\documentclass[reqno,11pt]{amsart}

\usepackage{graphicx}
\usepackage{amssymb}
\usepackage{amsmath}
\usepackage{amsfonts}
\usepackage{amsthm}
\usepackage{times}
\usepackage{xcolor}

\usepackage{hyperref}
\usepackage[mathscr]{eucal}
\usepackage{enumerate}

\usepackage{authblk}
\usepackage[foot]{amsaddr}

\numberwithin{equation}{section}

\newtheorem{Def}{Definition}[section]
\newtheorem{Thm}[Def]{Theorem}
\newtheorem{Prop}[Def]{Proposition}
\newtheorem{Lemma}[Def]{Lemma}

\newtheorem{Corollary}[Def]{Corollary}

\newcommand{\beq}{\begin{equation}}
\newcommand{\eeq}{\end{equation}}
\newcommand{\Proof}{\begin{proof}}
\newcommand{\QED}{\end{proof} \noindent}

\newcommand{\mm}{\hspace{-.08cm}\cdot \hspace{-.08cm}}
\newcommand{\M}{\mathcal{M}}

\newcommand{\R}{\mathbb{R}}
\newcommand{\1}{\mbox{\rm 1 \hspace{-1.05 em} 1}}

\newcommand{\Gammati}{\tilde{\Gamma}}

\DeclareMathOperator{\tr}{tr}

\newcommand{\cutoff}{\chi_{_\delta}}

\allowdisplaybreaks[1]

\title[Optimal Regularity for Connections on Tangent Bundles]{On the Optimal Regularity Implied by the Assumptions of Geometry I: Connections on Tangent Bundles}

\author[M.\ Reintjes]{Moritz Reintjes$^*$}
\address[*]{Department of Mathematics\\ City University of Hong Kong \\ Hong Kong}
\email{moritzreintjes@gmail.com}

\author[B.\ Temple]{Blake Temple$^{**}$ \\ \\ April 13, 2022}
\address[**]{Department of Mathematics\\ University of California\\ Davis, CA \\ USA}
 \email{temple@math.ucdavis.edu}

\begin{document}

\begin{abstract}
We resolve the problem of \emph{optimal regularity} and \emph{Uhlenbeck compactness} for affine connections in General Relativity and Mathematical Physics. First, we prove that any  affine connection $\Gamma$, with components $\Gamma \in L^{2p}$ and components of its Riemann curvature ${\rm Riem}(\Gamma)$ in $L^p$,  in some coordinate system,  can be smoothed by coordinate transformation to optimal regularity, $\Gamma \in W^{1,p}$ (one derivative smoother than the curvature), $p> \max\{n/2,2\}$,  dimension $n\geq 2$. For Lorentzian metrics in General Relativity this implies that shock wave solutions of the Einstein-Euler equations are non-singular---geodesic curves, locally inertial coordinates and the Newtonian limit, all exist in a classical sense, and the Einstein equations hold in the strong sense. The proof is based on an $L^p$ existence theory for the \emph{Regularity Transformation (RT) equations}, a system of {\it elliptic} partial differential equations (introduced by the authors) which determine the Jacobians of the regularizing coordinate transformations. Secondly, this existence theory gives the first extension of Uhlenbeck compactness from Riemannian metrics, to general affine connections bounded in $L^\infty$, with curvature in $L^{p}$, $p>n$, including semi-Riemannian metrics, and Lorentzian metric connections of relativistic Physics. We interpret this as a ``geometric'' improvement of the generalized Div-Curl Lemma. Our theory shows that Uhlenbeck compactness and optimal regularity are pure logical consequences of the rule which defines how connections transform from one coordinate system to another---what one could take to be the ``starting assumption of geometry''.    
\end{abstract}

\maketitle

\setcounter{tocdepth}{1}
\small
\tableofcontents
\normalsize

\section{Introduction} \label{Sec_intro}          

In this paper we resolve two problems in Mathematical Physics by extending the multi-dimensional existence theory for the \emph{Regularity Transformation (RT) equations}\footnote{The RT-equations, introduced in \cite{ReintjesTemple_ell1}, are referred to as the \emph{Reintjes-Temple equations} in \cite{ReintjesTemple_ell3}.} in \cite{ReintjesTemple_ell2} to affine connections at the low level of $L^p$ curvature regularity; specifically, to connections with components in $L^{2p}$ and with Riemann curvature  ${\rm Riem}(\Gamma)$ bounded component-wise in $L^p$, $p> \max\{n/2,2\}$, in dimension $n\geq 2$, (or equivalently $\Gamma \in L^{2p}$, $d\Gamma \in L^p$).\footnote{In this paper, all $L^p$ norms of tensors and connections are taken component-wise, local and coordinate dependent. That is, $L^p$ norms are taken on tensor components represented in coordinate systems on open and bounded neighborhoods of points. We choose $\Gamma \in L^{2p}$ and $d\Gamma \in L^{p}$ to place both terms in ${\rm Riem}(\Gamma) = d\Gamma + \Gamma \wedge \Gamma$ in $L^p$. For $\Gamma \in L^{2p}$, ${\rm Riem}(\Gamma)$ is in $L^p$ if and only if $d\Gamma$ is in $L^p$.} First, this existence theory establishes that any such connection, defined on an arbitrary manifold, including the Lorentzian metric connections of General Relativity (GR), can always be smoothed to optimal regularity $W^{1,p}$ by local coordinate transformations, any $p <\infty$. This extends the optimal regularity result of Kazdan and DeTurck \cite{DeTurckKazdan} from (positive definite) Riemannian metrics to general affine connections, including Lorentzian and semi-Riemannian metrics.\footnote{To emphasize this generality, we chose our title to conjure up Riemann's celebrated habilitation of 1854: ``{\it On the hypotheses which lie at the foundations of geometry}''.} Our optimal regularity result settles in the affirmative that spacetime singularities in the Lorentzian metrics of GR shock waves are removable. In particular, this establishes for the first time that (weak) shock wave solutions of the Einstein-Euler equations constructed by the Glimm scheme are one order more regular than previously known \cite{GroahTemple}. More generally, it implies that all multi-dimensional {\it weak} shock wave solutions of the Einstein equations $G=\kappa T$ with Lipschitz continuous gravitational metric and $L^\infty$ curvature are non-singular in the sense that they solve $G=\kappa T$ in the strong $L^p$-sense under coordinate transformation; and geodesic curves,\footnote{For $L^p$ as well as $L^\infty$ connections, the basic local existence theorem for ODE's at low regularities (Peano's Theorem) does not apply to construct geodesic curves or particle trajectories.} locally inertial coordinates and the Newtonian limit all exist in a classical sense.  

Secondly, our existence theory for the RT-equations suffices to extend Uhlenbeck compactness\footnote{By \emph{Uhlenbeck compactness} we mean compactness of a sequence of connections $\Gamma_i$ derived from a uniform bound on the un-differentiated connection and curvature components alone.}       from Riemannian to Lorentzian geometry.       To be precise, our theory extends Uhlenbeck compactness from the case of connections on {\it vector bundles} over fixed Riemannian manifolds, with optimal regularity $W^{1,p}$ and uniform $L^p$ curvature bounds ($p\geq n/2$) \cite{Uhlenbeck}, to the case of affine connections on {\it tangent bundles} over {\it arbitrary} manifolds (not endowed with any metric), uniformly bounded in $L^\infty$ with $L^{p}$ curvature bounds ($p>n$), allowing for non-optimal connection regularity at the start. That is, in \cite{Uhlenbeck} the part of the connection acting on the tangent bundle, its affine part, is assumed to be the Christoffel symbol of a fixed positive definite metric, and compactness is established for the part of the connection acting on the non-tangential fibre only, while our result addresses general affine connections, (not necessarily of a metric, but including Lorentzian and semi-Riemannian metric connections), acting on the tangent bundle without non-tangential fibres.  In \cite{ReintjesTemple_ell5}, we extend our results to connections on {\it vector bundles} (acting on tangent bundles and on non-tangential fibres), allowing again for a general affine part.        Uhlenbeck compactness establishes existence of a convergent subsequence of connections, weakly in $W^{1,p}$, from uniform bounds on the curvature alone, without the need to bound all connection derivatives, and the convergence is strong enough to pass limits through non-linear products.\footnote{Uhlenbeck's compactness theorem in \cite{Uhlenbeck} for Riemannian geometry was a topic of the 2019 Abel Prize and 2007 Steele Prize, and was a crucial step in the proof of fundamental results in geometry, including Donaldson's work in \cite{Donaldson}. See \cite{Taubes,Wehrheim} for a summary of the important applications of Uhlenbeck compactness.}      We interpret Uhlenbeck compactness as a ``geometric'' improvement of the generalized Div-Curl Lemma applicable to sequences of connections, (c.f. Section \ref{Sec_Div-Curl}). With the same assumptions, Uhlenbeck compactness provides strong convergence of a subsequence in transformed coordinates determined by the RT-equations, when the Div-Curl Lemma implies only weak continuity of wedge products in the original coordinates.         

The RT-equations are a system of nonlinear {\it elliptic} partial differential equations in matrix valued differential forms $(\Gammati,J,A)$. These equations determine the Jacobians $J$ of coordinate transformations which transform a given connection $\Gamma$ to optimal regularity. The unknown $\Gammati$ represents the regularized connection components, $A$ is an auxiliary variable introduced to impose the integrability condition $dJ=0$, and $\Gamma$, the connection components of an arbitrary given connection, appears as a source term on the right hand side of the RT-equations, along with a vector valued $0$-form $v$ free to be chosen. Our theory starts with no more than the component functions $\Gamma \equiv (\Gamma^k_{ij})$ defined on some open set $\Omega \subset \R^n$, and we view $\Gamma$ as the components of a connection in some given but arbitrary coordinate system $x$ on $\Omega$. The RT-equations, derived in \cite{ReintjesTemple_ell1} from the connection transformation law and first analyzed in \cite{ReintjesTemple_ell2}, are given by 
\begin{eqnarray} 
\Delta \Gammati &=& \delta d\Gamma - \delta \big(d J^{-1} \wedge dJ \big) + d(J^{-1} A ), \label{PDE1} \\
\Delta J &=& \delta ( J \mm \Gamma ) - \langle d J ; \tilde{\Gamma}\rangle - A , \label{PDE2} \\
d \vec{A} &=& \overrightarrow{\text{div}} \big(dJ \wedge \Gamma\big) + \overrightarrow{\text{div}} \big( J\, d\Gamma\big) - d\big(\overrightarrow{\langle d J ; \tilde{\Gamma}\rangle }\big),   \label{PDE3}\\
\delta \vec{A} &=& v,  \label{PDE4}
\end{eqnarray}
with boundary data $d\vec{J} =0$ on $\partial \Omega$. The unknowns $(\Gammati,J,A)$ in the RT-equations, together with the given connection components $\Gamma$, are defined by their components in $x$-coordinates on $\Omega$.\footnote{Here $\Gamma \equiv \Gamma^\mu_{\nu i} dx^i$ and $\Gammati \equiv \Gammati^\mu_{\nu i} dx^i$ are matrix valued $1$-forms, $J\equiv J^\mu_\nu$ and $A\equiv A^\mu_\nu$ are matrix valued $0$-forms, and $\vec{A} \equiv \vec{A}^\mu_i dx^i$ is a vector valued $1$-form.} The operations on the right hand side are defined in Section \ref{Sec_Prelim} in terms of the Cartan Algebra of differential forms in $x$-coordinates, and the RT-equations are introduced in detail in Section \ref{Sec_reducedRT}.   The vector field $v$, free to be chosen, constitutes the gauge freedom inherent to the RT-equations, (reflecting the multiplicity of coordinate maps to optimal connection regularity). The RT-equations admit a weak formulation, which is required for the low regularity of $L^p$ connections and curvature addressed here.\footnote{It is well known that weak formulations of equations which are equivalent for smooth solutions,  are not always equivalent for weak solutions. (For example, in conservation laws, different choices of conserved quantities lead to different weak formulations, and even Riemann entertained such confusions \cite{Smoller}.)   Our choice of the weak formulation is based on the geometric operators $d,\delta$ and $\overrightarrow{div}$ on vector and matrix valued differential forms, by introducing suitable $L^2$ adjoints for these differential operators. Our idea to base the weak formulation on these geometric operators is guided in part by the requirement that Jacobians $J$, which solve the reduced RT-equations in this weak sense, are automatically integrable to coordinates, c.f. Sections \ref{Sec_weak} and \ref{Sec_Proofs}.}

The RT-equations are elliptic regardless of metric signature, because $\Delta$ is the Laplacian of the Euclidean metric in $x$-coordinates. They determine the Jacobians of coordinate transformations to optimal regularity. Thus the problem of optimal regularity and Uhlenbeck compactness is reduced to an existence theory for the RT-equations, c.f. \cite{ReintjesTemple_ell1,ReintjesTemple_ell3}.  The RT-equations, and hence both Uhlenbeck compactness and optimal regularity, are mathematical consequences of just the rule which defines how connections transform from one coordinate system to another, logically independent of any additional structure on the geometry, like positive definiteness of the metric, or the Einstein equations.

Although the original formulation of the RT-equation \eqref{PDE1} - \eqref{PDE4} is amenable to a rigorous existence theory at the higher non-optimal regularity of connection and curvature in $W^{1,p}$, (we accomplished this existence theory in \cite{ReintjesTemple_ell2}, proving optimal connection regularity $W^{2,p}$), the non-linearities in the RT-equations as formulated in \eqref{PDE1} - \eqref{PDE4} are too fierce to extend the method of proof in \cite{ReintjesTemple_ell2} to the low regularity of non-optimal $L^p$ connections. In this paper we extend the existence theory for the RT-equations to the level of $L^p$ connections. For this we use the gauge freedom inherent in system \eqref{PDE1} - \eqref{PDE4} to introduce what we call the {\it reduced} RT-equations, (system \eqref{RT_withB_2_intro} - \eqref{RT_withB_4_intro} below), an elliptic system of equations equivalent to the original RT-equations \eqref{PDE1} - \eqref{PDE4}.  The reduced RT-equations simplify the nonlinearities in the problem of optimal regularity to a degree sufficient to extend our analysis of the RT-equations by one order, from the level of connection components $\Gamma\in W^{1,p}$ achieved in \cite{ReintjesTemple_ell2}, to $L^p$ connections, including the $L^\infty$ connections of GR shock waves.     

Our first main result, which follows from this existence theory, establishes that if in a given coordinate system the components of  $\Gamma$ are in $L^{2p}$ and those of its Riemann tensor are in $L^p$, $p> \max\{n/2,2\}$, $n\geq 2$, then in a neighborhood of every point there exists a $W^{2,2p}$ coordinate transformation, such that in the transformed coordinates the components of $\Gamma$ exhibit optimal regularity, $\Gamma\in W^{1,p}$, (i.e., the components of $\Gamma$ are one derivative more regular than the components of its Riemann curvature tensor, ${\rm Riem}(\Gamma) \in L^p$), c.f. Theorem \ref{Thm_Smoothing} below. This new existence theory for the reduced RT-equations provides uniform $W^{1,p}$ estimates for the connection in the transformed coordinates, and this directly implies our new Uhlenbeck compactness theorem, stated in Theorem \ref{Thm_compactness} below, which does not rely on any underlying Riemannian or Lorentzian metric. 

In \cite{ReintjesTemple_ell5} we extend our results on Uhlenbeck compactness and optimal regularity to connections on vector bundles, including Yang-Mills gauge theories over Lorentzian and semi-Riemannian manifolds, allowing for both compact and non-compact Lie groups. Our proofs are based on introducing the {\it RT-equations for vector bundles}. This further illuminates the generality of the new mathematical principle developed in this paper.  A conclusive summary of our results here, combined with the results in \cite{ReintjesTemple_ell5}, is provided in our forthcoming RSPA publication \cite{ReintjesTemple_ell6}.\footnote{Our exposition in \cite{ReintjesTemple_ell3} summarizes the earlier developments of the RT-equations in \cite{ReintjesTemple_ell1,ReintjesTemple_ell2}.}

\vspace{.2cm}\noindent{\bf The Reduced RT-equations:} 
The reduced RT-equations, which are central to the proof of our main results in this paper, are given by the following system,
\begin{eqnarray} 
\Delta J &=& \delta ( J \mm \Gamma ) - B , \label{RT_withB_2_intro} \\
d \vec{B} &=& \overrightarrow{\text{div}} \big(dJ \wedge \Gamma\big) + \overrightarrow{\text{div}} \big( J\, d\Gamma\big) ,   \label{RT_withB_3_intro} \\
\delta \vec{B} &=& w,  \label{RT_withB_4_intro}
\end{eqnarray}
where $J$ is the Jacobian of the transformation to optimal regularity,  $B$ is an auxiliary matrix valued differential form introduced to impose the integrability of $J$ to coordinates, and the new {\it gauge freedom} is the freedom to choose the vector valued function $w$.  The vectorization of $B$, ($\vec{B} \equiv \vec{B}^\mu_i dx^i$), is required to incorporate into the reduced RT-equations the condition that $J$ be integrable to coordinates, expressed in terms of the vectorization $\vec{J} = J^\mu_i dx^i$ through $d\vec{J} \equiv {\rm Curl}(J) =0$. The operations on the right hand side of (\ref{RT_withB_2_intro}) - (\ref{RT_withB_4_intro}) are defined in Section \ref{Sec_Prelim}.

The reduced RT-equations are derived from the original RT-equations \eqref{PDE1} - \eqref{PDE4} in Section \ref{Sec_reducedRT}, by using the gauge freedom $v$ in \eqref{PDE1} - \eqref{PDE4} to uncouple the equations for the Jacobian $J$ of the smoothing transformation from \eqref{PDE1}, the equation for $\Gammati$. This isolates the non-linearities in the $\Gammati$-equation \eqref{PDE1}, and thereby eliminates them from the iteration scheme for $J$ on which our existence theory in Section \ref{Sec_existence_theory} is based.  The reduced RT-equations consist of the resulting uncoupled system of equations \eqref{RT_withB_2_intro} - \eqref{RT_withB_4_intro} in $J$, a linear elliptic system in $(J,B)$ which is solvable at the level of $L^p$ connections.  The proof of our optimal regularity result, Theorem \ref{Thm_Smoothing}, is then completed by using the $\Gammati$-equation \eqref{PDE1} to show directly that the Jacobian determined by a solution $J$ of the reduced RT-equations, does indeed lift the original connection $\Gamma$ to optimal regularity $\Gamma\in W^{1,p}$, by proving this $J$ determines a $\Gammati$ such that $(J,\Gammati)$ solves the original RT-equations, c.f. Theorem \ref{Thm_gauge_existence}.

Although the derivation of the reduced RT-equations begins with the original RT-equations (we know of no independent derivation), the reduced RT-equations introduced in this paper represent a new starting point, and the subsequent proofs are self-contained and stand logically independent of arguments we gave for the original RT-equations.

\vspace{.2cm}\noindent{\bf Outline of the Paper:} 
In Section \ref{Sec_landscape} mathematical context is provided for our results. In Section \ref{Sec_MainResults} we state our main results, Theorems \ref{Thm_Smoothing} and \ref{Thm_compactness}. In Section \ref{Sec_Applications} we give three applications of Theorems \ref{Thm_Smoothing} and \ref{Thm_compactness}:  A new compactness result for the vacuum Einstein equations in GR at low regularities as an application of Uhlenbeck compactness, a proof of optimal regularity for GR shock waves constructed by Glimm's theorem, and a construction of locally inertial frames for general $L^\infty$ connections. In particular, the existence of locally inertial frames rules out regularity singularities at GR shock waves by establishing that shock wave solutions of the Einstein-Euler equations are locally inertial.   In Section \ref{Sec_Prelim} we introduce the Cartan calculus for matrix and vector valued differential forms required in this paper. Section \ref{Sec_reducedRT} contains the main idea of proof of our optimal regularity result, in particular we show that solutions of the reduced RT-equations determine solutions of the full RT-equations, Theorem \ref{Thm_gauge_existence}, and we state our main existence theorem for the reduced RT-equations, Theorem \ref{Thm_Existence_J}. The proof of Theorem \ref{Thm_gauge_existence} is given in Sections \ref{Sec_Proof1} and \ref{Sec_gauge}. In Section \ref{Sec_weak} we introduce the weak formulation of the RT-equations required at the level of regularity addressed in this paper. In Sections \ref{Sec_existence_theory} - \ref{Sec_Proofs} we give a self-contained presentation of the existence theory for weak solutions of the reduced RT-equations, (based only on two basic theorems from linear elliptic PDE theory, Theorems \ref{Thm_Poisson} and \ref{Thm_CauchyRiemann} recorded in the appendix), and we prove existence of solutions to the reduced RT-equations in the case of connection regularity $\Gamma, d\Gamma \in L^\infty$. We extend this proof to the sought after regularity $\Gamma \in L^{2p}, d\Gamma \in L^p$ in Section \ref{Sec_Lp-extension}, thereby completing the proof of our existence result Theorem \ref{Thm_Existence_J}.  The proof of our main results, Theorems \ref{Thm_Smoothing} and \ref{Thm_compactness}, is completed in Section \ref{Sec_Proof_optimal_regularity}.

\section{The mathematical landscape}   \label{Sec_landscape}

In the following subsections we place our results within a mathematical landscape, including the theory of GR shock waves. The reader interested in focusing in the mathematical development may skip these subsections and continue with Section \ref{Sec_MainResults}, where we state our main results, and Section \ref{Sec_reducedRT}, where we describe the main ideas of the method of proof.  

\subsection{The problem of optimal regularity}
The existence of coordinates in which connections are non-optimal is a fundamental feature of Riemann's curvature tensor, following directly from the fact that the Riemann curvature transforms as a tensor by contraction with undifferentiated Jacobians, while the transformation of a connection involves derivatives of the Jacobian. So any transformation by a Jacobian which has the {\it same regularity} as a given connection, will lower the regularity of a connection of optimal regularity (one derivative more regular than its curvature) by one order due to the terms containing derivatives of the Jacobian in the transformation law for connections.\footnote{Note, this principle directly carries over to metric tensors which, by Christoffel's formula,  are always exactly one derivative more regular than their connections.}    The result is a transformed connection with components one order less regular, and in the same regularity class as the curvature, because the Riemann curvature tensor would preserve its regularity under tensor transformation.     This holds for classical spaces of regularity like $C^k$, as well as weak regularity measured by Sobolev spaces $W^{m,p}$, and H\"older regularity $C^{m,\alpha}$.       To prove the reverse direction, that non-optimal connections can always be smoothed to optimal regularity by coordinate transformation, one needs to undo the above process, and this requires constructing a \emph{singular} transformation given only the information about the non-optimal connection and its curvature. For example, at the level of $L^{\infty}$ connections, such a coordinate transformation must be singular in the sense that jumps in derivatives of the Jacobian must be tuned to precisely cancel out the discontinuities in the given non-optimal connection in the transformation law for connections. 

The existence of coordinate transformations which smooth connections to optimal regularity, one derivative  more regular than the curvature, is surprising in light of the fact that the curvature, being a ``curl'' plus a ``commutator'',  does not directly control every derivative of $\Gamma$, only $d\Gamma\equiv {\rm Curl}(\Gamma)$.   That is, the complementary derivatives, $\delta\Gamma \equiv {\rm div}(\Gamma)$, are not controlled directly by assuming a given regularity of the curvature.   Since the basic compactness theorems for Sobolev spaces are based on controlling every derivative, it follows that optimal regularity is intimately connected to compactness.  This principle, as expressed through the exterior derivative $d$ and the co-derivative $\delta$ of the Cartan algebra of differential forms associated with an assumed positive definite metric, underlies Uhlenbeck's celebrated compactness result. 

We note that the regularity of metric, connection and curvature is not altered by sufficiently smooth coordinate transformations, so in this sense regularity is a geometric property of the manifold when one takes the smooth atlas. Thus one can view the RT-equations as providing a low regularity transformation which lifts regularity, but that regularity then becomes a geometric property of the resulting manifold when again the atlas of smooth coordinate transformations is taken, c.f. the discussion in \cite{ReintjesTemple_ell3}.

\subsection{Uhlenbeck compactness}
Uhlenbeck's compactness theorem, Theorem 1.5 of \cite{Uhlenbeck},  applies to Riemannian metrics, and is based on establishing a uniform bound on the components of a connection in Coulomb gauge, the Coulomb gauge providing a coordinate system arranged to satisfy $\delta\Gamma=0$ to bound the derivatives uncontrolled by the curvature through $d\Gamma$.  Compactness in Coulomb gauge then follows from a uniform bound on the curvature.  To illustrate the heart of the issue in \cite{Uhlenbeck}, taking $\delta$ of ${\rm Riem}(\Gamma)=d\Gamma+\Gamma\wedge\Gamma$, when $\delta\Gamma=0$, results in an equation of (essentially) the form $\Delta\Gamma=\delta\, {\rm Riem}(\Gamma)$, where $\Delta=d\delta+\delta d$ is the Laplacian of the underlying Riemannian metric; so by elliptic regularity, a sequence of connections $\Gamma_i\in W^{1,p}$ with ${\rm Riem}(\Gamma_i)$ uniformly bounded in $L^p$, will be uniformly bounded in $W^{1,p}$ in Coulomb gauge, for $p<\infty$. Sobolev compactness then implies a subsequence converges weakly in $W^{1,p}$ and strongly in $L^p$ in Coulomb gauge.           

In the case of Lorentzian metrics, $d\delta+\delta d$ is the hyperbolic D'Alembert (wave) operator, and since hyperbolic operators propagate irregularities from initial data surfaces along characteristics, deducing optimal regularity for Lorentzian metrics in Coulomb gauge is at best problematic, c.f. \cite[Ch. 9]{ReintjesTemple_ell3}. Our incoming point of view is that the Coulomb gauge condition $\delta\Gamma=0$ is too restrictive for general connections, and instead of trying to eliminate the uncontrolled $\delta$ derivatives of $\Gamma$ altogether, our idea is to bound them in the right space by the RT-equations, elliptic equations in $\tilde{\Gamma}$ and $J$.  Different from Uhlenbeck's argument, the RT-equations are formulated in terms of the Cartan algebra of differential forms associated with the Euclidean metric of an arbitrary coordinate system $x$ in which the components $\Gamma^k_{ij}$ of $\Gamma$ are given, not the invariant Cartan algebra of any underlying metric. Because they are based on the auxiliary Riemannian structure provided by the coordinate Euclidean metric, the RT-equations are {\it elliptic} regardless of any invariant metric structure for $\Gamma$. This allows us to obtain optimal regularity and Uhlenbeck compactness for arbitrary connections on the tangent bundle of arbitrary manifolds, without recourse to metric signature or even any underlying metric structure. Note that the RT-equations are not invariant in a tensorial sense, but a different version of them is given in each coordinate system. As a consequence the RT-equations have the same simple elliptic structure in every coordinate system, and this makes them inherently useful for analysis.\footnote{In contrast, the Einstein equations only take simpler forms in canonical coordinates like Standard Schwarzschild Coordinates \cite{GroahTemple} or wave coordinates \cite{Choquet}.}   

We now compare our compactness theorem to Uhlenbeck's result in \cite{Uhlenbeck}. Theorem 1.5 of \cite{Uhlenbeck} assumes a sequence of connections $\Gamma_i\in W^{1,p}$, with curvature ${\rm Riem}(\Gamma_i)$ uniformly bounded in $L^p$, and from this concludes that in Coulomb gauge,  the connection coefficients are uniformly bounded in $W^{1,p}$, with uniform bound provided by the original bound on the curvature in $L^p$.  The uniform bound on the extra derivative in $W^{1,p}$ then implies Uhlenbeck compactness, i.e., the convergence of a subsequence of $\Gamma_i$ weakly in $W^{1,p}$, and hence strongly in $L^p$, in Coulomb gauge.  In contrast, our Theorem \ref{Thm_compactness} stated below, assumes a sequence of connections $\Gamma_i \in L^\infty$, which need not lie in $W^{1,p}$ at the start, but assumes uniform bounds on $\Gamma_i$ in $L^\infty$ and ${\rm Riem}(\Gamma_i)$ in $L^{p}$, $n<p<\infty$, (or equivalently on $\Gamma_i$ in $L^\infty$ and $d\Gamma_i$ in $L^p$).  From this, Theorem \ref{Thm_compactness} concludes the existence of coordinate transformations $x\to y_i(x)$, (which play the role of Coulomb gauge), with Jacobians uniformly bounded in $W^{1,2p}$, such that in the new coordinates, the sequence of connection components $\Gamma_{y_i}$ are uniformly bounded in $W^{1,p}$, with bound given by our original $L^{\infty}$ bounds on $\Gamma_i$ and $d\Gamma_i$.  From this we again conclude Uhlenbeck compactness, i.e.,  weak convergence of a subsequence of $\Gamma_{y_i}$ in $W^{1,p}$, hence converging strongly in $L^p$, with $y_i$ converging in $W^{1,2p}$. Thus our theorem replaces the assumption $\Gamma_i\in W^{1,p}$, (which for us is the assumption of optimal regularity at the start), with the assumption that the components of $\Gamma_i$ are uniformly bounded in $L^{\infty}$.    We view, our assumption of a uniform $L^\infty$ bound on $\Gamma_i$ as a small concession considering that it just addresses the undifferentiated terms in the Riemann curvature tensor. 

\subsection{A refinement of the Div-Curl Lemma}    \label{Sec_Div-Curl}
Our version of Uhlenbeck compactness gives a ``geometric'' improvement of the generalized Div-Curl Lemma  for sequences of connections $\Gamma_i$. To see this, recall the generalized Div-Curl Lemma states that wedge products are weakly continuous when differential constraints take the form of exterior derivatives, \cite{RogersTemple}.    So consider the case when the components of a sequence of affine connections $\Gamma_i$ satisfy the constraint $\Gamma_i, d\Gamma_i$ uniformly bounded in, say, $L^\infty$, in a given coordinate system, c.f. \cite{Tartar}. The Banach-Alaoglu Theorem implies a subsequence of the connection components converges weakly in $L^p$, and the generalized Div-Curl Lemma then implies that the components of curvature ${\rm Riem}(\Gamma_i)=d\Gamma_i+\Gamma_i\wedge\Gamma_i$ are weakly continuous on this limit, that is, one can pass weak limits through both terms in ${\rm Riem}(\Gamma)$.  Uhlenbeck compactness assumes, in addition, ``geometry'' in the form of the transformation law for connections, and implies that connection components actually lie in $W^{1,p}$, (i.e., have one more derivative), and that a subsequence converges weakly in $W^{1,p}$, and hence {\it strongly} in $L^p$.  But this improved convergence occurs in a different coordinate system, a coordinate system obtained by solving the RT-equations.   One cannot overstate the importance of strong over weak convergence. Compactness theorems giving strong convergence are the starting point for validating approximation schemes in nonlinear problems.  For example, in Diperna's well known proof of the zero viscosity limit for $2\times2$ systems of conservation laws, an infinite family of entropy fields was required to bootstrap the weak convergence to strong convergence.  Uhlenbeck compactness extracts strong convergence from weak convergence from the theory of curvature, without need of any auxiliary estimates \cite{Diperna,Tartar}.

\subsection{GR shock waves}  \label{Sec_GR-shocks}
The authors' present multi-dimensional theory of optimal regularity began with the special case of shock wave solutions of the Einstein-Euler equations constructed by Glimm's random choice method  in \cite{GroahTemple}, (see also \cite{LeFlochStewart}). The Lorentzian metrics associated with these shock wave solutions are only Lipschitz continuous ($C^{0,1}$), a regularity too low to construct geodesic curves and locally inertial coordinates directly by classical ODE methods. This motivated the question as to whether one can raise the metric regularity by coordinate transformation, to recover these basic objects of geometry, or whether the Lorentzian metrics of GR shock waves are exhibiting essential non-removable spacetime singularities. A coordinate transformation to optimal regularity would remove these singularities.\footnote{This is a perspicatious warm-up problem for the multi-dimensional theory of GR shock waves because the role played by non-optimal coordinates in spherically symmetric spacetimes is no different than the role they play in general multi-dimensional spacetimes: They exist simply because the Riemann curvature tensor involves second derivatives of the metric, but transforms as a tensor by first derivative Jacobians.}  Thus, since shock waves form generically in the compressible Euler equations and correctly model gas dynamics, resolving the question whether these singularities can be removed, directly addresses the basic consistency of the Einstein-Euler system.\footnote{Interestingly, metrics of a similar low regularity arise in the recent study of  ``wild'' solutions of the non-relativistic Euler equations in \cite{ChenSlemrod}; and the problem of optimal metric regularity is also of interest in Conformal Geometry \cite{LassasLiimatainen}.}

In his classic 1966 paper \cite{Israel} Israel introduced the multi-dimensional theory of junction conditions and used it to prove that a metric $C^{0,1}$ across a \emph{single} smooth shock surface can be locally smoothed to optimal regularity $C^{1,1}$ by coordinate transformation to Gaussian normal coordinates. But the optimal regularity results in \cite{Israel} do not apply to shock wave interactions, and thus not to the $C^{0,1}$ metrics in \cite{GroahTemple}, because the underlying Gaussian normal coordinate construction cannot be associated to intersecting shock surfaces. The only extension of Israel's result to shock wave interactions (before this paper) was accomplished for the special case of spherically symmetric shock wave interactions in \cite{Reintjes,ReintjesTemple1}. But it remained out of reach how to address these apparent singularities in shock wave solutions constructed in \cite{GroahTemple}, (or constructed in multi-dimensions by the junction conditions), when they contain more complex shock wave interactions.       

Authors' paper \cite{ReintjesTemple_geo} was a first step for the general problem of smoothing metrics and connections.   In  \cite{ReintjesTemple_geo}, we introduced a necessary and sufficient condition for the general problem of smoothing metrics with connection and Riemann curvature tensor in $L^{\infty}$, the so-called {\it Riemann-flat condition}, which is the condition that there should exist a tensor $\tilde{\Gamma}\in C^{0,1}$ such that ${\rm Riem}(\Gamma-\tilde{\Gamma})=0$. Our main theorem in \cite{ReintjesTemple_geo} then states that there exists a $C^{1,1}$ coordinate transformation which smooths an $L^{\infty}$ connection $\Gamma$ by one order to Lipschitz continuous $C^{0,1}=W^{1,\infty}$ (hence optimal) if and only if the Riemann-flat condition holds.\footnote{This equivalence extends easily to $\Gamma \in L^\infty$ and $\Gammati \in W^{1,p}$, the case address in this paper.} The tensor $\Gammati$ gives rise to a coordinate system $y$ in which $\Gamma$ exhibits optimal regularity, and the components of $\Gammati$ and $\Gamma$ agree in $y$-coordinates. However, even though the Riemann-flat condition gives a new geometric point of view on the problem of optimal regularity, it was entirely unclear how to construct such a tensor $\Gammati$, or whether this is always possible. The breakthrough in our research program came about in \cite{ReintjesTemple_ell1}, when we derived, from two equivalent forms of the Riemann-flat condition, the RT-equations \eqref{PDE1} - \eqref{PDE4}, a system of solvable elliptic equations in the sought after tensor $\Gammati$ and Jacobian $J$.  

In this paper we extend our current existence theory for the RT-equations in \cite{ReintjesTemple_ell2} by one order of regularity to $\Gamma \in L^{2p}$, ${\rm Riem}(\Gamma) \in L^p$, (or equivalently $\Gamma \in L^{2p}$, $d\Gamma \in L^p$), and prove that any such connection can be locally smoothed to optimal regularity $\Gamma\in W^{1,p}$. This resolves the problem of optimal regularity at GR shock waves by establishing that for any weak solution of the Einstein equations satisfying $\Gamma, d\Gamma\in L^{\infty}$ in $x$-coordinates, there always exist local coordinate transformations $x\to y$ with Jacobian $J\in W^{1,2p}$, such that $\Gamma\in W^{1,p}$ in $y$ coordinates, for any $n/2 < p<\infty$. Here $p > n/2$ can be taken to be arbitrarily large, but not yet $p=\infty$.  So we do not obtain $\Gamma\in C^{0,1},\ g\in C^{1,1}$ as Israel did for smooth shock surfaces, but we are arbitrarily close in the sense that $p$ can be arbitrarily large. For $p>n$,  any $\Gamma\in W^{1,p}$ is H\"older continuous by Morrey's inequality, a regularity sufficient for geodesic curves to exist (by Peano's Theorem), for spacetime to admit locally inertial frames,  and hence for the Newtonian limit to exist at each point in spacetime.   An explicit construction of locally inertial frames is given in Corollary \ref{Cor_locally_inertial_frames} below.

\subsection{Prior results}
It was shown by DeTurck and Kazdan in \cite{DeTurckKazdan} that for (positive definite) Riemannian metrics, optimal regularity can always be achie\-ved in harmonic coordinates. The first optimal regularity result in Lorentzian geometry is due to Anderson \cite{Anderson}.  Anderson's results are based on using harmonic coordinates on the Riemannian hypersurfaces of a given foliation of spacetime, and establish curvature bounds for vacuum spacetimes and certain matter fields when the Riemann curvature is in $L^\infty$, under some technical assumptions. A similar result for vacuum spacetimes was proven in \cite{ChenLeFloch}. As far as we can tell, these results do not apply to GR shock waves, and our result cannot be obtained from these prior methods, free of additional assumptions, even in the special case of vacuum spacetimes. (Keep in mind that the setting of vacuum excludes fluid dynamical shock waves, and so is a warm-up problem from the point of view of shock wave theory. Historically shock waves are one of the main motivations for the study of low regularity solutions.) The results in \cite{Anderson,ChenLeFloch} require applying sophisticated analytical and geometric machinery on top of the classical harmonic coordinate construction in \cite{DeTurckKazdan}, and suggest strongly that metric signature is a central issue. Our results show that optimal regularity is entirely independent of metric and metric signature.\footnote{We note that the recent resolution of the bounded $L^2$ curvature conjecture for vacuum spacetimes, (c.f. Theorem 1.6 of \cite{Klainermann}), does not address the issue of optimal regularity, essentially because initial data taken from non-optimal connections is one order less regular than the data assumed in \cite{Klainermann}, see the discussion in \cite{ReintjesTemple_ell3}.}

\section{Statement of our main results}      \label{Sec_MainResults}

\subsection{Optimal Regularity}
Let $\Gamma$ denote a connection on the tangent bundle $T\mathcal{M}$ of an arbitrary $n$-dimensional differentiable manifold $\mathcal{M}$, $n\geq 2$.  Since the problem of optimal regularity is local, we assume at the start a given coordinate system $x$ defined on an open set $\Omega \subset\mathcal{M}$, such that $\Omega_x \equiv x(\Omega)\subset \R^n$ is bounded. That is, we work in a fixed chart $(x,\Omega)$ on $\mathcal{M}$. Without loss of generality we assume $\Omega_x$ has a smooth boundary. We use the notation $\Gamma_x$ to denote the components of $\Gamma$ in $x$-coordinates, $\Gamma_x\equiv\Gamma^k_{ij}(x)$. We say $\Gamma_x \in L^{p}(\Omega_x)$, and likewise $d\Gamma_x \in L^{p}(\Omega_x)$, if all component functions are in $L^p(\Omega_x)$ in $x$-coordinates. Here $d\Gamma_x$ denotes the exterior derivative of $\Gamma_x$ viewed as a matrix valued $1$-form in $x$-coordinates, a non-invariant object in the sense that it transforms neither as a tensor nor as a connection. Given a coordinate transformation $x \to y$, we let $\Gamma_y\equiv \Gamma^\gamma_{\alpha\beta}(y)$ denote the connection components in $y$-coordinates defined on $\Omega_y \equiv y(\Omega)$.

In this paper we assume $\Gamma_x \in L^{2p}(\Omega_x)$ and $d\Gamma_x \in L^{p}(\Omega_x)$ for $p>\max\{ n/2, 2\}$, a balance of $L^p$ spaces consistent with solutions $J \in W^{1,2p}(\Omega_x)$ of the RT-equations \eqref{PDE1} - \eqref{PDE4}. The two theorem which direct our choice of norms and solution spaces are, first, the product of two functions in $L^{2p}$ is always in $L^p$ by H\"older's inequality, and second, a function in $W^{1,2p}$ is H\"older continuous by Morrey's inequality when $p>n/2$. (The norms we use in this paper are recorded in Appendix \ref{Sec_appendix_norms}.) Note, since the Riemann curvature tensor can be written as      
\beq \label{Riem_intro}
{\rm Riem}(\Gamma_x) = d\Gamma_x +\Gamma_x \wedge \Gamma_x,
\eeq 
(c.f. \eqref{Riem} below), assuming $\Gamma_x \in L^{2p}(\Omega_x)$ and $d\Gamma_x \in L^p(\Omega_x)$ is equivalent to assuming $\Gamma_x \in L^{2p}(\Omega_x)$ and ${\rm Riem}(\Gamma_x) \in L^p(\Omega_x)$. For coordinate transformations with Jacobians $J \in W^{1,2p}(\Omega_x)$, $p>n/2$, (always assumed here), the assumption $\Gamma_x \in L^{2p}(\Omega_x)$ with $d\Gamma_x \in L^p(\Omega_x)$ is an invariant statement,\footnote{This holds since for such coordinate transformations ${\rm Riem}(\Gamma)$ transforms as a tensor, and contraction by H\"older continuous Jacobians does not lower the $L^p$ regularity. The $L^p$ regularity of $d\Gamma_y$ then follows from the regularity of ${\rm Riem}(\Gamma_y)$.} so one can omit the subscript $x$ on $\Gamma$ and $d\Gamma$ without confusion. Note, the statement that $\Gamma_x$ is in $W^{1,p}$, (i.e., $\Gamma_x$ has optimal regularity), is not an invariant statement for Jacobians at the low regularity $J \in W^{1,2p}(\Omega_x)$. 

The main norm on which our estimates are based is the following coordinate dependent norm,
\beq \label{L-infty of Gamma}
\|(\Gamma,d\Gamma) \|_{L^{q,p}(\Omega)} \equiv \|\Gamma_x \|_{L^q(\Omega_x)} + \|d\Gamma_x \|_{L^p(\Omega_x)},
\eeq  
which we used for either $q=2p$ or $q=\infty$ with $p>n/2$, where we always base $L^p$-norms on a coordinate system (c.f. Appendix \ref{Sec_appendix_norms}). Note that, for the purpose of this paper, we use $\|d\Gamma\|_{L^p}$ in the definition of the norm in \eqref{L-infty of Gamma}, but we would get an equivalent norm using $\| {\rm Riem}(\Gamma) \|_{L^{p}}$ in its place, because  \eqref{Riem_intro} implies
\begin{eqnarray} \label{RiemEquiv}
 \|d\Gamma\|_{L^p}  \ \leq \ \| {\rm Riem}(\Gamma) \|_{L^{p}} + 2 \|\Gamma\|_{L^{2p}}^2
 \ \leq \ \|d\Gamma\|_{L^p} + 4 \|\Gamma\|_{L^{2p}}^2  
\end{eqnarray}
by application of the H\"older inequality, c.f. \eqref{Holder_L2p-trick}.  From now on we always use $\Omega$ and $\Gamma$ to denote $\Omega_x$ and $\Gamma_x$ respectively, so without loss of generality $\Omega \equiv \Omega_x \subset \R^n$ and $\Gamma \equiv \Gamma_x$ denotes a collection of functions defined on $\Omega \subset \R^n$.

Our main theorem regarding optimal regularity at GR shock waves is the following:\vspace{.2cm}

\noindent {\bf Theorem \ref{Thm_Smoothing}$^\prime$} 
{\it Assume $\Gamma, d\Gamma \in L^{\infty}(\Omega)$ in $x$-coordinates and let $M>0$ be a constant such that
\beq 
 \|(\Gamma,d\Gamma)\|_{L^\infty(\Omega)}  \equiv \|\Gamma_x \|_{L^\infty(\Omega_x)} + \|d\Gamma_x \|_{L^\infty(\Omega_x)} \; \leq \; M.
\eeq   
Then for any $n < p < \infty$ and any point $q\in \Omega$ there exists a neighborhood $\Omega' \subset \Omega$ of $q$ and a coordinate transformation $x \to y$ with Jacobian $J=\frac{\partial y}{\partial x}\, \in W^{1,2p}(\Omega'_x),$ such that  
\beq \label{curvature_simple}
\|\Gamma_y \|_{W^{1,p}(\Omega'_y)}  \leq C\;  \|(\Gamma,d\Gamma)\|_{L^\infty(\Omega)},
\eeq
for some constant $C > 0$ depending only on $\Omega_x, p, n, q$ and $M$. That is, the connection components $\Gamma_y$ in $y$-coordinates exhibit optimal regularity, (i.e., one derivative above its curvature), in the sense that $\Gamma_y \in W^{1,p}(\Omega'_y)$.}
\vspace{.2cm}

Theorem \ref{Thm_Smoothing}$^\prime$ tells us that we can raise the connection regularity by essentially one derivative to $\Gamma_y \in W^{1,p}$, arbitrarily close to $W^{1,\infty}$ as $p \rightarrow \infty$.\footnote{Recall that the Sobolev space $W^{1,\infty}$ can be identified with the space of Lipschitz continuous functions, and $W^{1,p}$ can be identified with the space of H\"older continuous functions with H\"older coefficient $\alpha=\frac{n}{p}$ as long that $p>n$, \cite{Evans}.} Note, since $\Gamma$ and $d\Gamma$ are assumed in $L^\infty(\Omega)$, the statement of the theorem is sharper the larger $p$ is (and extends trivially to $1\leq p < \infty$ since $\Omega$ is bounded), and we can choose any $p<\infty$ but not yet $p=\infty$, a singular case in elliptic regularity theory, c.f. the discussion in \cite{ReintjesTemple_ell1}. By Morrey's inequality, $\Gamma_y$ is H\"older continuous when $p>n$, and this is sufficient regularity to construct classical geodesic curves and locally inertial coordinates, as we prove in Corollary \ref{Cor_locally_inertial_frames} below. Taken together, this resolves the open problem as to whether the spacetime singularities at GR shock waves are removable in the positive, establishing that every Lipschitz continuous metric of GR shock wave theory is regular enough to meet the physical requirements of spacetime.

Theorem \ref{Thm_Smoothing}$^\prime$ is the simplest statement of our optimal regularity result, because $\Gamma$ and $d\Gamma$ and $\Gamma \wedge \Gamma$ all lie in the same space. Theorem \ref{Thm_Smoothing}$^\prime$ follows directly from the following more refined theorem for $L^p$ connections---our main result---, establishing the improved version \eqref{curvature} of estimate \eqref{curvature_simple}. 

\begin{Thm} \label{Thm_Smoothing}  
Assume $\Gamma \in L^{2p}(\Omega)$ and $d\Gamma \in L^{p}(\Omega)$ in $x$-coordinates, for some $p>\max\{ n/2, 2\}$,\footnote{The assumption $p>2$ takes effect only in dimensions $n=2$ and $n= 3$, and is only required by our use of the Sobolev embedding Theorem in Section \ref{Sec_gauge_lowreg}.} $n\geq 2$,  and let $M>0$ be a constant such that
\beq \label{bound_incoming_ass}
 \|(\Gamma,d\Gamma)\|_{L^{2p,p}(\Omega)}  \equiv \|\Gamma_x \|_{L^{2p}(\Omega_x)} + \|d\Gamma_x \|_{L^p(\Omega_x)} \; \leq \; M.
\eeq    
Then for any point $q\in \Omega$ there exists a neighborhood $\Omega' \subset \Omega$ of $q$ and a coordinate transformation $x \to y$ with Jacobian $J=\frac{\partial y}{\partial x}\, \in W^{1,2p}(\Omega'_x),$ such that the connection components $\Gamma_y$ in $y$-coordinates exhibit optimal regularity $\Gamma_y \in W^{1,p}(\Omega''_y)$, (connection one derivative above its curvature), on every open set $\Omega''$ compactly contained in $\Omega'$, where $\Omega''_y \equiv y(\Omega'')$.  Moreover, for each $\Omega''$ compactly contained in $\Omega'$, $\Gamma_y$ satisfies the uniform bound   
\beq \label{curvature} 
\|\Gamma_y \|_{W^{1,p}(\Omega''_y)}  \leq C_1(M)\; \|(\Gamma,d\Gamma)\|_{L^{2p,p}(\Omega')} , 
\eeq
and the Jacobian $J$ satisfies 
\beq \label{curvature_on_J}
\|J\|_{W^{1,2p}(\Omega''_x)}  + \|J^{-1}\|_{W^{1,2p}(\Omega''_x)} \leq C_1(M)\; \|(\Gamma,d\Gamma)\|_{L^{2p,p}(\Omega')} ,  
\eeq          
for some constant $C_1(M) > 0$ depending only on $\Omega''_x, \Omega'_x, p, n, q$ and $M$. The neighborhood $\Omega'$ can be taken as $\Omega'_x = \Omega_x \cap B_r(q)$, for $B_r(q)$ the Euclidean ball of radius $r$ in $x$-coordinates, where $r$ depends only on $\Omega_x, p, n$ and $\Gamma$ near $q$; if $\|(\Gamma,d\Gamma)\|_{L^{\infty,2p}(\Omega)}~\leq~M$, then $r$ depends only on $\Omega_x, p, n$ and $M$.
\end{Thm}

\noindent The refinement \eqref{curvature} of estimate \eqref{curvature_simple} provides the extension to $L^p$ connections. The proofs of Theorems \ref{Thm_Smoothing}$^\prime$ and \ref{Thm_Smoothing} are based on developing an existence theory for the RT-equations, as summarized in Section \ref{Sec_Proof_optimal_regularity}. The strategy of our proof is to first prove existence to the RT-equations in the simpler $L^\infty$ case addressed in Theorem \ref{Thm_Smoothing}$^\prime$ (in Sections \ref{Sec_existence_theory} - \ref{Sec_Proofs}), where the analysis is cleanest, and then extend this proof to the $L^p$ case of Theorem \ref{Thm_Smoothing} in Section \ref{Sec_Lp-extension}.

\subsection{Uhlenbeck compactness}
To introduce our compactness theorem, let us recall  briefly the relation between weak and strong convergence in Banach spaces $L^p$.  Recall that whenever we have a uniform bound on a sequence of functions in $L^p$, there always exists a weakly convergent subsequence whose limit satisfies the same uniform bound $L^p$ as the original sequence.  (By the Banach-Alaoglu Theorem, the closed unit ball is weakly compact in $L^p$ \cite{Evans}.)  But the compactness we seek here is the statement that this weak limit is actually a strong limit in $L^p$.  For this it suffices to have a uniform bound on the sequence of functions in $W^{1,p}$, the resulting weak convergence in $W^{1,p}$ then implies also strong convergence in $L^p$. Weak convergence is generally not sufficient for non-linear problems because products are generally not continuous under weak limits, but are always continuous under strong limits, and weak limits cannot be estimated as close to the weakly convergent subsequence in the norms in which the global bounds are obtained.

We now develop some notation required to state our extension of Uhlenbeck's compactness result, Theorem 1.5 in \cite{Uhlenbeck}. Let $\{\Gamma_i\}_{i\in \mathbb{N}}$ be a sequence of connections $\Gamma_i$ defined on the tangent bundle $T\mathcal{M}$, and let $(\Gamma_i)_x$ denote their components in fixed $x$-coordinates defined on $\Omega_x\subset\mathbb{R}^n$, bounded and open. Assume $d\Gamma_i \in L^p(\Omega_x)$ for some $p>n$, and assume $\Gamma_i \in L^\infty(\Omega_x)$, in fixed $x$-coordinates.\footnote{The assumption $\Gamma_i \in L^{2p}(\Omega_x)$ for $p<\infty$, as well as $d\Gamma_i \in L^{p}(\Omega_x)$ for $n/2 < p \leq n$, would currently not suffice, because of our $\epsilon$-scaling argument for the existence theory in Section \ref{Sec_existence_theory}.}   Again, from now on we use $\Omega$ and $\Gamma_i$ to denote $\Omega_x$ and $(\Gamma_i)_x$ respectively. Our compactness theorem states the existence of a \emph{strongly} convergent subsequence of $\{\Gamma_i\}_{i\in \mathbb{N}}$ in $L^p$ under coordinate transformation, assuming only the bound 
\beq \label{uniform_bound}
\|(\Gamma_i,d\Gamma_i)\|_{L^{\infty,p}(\Omega_x)} \leq M,
\eeq
for some constant $M>0$ independent of $\Gamma_i$.   More precisely, assuming \eqref{uniform_bound}, Theorem \ref{Thm_Smoothing} implies that for each $i\in\mathbb{N}$, there exists a coordinate transformation $x\to y_i(x)$, such that the connection components $(\Gamma_i)_{y_{i}}\equiv \Gamma_{y_{i}}$ in $y_i$-coordinates are one order more regular and satisfy the uniform bound $\|\Gamma_{y_{i}}\|_{W^{1,p}(\Omega'_{y_i})}\leq C(M)$ for some constant $C(M)>0$ and some open set $\Omega'$, both depending only on $M$, independent of $i$, (taking $p,n$ and $\Omega$ to be fixed). To establish the existence of a convergent subsequence of $\{\Gamma_{y_i}\}_{i \in \mathbb{N}}$, we express the components of each $\Gamma_{y_i} \equiv \Gamma_{y_i}(y_i)$ as functions of the original $x$-coordinates $\Gamma_{y_i}(x) \equiv \Gamma_{y_i}(y_i(x))$, so that $\Gamma_{y_i}(x)$ has the same (optimal) regularity as $\Gamma_{y_i}$, since the mapping from $x \rightarrow y_i(x)$ is one derivative more regular than $\Gamma_{y_i}$.  (That is, we transform the $y$-components back to $x$-components as scalars, in contrast to the connection transformation from $\Gamma_{y_{i}}$ to $(\Gamma_i)_x$ which looses one derivative of regularity.) Our proof of Uhlenbeck compactness establishes by use of the uniform estimate \eqref{curvature_on_J} on $J$ that the resulting components $\Gamma_{y_i}(x) \equiv \Gamma_{y_i}(y_i(x))$ will again meet the uniform $W^{1,p}$-bound $\|\Gamma_{y_{i}}\|_{W^{1,p}(\Omega'_x)}\leq C(M)$, but over the fixed region $\Omega'_x$ in $x$-coordinates and for a different constant $C(M)>0$ which accounts for the Jacobian of the transformation from $x$ to $y_i$. By the Banach-Alaoglu Theorem, we then conclude the existence of a subsequence of $\Gamma_{y_i}(x)$ which converges weakly in $W^{1,p}(\Omega'_x)$ and hence strongly in $L^p(\Omega'_x)$.\footnote{The implication that weak $W^{1,p}$ convergence implies strong $L^p$ convergence can be found in \cite[Chapter 5.7]{Evans}. In fact, our assumption $p>n$ implies that even strong convergence in $L^\infty$ and the supremums norm holds, c.f. the Rellich Kondrashov Theorem \cite[Theorem 8.9(iii)]{LiebLoss}, but for sake of presentation we only state strong $L^p$ convergence in Theorem \ref{Thm_compactness}.} This is our compactness theorem. The proof is given in Section \ref{Sec_Proof_optimal_regularity}.   

\begin{Thm}   \label{Thm_compactness}
Assume $\{(\Gamma_i)_x\}_{i\in \mathbb{N}}$ are the $x$-components of a sequence of connections $\Gamma_i$ on the tangent bundle $T\mathcal{M}$ of an $n$-dimensional manifold $\mathcal{M}$ in a fixed coordinate system $x$ on $\Omega$. Assume $\Gamma_i \in L^{\infty}(\Omega)$ and $d\Gamma_i \in L^{p}(\Omega)$ in $x$-coordinates, $n < p <\infty$, and assume the uniform bound
\beq \label{uniform_bound2}
\|(\Gamma_i,d\Gamma_i)\|_{L^{\infty,p}(\Omega_x)} \equiv \|(\Gamma_i)_x\|_{L^\infty(\Omega_x)} + \|(d\Gamma_i)_x\|_{L^p(\Omega_x)} \leq M
\eeq
holds for some constant $M>0$ independent of $i\in \mathbb{N}$. Then for any $q \in \Omega$ there exists a neighborhood $\Omega' \subset \Omega$ of $q$, and a subsequence of $\Gamma_i$, (also denoted by $\Gamma_i$), for which the following holds: 

\noindent{\bf (i)} There exists for each $(\Gamma_i)_x$ a coordinate transformation $x \to y_i(x)$ taking $\Omega'_x$ to $\Omega'_{y_i}$, such that the components $(\Gamma_i)_{y_i}\equiv\Gamma_{y_i}$ of $\Gamma_i$ in $y_i$-coordinates exhibit optimal regularity $\Gamma_{y_i}\in W^{1,p}(\Omega'_{y_i})$, with uniform  bound \eqref{curvature_simple} in $W^{1,p}(\Omega'_{y_i}).$ 

\noindent{\bf (ii)} The $y_i$-components $\Gamma_{y_i}$, taken as functions of $x$, also exhibit optimal regularity $\Gamma_{y_i}(x) \equiv \Gamma_{y_i}(y_i(x)) \in W^{1,p}(\Omega'_{x})$, with uniform bound \eqref{curvature} in $W^{1,p}(\Omega'_{x})$. 

\noindent{\bf (iii)} The transformations $x\to y_i(x)$ are uniformly bounded in $W^{2,2p}(\Omega'_x)$ in light of \eqref{curvature_on_J}, and converge to a transformation $x\to y(x)$, weakly in $W^{2,2p}(\Omega'_x)$ and strongly in $W^{1,2p}(\Omega'_x)$.

\noindent{\bf (iv)} {\rm Main Conclusion:}   There is a subsequence on which the $y_i$-components $\Gamma_{y_i}(x)$ converge to some $\Gamma_y(x)$, weakly in $W^{1,p}(\Omega_x')$, strongly in $L^p(\Omega_x')$, and $\Gamma_y \equiv \Gamma_y(x^{-1}(y))$ are the connection coefficients of $\Gamma_x$ in $y$-coordinates, where $\Gamma_x$ is the weak limit of $(\Gamma_i)_x$ in $L^p(\Omega'_x).$
\end{Thm}

Theorem \ref{Thm_compactness} extends Uhlenbeck's compactness result, Theorem 1.5 of \cite{Uhlenbeck}, to Lorentzian geometry and beyond as follows:  Theorem 1.5 of \cite{Uhlenbeck} applies to connections on vector bundles (with compact gauge groups) over {\it Riemannian} manifolds, and our Theorem \ref{Thm_compactness} applies to connections on tangent bundles, but for arbitrary manifolds, including {\it Lorentzian} manifolds of General Relativity. The extension of our results in this paper to connections on vector bundles over arbitrary base manifolds with compact and non-compact gauge groups is accomplished in \cite{ReintjesTemple_ell5}. As we mentioned in Section \ref{Sec_intro}, our extension to general affine connections requires a small modification in the regularity assumptions for the connection and curvature. Namely, Theorem 1.5 of \cite{Uhlenbeck} assumes a sequence of connections $(\Gamma_x)_i\in W^{1,p}$, with curvature ${\rm Riem}(\Gamma_i)$ uniformly bounded in $L^p$, $p>n$, and from this concludes with a uniform $W^{1,p}$ bound on connection coefficients in Coulomb gauge (where $\delta\Gamma=0$), with resulting compactness in $L^p$. In contrast, our Theorem \ref{Thm_compactness} assumes a sequence of connections $(\Gamma_x)_i$ which need not lie in $W^{1,p}$ at the start, but assumes the same uniform bound on the curvature in $L^p$, replacing only the assumption $\Gamma_i\in W^{1,p}$ with the assumption of a  uniform bound on $\Gamma_i$ in $L^\infty.$  That is, we need not assume differentiability of the sequence of connections, but require a uniform bound in the less regular space $L^\infty$.   Since the assumption of $L^p$ regularity of the curvature is the essential part, we view our assumptions as being essentially equivalent to the assumptions of Uhlenbeck.   Our assumptions $\Gamma_i$ in $L^\infty$ and ${\rm Riem}(\Gamma_i)\in L^p$, ($p>n$), are natural for the setting of the RT-equations, as we now discuss (setting for simplicity $p=\infty$ as in Theorem \ref{Thm_Smoothing}$^\prime$). First note that a uniform bound on $\|{\rm Riem}(\Gamma_i)\|_{L^\infty}$ is implied by, but does not imply a uniform bound on $\|(\Gamma_i,d\Gamma_i)\|_{L^\infty}=\|\Gamma_i\|_{L^\infty}+\|d\Gamma\|_{L^\infty}$, since uncontrolled terms in $\Gamma$ could cancel in the wedge-product in ${\rm Riem}(\Gamma)=d\Gamma+\Gamma\wedge\Gamma$. In light of \eqref{RiemEquiv}, for the bound on $\|(\Gamma_i,d\Gamma_i)\|_{L^\infty}$ to imply a bound on the curvature tensor would require starting in a coordinate system $x$ in which $\|\Gamma_i\|_{L^\infty}$ is bounded by $\|d\Gamma_i\|_{L^\infty}$, or alternatively by $\|{\rm Riem}(\Gamma_i)\|_{L^\infty}$.  For this one could take the locally inertial coordinate frames proven in Corollary \ref{Cor_locally_inertial_frames} to exist for $W^{1,p}$ connections. This shows that our assumption of a uniform bound on $\|\Gamma_i\|_{L^\infty}$ is implied by an $L^\infty$ bound on the curvature alone, in natural coordinates, but not necessarily in all coordinate systems, which could involve transformations with arbitrarily large Jacobians. 

Theorems \ref{Thm_Smoothing} and \ref{Thm_compactness} are based on extending the existence theory for the \emph{RT-equations} to the setting of weak (distributional) solutions.  The RT-equations, introduced by the authors in \cite{ReintjesTemple_ell1},  are a system of elliptic partial differential equations which determine whether coordinate systems exist in which the connection exhibits optimal regularity. The RT-equations are elliptic independent of any underlying metric structure on the tangent bundle, hence our methods do not require the ellipticity of the Laplace-Beltrami operator of a metric, and by this we can extend Uhlenbeck's results to  tangent bundles of arbitrary manifolds. (Again, using the Coulomb gauge method in the case of Lorentzian metrics would entail hyperbolic estimates, which are problematic, c.f. \cite{ReintjesTemple_ell3}.) To formulate the RT-equations, we require an Euclidean Cartan algebra for matrix valued differential forms, the topic of Section \ref{Sec_Prelim}.

\section{Applications of the main results} \label{Sec_Applications}

We now present some applications of Theorems \ref{Thm_Smoothing} and \ref{Thm_compactness}. In particular, we show that the GR shock waves solutions constructed in \cite{GroahTemple} can be regularized, and we formulate a version of Uhlenbeck compactness aimed at constructing low regularity solutions of the vacuum Einstein equations. The reader interested only in the proof of Theorems \ref{Thm_Smoothing} and \ref{Thm_compactness} and the RT-equations may skip this section.

\subsection{Application of Uhlenbeck compactness to the vacuum Einstein equations} \label{Sec_appl_Einstein-eqns}

As an application of Uhlenbeck compactness,  we prove the following corollary of Theorem \ref{Thm_compactness} which provides a new compactness theorem applicable to vacuum solutions of the Einstein equations. The main difficulty in a convergence proof for a PDE in an existence theory, is typically the problem of establishing a uniform bound on the highest order derivatives, suitable to apply Sobolev compactness. Uhlenbeck compactness tells us that it suffices to establish a bound on just the Riemann curvature, not all highest order derivatives of a connection, in order to imply subsequential convergence of connection and metric.                    

\begin{Corollary} \label{Cor_compactness}  
Let $g_i$ be a sequence of Lipschitz continuous metrics given on a manifold $\M$, and let $\Gamma_i$ denote the Christoffel symbols of $g_i$ for each $i \in \mathbb{N}$. Assume that $(g_i)_{i\in\mathbb{N}}$ is a sequence of approximate solutions of the vacuum Einstein equations such that, in a neighborhood of each point, there exists a coordinate system $x$ in which ${\rm Ric}(g_i) \to 0$ weakly in $L^p$, for $n<p<\infty$, 
and $g_i$ satisfies the uniform bound
\beq \label{Cor_compct_eqn1}
\|g_i\|_{L^\infty} + \|\Gamma_i\|_{L^\infty} + \|{\rm Weyl}(g_i)\|_{L^p} \leq M
\eeq
for some constant $M>0$, together with the non-degeneracy condition that $|\det(g_i)|$ is uniformly bounded away from zero. (Norms are taken in coordinate systems and ${\rm Weyl}(g_i)$ denotes the Weyl curvature of $g_i$.) Then, in each such coordinate system, there exists a subsequence of $(g_i)_{i\in \mathbb{N}}$ which converges component-wise and weakly in $W^{1,p}(\Omega)$ to some metric $g$ which satisfies \eqref{Cor_compct_eqn1} and solves the vacuum Einstein equations ${\rm Ric}(g)=0$. Furthermore, according to Theorem \ref{Thm_Smoothing}, there exists locally, (i.e., in a neighborhood of each point), a $W^{2,2p}$ coordinate transformation $x\to y$ which lifts the components of $g$ to $W^{2,p}$ and these are the $W^{2,p}$-limits of $g_{y_i}$, the components of $g_i$ in optimal coordinates $y_i$, as in (ii) of Theorem \ref{Thm_compactness}.
\end{Corollary} 

Note that if $\M$ is a compact manifold, one can cover $\M$ with a finite number of  such optimal coordinate patches, and by a diagonal argument, extract a subsequence which converges to a solution of the vacuum Einstein equations in each coordinate system of the finite covering, and hence in all of $\M$.  

\Proof 
To begin, note that the $L^\infty$ bound on $\Gamma_i$ provides an $L^\infty$ bound on the derivatives of $g_i$, so the sequence $(g_i)_{i\in \mathbb{N}}$ is uniformly bounded in $W^{1,p}$, any $p \in (1,\infty)$. Thus the weak $W^{1,p}$ convergence of a subsequence of $(g_i)_{i\in \mathbb{N}}$ to some metric $g \in W^{1,p}$ in $x$-coordinates follows by the Banach Alaoglu Theorem for any $p\leq \infty$; (that $g$ is a metric follows by our non-degeneracy assumption). But this is not enough to conclude that ${\rm Ric}(g)=0$, because the convergence is not strong enough to pass weak limits through ${\rm Ric}(g_i)$. To prove that ${\rm Ric}(g)=0$ we apply now Uhlenbeck compactness of Theorem \ref{Thm_compactness}.  Since ${\rm Ric}(g)=0$ is a point-wise condition, we assume without loss of generality that the coordinate transformations to optimal regularity ($x\to y_i$), asserted to exist by Theorem \ref{Thm_compactness}, are defined on the entire coordinate patch in which each $g_i$ is given. 

Note first that a uniform $L^p$ bound on ${\rm Ric}(g_i)$ and ${\rm Weyl}(g_i)$ implies that ${\rm Riem}(g_i)$ is uniformly bounded in $L^p$, since the Ricci tensor together with the Weyl tensor comprise the Riemann curvature tensor \cite{Choquet}. Moreover, by assumption ${\rm Ric}(g_i) \to 0$ weakly in $L^p$, which implies that ${\rm Ric}(g_i)$ is uniformly bounded in $L^p$.  Thus, also taking into account the bound on the metric and connection in \eqref{Cor_compct_eqn1}, Theorem \ref{Thm_compactness} applies and yields the existence of a convergent subsequence of $(g_i)_{i\in \mathbb{N}}$, asserting weak $W^{2,p}$ and strong $W^{1,p}$ convergence, ($n<p<\infty$). Namely, let $y_i$ be a coordinate system in which $\Gamma_i$ and hence $g_i$ has optimal regularity, and denote by $g_{y_i}$ the metric $g_i(x)$ in $y_i$ coordinates but with its components expressed as functions over $x$-coordinates, c.f. (ii) of Theorem \ref{Thm_compactness}. Then $g_{y_i} \in W^{2,p}$ in $x$-coordinates and 
\beq \nonumber
\|g_{y_i}\|_{W^{2,p}} \leq  \|g_{y_i}\|_{L^p} + \|\Gamma_{y_i}\|_{W^{1,p}}
\eeq
is bounded uniformly by some constant $C(M)>0$, since $\|\Gamma_{y_i}\|_{W^{1,p}}$ and the Jacobians $\big\|\frac{\partial y_i}{\partial x}\big\|_{W^{1,p}}$ are both uniformly bounded, c.f. \eqref{curvature} and \eqref{curvature_on_J}. The asserted convergence of a subsequence now follows by the Banach Alaoglu Theorem. We denote this convergent subsequence by $(g_{y_i})_{i\in \mathbb{N}}$, where $g_{y_i}\equiv g_{y_i}(x)$ is to be understood as the metric in $y_i$-coordinates with components expressed in $x$-coordinates. 

The main point then is that the curvature is linear in derivatives, and one can pass weak limits through such derivatives.   That is, by assumption ${\rm Ric}(g_i) \rightarrow 0$ weakly in $L^p$ for some $p \in (n,\infty)$ as $i\rightarrow \infty$, and to prove that the limit metric $g \equiv \lim\limits_{i\to \infty} g_{i}$ solves the vacuum Einstein equations, we need only show that  ${\rm Ric}(g_i)$ converges to ${\rm Ric}(g)$ weakly in $L^p$ as $i\rightarrow \infty$. For this, observe that the weak $W^{1,p}$ convergence of $\Gamma_i$ implies weak $L^p$ convergence of $d\Gamma_i$ to $d\Gamma$, where $\Gamma$ denotes the connection of $g$. Moreover, the strong $L^p$ convergence of $\Gamma_i$ implies strong convergence of $\Gamma_i \wedge \Gamma_i$ to $\Gamma \wedge \Gamma$ in $L^{\frac{p}{2}}$. This implies weak convergence of the Riemann curvature, namely for any matrix valued $2$-form $\Psi \in W^{1,(p/2)^*}_0 \subset W^{1,p^*}_0$ we have
\begin{eqnarray} \nonumber
\langle {\rm Riem}(\Gamma_i),\Psi \rangle_{L^2} 
&=& - \langle \Gamma_i,\delta\Psi \rangle_{L^2} +\langle \Gamma_i\wedge \Gamma_i,\Psi \rangle_{L^2} \cr
& \overset{i\to\infty}{\longrightarrow} & - \langle \Gamma,\delta\Psi \rangle_{L^2} +\langle \Gamma\wedge \Gamma,\Psi \rangle_{L^2} 
= \langle {\rm Riem}(\Gamma),\Psi \rangle_{L^2} ,
\end{eqnarray}
which implies weak convergence in $L^\frac{p}{2}$ by denseness of $W^{1,(p/2)^*}_0$ in $L^{p^*}$. Since this applies to any $p>n$, we conclude that ${\rm Riem}(\Gamma_i)$ converges weakly to ${\rm Riem}(\Gamma)$ in $L^p$ which implies the sought after convergence, ${\rm Ric}(g_i) \rightarrow {\rm Ric}(g)$ weakly in $L^p$. This proves $g$ solves the vacuum Einstein equations  ${\rm Ric}(g)=0$ and this holds in any coordinate system by tensor transformation. 
\QED

Note that without Uhlenbeck compactness Theorem \ref{Thm_compactness}, the uniform $L^\infty$ bound on a sequence of metric connections and their curvatures would not in general imply that the limit metric solves the vacuum Einstein equations.  Indeed, weak $L^p$ convergence of a subsequence of the metric connections is not  in general sufficient to pass weak limits through nonlinear functions like products  \cite[Chapter 16]{Dafermos}. As a result, even though the Ricci tensor would correctly converge to zero, the limit Ricci tensor would in general fail to be the Ricci tensor of the limit connection.

\subsection{A generalization of the Israel junction conditions}

As a corollary of Theorem \ref{Thm_Smoothing}$^\prime$, in the spirit of Uhlenbeck's earlier paper \cite{Uhlenbeck_2} for positive definite metrics, we immediately obtain that $d\Gamma\in L^\infty$ implies that singularities on sets of measure zero in non-optimal connections are always removable. The condition $d\Gamma\in L^\infty$ plays the role of a generalized Rankine Hugoniot jump condition \cite{Smoller}, or ``Junction Condition'' \cite{Israel}, and it gives general expression to the condition that the curvature be free of ``delta function sources'', necessary and sufficient conditions introduced by Israel for smoothing discontinuous connections across \emph{single} shock surfaces \cite{Israel}.  

\begin{Corollary} \label{Cor_singularity_removal}
Assume $\Gamma, d\Gamma$ are bounded and continuous off a set of measure zero in some open set $\Omega$ in $x$-coordinates. Then the additional condition that the $L^\infty$ extension of $\Gamma$ to $\Omega$ satisfies $d\Gamma \in L^{\infty}(\Omega),$ is sufficient to imply that for any point $q\in \Omega$,  there exists a neighborhood $\Omega' \subset \Omega$ of $q,$ and a coordinate transformation $x \to y$ on $\Omega',$ such that the connection components $\Gamma_y$ in $y$-coordinates can be extended as H\"older continuous functions to $\Omega'_y$ with $\Gamma_y \in W^{1,p}(\Omega'_y)$. 
\end{Corollary}

\noindent This is a direct consequence of Theorem \ref{Thm_Smoothing}$^\prime$, keeping in mind that $W^{1,p}$ is embedded in the space of H\"older continuous functions for $p>n$ by Morrey's inequality, c.f. \eqref{Morrey_textbook}.

\subsection{Optimal regularity in spherically symmetric spacetimes} 

The following corollary of Theorem \ref{Thm_Smoothing} establishes for the first time that solutions of the Einstein equations constructed in Standard Schwarzschild Coordinates, including the Lipschitz continuous metrics associated with shock waves in \cite{GroahTemple}, can always be smoothed to optimal regularity by coordinate transformation.  Solutions of the Einstein equations in SSC have a long history in General Relativity going back to Schwarzschild and Birkhoff.   The existence theory in \cite{GroahTemple} establishes (weak) shock wave solutions of the Einstein-Euler equations by Glimm's method, (see also \cite{LeFlochStewart}).  The Einstein-Euler system couples the unknown metric $g_{ij}$ to the unknown density $\rho$, pressure $p$ and velocity $u$ of a perfect fluid via $T^{ij}=(\rho+p)u^iu^j+pg^{ij}$ in $G=\kappa T$. The spacetime metrics of these solutions are non-optimal with curvature in $L^\infty$, but optimal metric regularity would be required to introduce locally inertial frames and geodesic curves by standard methods.  

For this consider a metric in Standard Schwarzschild Coordinates (SSC)
\begin{eqnarray}
ds^2=-B(t,r)dt^2+\frac{dr^2}{A(t,r)}+r^2d\Omega^2.
\end{eqnarray}
This represents the coordinates in which the Einstein equations for a spherically symmetric spacetime metric (arguably) take their simplest form. Since the first three Einstein equations in SSC are 
\begin{eqnarray}\label{firstorder1}
-r A_r +(1-A)&=&\kappa B T^{00}r^2 \\ \label{firstorder2}
A_t &=&\kappa B T^{01}r \\ \label{firstorder3}
r\frac{B_r}{B}-\frac{1-A}{A}&=&\frac{\kappa }{A^2}T^{11}r^2
\end{eqnarray}
the metric can generically be only one level more regular than the curvature tensor, at every level of regularity, and is hence non-optimal. (See \cite{GroahTemple} for the full system of equations.) As an application of Theorem \ref{Thm_Smoothing}, we have the following result which establishes that shock wave solutions of the Einstein equations constructed by the Glimm scheme are one order more regular than previously known \cite{GroahTemple}.   (The result here extends to every level of regularity, c.f. \cite{ReintjesTemple_ell3}.) 

\begin{Corollary} \label{Cor_opt}
Let $T\in L^\infty$ and assume $g\equiv(A,B)$ is a (weak) solution of the Einstein equations in SSC satisfying $g\in C^{0,1}$ and hence $\Gamma\in L^\infty$ in an open set $\Omega$. Then for any $p>4$ and any $q\in\Omega$ there exists a coordinate transformation $x\to y$ defined in a neighborhood of $q$, such that, in $y$-coordinates, $g\in W^{2,p},$ $\Gamma\in W^{1,p}$.
\end{Corollary} 

\Proof
In Standard Schwarzschild coordinates the Ricci and Riemann curvature tensor have the same regularity (as can be verified using Mathematica). So assuming $T$ in $L^\infty$ implies $d\Gamma$ in $L^\infty$, and Theorem \ref{Thm_Smoothing}  implies the corollary. 
\QED

\subsection{Construction of locally inertial coordinates} \label{Sec_loc_inertial}  
The standard method for constructing locally inertial frames does not apply to connections $\Gamma\in L^\infty(\Omega)$ because the classical Riemann normal coordinate construction requires Lipschitz continuity for a connection, and regularity $C^{1,1}$ for a metric, \cite{ReintjesTemple1}.  The following corollary of Theorem \ref{Thm_Smoothing} establishes that locally inertial coordinates always exist in a H\"older sense, for any $L^\infty$ connection with $Riem(\Gamma)\in L^\infty(\Omega).$

\begin{Corollary} \label{Cor_locally_inertial_frames}
Assume $\Gamma, {\rm Riem}(\Gamma) \in L^{\infty}(\Omega)$ on a bounded spacetime domain $\Omega \subset \R^{n}$. Then for any $p \geq 1$ and any point $q\in \Omega$ there exists a neighborhood $\Omega' \subset \Omega$ of $q$ and a coordinate transformation with Jacobian $J\in W^{1,2p}(\Omega')$ such that the connection in the resulting coordinates $z$ has regularity $\Gamma \in W^{1,p}(\Omega')$ and satisfies
\begin{eqnarray}
\Gamma^\gamma_{\alpha\beta}(q) &=& 0 \label{loc_inertial_prop1} \\
\big|\Gamma^\gamma_{\alpha\beta}(\hat{q})\big| &\leq & C\, |q-\hat{q}|^\alpha, \label{loc_inertial_prop2}
\end{eqnarray} 
where $\alpha \in (0,1)$ is the H\"older coefficient associated with $2p>n$ by Morrey's inequality and $|\cdot|$ is the Euclidean norm on $\R^n$ applied to $q-\hat{q}$ in $z$-coordinates.
\end{Corollary}

\noindent We call a coordinate system $y$ in which the connection is in $W^{1,p}$ for $p>n$ and satisfies \eqref{loc_inertial_prop1} and \eqref{loc_inertial_prop2}, a locally inertial coordinate system with H\"older corrections to the gravitational field.  The case $\alpha =1$ in \eqref{loc_inertial_prop2} would give the standard second order correction due to the gravitational field. For Lorentz metrics one can in addition arrange for the metric to be equal to the Minkowski metric at $q$ by suitable multiplication with a constant Jacobian.\footnote{This Jacobian is the unique composition of the orthogonal matrix diagonalizing the metric at $q$ multiplied with the diagonal matrix that has the inverse of the square root of each eigenvalue of the metric on its diagonal. See the construction in \cite{ReintjesTemple_geo} for details. Since the Jacobian is constant, properties \eqref{loc_inertial_prop1} - \eqref{loc_inertial_prop2} are preserved.} \vspace{.2cm}

\noindent {\it Proof of Corollary \ref{Cor_locally_inertial_frames}.}
The assumptions of Corollary \ref{Sec_loc_inertial} are identical to those of Theorem \ref{Thm_Smoothing}. Applying Theorem \ref{Thm_Smoothing} gives us a Jacobian $J\in W^{1,2p}(\Omega')$, as determined by the RT-equations, defined in some neighborhood $\Omega' \subset \Omega$ of $q$, such that the connection in the resulting coordinates $y^\alpha$ has regularity $\Gamma \in W^{1,p}(\Omega')$. Without loss of generality, we assume that $y(q)=0$.

To arrange for condition \eqref{loc_inertial_prop1}, following the development in Chapter 8 of \cite{ReintjesTemple_geo}, we introduce a smooth coordinate transformation $y \rightarrow z$, (hence  preserving the regularity $W^{1,p}$ of $\Gamma$), such that $\Gamma$ satisfied the sought after properties \eqref{loc_inertial_prop1} - \eqref{loc_inertial_prop2} in $z$-coordinates. The H\"older continuity of $\Gamma$, (implied by Morrey's inequality \eqref{Morrey_textbook}), allows us to evaluate the connection in $y$-coordinates at the point $q$, $\Gamma^\alpha_{\beta\gamma}\big|_q$, and hence introduce the coordinate transformation
\beq \label{loc_inertial_transfo_z}
z^\mu(y) \equiv  \delta^\mu_\alpha \Gamma^\alpha_{\beta\gamma}\big|_q \: y^\beta y^\gamma + \delta^\mu_\alpha \: y^\alpha  ,
\eeq
where $\delta^\mu_\alpha$ denotes the Kronecker symbol. Clearly \eqref{loc_inertial_transfo_z} defines a smooth coordinate transformation and, by our incoming assumption $y(q)=0$, it follows that 
\beq \label{loc_inertial_eqn1}
z(y(q))=0 \hspace{1cm} \text{and} \hspace{1cm} \frac{\partial z^\mu}{\partial y^\alpha}\Big|_q = \delta^\mu_\alpha. 
\eeq
Moreover, and this is the main point of definition \eqref{loc_inertial_transfo_z}, we have
\beq \label{loc_inertial__eqn2}
\frac{\partial^2 z^\mu}{\partial y^\beta \partial y^\gamma}\Big|_q =\delta^\mu_\alpha \Gamma^\alpha_{\beta\gamma}\big|_q,
\eeq
which implies that the connection $\Gamma^\sigma_{\mu\nu}$ in $z$-coordinates vanishes at $q$.  Indeed, from the transformation law of connections we find that
\begin{eqnarray} \nonumber
\frac{\partial z^\sigma}{\partial y^\alpha} \Gamma^\alpha_{\beta\gamma} 
=\frac{\partial^2 z^\sigma}{\partial y^\beta \partial y^\gamma} +\Gamma^\sigma_{\mu\nu}\frac{\partial z^\mu}{\partial y^\beta}\frac{\partial z^\nu}{\partial y^\gamma} ,
\end{eqnarray} 
so using \eqref{loc_inertial_eqn1} and \eqref{loc_inertial__eqn2} to evaluate $\Gamma^\sigma_{\mu\nu}$ at $q$ gives
\begin{eqnarray} \nonumber
\delta^\sigma_\alpha \Gamma^\alpha_{\beta\gamma}\big|_q 
=\delta^\sigma_\alpha \Gamma^\alpha_{\beta\gamma}\big|_q + \Gamma^\sigma_{\mu\nu}\big|_q \delta^\mu_\beta \delta^\nu_\gamma
\end{eqnarray} 
and this implies that in $z$-coordinates $\Gamma^\sigma_{\mu\nu}\big|_q =0$ for all $\sigma,\mu,\nu \in \{1,...,n\}$. This proves property \eqref{loc_inertial_prop1} of Corollary \ref{Cor_locally_inertial_frames}.

Now property \eqref{loc_inertial_prop2} follows directly from \eqref{loc_inertial_prop1} together with the H\"older continuity of $\Gamma$ in $z$-coordinates. Namely, since the coordinate transformation $y \rightarrow z$ is in $C^\infty(\Omega')$, we again have $\Gamma \in W^{1,p}(\Omega')$ in $z$-coordinates, so Morrey's inequality implies that $\Gamma \in C^{0,\alpha}$ for $\alpha = 1 - \frac{n}{p}$. This completes the proof of Corollary \ref{Cor_locally_inertial_frames}. \hfill $\Box$

\section{Preliminaries - The Euclidean Cartan algebra} \label{Sec_Prelim}

We now summarize the Cartan Calculus which we require to formulate the RT-equations and refer the reader to Section 2 in \cite{ReintjesTemple_ell1} for further detail and proofs. We work again in fixed $x$-coordinates defined on a open set $\Omega \equiv \Omega_x \subset \R^n$.  By a matrix valued differential $k$-form $\omega$ we mean an $(n\times n)$-matrix whose components are $k$-forms, and we write
\beq \label{def_matrixvalued_diff-form}
\omega = \omega_{[i_1...i_k]} dx^{i_1} \wedge ... \wedge dx^{i_k} \equiv \sum_{i_1< ... < i_k} \omega_{i_1...i_k} dx^{i_1} \wedge ... \wedge dx^{i_k},
\eeq 
for $(n\times n)$-matrices  $\omega_{i_1...i_k}$  such that total anti-symmetry holds in the indices $i_1,...,i_k \in \{1,...,n\}$. (We always sum over repeated indices, following Einstein's convention, but we never ``raise'' or ``lower'' indices.) We define the wedge product of a matrix valued $k$-form $\omega$ with a matrix valued $l$-form $u = u_{j_1...j_l} dx^{j_1} \wedge ... \wedge dx^{j_l}$ as  
\begin{eqnarray} \label{def_wedge}
\omega \wedge u  
&\equiv & \frac{1}{l!k!} \omega_{i_1...i_k} \cdot u_{j_1...j_l} \; dx^{i_1} \wedge ... \wedge dx^{i_k} \wedge dx^{j_1} \wedge ... \wedge dx^{j_l}, 
\end{eqnarray}
where ``$\cdot$'' denotes standard matrix multiplication. In contrast to scalar valued differential forms, $\omega\wedge \omega$ can be non-zero, because matrices do in general not commute. The exterior derivative $d$ is defined component wise on matrix-components,
\beq \label{exterior_derivative}
d \omega \equiv \partial_l \omega_{[i_1...i_k]} dx^l \wedge dx^{i_1} \wedge ... \wedge dx^{i_k} ,
\eeq
and we define the co-derivative $\delta$ on a matrix valued $k$-form $\omega$ as 
$$
\delta \omega  \equiv (-1)^{(k+1)(n-k)} *d* \omega,
$$ 
where $*$ is the Hodge star introduced in terms of the Euclidean metric in $x$-coordinates. That is, $*$ satisfies the orthogonality condition 
\beq \label{Hodge_ortho}
dx^{[i_1} \wedge ... \wedge dx^{i_k]} \wedge\, * \big( dx^{[j_1} \wedge ... \wedge dx^{j_k]} \big) 
= \begin{cases} dx^1 \wedge ... \wedge dx^n, \hspace{.3cm} \text{if} \ \ i_1=j_1,...,i_k=j_k, \cr 0 \hspace{.3cm} \text{otherwise},    \end{cases}
\eeq 
where indices are taken to be increasing. So $\delta$ is defined via the Euclidean metric in $x$-coordinates, while $d$ requires no metric.  Both $d$ and $\delta$ act component wise on matrix components, so all properties of $d$ and $\delta$ for scalar valued differential forms carry over to matrix valued forms. The Laplacian $\Delta \equiv d \delta + \delta d$ acts component wise on matrix-components and also on differential form components. By \eqref{Hodge_ortho}, one can show that $\Delta$ is in fact identical to the Laplacian of the Euclidean metric in $x$-coordinates, 
$$
\Delta = \partial^2_{x^1} + ... + \partial^2_{x^n},
$$ 
c.f. \cite{Dac,ReintjesTemple_ell1} for more detail. 

By \eqref{def_wedge} and \eqref{exterior_derivative}, the Riemann curvature tensor can be written as
\beq \label{Riem}
{\rm Riem}(\Gamma_x) = d\Gamma_x +\Gamma_x \wedge \Gamma_x,
\eeq 
in $x$-coordinates. The exterior derivative satisfies the product rule  
\beq \label{ Leibniz-rule-d}
d(\omega\wedge u) = d\omega \wedge u + (-1)^k \omega \wedge du  ,
\eeq
where $\omega\in W^{1,p}(\Omega)$ is a matrix valued $k$-form and $u \in W^{1,p}(\Omega)$ is a matrix valued $j$-form, (c.f. Lemma 3.3 of \cite{ReintjesTemple_ell1}). Since the wedge product \eqref{def_wedge} for matrix valued $0$-forms $J$ is identical to matrix multiplication, and since $dJ^{-1} = - J^{-1}\cdot dJ\cdot J^{-1}$, the  Leibniz rule \eqref{ Leibniz-rule-d} implies that    
\beq  \label{Leibniz_rule_J-application} 
d\big( J^{-1} \cdot dJ \big) =  d(J^{-1}) \wedge  dJ  = -  J^{-1} d J \wedge J^{-1} dJ,
\eeq
c.f. Lemma 4.3 in \cite{ReintjesTemple_ell1}. Regarding the co-derivative $\delta$, we require the following product rule
\beq \label{ Leibniz-rule-delta}
\delta (J\mm w ) = J \mm \delta w  + \langle d J ;  w \rangle
\eeq
where $J\in W^{1,p}(\Omega)$ is a matrix valued $0$-form, $w \in W^{1,p}(\Omega)$ a matrix valued $1$-form, and where $\langle \cdot\; ; \cdot \rangle $ is the matrix valued inner product defined on matrix valued $k$-forms $\omega$ and $u$ by
\beq \label{def_inner-product}
\langle \omega\; ; u \rangle^\mu_\nu \equiv \sum_{\sigma=1}^n \sum_{i_1<...<i_k} \omega^\mu_{\sigma\: i_1...i_k} u^\sigma_{\nu\: i_1...i_k}.
\eeq
So $\langle \omega\; ; u \rangle$ converts two matrix valued $k$-forms into a matrix valued $0$-form. For multiplication by a matrix valued $0$-form $J$ we have the following multiplication property
\beq \label{inner-product_muliplications}
 J \cdot \langle \omega\; ; u \rangle =  \langle  J \cdot \omega\; ; u \rangle,  \hspace{.5cm} 
\langle \omega\cdot J \; ; u \rangle =  \langle \omega\; ;  J \cdot u \rangle, \hspace{.5cm}
\langle \omega \; ; u \cdot J \rangle =  \langle \omega\; ; u \rangle \cdot J  .
\eeq

We also need to interpret matrix valued forms as vector valued differential forms. The two operations which convert matrix valued differential forms to vector valued forms on the right hand side of the RT-equations are $\vec{\cdot}$ and $\overrightarrow{\text{div}}(\cdot)$.   First, $\vec{\cdot}$ converts matrix valued $k$-forms $\omega$ into vector valued $(k+1)$-forms $\vec{\omega}$ by 
\beq \label{Def_vec-div}
\vec{\omega}^\mu \equiv \omega^\mu_{\nu\, i_1...i_k} dx^\nu \wedge dx^{i_1} \wedge ... \wedge dx^{i_k},
\eeq
with $\omega$ taken as in \eqref{def_matrixvalued_diff-form}, c.f. (2.20) in \cite{ReintjesTemple_ell1} for the case $k=0$, most relevant to us. Secondly, the operation $\overrightarrow{\text{div}}(\cdot)$ converts matrix valued $k$-forms $\omega$ into vector valued $k$-forms $\overrightarrow{\text{div}}(\omega)$ by the operation  
\beq \nonumber
\overrightarrow{\text{div}}(\omega)^\alpha \equiv \sum_{l=1}^n \partial_l \big( (\omega^\alpha_l)_{i_1...i_k}\big) dx^{i_1}\wedge . . . \wedge dx^{i_k}.
\eeq
Finally, for a matrix valued $1$-form $w$ and a matrix valued $0$-form $J$, Lemma 2.4 of \cite{ReintjesTemple_ell1} gives the important identity
\beq \label{regularity-miracle}
d \big(\overrightarrow{\delta ( J \mm w )}\big) 
= \overrightarrow{\text{div}} \big(d(J \cdot w)\big)
= \overrightarrow{\text{div}} \big(dJ \wedge w\big) + \overrightarrow{\text{div}} \big( J\mm dw\big) ,
\eeq                
which is crucial for the regularity to close in the RT-equations.

\section{The reduced RT-equations and resulting optimal regularity} \label{Sec_reducedRT}

In this section we derive the reduced RT-equations from the RT-equations, the system of elliptic PDE's introduced in \cite{ReintjesTemple_ell1} which determines whether a connection $\Gamma$ can be mapped to optimal regularity, and prove their equivalence. We then state the main theorems concerning the existence of solutions of the reduced RT-equations and resulting optimal regularity for $L^p$ connections, Theorems \ref{Thm_Existence_J} and \ref{Thm_gauge_existence} respectively, which are proven in Sections \ref{Sec_weak} - \ref{Sec_Proofs}. In Section \ref{Sec_Proof_optimal_regularity} we apply Theorems \ref{Thm_gauge_existence} and \ref{Thm_Existence_J} to give the proof of our main results,  Theorems \ref{Thm_Smoothing} and \ref{Thm_compactness}. 

We begin by reviewing the RT-equations derived in \cite{ReintjesTemple_ell1}. The RT-equations consist of the following nonlinear elliptic system of PDE's
\begin{eqnarray} 
\Delta \Gammati &=& \delta d \Gamma - \delta \big( dJ^{-1} \wedge dJ \big) + d(J^{-1} A ), \label{RT_1} \\
\Delta J &=& \delta ( J \mm \Gamma ) - \langle d J ; \tilde{\Gamma}\rangle - A , \label{RT_2} \\
d \vec{A} &=& \overrightarrow{\text{div}} \big(dJ \wedge \Gamma\big) + \overrightarrow{\text{div}} \big( J\, d\Gamma\big) - d\big(\overrightarrow{\langle d J ; \tilde{\Gamma}\rangle }\big),   \label{RT_3}\\
\delta \vec{A} &=& v,  \label{RT_4}
\end{eqnarray}
together with boundary data 
\begin{eqnarray}   
d\vec{J} =0 \ \ \text{on} \ \partial \Omega.  \label{RT_data}
\end{eqnarray}
The connection $\Gamma\equiv\Gamma^\mu_{\nu k}dx^k$ is given and interpreted as a matrix valued $1$-form, and $\Gamma$ on the right hand side of \eqref{RT_1} - \eqref{RT_4} always denotes the components $\Gamma_x$ in $x$-coordinates. The unknowns in the RT-equations are $(\Gammati,J,A)$ which are matrix valued differential forms as follows:  $J\equiv J^\mu_\nu$ is the Jacobian of the sought after coordinate transformation which smooths the connection, viewed as a matrix-valued $0$-form; $\Gammati\equiv \Gammati^\mu_{\nu k}dx^k$ represents the unknown tensor one order smoother than $\Gamma$ such that ${\rm Riem}(\Gamma-\Gammati)=0$, viewed as a matrix-valued $1$-form; and $A\equiv A^\mu_\nu$ is an auxiliary matrix valued $0$-form introduced together with boundary data \eqref{RT_data} to impose $d\vec{J} \equiv {\rm Curl}( J)=0$, the condition for the Jacobian $J$ that guarantees it is integrable to a coordinate system, c.f. Theorem \ref{Thm_CauchyRiemann} and \cite{ReintjesTemple_ell1}. See Section \ref{Sec_Prelim} for definitions of the remaining operations in \eqref{RT_1} - \eqref{RT_4}.   

The RT-equations (\ref{RT_1}) and (\ref{RT_2}) were derived by constructing Laplacians out of two equivalent forms of the Riemann-flat condition, a condition introduced in \cite{ReintjesTemple_geo} equivalent to the existence of coordinates in which the connection has optimal regularity.  These two starting conditions were that ${\rm Riem}(\Gamma-\Gammati)=0$, or alternatively, that $\Gammati=\Gamma-J^{-1}dJ$ for some tensor $\Gammati$ one order smoother than $\Gamma$.  If $\Gamma_x$ can be smoothed to optimal regularity by the transformation $x \to y$ with Jacobian $J$, then, defining  
\beq  \label{Gammati'} 
\Gammati_J \equiv \Gamma - J^{-1} dJ,
\eeq
the connection components $\Gamma_y$ of optimal regularity are given by the tensor transformation rule
\beq \label{Gamma_y}
(\Gammati_J)^k_{ij} = (J^{-1})^k_\gamma J^\alpha_i J^\beta_j (\Gamma_y)^\gamma_{\alpha\beta},
\eeq
and $\Gammati_J$ will solve the RT-equations \eqref{RT_1} - \eqref{RT_4} as well as the Riemann-flat condition ${\rm Riem}(\Gamma-\Gammati_J)=0$.   That is, $\Gammati_J,$ the tensor transformation by $J$ of the components of the connection $\Gamma_y$ in $y$-coordinates, will solve the RT-equations for some $(A,v)$, when $J$ is paired with $\Gammati_J$. Conversely, one can recover the connection of optimal regularity $\Gamma_y$ via equation \eqref{Gamma_y} from a general solution $\Gammati,J$ of the RT-equations, but this requires an existence theory which establishes the relationship between the solution $\Gammati$ and $\Gammati_J$, as we now explain.

In \cite{ReintjesTemple_ell1} we prove that if $\Gamma_x, d\Gamma_x \in W^{m,p}$, for $m\geq 1$, $p>n$, then there exists a coordinate transformation $x \to y$ which raises the regularity by one order to $\Gamma_y \in W^{m+1,p}$ if and only if there exists a solution $(J,\Gammati,A)$ of the RT-equations \eqref{RT_1} - \eqref{RT_4}, (taking $\Gamma=\Gamma_x$ on the right hand side), with boundary data \eqref{RT_data}, and regularity $J, \Gammati \in W^{m+1,p}$, $A \in W^{m,p}$. In \cite{ReintjesTemple_ell2}  we proved that such a solution $(J,\Gammati,A)$ exists for any such connection $\Gamma \in W^{m,p}$ with $d\Gamma \in W^{m,p}$, when $m\geq 1$, $p>n$.    Extending this theory to the case of $\Gamma, d\Gamma \in L^{\infty}$ as well as the case $\Gamma \in L^{2p}, d\Gamma\in L^p$, ($p>n/2$), when the RT-equations only have meaning in a weak sense, is accomplished in the present paper. This was not possible with the methods used in our previous paper \cite{ReintjesTemple_ell2}. The main obstacle is proving an existence theory for the RT-equations with $J\in W^{1,2p}$ at the low level of regularity $\Gamma, d\Gamma \in L^{\infty}$ (or $\Gamma \in L^{2p}, d\Gamma\in L^p$).   The problem is that the iteration scheme in \cite{ReintjesTemple_ell2} does not close because the gradient product 
$dJ^{-1} \wedge dJ$  on the right hand side of equation \eqref{RT_1} fails to stay in a fixed $L^{p}$ space under iteration. Alternatively, trying to construct solutions $J \in W^{1,\infty}$ is problematic as well, because $p=\infty$ is a singular case in elliptic regularity theory, and our iteration scheme in \cite{ReintjesTemple_ell2} would not close in $L^\infty$ for this different reason.  We here extend the existence theory and consequent optimal regularity theory first to the case $\Gamma, d\Gamma \in L^{\infty}$, (in Sections \ref{Sec_existence_theory} - \ref{Sec_Proofs}), and then to the case $\Gamma \in L^{2p}, d\Gamma\in L^p$, (in Section \ref{Sec_Lp-extension}), by a serendipitous modification of the RT-equations.  

In this paper we employ the gauge freedom of the RT-equations to circumvent the problem of incorporating the nonlinear product $dJ^{-1} \wedge dJ$ in \eqref{RT_1} into an iteration scheme which closes in $L^p$ spaces. The idea is to separate this term from the iteration scheme by using the gauge freedom $v$ in the $A$ equation \eqref{RT_4} to consolidate $\tilde{\Gamma}$ and $A$ into a single variable $B$, and thereby uncouple equations \eqref{RT_2} - \eqref{RT_4} for $J$ from equation \eqref{RT_1} for $\tilde{\Gamma}$.  Defining 
\begin{eqnarray} \label{def_B}
B &\equiv& A +\langle d J ; \tilde{\Gamma}\rangle,\\
w&\equiv& v + \delta\overrightarrow{\langle d J ; \tilde{\Gamma}\rangle}, \label{def_w}
\end{eqnarray} 
observe now that we can write \eqref{RT_2} - \eqref{RT_4} as 
\begin{eqnarray} 
\Delta J &=& \delta ( J \mm \Gamma ) - B , \label{RT_withB_2} \\
d \vec{B} &=& \overrightarrow{\text{div}} \big(dJ \wedge \Gamma\big) + \overrightarrow{\text{div}} \big( J\, d\Gamma\big) ,   \label{RT_withB_3} \\
\delta \vec{B} &=& w.  \label{RT_withB_4}
\end{eqnarray}
Equations (\ref{RT_withB_2}) - (\ref{RT_withB_4}) are the reduced RT-equations, c.f. (\ref{RT_withB_2_intro}) - (\ref{RT_withB_4_intro}). Since the transformation from $v$ to $w$ can be viewed as a gauge transformation, the gauge freedom of the RT-equations implies that we can ignore the dependence of $\Gammati$ on $w$, and view $w$ as the independent gauge freedom in the RT-equations; (this is made rigorous in Theorem \ref{Thm_gauge_existence} below).  Therefore equations \eqref{RT_withB_2} - \eqref{RT_withB_4} decouple the equations for $J$ and $B$ from $\tilde{\Gamma}$, and hence from the first RT-equation \eqref{RT_1}, which in terms of $B$ becomes
\beq \label{RT_withB_1}
\Delta \Gammati = \delta d \Gamma - \delta \big( dJ^{-1} \wedge dJ \big) + d\big(J^{-1}(B - \langle d J ; \tilde{\Gamma}\rangle) \big).
\eeq
By this decoupling the Jacobians $J$ which map $\Gamma$ to optimal regularity can be constructed independently of $\Gammati$, (which we can now discard), by solving the reduced RT-equations \eqref{RT_withB_2} - \eqref{RT_withB_4} alone. 

So, discard the original $\Gammati$, and assume $(J,B)$ is a solution of the reduced RT-equations \eqref{RT_withB_2} - \eqref{RT_withB_4} with $J \in W^{1,2p}(\Omega'_x)$ and $B \in L^{p}(\Omega'_x)$. The goal now is to use the solutions $J,B$ of the reduced RT-equations to establish optimal regularity without reference to the original $\Gammati$.   To show such a Jacobian $J$ indeed maps $\Gamma$ to optimal regularity, it suffices to prove that $\Gammati=\Gammati_J$ provides {\it another} solution of (\ref{RT_withB_1}).  In our next theorem we show that, incredibly enough, this is true: 
$\Gammati=\Gammati_J$ is an exact solution of the elliptic equation (\ref{RT_withB_1}), an equation we could not solve by our previous methods at the low regularity $\Gamma,d\Gamma\in L^\infty$.
Equation (\ref{RT_withB_1}) then establishes the requisite smoothness $\Gammati_J\in W^{1,p}$.

To complete the circle, we now explain how to recover a solution of the full RT-equations form \eqref{RT_withB_2} - \eqref{RT_withB_4}, when $\Gammati$ is replaced by $\Gammati_J$.   For this, we need only show that $J,\Gammati_J$ solve the original RT-equations with a different choice of gauge $A',v'.$   Reversing the above steps using $\Gammati_J$ in place of $\Gammati$, it follows that the back change of gauge 
 \begin{eqnarray} \label{A'}
A' &\equiv& B - \langle d J ; \Gammati_J \rangle ,\\\label{v'}
v' &\equiv& w - \delta \overrightarrow{\langle d J ; \Gammati_J \rangle},
\end{eqnarray} 
takes a solution $(J,B)$ of the reduced RT-equations back to a solution of the original RT-equations with the same $J$, but with $\Gammati$ replaced by $\Gammati_J$.  These are recorded in parts  $(i)$ and $(ii)$ of Theorem \ref{Thm_gauge_existence} below, which states that $(J,\Gammati_J,A')$ defined in \eqref{Gammati'} - \eqref{A'} indeed solves the full RT-equations \eqref{RT_1} - \eqref{RT_4}, and, by this, $\Gammati_J$ has the requisite smoothness $\Gammati_J \in W^{1,p}(\Omega)$. Part $(iii)$ of Theorem \ref{Thm_gauge_existence} establishes an estimate for $\Gammati_J$ from which we deduce the uniform $W^{1,p}$-bound \eqref{curvature} on $\Gamma_y$ in Theorem \ref{Thm_Smoothing}, the bound that underlies Uhlenbeck compactness. The existence of solutions $(J,B)$ of the reduced RT-equations, satisfying estimate \eqref{curvature_estimate_soln}, which are assumed in Theorem \ref{Thm_gauge_existence}, are shown to exist in Theorem \ref{Thm_Existence_J} below.  For the low regularities considered in this paper, we need to establish the above equivalence and existence theory in a weak sense, accomplished in Sections \ref{Sec_gauge} - \ref{Sec_Lp-extension}. Serendipitously, the RT-equations allow for a weak formulation because all lowest regularity terms on the right hand side have derivatives $\delta$ or $d$ on them, making them amenable to integration by parts, (as in the theory of conservation laws \cite{Smoller}), c.f. Section \ref{Sec_weak}. 

\begin{Thm} \label{Thm_gauge_existence} 
Assume $\Gamma \in L^{2p}(\Omega_x)$ and $d\Gamma \in L^{p}(\Omega_x)$, in $x$-coordinates, where $p>\max\{ n/2, 2\}$, $p < \infty$, $n\geq 2$. Assume $(J,B)$ solves the reduced RT-equations \eqref{RT_withB_2} - \eqref{RT_withB_4} weakly for some $w$ on an open set $\Omega' \subset \Omega$, such that $J, J^{-1}\, \in W^{1,2p}(\Omega'_x)$ and $B \in L^{p}(\Omega'_x)$. Then the following holds: 

\noindent {\bf (i)} $\Gammati_J$ solves (\ref{RT_withB_1}), where $\Gammati_J$ is defined in \eqref{Gammati'}. The tuple $(J,\Gammati_J,A')$ solves the full RT-equations \eqref{RT_1} - \eqref{RT_4} in $\Omega'$ for $v=v',$ where $A'$ and $v'$ are defined in \eqref{A'} and \eqref{v'}. 

\noindent {\bf (ii)} $\Gammati_J$ is one derivative more regular than the terms constituting $\Gammati_J$ in its defining equation \eqref{Gammati'} are separately, that is, $\Gammati_J\in W^{1,p}(\Omega_x'')$ for any open set $\Omega''$ compactly contained in $\Omega'$.

\noindent {\bf (iii)} Assume the initial bound \eqref{bound_incoming_ass} of Theorem \ref{Thm_Smoothing} holds, i.e., 
\beq \nonumber
\|(\Gamma,d\Gamma) \|_{L^{2p,p}(\Omega)} \equiv \|\Gamma_x \|_{L^{2p}(\Omega_x)} + \|d\Gamma_x \|_{L^p(\Omega_x)} \leq M
\eeq
for some constant $M>0$, and assume that $(J,B)$ satisfies the estimate  
\small
\beq \label{curvature_estimate_soln}
\|I-J\|_{W^{1,2p}(\Omega'_x)} + \|I-J^{-1}\|_{W^{1,2p}(\Omega'_x)} + \|B\|_{L^{p}(\Omega'_x)} 
\ \leq \ C_2(M)\; \|(\Gamma,d\Gamma)\|_{L^{2p,p}(\Omega')} , 
\eeq
\normalsize
for some constant $C_2(M)>0$ depending only on $\Omega'_x, n, p$ and $M$.\footnote{Estimate \eqref{curvature_estimate_soln} bounds $J$ and $J^{-1}$, but is expressed in terms of $I-J$ and $I-J^{-1}$ to reflect the fact that $J$ typically tends to the identity as $M$ tends to zero.}   Then, on any open set $\Omega''$ compactly contained in $\Omega'$, $\Gammati$ satisfies the uniform bound                
\beq  \label{Gammati'_estimate}
\| \Gammati_J \|_{W^{1,p}(\Omega''_x)} \leq  C_3(M) \; \|(\Gamma,d\Gamma)\|_{L^{2p,p}(\Omega')}
\eeq
where $C_3(M)>0$ is some constant depending only on $\Omega''_x, \Omega'_x, n, p$ and $M$.\footnote{Theorem \ref{Thm_gauge_existence} also hold for $\|(\Gamma,d\Gamma)\|_{L^{2p,p}}$ replaced everywhere by $\|(\Gamma,d\Gamma)\|_{L^{\infty}}$, essentially since $\Gamma, d\Gamma \in L^{\infty}$ implies $\Gamma \in L^{2p}$ and $d\Gamma\in L^p$ for any $p<\infty$ by boundedness of $\Omega$.}  
\end{Thm}

The key step in the proof of Theorem \ref{Thm_gauge_existence} is establish in Lemma \ref{Thm_gauge_informal} below, by proving that \eqref{Gammati'} is an exact formula for the solution $\Gammati = \Gammati_J$ of the first RT-equation \eqref{RT_1}, from which the regularity gain of $\Gammati_J$ in $(ii)$ can be deduced. To give the argument in its essence, we assume one more level of smoothness in Lemma \ref{Thm_gauge_informal}. More care is required to extend the argument of Lemma \ref{Thm_gauge_informal} to the low regularities of Theorem \ref{Thm_gauge_existence} and prove the theorem rigorously, which is the subject of Section \ref{Sec_gauge}.  Assuming only that $(ii)$ of Theorem \ref{Thm_gauge_existence} holds, the equivalence of optimal regularity and the reduced RT-equations, in the spirit of our previous paper \cite{ReintjesTemple_ell1}, can now be established as a corollary. This reduces the problem of optimal regularity to an existence theorem for the reduced RT-equations. \footnote{Although equation \eqref{RT_1} can be bypassed for constructing solutions, equation \eqref{RT_1} is required to prove optimal connection regularity in the coordinate system introduced by $J$ and is therefore a vital part of the RT-equations. Note also that one can use an underlying Cauchy-Riemann-type equation for $\Gammati$ instead of \eqref{RT_1} and establish optimal regularity by applying Gaffney's inequality, but we prefer the Poisson type equation \eqref{RT_1}.} 

\begin{Corollary} \label{Thm_equivalence}
Assume $\Gamma \in L^{2p}(\Omega_x)$ and $d\Gamma \in L^{p}(\Omega_x)$ in $x$-coordinates, $p>\max\{ n/2, 2\}$, $p < \infty$, $n\geq 2$. Then for any $q\in \Omega$ there exists a neighborhood $\tilde{\Omega} \subset \Omega$ of $q$ and a coordinate transformation $x \to y$ such that the connection components $\Gamma_y$ in $y$-coordinates have optimal regularity $\Gamma_y \in W^{1,p}(\tilde{\Omega}_y)$ if and only if there exists a weak solution $(J,B)$ of the reduced RT-equations \eqref{RT_withB_2} - \eqref{RT_withB_4}, defined on some neighborhood $\Omega'$ of $q$, with $J, J^{-1}\, \in W^{1,2p}(\Omega'_x)$, $B \in L^{p}(\Omega'_x)$ and $d\vec{J}=0$ in $\Omega_x'$. The Jacobian of the coordinate transformation $x \to y$ is $dy=J \in W^{1,2p}(\Omega'_x)$.
\end{Corollary}       

\Proof 
The forward implication of Corollary \ref{Thm_equivalence} is straightforward because the reduced RT-equations are derived from the full RT-equations which are in turn deduced from the Riemann-flat condition, a condition equivalent to optimal regularity, c.f. \cite{ReintjesTemple_ell1}. That is, given the Jacobian $J$ and resulting connection $\Gamma_y$ of optimal regularity, and defining
\beq \nonumber
\Gammati^k_{ij} = (J^{-1})^k_\gamma J^\alpha_i J^\beta_j (\Gamma_y)^\gamma_{\alpha\beta},
\eeq
then $\Gammati$ satisfies the Riemann-flat condition. It is now straightforward to check that $(\Gammati,J)$ solves the RT-equation for some $A$, and, defining $B$ by \eqref{def_B}, that $(J,B)$ solves the reduced RT-equations \eqref{RT_withB_2} - \eqref{RT_withB_4}. Recall that $d\vec{J}=0$ is equivalent to $J$ being the Jacobian of a coordinate transformation, c.f. Theorem \ref{Thm_CauchyRiemann}. 

To prove the reverse implication we apply Theorem \ref{Thm_gauge_existence}, which we assume to be valid here. By part $(ii)$ of  Theorem \ref{Thm_gauge_existence}, $\Gammati_J$ defined by \eqref{Gammati'} is in $W^{1,p}(\Omega'_x)$ and $\Gammati_J$ solves the first RT-equation \eqref{RT_withB_1} in terms of the solution $(J,B)$ of the reduced RT-equations \eqref{RT_withB_2} - \eqref{RT_withB_4}. Let $x \to y$ be a coordinate transformation with Jacobian $dy=J$, which exists since we assumed that $d\vec{J}=0$ in $\Omega_x'$. Now define the connection $\Gamma_y$ in terms of $\Gammati_J$ in $x$-coordinates by \eqref{Gamma_y}, that is,
\beq \label{Gamma_y_reverse}
(\Gamma_y)^\gamma_{\alpha\beta} = J_k^\gamma (J^{-1})^i_\alpha  (J^{-1})^j_\beta   (\Gammati_J)^k_{ij}.
\eeq
Since $\Gammati_J \in W^{1,p}(\Omega'_x)$ and $J, J^{-1} \in W^{1,2p}(\Omega'_x)$, $p>n/2$, it follows that $\Gamma_y(x) \in W^{1,p}(\Omega'_x)$ in $x$-coordinates, and therefore also when expressed in $y$-coordinates $\Gamma_y \in W^{1,p}(\Omega'_y)$, as can be shown using Morrey's and H\"older's inequalities in combination with Sobolev embedding, (see Section \ref{Sec_Proof_optimal_regularity} for details). Substituting the definition of $\Gammati_J$ in \eqref{Gammati'} into \eqref{Gamma_y_reverse} implies that 
\begin{eqnarray} \nonumber
(\Gamma_y)^\gamma_{\alpha\beta} 
&=& J_k^\gamma (J^{-1})^i_\alpha  (J^{-1})^j_\beta   \big(\Gamma_x - J^{-1} dJ\big)^k_{ij}  \cr
&=& J_k^\gamma (J^{-1})^i_\alpha  (J^{-1})^j_\beta (\Gamma_x)^k_{ij}  - (J^{-1})^i_\alpha  (J^{-1})^j_\beta \partial_i J^\gamma_j ,
\end{eqnarray}
from which we conclude that $\Gamma_y$ are the connection components $\Gamma_x$ transformed to a coordinate system $y$ in which $\Gamma$ exhibits optimal regularity, $\Gamma_y \in W^{1,p}(\Omega)$. This completes the proof.
\QED

Finally, to obtain the optimal regularity result stated in Theorem \ref{Thm_Smoothing}, together with the uniform estimate \eqref{curvature}, we require the following theorem which establishes the existence of solutions to the reduced RT-equations satisfying the assumptions $J, \in W^{1,2p}(\Omega'_x)$, $B \in L^{p}(\Omega'_x)$ of Theorem \ref{Thm_gauge_existence}, together with the additional estimate \eqref{curvature_estimate_soln}. (The existence theory is worked out in fixed $x$-coordinates, so we omit subscript $x$ on $\Gamma$ and $\Omega$.)

\begin{Thm} \label{Thm_Existence_J}
Assume $\Gamma \in L^{2p}(\Omega)$ and $d\Gamma \in L^{p}(\Omega)$ in $x$-coordinates, where $ n/2 <p < \infty$, $n\geq 2$, and assume the initial bound \eqref{bound_incoming_ass} holds for some constant $M>0$. Then for any $q\in \Omega$ there exists a neighborhood $\Omega' \subset \Omega$ of $q$, and there exists $J \in W^{1,2p}(\Omega')$ and $B \in L^{p}(\Omega')$ such that $(J,B)$ solves the reduced RT-equations \eqref{RT_withB_2} - \eqref{RT_withB_4} in a weak sense and satisfies the uniform bound \eqref{curvature_estimate_soln}. Moreover, $J$ is invertible with $J^{-1} \in W^{1,2p}(\Omega')$ and integrable to coordinates ($d\vec{J}=0$ in $\Omega'$). One can take $\Omega'_x = \Omega_x \cap B_r(q)$, for $B_r(q)$ the Euclidean ball of radius $r$ in $x$-coordinates, where $r$ depends only on $\Omega_x, p, n$ and $\Gamma$ near $q$;  if $\|(\Gamma,d\Gamma)\|_{L^{\infty,2p}(\Omega)}~\leq~M$, then $r$ depends only on $\Omega_x, p, n$ and $M$.
\end{Thm}                                         

Theorem \ref{Thm_Smoothing}, our main result on optimal regularity, is now a rather direct consequence of Theorem \ref{Thm_Existence_J} in combination with Corollary \ref{Thm_equivalence} and estimate \eqref{Gammati'_estimate} of Theorem \ref{Thm_gauge_existence}. This is shown in detail in Section \ref{Sec_Proof_optimal_regularity} below. The proof of Theorems \ref{Thm_gauge_existence} and \ref{Thm_Existence_J} is the main technical effort in this paper. The proof of Theorem \ref{Thm_gauge_existence} is given in Sections \ref{Sec_Proof1} and \ref{Sec_gauge}. The proof of Theorem \ref{Thm_Existence_J} is established for connections $\Gamma, d\Gamma \in L^{\infty}(\Omega)$ in Sections \ref{Sec_existence_theory} and \ref{Sec_Proofs}, and extended to connections $\Gamma \in L^{2p}(\Omega)$, $d\Gamma \in L^{p}(\Omega)$ in Section \ref{Sec_Lp-extension}. Note that the boundary data \eqref{RT_data} is ill-defined at the low regularity $J \in W^{1,2p}$. So to make sense of this, we augment the reduced RT-equations with auxiliary elliptic PDE's for $y$, which allows us to replace \eqref{RT_data} by Dirichlet data for $J=dy$, data one degree more regular than \eqref{RT_data} and thus well-defined.  To summarize, the RT-equations reduce the nonlinear problem of regularizing connections to a linear existence problem for constructing the Jacobian $J$ via the reduced RT-equations, followed by a regularity boost for $\Gammati_J$ provided by the non-linear first RT-equation \eqref{RT_1}. So for applications it suffices to solve the linear reduced RT-equations for $J$ to obtain the regularizing coordinate transformation, and our iteration scheme in Section \ref{Sec_existence_theory} provides an algorithm for doing this.

\section{How to recover the full from the reduced RT-equations by gauge transformation} \label{Sec_Proof1}

\subsection{Conceptual overview}

We start by describing, more carefully, the logical connection between the full RT-equations (\ref{RT_1}) - (\ref{RT_4}) and the reduced RT-equations (\ref{RT_withB_2}) - (\ref{RT_withB_4}).  Recall from Section \ref{Sec_reducedRT} that the original RT-equations were derived by constructing the Laplacian $d\delta+\delta d$ starting from two equivalent formulations of the Riemann-flat condition, one involving $dJ$ and one involving $d\Gammati;$ and the first order $A$ equation came by replacing $A=J\delta\Gammati$ in the $J$ equation, setting $d$ of the right hand side equal to zero, and imposing $dJ=0$ on the boundary in (\ref{RT_data}).  Now in the existence theory set out in \cite{ReintjesTemple_ell2} for the case $\Gamma,d\Gamma\in W^{1,p}$, we saw that not every solution of the original RT-equations determines a solution in which $J$ is paired with $\Gammati=\Gammati_J\equiv\Gamma-J^{-1}dJ$ given in (\ref{Gammati'}).  Here $\Gammati_J$ is the tensor transformation (\ref{Gamma_y}) of the connection coefficients obtained by transforming the original $\Gamma$ by $J$.   To complete the argument in \cite{ReintjesTemple_ell1}, we proved that given a solution $(\Gammati,J,A,v)$ of the full RT-equations, $\Gammati_J$ will solve a modified version of the first RT-equation (\ref{RT_1}).   In \cite{ReintjesTemple_ell1}, the role of this modified elliptic equation was to establish that $\Gammati_J$, (and hence also $\Gamma_y$ by (\ref{Gamma_y})), is of optimal regularity.  This was established rigorously in \cite{ReintjesTemple_ell2} at the smoothness level $\Gamma,d\Gamma\in W^{1,p}$. 

We now understand this more conceptually as follows.  The variables for the original RT-equations are $(\Gammati,J,A,v)$.   The transformation $(\Gammati,J,A,v)\to (\Gammati,J,B,w)$ effected by the change of variables (\ref{def_B}) - (\ref{def_w}) given by
\begin{eqnarray} \nonumber
B &=& A - \langle d J ; \Gammati \rangle ,  \cr
w &=& v - \delta \langle d J ; \Gammati_J \rangle,
\end{eqnarray}
transforms the last three RT-equations (\ref{RT_2})-(\ref{RT_4}) into the reduced RT-equations (\ref{RT_withB_2})-(\ref{RT_withB_4}), and transforms the first RT-equation (\ref{RT_1}) into equation (\ref{RT_withB_1}), an elliptic equation for $\Gammati$ involving $(\Gammati,J,B)$ on the right hand side. Thus the original four RT-equations (\ref{RT_1})-(\ref{RT_4}) are equivalent to the three reduced RT-equations (\ref{RT_withB_2})-(\ref{RT_withB_4}) together with (\ref{RT_withB_1}). Now the rather remarkable discovery, which is the basis for the present paper, is that $\Gammati=\Gammati_J$ turns out to exactly solve equation (\ref{RT_withB_1}), but only on solutions $(J,B)$ of the reduced RT-equations.  That is,  $\Gammati_J$ does not in general solve the first RT-equation (\ref{RT_1}), but the transformation (\ref{def_B}) - (\ref{def_w}), which uncouples $\Gammati$ from the last three equations, also produces the elliptic equation (\ref{RT_withB_1}) satisfied by $\Gammati_J.$    The result then, is that we no longer need the original RT-equations, because optimal regularity is determined entirely from the reduced RT-equations (\ref{RT_withB_2}) - (\ref{RT_withB_4}) for $(J,B),$ together with the elliptic equation (\ref{RT_withB_1}) for the gain in regularity of $\Gammati_J$.   At the end, the original $\Gammati$ is out of the picture.    To borrow words from Ludwig Wittgenstein (regarding his private language argument), the original RT-equations are a ``ladder we climb'' to obtain the reduced RT-equations within the gauge freedom of the original RT-equations,  but that ladder can then be thrown away once we find them.  But still, to complete the picture, it is interesting to understand the sense in which solutions $(\Gammati_J,J,B,w)$ of the reduced RT-equations correspond to solutions $(\Gammati,J,A,v)$ of the original RT-equations.

To clarify this, recall that if we are given a coordinate transformation $x\to y$ and $J$ is its Jacobian, applying the tensor transformation law  (\ref{Gamma_y}) to $\Gammati_J$ produces the coefficients of the connection $\Gamma_y$ in $y$ coordinates.   Now we know from \cite{ReintjesTemple_ell2} that the solution space of the original RT-equations (\ref{RT_1})-(\ref{RT_data}) is larger than we want, because it contains solutions $(\Gammati,J,A,v)$ for which $\Gammati\neq\Gammati_J$.  That is, $\Gammati$ need not have anything to do with the transformation of our starting non-optimal connection $\Gamma.$  In fact, we have discovered that there exist solutions of the RT-equations with the same $J$, but different $(\Gammati,A,v).$  Define then the equivalence class ${\mathcal E}(J)$ of all solutions of the original  RT-equations (\ref{RT_1})-(\ref{RT_data}) which share the same Jacobian field $J$.  (It could be empty).   Recall that only the $J$-equation (\ref{RT_2}) comes with a boundary condition (\ref{RT_data}), so the ``gauge freedom'' in the RT-equations is the freedom to choose the free function $v$, and the freedom to impose boundary conditions for $\Gammati$ and $A$ in equations (\ref{RT_1}) and (\ref{RT_3}), (\ref{RT_4}), respectively.   Thus the equivalence class ${\mathcal E}(J)$ associated with a given Jacobian $J$ is the set of all solutions $(\Gammati,J,A,v)$ of the RT-equations (\ref{RT_1})-(\ref{RT_data}), solutions determined by $v$ and all the possible boundary conditions we might impose to determine $\Gammati$ and $A$ from equations (\ref{RT_1}) and  (\ref{RT_3}), (\ref{RT_4}).    Now once we have established that $\Gammati_J$ solves (\ref{RT_withB_1}), reversing the argument leading to (\ref{def_B}) - (\ref{def_w}) in Section \ref{Sec_reducedRT}, (which entails simply replacing $\Gammati$ by $\Gammati_J$ in the formulas for $B$ and $w$), shows directly that $(\Gammati_J,J,A',v')$ will solve the original original RT-equations whenever $(\Gammati,J,A,v)$ does,  where
\begin{eqnarray} \nonumber
A' & = & B - \langle d J ; \Gammati_J \rangle ,  \cr
v' &= & w - \delta \langle d J ; \Gammati_J \rangle.
\end{eqnarray}
Thus conceptually,  starting with a solution $(\Gammati,J,A,v)$ of the original RT-equations, $(J,B)$ will solve the reduced RT-equations, $\Gammati_J$ will solve the modified first RT-equation (\ref{RT_withB_1}), and the back transformation (\ref{A'}), (\ref{v'}) determines a new solution of the original RT-equations within the equivalence class ${\mathcal E}(J)$, this new solution having as its first component $\Gammati=\Gammati_J$.  That is, through a change of gauge, we can substitute $\Gammati$ for $\Gammati_J$ in any solution of the original RT-equations.  This gives expression to the content of what is claimed in (i),(ii) of Theorem \ref{Thm_gauge_existence}.    In summary, we write this as a direct corollary of Theorem \ref{Thm_gauge_existence}:
\begin{Corollary}\label{CorEquiv}
If $(\Gammati,J,A,v)$ lies within the equivalence class ${\mathcal E}(J)$ of the RT-equations, then $(\Gammati_J,J,A',v')\in{\mathcal E}(J)$ as well. 
\end{Corollary} 
That $\Gammati_J$ given by (\ref{Gammati'}) actually solves (\ref{RT_withB_1}),  the first RT-equation (\ref{RT_1}) modified by the substitution (\ref{def_B})), on solutions of the reduced RT-equations, is crucial because it is this equation which gives the requisite optimal regularity of $\Gammati_J$, the connection coefficients of $\Gamma$ transformed by $J$.   As laid out above, it really is quite remarkable that a change of gauge simultaneously eliminates $\Gammati$ from the last three RT-equations, and then on top of that, transforms the first RT-equations into a new elliptic equation satisfied by $\Gammati_J$. So as a preliminary to the proof of Theorem \ref{Thm_gauge_existence},  we now go through the key idea in the proof of parts (i) and (ii) leading to $\Gamma_J$ being a solution of (\ref{RT_withB_1}), in the case when $\Gamma$ is smooth, thereby displaying how it works without being distracted by the weak formulation of the equations.

\subsection{Recovering solutions of the RT-equations from the reduced RT-equations} 

A complete proof of Theorem \ref{Thm_gauge_existence} is the subject of Section \ref{Sec_gauge}. As a preliminary, we explain in this section the key step in the equivalence between the reduced and full RT-equations in its simplest setting, when $\Gamma$ is a smooth connection. That is, we explain why $\Gammati = \Gammati_J \equiv\Gamma - J^{-1} dJ$ automatically solves the first RT-equation when defined in terms of a solution $(J,B)$ of the reduced RT-equations. On the face of it, this is quite remarkable, because the $J$-equation appears to have lost all knowledge of $\Gammati$ once we gauge transform from $A$ to $B$.  The point, which we establish here, is that when we take the $B$ gauge, the formula \eqref{Gammati'}  for  $\Gammati_J$ in terms of $J$ and $\Gamma$ alone gives an exact solution of the first RT-equation \eqref{RT_1},  provided $J$ solves the reduced RT-equations. Thus the regularity of $\Gammati_J$ is determined by the first RT-equation, but is ultimately encoded in the reduced RT-equations. 

\begin{Lemma} \label{Thm_gauge_informal} 
Let $\Gamma$ be smooth and assume $(J,B)$ is a smooth solution of the reduced RT-equations \eqref{RT_withB_2} - \eqref{RT_withB_4} for some given $w$, such that $J$ is invertible. Then $\Gammati=\Gammati_J$ and $A=A'$, defined in \eqref{Gammati'} and \eqref{A'}, satisfy the first RT-equation \eqref{RT_1}.
\end{Lemma} 

\Proof
To prove Lemma \ref{Thm_gauge_informal}, we first take the exterior derivative $d$ of equation $\Gammati_J = \Gamma - J^{-1} dJ$, c.f. \eqref{Gammati'}, to obtain 
\begin{eqnarray} \label{Proof_Thm_informal_eqn1}
d \Gammati_J &=& d\Gamma -d\big( J^{-1}dJ \big) \cr
&=& d\Gamma - d\big(J^{-1}\big) \wedge dJ,
\end{eqnarray}
where we use $d\big( J^{-1} dJ \big) = d\big(J^{-1}\big) \wedge dJ$ by the Leibniz rule \eqref{Leibniz_rule_J-application}. Taking now the co-derivative $\delta$ of \eqref{Proof_Thm_informal_eqn1} gives
\beq \label{Proof_Thm_informal_eqn2}
\delta d \Gammati_J = \delta d \Gamma - \delta \big( dJ^{-1} \wedge dJ \big) ,
\eeq
thus giving the first term of the Laplacian $\Delta \Gammati_J \equiv \delta d \Gammati_J +d \delta \Gammati_J$.  

To determine the second term of $\Delta\Gammati_J$, we take $\delta$ of equation $\Gammati_J = \Gamma - J^{-1} dJ$, c.f. \eqref{Gammati'},  to compute
\begin{eqnarray} \label{Proof_Thm_informal_eqn3a}
\delta \Gammati_J 
= \delta \Gamma - \delta \big( J^{-1} dJ \big).  
\end{eqnarray}
Using now the Leibniz rule for co-derivatives \eqref{ Leibniz-rule-delta} we have
\begin{eqnarray} \label{Proof_Thm_informal_eqn3}
\delta \Gammati_J 
&=& \delta \Gamma  -  \langle d (J^{-1}) ;  dJ \rangle   -  J^{-1} \delta d J   \cr
&=& \delta \Gamma  -  \langle d (J^{-1}) ;  dJ \rangle   -  J^{-1} \Delta J,  
\end{eqnarray}
since $\Delta J = \delta d J$ by $\delta J=0$, because the co-derivative $\delta$ vanishes on $0$-forms. Substituting now the reduced RT-equation \eqref{RT_withB_2} for $\Delta J$ into \eqref{Proof_Thm_informal_eqn3} gives
\begin{eqnarray}  \label{Proof_Thm_informal_eqn4a}
\delta \Gammati_J  
&=& \delta \Gamma  -  \langle d (J^{-1}) ;  dJ \rangle   -  J^{-1} \big( \delta ( J \mm \Gamma ) -  B  \big) ,
\end{eqnarray}
and since $\delta ( J \mm \Gamma ) = J \delta \Gamma + \langle d J ; \Gamma \rangle$ by \eqref{ Leibniz-rule-delta}, we obtain that
\begin{eqnarray} \label{Proof_Thm_informal_eqn4}
\delta \Gammati_J &=&   J^{-1}  B  - \langle d (J^{-1}); dJ \rangle -  J^{-1} \langle d J ; \Gamma \rangle  .
\end{eqnarray} 
The cancellation of the lowest regularity term $\delta\Gamma$ in the step from \eqref{Proof_Thm_informal_eqn4a} to \eqref{Proof_Thm_informal_eqn4} is the essence of the gain of regularity implied by the RT-equations, c.f. Theorem \ref{Thm_gauge_existence}. (To establish this cancellation for weak solutions takes some work, see Section \ref{Sec_gauge_lowreg}.) Now the right hand side of \eqref{Proof_Thm_informal_eqn4} is equal to $J^{-1} A$ as a consequence of the definition of $A$ in \eqref{A'}, as proven in Lemma \ref{Lemma_first} below. So assuming Lemma \ref{Lemma_first} to be true for the moment, we find equation \eqref{Proof_Thm_informal_eqn4} to be identical to
\begin{eqnarray} \label{Proof_Thm_informal_eqn5}
\delta \tilde{\Gamma}    &=&   J^{-1}  A.
\end{eqnarray} 
Taking now the exterior derivative $d$ of \eqref{Proof_Thm_informal_eqn5} and adding the resulting equation to \eqref{Proof_Thm_informal_eqn2} gives us
\begin{eqnarray} \label{Proof_Thm_informal_eqn8}
\Delta \Gammati_J  
&=&   \delta d \Gamma - \delta \big( dJ^{-1} \wedge dJ \big) + d\big( J^{-1}A \big)  
\end{eqnarray}
which proves that $\Gammati_J$ solves the sought after equation \eqref{RT_1}. 
\QED

To complete the proof of Lemma \ref{Thm_gauge_informal}, it remains to prove Lemma \ref{Lemma_first}.

\begin{Lemma} \label{Lemma_first}
For $A$ defined in \eqref{A'}, we have
\beq \label{Lemma_first_eqn}
A = B  - J \langle d (J^{-1}); dJ \rangle -  \langle d J ; \Gamma \rangle,
\eeq
so the right hand side of \eqref{Proof_Thm_informal_eqn4} is equal to $J^{-1} A$.
\end{Lemma} 

\Proof
To verify \eqref{Lemma_first_eqn}, we substitute $\Gammati_J = \Gamma - J^{-1} dJ$, c.f. \eqref{Gammati'}, into the definition $A= B - \big\langle dJ ; \Gammati_J \big\rangle$, c.f. \eqref{A'}, to compute 
\begin{eqnarray} \label{Proof_Thm_informal_eqn7}
A &=&  B - \big\langle dJ ; (\Gamma - J^{-1} dJ ) \big\rangle  \cr 
&=&   B+ \langle dJ ; J^{-1} dJ\rangle - \langle d J ; \Gamma \rangle .
\end{eqnarray} 
We now use the multiplication property \eqref{inner-product_muliplications} of the matrix valued inner product $\langle \cdot \: ; \cdot \rangle$ twice, to write the second term in \eqref{Proof_Thm_informal_eqn7} as
\begin{eqnarray} \label{Proof_Thm_informal_eqn6}
\langle dJ ; J^{-1} dJ \rangle
 &=& \langle  dJ \mm J^{-1}; dJ \rangle  \cr
 &=&  J  \langle J^{-1} dJ \mm J^{-1}; dJ \rangle \cr
&=&  - J  \langle d (J^{-1}); dJ \rangle ,
\end{eqnarray}
where the last equality holds since $d(J^{-1}) = - J^{-1} \, dJ \cdot  J^{-1}$ by the Leibniz rule for gradients, (since $J$ is a $0$-form so $d$ is the gradient). Substituting \eqref{Proof_Thm_informal_eqn6} into \eqref{Proof_Thm_informal_eqn7} gives the sought after identity \eqref{Lemma_first_eqn} and proves Lemma \ref{Lemma_first}. 
\QED

\noindent This completes the proof of Lemma \ref{Thm_gauge_informal}, the case of smooth solutions. Accomplishing this for weak solutions is subject of Section \ref{Sec_gauge}.

\section{Weak formulation of the RT-equations} \label{Sec_weak}

We now begin the existence theory for weak solutions of the reduced RT-equations (\ref{RT_withB_2})-(\ref{RT_withB_4}) on bounded domains $\Omega \subset \mathbb{R}^n$.   This provides an existence theory for weak solutions of the full RT-equations (\ref{RT_1})-(\ref{RT_data}) by using the change of gauge $(B,w)\to(A',v')$ given in (\ref{A'}) - (\ref{v'}).  

The RT-equations are a nonlinear elliptic system of equations in unknowns $J,A$ and $\Gammati$ determined by the assumed given connection $\Gamma$, and they allow for the freedom to choose the arbitrary function $v$ in the second $A$-equation (\ref{RT_4}), together with boundary conditions for $\Gammati$ and $A$ in (\ref{RT_1}) and (\ref{RT_3}), (\ref{RT_4}), within the appropriate regularity class. We refer to this as the gauge freedom in the full RT-equations, and loosely refer to $v$ or $A$ as the choice of gauge.  Note first that because the right hand sides involve the derivatives $D\Gamma$ of $\Gamma$,  the RT-equations are consistent with the regularity $J,\Gammati$ being one order more regular than $\Gamma, d\Gamma$ because this puts the right hand sides of the $\Gammati$ and $J$ equations (\ref{RT_2}) and (\ref{RT_1}) at the same regularity as $D\Gamma$, so long as $A$ has the same regularity as $\Gamma$;  and this is consistent with the regularity of the right hand side of the $A$ equations (\ref{RT_3}) - (\ref{RT_4}) being one order less regular than $\Gamma$.   Thus, the RT-equations are consistent with elliptic PDE theory in the sense that Laplacians in (\ref{RT_1}) - (\ref{RT_2}) should raise the regularity of $\Gammati$ and $J$ two orders above the right hand side, which is one order above the regularity of $\Gamma$; and the first order Cauchy Riemann system (\ref{RT_3}) - (\ref{RT_4}) should raise the regularity of $A$ to the same regularity as $\Gamma$.  In \cite{ReintjesTemple_ell2}  authors proved that this consistency of the RT-equations is correct in the classical sense by giving an existence theory for the full RT-equations in the case $\Gamma,d\Gamma\in W^{1,p}$,  and in Sections \ref{Sec_gauge} - \ref{Sec_Lp-extension} we extend this classical theory to the case of weak solutions, when $\Gamma, d\Gamma\in L^{\infty}.$   We assume $\Gamma \in L^{2p}$ with $d\Gamma\in L^{p}$, for $p>n/2$, $p<\infty$, and our goal is to establish existence of weak solutions $J,\Gammati\in W^{1,p},$ $A\in L^p,$ on sufficiently small domains.\footnote{Recall that $L^\infty(\Omega) \subset L^p(\Omega)$ for any $p\leq \infty$, since $\Omega$ is be bounded here, but standard elliptic theory does not suffice to give optimal regularity in $W^{1,\infty}$ even when assuming $\Gamma, d\Gamma \in L^\infty$, c.f. Theorem \ref{Thm_Smoothing}'.}  Note that in this case, the right hand sides of the RT-equations  (\ref{RT_1}) - (\ref{RT_4}) are at the regularity $W^{-1,p}$, a regularity too weak for classical solutions.  So an existence theory requires a weak formulation of the RT-equations.         

The RT-equations do indeed admit a weak formulation because all of the lowest order terms on the right hand side of (\ref{RT_1})-(\ref{RT_4}) are matrix valued differential forms with ``geometric'' total derivatives ($d,\delta$ or $\overrightarrow{div}$) operating on them, so integration by parts will raise the regularities by one order.  To accomplish the integration by parts and express a rigorous weak formulation, we need to introduce a suitable inner product together with adjoint operators associated with $d,\delta$ or $\overrightarrow{div}$ on the right hand side of the RT-equations.   This is all accomplished in this section. It is interesting at this point to comment that it is well known that weak formulations are not always equivalent.  Our choice of the weak formulation is based on the geometric operators $d,\delta$ and $\overrightarrow{div}$ which appear on the right hand side of our formulation of the reduced RT-equations.   The idea to base the RT-equation and their weak formulation on these geometric operators instead of on pure partial derivatives separately, is guided by the requirement that Jacobians $J$ which solve the reduced RT-equations in this weak sense are indeed integrable to coordinates. 

The existence theory for weak solutions of the RT-equations is accomplished in Sections \ref{Sec_gauge} - \ref{Sec_Lp-extension}.   A few preliminary comments are in order.  First, the iteration scheme used in \cite{ReintjesTemple_ell2} only closes in $L^p$ spaces for classical solutions, because of the bad nonlinear term $dJ^{-1} \wedge dJ$  on the right hand side of the $\Gammati$ equation (\ref{RT_1}).    The problem is that products of functions in $L^p$ are not in $L^p$, so the iteration scheme does not close in any $L^p$ space, (working alternatively with $L^\infty$ is of no use, since the Laplacian does not lift $L^\infty$ to $C^{1,1}$).   We overcome this problem by showing that for solutions $(\Gammati,J,A,v)$ with $v\in W^{-1,p}$, the change of gauge $(A,v)\to(B,w)$, given in (\ref{def_B}) - (\ref{def_w}), uncouples the $(J,B)$ equation from the $\Gammati$ equation which contains the bad nonlinear term.  We named the resulting  system (\ref{RT_withB_2}) - (\ref{RT_withB_4}) in $(J,B)$ the reduced RT-equations, a system of {\it linear} elliptic equations. We prove in Section \ref{Sec_existence_theory} that our iteration scheme, modified to the weak formulation of the equations, does converge when $\Gamma,d\Gamma\in L^\infty$, for sufficiently small bounded sets $\Omega$, and in Section \ref{Sec_Lp-extension} we extend this result to the low regularity $\Gamma \in L^{2p}$ with $d\Gamma\in L^{p}$, $p>n/2$.  Even though the reduced RT-equations are linear, it is a system and the coefficients are at critical low regularity, so an iteration scheme is still required to handle the lower order terms. The smallness of the neighborhood $\Omega$ is used to rule out complications with the Fredholm alternative.  Once we obtain a solution $(J,B)$ of the reduced RT-equations, we then prove that (\ref{Gammati'}) provides an exact solution $\Gammati=\Gammati_J$ of the  first RT-equation (\ref{RT_1}) (with $B$ substituted for $A$), and by this we obtain the desired regularity of the transformed connection coefficients $\Gammati_J$ from the now classical linear $L^p$ theory of the Laplacian.   In this way the bad nonlinear product in (\ref{RT_1}) can be handled by simply using solutions of the reduced RT-equations with larger $p$, $J\in W^{1,2p})$, so the bad nonlinear term $dJ^{-1} \wedge dJ$  on the right hand side of (\ref{RT_1}) is in $W^{-1,p}$, thereby placing the solution $\Gammati=\Gammati_J\in W^{1,p}.$  Once we have a solution to the RT-equation in the $(B,w)$ gauge, we no longer require a solution of the original RT-equations, but to complete the circle we show in Section \ref{Sec_gauge} that the transformation back to $(A',v')$ in (\ref{A'}), (\ref{v'}) provides a weak solution in the original gauge, thereby demonstrating the consistency of the whole theory for every gauge.  

We finish this introduction of the existence theory to follow, by pointing out some of the obstacles our theory faces in the weak formulation required for the regularities here. One central step in the argument is to prove that a weak solution of the $B$ equation really does impose the integrability equation $dJ=0$. That is, the boundary condition $dJ=0$ (\ref{RT_data}), is not a classical Dirichlet boundary condition, and when $J\in W^{1,p}$, $dJ=0$ is too weak to impose on a boundary.  Fortunately, the way we handled this boundary condition in the iteration scheme introduced in \cite{ReintjesTemple_ell2}, can be modified to the weak setting.   The idea is to introduce an auxiliary elliptic equation for $y$ satisfying $dy=J$ in the iteration.    Then we can use Dirichlet boundary conditions for $J$ which make sense at this low regularity, and thereby obtain the integrability from $dy=J$ which implies $dJ=0$. This provides a very clean way to handle the boundary condition since we can then apply classical linear $L^p$-elliptic theory for the Dirichlet problem at each stage of the iteration. However, for the low regularity considered here, this procedure is more technical because it involves two different version of the weak Laplacian combined with operations on the Cartan algebra of differential forms. This is accomplished in Sections \ref{Sec_existence_theory} - \ref{Sec_Proofs}. Finally, the proof that $\Gammati_J$ solves the first RT-equation in a weak sense is more involved, because the weak reduced RT-equation for $J$ cannot be used in a straightforward way to achieve the cancellation of the lowest regularity term $\delta \Gamma$ in equation \eqref{Proof_Thm_informal_eqn4}. This is achieved in Section \ref{Sec_gauge_lowreg}.

\subsection{Integration by parts for matrix valued differential forms}

To introduce the weak formulation of the RT-equations, we first define the following inner products over matrix and vector valued differential forms. On matrix valued $k$-forms $A$ and $B$, we define the point-wise inner product
\begin{eqnarray} \label{inner-product_pointwise}
\langle A, B\rangle 
&\equiv & \tr \langle A ; B^T \rangle   \cr
&\overset{\eqref{def_inner-product}}{=}&  \sum_{\nu, \sigma=1}^n  \sum_{i_1<...<i_k} A^\nu_{\sigma\: i_1...i_k} B^\nu_{\sigma\: i_1...i_k}, 
\end{eqnarray}
where the matrix valued inner product $\langle A ; B \rangle$ is defined in \eqref{def_inner-product}.  We further introduce the $L^2$-inner product                    
\begin{eqnarray} \label{inner-product_L2}
\langle A, B\rangle_{L^2} 
&\equiv & \int_\Omega \langle A , B\rangle dx
\end{eqnarray}
where $dx$ denotes the Lebesgue measure in $\R^n$. For vector valued $k$-forms $v$ and $w$ we define the point-wise inner product
\begin{eqnarray} \label{inner-product_pointwise_vec}
\langle v, w\rangle &\equiv &  \sum_{\alpha =1}^n  \sum_{\; i_1<...<i_k} v^\alpha_{\; i_1...i_k} w^\alpha_{i_1...i_k},   
\end{eqnarray}
in terms of the Euclidean inner product, and we introduce the $L^2$-inner product by
\begin{eqnarray} \label{inner-product_L2_vec}
\langle v, w\rangle_{L^2} 
&\equiv & \int_\Omega \langle v , w\rangle dx .
\end{eqnarray}
The inner products on matrix valued $0$-forms and vector valued $1$-forms are in fact identical, 
\beq \label{tracestuff}
\langle A, B\rangle = \langle \vec{A}, \vec{B}\rangle,
\eeq
where $A$ and $B$ are matrix valued $0$-forms. For $k\geq 1$, vectorization of matrix valued $k$-forms generally results in a loss of information due to cancellation of symmetries, and one can not expect \eqref{tracestuff} to be valid.

To introduce the weak formalism of the RT-equations below, we further require the following well-known partial integration formula for scalar valued differential forms,
\beq \label{partial_integration_no_bdd}
\langle d \alpha, \beta \rangle_{L^2} + \langle  \alpha, \delta \beta \rangle_{L^2} =0,
\eeq
where $\alpha$ is a $k$-form and $\beta$ a $(k+1)$-form, such that either $\alpha|_{\partial\Omega}=0$ or $\beta|_{\partial\Omega}=0$, c.f. Theorem 1.11 in \cite{Dac}. Equation \eqref{partial_integration_no_bdd} holds for regularity $\alpha \in W^{1,p}(\Omega)$ and $\beta \in W^{1,p^*}(\Omega)$, (where $p,p^*$ are conjugate exponents, $\frac{1}{p}+\frac{1}{p^*}=1$), and the condition $\alpha|_{\partial\Omega}=0$ or $\beta|_{\partial\Omega}=0$ is understood in the sense that $\alpha \in W^{1,p}_0(\Omega)$ or $\beta \in W^{1,p^*}_0(\Omega)$. Here $W^{1,p}_0(\Omega)$ is the closure of the space of smooth functions $C_0^\infty$ with respect to the $W^{1,p}$-norm, so for $p>n$ functions in $W^{1,p}_0(\Omega)$ vanish on $\partial\Omega$ in the sense of continuous functions. In our first lemma, we extend \eqref{partial_integration_no_bdd} to matrix and vector valued differential forms. 

\begin{Lemma} \label{Lemma_partial_integration_matrix}
Let $u$ be a matrix valued $k$-form and $\omega$ be a matrix valued $(k+1)$-form, $k\geq 0$, such that $u \in W^{1,p}_0(\Omega)$ and $\omega \in W^{1,p^*}_0(\Omega)$, where $\frac{1}{p} + \frac{1}{p^*} =1$, then
\beq \label{partial_integration_matrix}
\langle d u, \omega \rangle_{L^2} + \langle  u, \delta \omega \rangle_{L^2} =0,
\eeq
and \eqref{partial_integration_matrix} continues to hold if only one of the forms $u$ and $\omega$ vanishes on the boundary, i.e., only $u \in W^{1,p}_0(\Omega)$ or $\omega \in W^{1,p^*}_0(\Omega)$. Moreover, \eqref{partial_integration_matrix} holds as well when $u \in W^{1,p}_0(\Omega)$ is a vector valued $k$-form and $\omega \in W^{1,p^*}_0(\Omega)$ a vector valued $(k+1)$-form, (or if either $\omega|_{\partial\Omega}=0$ or $u|_{\partial\Omega}=0$ in the above sense).
\end{Lemma}     

\Proof
The Lemma follows directly from \eqref{partial_integration_no_bdd} together with the fact that the exterior derivative and co-derivative act on matrix-components separately, c.f. \cite{ReintjesTemple_ell1}. Namely, assuming the case that $u$ and $v$ are matrix valued differential forms, we find from definition \eqref{inner-product_L2} that 
\begin{eqnarray} \nonumber 
\langle du, \omega \rangle_{L^2} 
=  \sum_{\nu, \sigma=1}^n \int_\Omega \sum_{i_1<...<i_{k+1}} (du)^\nu_{\sigma\: i_1...i_{k+1}} \omega^\nu_{\sigma\: i_1...i_{k+1}} dx .
\end{eqnarray}
Applying now the partial integration formula for scalar valued forms \eqref{partial_integration_no_bdd} to the above right hand side for fixed $\nu$ and $\sigma$, we obtain that
\begin{eqnarray} \nonumber
\langle du, \omega \rangle_{L^2} 
&=& - \sum_{\nu, \sigma=1}^n  \int_\Omega \sum_{i_1<...<i_k} u^\nu_{\sigma\: i_1...i_k} (\delta\omega)^\nu_{\sigma\: i_1...i_k} dx \cr
&=& - \langle  u, \delta \omega \rangle_{L^2}.
\end{eqnarray}
This is the sought after identity \eqref{partial_integration_matrix} for matrix valued forms. The case for vector valued forms follows analogously.
\QED

Before we define the weak formulation of the RT-equations, we first introduce the weak formulations of Cauchy Riemann type and Poisson equations. So consider an equation of Cauchy Riemann type, 
\beq \label{Cauchy_Riemann}
\begin{cases}
du = f \cr
\delta u = g,
\end{cases} 
\eeq
where $u$, $f$ and $g$ are vector valued differential forms. In light of \eqref{partial_integration_matrix}, we say that a vector valued $k$-form $u \in W^{1,p}(\Omega)$ solves \eqref{Cauchy_Riemann} weakly, if 
\beq \label{Cauchy_Riemann_weak}
\begin{cases}
\langle u, \delta \psi \rangle_{L^2} = - f(\psi) \cr
\langle u, d \varphi \rangle_{L^2} = - g(\varphi) ,
\end{cases} 
\eeq
for any vector valued $(k+1)$-form $\psi \in W^{1,p^*}_0(\Omega)$ and any vector valued $(k-1)$-form $\varphi \in W^{1,p^*}_0(\Omega)$ both vanishing on $\partial\Omega$, where we assume that $f$ is a scalar valued linear functional on the space of vector valued $(k+1)$-forms in $W^{1,p^*}_0(\Omega)$ and $g$ is a scalar valued linear functional on the space of vector valued $(k-1)$-forms in $W^{1,p^*}_0(\Omega)$. (See Section \ref{Sec_A} for a complete list of consistency conditions on $f$ and $g$ required for existence of solutions.) Weak solutions of \eqref{Cauchy_Riemann} for matrix valued differential forms can be introduced in a similar way, but are not required in this paper since the first order system of the RT-equations \eqref{RT_3} - \eqref{RT_4} is vector valued.

Now consider the Poisson equation 
\beq \label{Poisson_eqn}
\Delta u = f.
\eeq
We say that a matrix valued $k$-form $u \in W^{1,p}(\Omega)$ solves \eqref{Poisson_eqn} weakly, if
\beq \label{Poisson_eqn_weakly}
\Delta u[\phi]  =  f(\phi),
\eeq
for any matrix valued $k$-form $\phi \in W^{1,p^*}_0(\Omega)$, where 
\beq \label{Laplacian_weak}
\Delta u[\phi] \equiv - \langle d u, d\phi \rangle_{L^2} - \langle \delta u, \delta \phi \rangle_{L^2}
\eeq
and $f$ is a scalar valued linear functional on the space of matrix valued $k$-forms $\phi \in W^{1,p^*}_0(\Omega)$. To extend the notion of weak solutions of the Poisson equation \eqref{Poisson_eqn} to vector valued differential forms, simply use vector valued test $k$-forms $\phi \in W^{1,p^*}_0(\Omega)$ and the corresponding inner product \eqref{inner-product_L2_vec}, which suffices since the exterior derivative $d$ and the co-derivative $\delta$ act only on form-indices, but not on matrix or vector indices. 

The following lemma clarifies that the weak formulation of the Laplacian  in \eqref{Laplacian_weak} for differential forms is identical to the standard weak form of the Laplacian, taken component wise by restricting to single matrix component of the test functions (an orthogonal decomposition of the space of test functions with respect to the Hilbert-Schmidt inner product).  As a consequence, it is possible to employ standard theorems of elliptic regularity theory for the analysis in this paper, c.f. Section \ref{Sec_Prelimiaries-elliptic}.

\begin{Lemma} \label{Lemma_weak_Poisson_equivalence}
Let $u \in W^{1,p}(\Omega)$ be a matrix (or vector) valued $k$-form, then 
\beq \label{weak_Laplacian_equivalence}
\Delta u[\phi] = - \langle \nabla u, \nabla \phi \rangle_{L^2} ,
\eeq
for any matrix (or vector) $k$-form $\phi \in W^{1,p^*}_0(\Omega)$, where $\Delta u[\phi]$ is the weak Laplacian defined in \eqref{Laplacian_weak}, $\nabla$ is the Euclidean gradient in $x$-coordinates taken on each component\footnote{So $\nabla$ is taken on each matrix, vector or differential form component separately, e.g., $\nabla u = (\nabla u)_{i_1...i_k} dx^{i_1} \wedge ... \wedge dx^{i_k}$ for $u = u_{i_1...i_k} dx^{i_1} \wedge ... \wedge dx^{i_k}$.}, and we set 
\begin{eqnarray} \label{inner-product_Du_matrix}
\langle \nabla u, \nabla \phi \rangle_{L^2} 
&\equiv & \sum_{j=1}^n \int_\Omega \langle \partial_j u, \partial_j \phi \rangle dx  .
\end{eqnarray}
Moreover, \eqref{weak_Laplacian_equivalence} holds assuming only $u \in L^p(\Omega)$ with $du, \delta u \in L^p(\Omega)$, (the low regularity we encounter in the proof of Theorem \ref{Thm_gauge_existence}).
\end{Lemma}            

\Proof
By compactness of $C^\infty_0(\Omega)$ in $W^{1,p^*}_0(\Omega)$ it suffices to prove \eqref{weak_Laplacian_equivalence} for $\phi \in C^\infty_0(\Omega)$. So let $\phi \in C^\infty_0(\Omega)$ and use partial integration component wise to compute
\beq \nonumber
\langle \nabla u, \nabla \phi \rangle_{L^2}  = - \langle u, \nabla\cdot (\nabla \phi) \rangle_{L^2} .
\eeq
Substituting now that $\nabla\cdot (\nabla \phi) = \Delta \phi = d\delta \phi + \delta d \phi$ and using partial integration for differential forms \eqref{partial_integration_matrix}, we obtain that
\begin{eqnarray}
\langle \nabla u, \nabla \phi \rangle_{L^2}  
&=& - \langle u, d\delta \phi \rangle_{L^2} - \langle u, \delta d \phi \rangle_{L^2}  \cr
&=&  \langle\delta u, \delta \phi \rangle_{L^2} + \langle d u, d \phi \rangle_{L^2} \cr
&=& - \Delta u[\phi]. 
\end{eqnarray}
This proves \eqref{weak_Laplacian_equivalence} in the case $u \in W^{1,p}(\Omega)$.

To prove the supplement, assume that $u \in L^p(\Omega)$ with $du, \delta u \in L^p(\Omega)$. For this regularity the weak Laplacian \eqref{Laplacian_weak} is well-defined, i.e.,
\beq \label{Laplacian_weak_techeqn1}
\Delta u[\phi] \equiv - \langle d u, d\phi \rangle_{L^2} - \langle \delta u, \delta \phi \rangle_{L^2},
\eeq
exists for any matrix valued $k$-form $\phi \in W^{1,p^*}_0(\Omega)$. Again, considering $\phi$ in the dense subspace $C^\infty_0(\Omega)$, $\Delta \phi = \delta d \phi + d \delta \phi$ is well-defined. We now apply partial integration \eqref{partial_integration_matrix} to write  \eqref{Laplacian_weak_techeqn1} as
\beq \label{Laplacian_weak_techeqn2}
\Delta u[\phi] =  \langle u, \Delta \phi \rangle_{L^2} = \lim\limits_{h\to 0} \langle u, \nabla_h \cdot \nabla \phi \rangle_{L^2},
\eeq
where the last equality holds by convergence of the difference quotient  $\nabla_h \phi$ to $\nabla\phi$, so that we have $\nabla_h \cdot \nabla \phi \to \Delta \phi$ as $h \to 0$. By partial integration for $\nabla_h$, we find that
\beq \label{Laplacian_weak_techeqn3}
- \langle \nabla_h u, \nabla \phi \rangle_{L^2} = \langle u, \nabla_h \cdot \nabla \phi \rangle_{L^2}.
\eeq
By \eqref{Laplacian_weak_techeqn2} the right hand side in \eqref{Laplacian_weak_techeqn3} converges, which implies that the left hand side in \eqref{Laplacian_weak_techeqn3} converges as $h \to 0$ as well, and we have 
\beq
\lim\limits_{h\to 0} \langle \nabla_h u, \nabla \phi \rangle_{L^2} = \langle \nabla u, \nabla \phi \rangle_{L^2}.
\eeq 
Combing this with \eqref{Laplacian_weak_techeqn2}, we find that
\beq \nonumber
 \Delta u[\phi] = - \langle \nabla u, \nabla \phi \rangle_{L^2} 
\eeq
for any $\phi \in C^\infty_0(\Omega)$. By denseness of $C^\infty_0(\Omega)$ in $W^{1,p^*}_0(\Omega)$ this establishes \eqref{weak_Laplacian_equivalence} for the low regularity $u \in L^p(\Omega)$ with $du, \delta u \in L^p(\Omega)$. This completes the proof of Lemma \ref{Lemma_weak_Poisson_equivalence}.
\QED

\subsection{The weak RT-equations and weak reduced RT-equations}

We are now prepared to derive the weak formulation of the RT-equations \eqref{RT_1} - \eqref{RT_4}, that is, of the system
\begin{eqnarray} \nonumber
\Delta \Gammati &=& \delta d \Gamma - \delta \big( dJ^{-1} \wedge dJ \big) + d(J^{-1} A ),  \cr
\Delta J &=& \delta ( J \mm \Gamma ) - \langle d J ; \tilde{\Gamma}\rangle - A , \cr
d \vec{A} &=& \overrightarrow{\text{div}} \big(dJ \wedge \Gamma\big) + \overrightarrow{\text{div}} \big( J\, d\Gamma\big) - d\big(\overrightarrow{\langle d J ; \tilde{\Gamma}\rangle }\big),  \cr
\delta \vec{A} &=& v.  
\end{eqnarray}
To begin, let $\Gamma \in L^{2p}(\Omega)$ and $d\Gamma \in L^p(\Omega)$ for $p>n/2$, let $v\in L^p(\Omega)$ be a vector valued $0$-form, and assume we are given a smooth solution $(\Gammati,J,A)$ of the RT-equations \eqref{RT_1} - \eqref{RT_4}. We now introduce a weak formulation for solutions $\Gammati \in W^{1,p}(\Omega)$, $J \in W^{1,2p}(\Omega)$ and $A \in L^{p}(\Omega)$.

For any matrix valued $1$-form $\Phi \in W^{1,p^*}_0(\Omega)$, $\frac{1}{p^*} + \frac{1}{p} =1$, we write the left hand side of the first RT-equation \eqref{RT_1} as       
\begin{eqnarray} \label{weak_Delta_Gammati}
\Delta \Gammati [\Phi]   
\equiv  - \langle \delta \Gammati, \delta \Phi \rangle_{L^2} - \langle d \Gammati, d\Phi \rangle_{L^2},
\end{eqnarray}
by applying the  Leibniz-rule \eqref{partial_integration_matrix} to $\Delta = \delta d + d \delta$, c.f. \eqref{Laplacian_weak}.  Using \eqref{partial_integration_matrix} to rewrite the right hand side of \eqref{RT_1} in an analogous way, we find that the first RT-equation \eqref{RT_1} is equivalent to
\beq \label{weak_RT_1}
- \Delta \Gammati [\Phi] 
= \big\langle (d \Gamma  -  dJ^{-1} \wedge dJ ) , \: d\Phi \big\rangle_{L^2} + \langle J^{-1} A  ,\: \delta \Phi \rangle_{L^2}
\eeq
to hold for any matrix valued $1$-form $\Phi \in W^{1,p^*}_0(\Omega)$. This is our weak formulation of \eqref{RT_1}. 

Similarly, we find the weak formulation of \eqref{RT_2} to be 
\beq \label{weak_RT_2}
- \Delta J [\phi] 
= \big\langle J \mm \Gamma , \: d\phi \big\rangle_{L^2} + \big\langle \big(\langle d J ; \tilde{\Gamma}\rangle + A\big)  ,\: \phi \big\rangle_{L^2},
\eeq
for all matrix valued $0$-form $\phi \in W^{1,(2p)^*}_0(\Omega)$, where 
$$
\Delta J[\phi] = - \langle dJ, d\phi \rangle_{L^2}.
$$ 
Since we address the solution space $J \in W^{1,2p}(\Omega)$, we require the test space $ W^{1,(2p)^*}_0(\Omega)$, where $(2p)^*$ denotes the conjugate exponent to $2p$, $\frac{1}{(2p)^*} + \frac{1}{2p} =1$, (and $(2p)^* \neq 2p^*$ in general).  Note that $\langle  A,\: \phi \rangle_{L^2}$ is finite for $A\in L^p(\Omega)$ by Sobolev embedding, as proven in Lemma \ref{Lemma_Sobolev_embedding} below.

To derive the weak formulation of \eqref{RT_3} - \eqref{RT_4}, we first clarify how to shift $\overrightarrow{\text{div}}$ over to test functions, i.e., we have to define a suitable adjoint to the operation defined in  \eqref{Def_vec-div}. For this, we introduce $\underleftarrow{\text{div}}$ as the mapping from vector valued $2$-forms $\psi \in W^{1,p^*}_0(\Omega)$ to matrix valued $2$-forms in $L^{p^*}(\Omega)$ defined by
\beq \label{vec-div_adjoint}
\big(\underleftarrow{\text{div}}(\psi)\big)^\mu_{\nu}  = \partial_\nu \psi^\mu_{ij} \; dx^i \wedge dx^j.
\eeq
By applying partial integration component wise, it is straightforward to verify the following lemma.

\begin{Lemma}
For any matrix valued $2$-form $\omega$, it is
\beq \label{vec-div_adjoint_eqn1}
\langle \overrightarrow{div}(\omega),\psi \rangle_{L^2} = - \langle \omega, \underleftarrow{div}(\psi) \rangle_{L^2}
\eeq
for any vector valued $2$-form $\psi \in W^{1,p^*}_0(\Omega)$.
\end{Lemma} 

\Proof
By \eqref{Def_vec-div} we have $\overrightarrow{\text{div}}(\omega)^\mu \equiv \sum_{\nu=1}^n \partial_\nu (\omega^\mu_\nu)_{ij} dx^{i}\wedge dx^{j}$, so components-wise partial integration gives us
\begin{eqnarray}
\langle \overrightarrow{div}(\omega),\psi \rangle_{L^2} 
&=& \sum_{\mu} \sum_{i<j} \int_\Omega \sum_\nu \partial_\nu (\omega^\mu_\nu)_{ij} \psi^\mu_{\, ij} dx   \cr
&=& \sum_{\mu,\nu} \sum_{i<j} \int_\Omega (\omega^\mu_\nu)_{ij} \partial_\nu \psi^\mu_{\, ij} dx  \cr
&=& - \langle \omega, \underleftarrow{div}(\psi) \rangle_{L^2},
\end{eqnarray}  
which proves \eqref{vec-div_adjoint_eqn1} and the lemma.
\QED

\noindent Applying now \eqref{vec-div_adjoint_eqn1} together with \eqref{partial_integration_matrix} for vector valued differential forms, we find that \eqref{RT_3} - \eqref{RT_4} can be written equivalently as
\beq
\begin{cases} \label{weak_RT_A}
\langle \vec{A}, \delta \psi \rangle_{L^2} 
= \big\langle ( dJ\wedge \Gamma + J\, d\Gamma), \underleftarrow{\text{div}}(\psi) \big\rangle_{L^2} - \langle \overrightarrow{\langle d J ; \tilde{\Gamma}\rangle} ,\delta \psi \rangle_{L^2} \cr
\langle \vec{A}, d \varphi \rangle_{L^2}  = -\langle v,  \varphi \rangle_{L^2} ,  
\end{cases}
\eeq
for all vector valued $2$-form $\psi \in W^{1,p^*}_0(\Omega)$ and all vector valued $0$-forms $\varphi \in W^{1,p^*}_0(\Omega)$. This is the weak formulation of \eqref{RT_3} - \eqref{RT_4}. 

\begin{Def} \label{Definition_weak_RT}
Let $\Gamma \in L^{2p}(\Omega)$ and $d\Gamma \in L^{p}(\Omega)$ for $p>n/2$, and assume $\Gammati \in W^{1,p}(\Omega)$, $J \in W^{1,2p}(\Omega)$ and $A \in L^{p}(\Omega)$. We say $(\Gammati,J,A)$ is a weak solution of the RT-equations if \eqref{weak_RT_1}, \eqref{weak_RT_2} and \eqref{weak_RT_A} hold for all matrix valued $1$-forms $\Phi \in W^{1,p^*}_0(\Omega)$, all matrix valued $0$-forms $\phi \in W^{1,(2p)^*}_0(\Omega)$, all vector valued $2$-forms $\psi \in W^{1,p^*}_0(\Omega)$ and all vector valued $0$-forms $\varphi  \in W^{1,p^*}_0(\Omega)$,  where $p^*$ and $(2p)^*$ are conjugate exponents defined by $\frac{1}{p^*} +\frac{1}{p} =1$ and $\frac{1}{(2p)^*} + \frac{1}{2p} =1$.   
\end{Def}

The weak formulation of the RT-equations in Definition \ref{Definition_weak_RT} can be easily adapted to the reduced RT-equations, subject of the next definition. Recall first that the reduced RT-equations \eqref{RT_withB_2} - \eqref{RT_withB_4} are  
\begin{eqnarray} 
\Delta J &=& \delta ( J \mm \Gamma ) - B ,    \label{RT_red1}  \\
d \vec{B} &=& \overrightarrow{\text{div}} \big(dJ \wedge \Gamma\big) + \overrightarrow{\text{div}} \big( J\, d\Gamma\big) , \label{RT_red2} \\
\delta \vec{B} &=& w,  \label{RT_red3}
\end{eqnarray}
with unknowns $J$ and $B$.

\begin{Def} \label{Def_weak_reduced_RT}
Let $\Gamma \in L^{2p}(\Omega)$ and $d\Gamma \in L^{p}(\Omega)$ for $p>n/2$, and assume $J\in W^{1,2p}(\Omega)$ and $B \in L^{p}(\Omega)$, we say that $(J,B)$ is a weak solution of the reduced RT-equations \eqref{RT_withB_2} - \eqref{RT_withB_4}, if 
\beq \label{RT_reduced_J_weak}
-\Delta J[\phi] =  \langle  J \mm \Gamma , d \phi \rangle_{L^2} + \langle B , \phi \rangle_{L^2} 
\eeq
holds for any matrix valued $0$-form $\phi \in W^{1,(2p)^*}_0(\Omega)$, (where $\Delta J[\phi] =- \langle d J, d \phi \rangle_{L^2}$), and if
\beq
\begin{cases} \label{RT_reduced_A_weak}
\langle \vec{B}, \delta \psi \rangle_{L^2} = \big\langle ( dJ\wedge \Gamma + J\, d\Gamma), \underleftarrow{\text{div}}(\psi) \big\rangle_{L^2} \cr
\langle \vec{B}, d \varphi \rangle_{L^2}  = w ,  
\end{cases}
\eeq
holds for any vector valued $2$-form $\psi \in W^{1,p^*}_0(\Omega)$ and any vector valued $0$-form $\varphi \in W^{1,p^*}_0(\Omega)$. 
\end{Def}

\subsection{Consistency of the RT-equations at low regularity}  \label{Sec_consistency}

We now show consistency of the reduced RT-equations (by which we mean that the operations on the right and left hand side of the RT-equations produce the same regularity) for the solutions space $J \in W^{1,2p}(\Omega)$ and $B \in L^{p}(\Omega)$, when $\Gamma \in L^{2p}(\Omega)$ and $d\Gamma \in L^p(\Omega)$, $p>n/2$. For this we apply the Sobolev embedding theorem to prove for completeness that $L^{p}(\Omega) \subset W^{-1,2p}(\Omega)$ in the next lemma. Given this embedding $L^{p}(\Omega) \subset W^{-1,2p}(\Omega)$, we obtain consistency of the reduced RT-equations as follows:  For $J \in W^{1,2p}(\Omega)$ and $\Gamma \in L^{2p}(\Omega)$, applying the H\"older inequality as in \eqref{Holder_L2p-trick} implies $dJ \wedge \Gamma \in L^p$, by which the right hand side of \eqref{RT_red2} is placed in $W^{-1,p}(\Omega)$, (the remaining term is more regular by Morrey's inequality). This places $B$ in the sought after solution space $L^p(\Omega)$. Using now the embedding $L^{p}(\Omega) \subset W^{-1,2p}(\Omega)$, we find that $B \in W^{-1,2p}(\Omega)$, which in turn implies that the right hand side of \eqref{RT_red1} is in $W^{-1,2p}(\Omega)$. Namely, since $J\Gamma \in W^{1,2p}(\Omega)$ implies that $\delta(J\Gamma)[\phi] = - \langle J\Gamma, d\phi\rangle_{L^2}$ is finite for any $\phi \in W^{1,(2p)^*}(\Omega)$, it follows that also $\delta(J\Gamma) \in W^{-1,2p}(\Omega)$. This places $J$ in the sought after solution space $W^{1,2p}(\Omega)$. Taken on whole, this gives the consistency of the reduced RT-equations once the embedding $L^{p}(\Omega) \subset W^{-1,2p}(\Omega)$ of the next lemma is proven.       

\begin{Lemma} \label{Lemma_Sobolev_embedding}
For $p > n/2$ the embedding $L^{p}(\Omega) \subset W^{-1,2p}(\Omega)$ holds, together with the estimate  
\beq \label{Sobolev_embedding_estimate}
\| B \|_{W^{-1,2p}(\Omega)} \leq C \| B \|_{L^p(\Omega)} ,
\eeq
where $C>0$ is some constant depending only on $\Omega, p, n$.
\end{Lemma}

\Proof
To prove the embedding $L^p(\Omega) \subset W^{-1,2p}(\Omega)$, given some fixed function $B\in L^p(\Omega)$, we need to show that the dual pairing $B(\phi)\equiv \langle B, \phi \rangle_{L^2}$ is finite for any $\phi \in W^{1,(2p)^*}_0(\Omega)$, i.e., that $B$ defines a functional over $W^{1,(2p)^*}_0(\Omega)$. By H\"older's inequality we have
\beq \label{Sobolev_emb_eqn0}
|\langle B, \phi \rangle_{L^2}| \leq \|B\|_{L^p} \|\phi\|_{L^{p^*}},
\eeq
so it suffices to show that 
\beq \label{Sobolev_emb_eqn1}
W^{1,(2p)^*}_0(\Omega) \subset L^{p^*}(\Omega).
\eeq 
We show \eqref{Sobolev_emb_eqn1} via the Sobolev embedding Theorem \cite[Thm 2, Ch. 5.6]{Evans}, which asserts that $W^{1,q}(\Omega) \subset L^{q'}(\Omega)$ for the Sobolev conjugate $q' \equiv \frac{n q}{n-q}$ whenever $1 \leq q <n$, together with the estimate
\beq \label{Sobolev_emb_eqn2}
\| f \|_{L^{q'}(\Omega)} \leq C_S \| f \|_{W^{1,q}(\Omega)},
\eeq 
for some constant $C_S>0$ depending only on $n, q$ and $\Omega$, (where $f\in W^{1,q}(\Omega)$).  Here we have $q=(2p)^*$, and it is straightforward to verify $1 \leq (2p)^* <n$ using $(2p)^* = \frac{2p}{2p-1} < \frac{n}{n-1} \leq 2$, so Sobolev embedding applies and gives $W^{1,(2p)^*}_0(\Omega) \subset L^{q'}(\Omega)$ for $q' \equiv \frac{n (2p)^*}{n-(2p)^*}$. Thus to establish the sought after embedding \eqref{Sobolev_embedding_estimate}, it only remains to verify that $L^{q'}(\Omega) \subset L^{p^*}(\Omega)$. This inclusion follows directly from our assumption $p> n/2$,  by verifying the inequality $q' \geq p^*$ by substitution of $q' = \frac{n (2p)^*}{n-(2p)^*}$, $(2p)^* = \frac{2p}{2p-1}$ and $p^* = \frac{p}{p-1}$, and since $q' \geq p^*$, $L^{q'}(\Omega) \subset L^{p^*}(\Omega)$ follows by the boundedness of $\Omega$. This proves the sought after embedding \eqref{Sobolev_emb_eqn1}.

We now prove estimate \eqref{Sobolev_embedding_estimate}. By applying H\"older's inequality to $\|f\|_{L^p}^p = \|f^p\cdot 1\|_{L^1}$, to show that $\|f\|_{L^p}^p = \|f^p\|_{L^\frac{q}{p}} \| 1\|_{L^{(q/p)^*}}$, we find
\beq \label{Sobolev_emb_eqn3}
\| f \|_{L^{p}(\Omega)} \leq {\rm vol}(\Omega)^\frac{q-p}{pq} \: \| f \|_{L^{q}(\Omega)} 
\eeq
whenever $q\geq p$. Combining now \eqref{Sobolev_emb_eqn3} with estimate \eqref{Sobolev_emb_eqn2} of the Sobolev embedding theorem, gives us
\beq \label{Sobolev_emb_eqn4}
\| \phi \|_{L^{p^*}(\Omega)} \leq C\: \| \phi \|_{W^{1,(2p)^*}(\Omega)}
\eeq 
for $C\equiv C_S\cdot  {\rm vol}(\Omega)^\frac{q'-p^*}{p^* q'}$ with $q' = \frac{n (2p)^*}{n-(2p)^*}$. From \eqref{Sobolev_emb_eqn4} together with H\"older's inequality we find that $B(\phi) \equiv \langle B,\phi\rangle_{L^2}$ is bounded by 
\beq \label{Sobolev_emb_eqn5}
|B(\phi)| \leq \|B\|_{L^p(\Omega)} \|\phi \|_{L^{p^*}(\Omega)} 
\leq C\: \|B\|_{L^p(\Omega)} \| \phi \|_{W^{1,(2p)^*}(\Omega)},
\eeq 
which implies the sought after estimate \eqref{Sobolev_embedding_estimate} by taking the supremum over all $\phi \in W^{1,(2p)^*}_0(\Omega)$ with $\| \phi \|_{W^{1,(2p)^*}(\Omega)}=1$.
\QED

The proof of Lemma \ref{Lemma_Sobolev_embedding} establishes consistency of the reduced RT-equations at the level of regularity, $\Gamma \in L^{2p}(\Omega)$,  $d\Gamma \in L^p(\Omega)$, $J \in W^{1,2p}(\Omega)$ and $B \in L^{p}(\Omega)$, for  $p>n/2$. To show consistency of the full RT-equations, it suffices to address only the first RT-equation \eqref{PDE1}, 
\beq  
\Delta \Gammati = \delta d\Gamma - \delta \big(d J^{-1} \wedge dJ \big) + d(J^{-1} A ),
\eeq
for $A \in L^p(\Omega)$. Consistency of the first RT-equation holds, since $d J^{-1} \wedge dJ \in L^p(\Omega)$ by H\"older's inequality \eqref{Holder_L2p-trick}, since $J^{-1}A \in L^p(\Omega)$ by Morrey's inequality, and since $d\Gamma \in L^p(\Omega)$ by assumption, which taken on whole places $\Gammati'$ in the desired space $\Gammati' \in W^{1,p}(\Omega)$, as proven in Section \ref{Sec_gauge}. Note that the assumption $d\Gamma \in L^p(\Omega)$ is only required for consistency of the first RT-equation\eqref{PDE1}, but not for the reduced RT-equations.

\section{Recovering the full from the reduced RT-equations - Proof of Theorem \ref{Thm_gauge_existence}}   \label{Sec_gauge}

In this section we prove that, given a solution $(J,B)$ of the reduced RT-equations \eqref{RT_withB_2} - \eqref{RT_withB_4}, then $(J,\Gammati,A)$ solves the full RT-equations \eqref{RT_1} - \eqref{RT_4} with $\Gammati = \Gammati_J$ and $A=A'$, where $\Gammati_J$ and $A'$ are defined in \eqref{Gammati'} and \eqref{A'} as
\begin{eqnarray}\nonumber
\Gammati_J & \equiv &\Gamma - J^{-1}dJ, \cr
A'  &\equiv &  B - \langle d J ; \Gammati_J \rangle .
\end{eqnarray} 
From this, using the first RT-equation \eqref{RT_1} in terms of $A=A'$ (which is equivalent to equation \eqref{RT_withB_1}), we then prove that $\Gammati_J$ gains one derivative over the regularity of the terms separately on the right hand side of its definition in \eqref{Gammati'}. This is asserted by Theorem \ref{Thm_gauge_existence}, and here we give the proof. Throughout the rest of this paper we only address the case $\Gammati = \Gammati_J$ and $A=A'$, so for ease of notation from here on, we denote $\Gammati_J$ by $\Gammati$ and $A'$ by $A$.    For completeness, we first prove a version of Theorem \ref{Thm_gauge_existence} at the higher level of connection regularity, $\Gamma, d\Gamma \in W^{m,p}(\Omega)$, $m\geq 1$, $p>n$, a regularity sufficient to take point-wise estimates by Morrey's inequality \eqref{Morrey_textbook}.\footnote{This also connects the use of the reduced RT-equation here with the approach in \cite{ReintjesTemple_ell3} based on the original RT-equations, in particular in Section \ref{Sec_gauge_lowreg} of \cite{ReintjesTemple_ell3}.} In Section \ref{Sec_gauge_lowreg} we then prove the sought after low regularity case $\Gamma \in L^{2p}(\Omega)$, $d\Gamma \in L^p(\Omega)$, $p>n/2$.

\subsection{A smooth version of Theorem \ref{Thm_gauge_existence}} \label{Sec_gauge_smooth}

We now prove the following proposition which is a version of  Theorem \ref{Thm_gauge_existence} in the case $\Gamma, d\Gamma \in W^{m,p}(\Omega)$, $m\geq 1$, $p>n$. This proposition is not needed to prove the results in this paper, but helpful for understanding the main steps in the proof of  Theorem \ref{Thm_gauge_existence} given in Section  \ref{Sec_gauge_lowreg}. For ease of notation we let $\Omega \equiv \Omega_x$ denote the neighborhood $\Omega'_x$ of Theorem \ref{Thm_gauge_existence} in $x$-coordinates, and we denote the compactly contained subset $\Omega''_x$ by $\Omega'$.

\begin{Prop}   \label{Thm_gauge-invariance} 
Assume $\Gamma, d\Gamma \in W^{m+1,p}(\Omega)$ in $x$-coordinates, for $m\geq 1$, $p> n$ and $p < \infty$. Assume $(J,B)$ solves the reduced RT-equations \eqref{RT_withB_2} - \eqref{RT_withB_4} in a weak sense for some $w \in W^{m-1,p}(\Omega)$, such that $J, J^{-1}\, \in W^{m+1,p}(\Omega)$ and $B \in W^{m,p}(\Omega)$.\footnote{The existence of such a solution follows from our existence theory in \cite{ReintjesTemple_ell2,ReintjesTemple_ell3}.} Then the following holds: 

\noindent {\bf (i)} Defining $\Gammati_J$ by \eqref{Gammati'}, $\Gammati\equiv\Gammati_J$ solves (\ref{RT_withB_1}), and the tuple $(J,\Gammati,A)$ solves the full RT-equations \eqref{RT_1} - \eqref{RT_4} in $\Omega$ for $v\equiv v'$ and $A \equiv A'$, where $A'$ and $v'$ are defined in \eqref{A'} and \eqref{v'}. 

\noindent {\bf (ii)} The regularity of $\Gammati\equiv \Gammati_J$ is given by $\Gammati_J \in W^{m+1,p}(\Omega')$ for any open set $\Omega'$ compactly contained in $\Omega$, i.e., $\Gammati_J \equiv \Gamma - J^{-1}dJ$ is one order more regular than the two terms on the right hand side are separately.

\noindent {\bf (iii)} Let $M>0$ be a constant such that 
$$
\|(\Gamma,d\Gamma)\|_{W^{m,p}(\Omega)} \equiv \|\Gamma \|_{W^{m,p}(\Omega)} + \| d\Gamma \|_{W^{m,p}(\Omega)}   \leq M. 
$$ 
Assume that $(J,B)$ satisfies further the estimate
\small
\beq \label{Gammati_estimate_smooth_assump}
\|I-J\|_{W^{m+1,p}(\Omega)} + \|I-J^{-1}\|_{W^{m+1,p}(\Omega)} + \|B\|_{W^{m,p}(\Omega)} 
\ \leq \ C_1(M)\; \|(\Gamma,d\Gamma)\|_{W^{m,p}(\Omega)} , 
\eeq
\normalsize
for some constant $C_1(M)>0$ depending only on $\Omega, n, p$ and $M$. Then, on any open set $\Omega'$ compactly contained in $\Omega$, $\Gammati\equiv\Gammati_J$ satisfies the uniform bound                
\beq  \label{Gammati'_estimate_smooth}
\| \Gammati \|_{W^{m+1,p}(\Omega'_x)} \leq  C_2(M) \; \|(\Gamma,d\Gamma)\|_{W^{m,p}(\Omega)}
\eeq
where $C_2(M)>0$ is some constant depending only on $\Omega', \Omega, n, p$ and $M$. 
\end{Prop}                 

\Proof
Let $\Gamma, d\Gamma \in W^{m,p}(\Omega)$, for $m\geq 1$, $p>n$, and assume $(J,B)$ is a solution of the reduced RT-equations \eqref{RT_withB_2} - \eqref{RT_withB_4} with $J, J^{-1} \in W^{m+1,p}(\Omega)$ and $B \in W^{m,p}(\Omega)$. For this regularity Lemma \ref{Thm_gauge_informal} applies and yields that $(J,\Gammati,A)$ solves the first RT-equation \eqref{RT_1}.

We now prove that $(J,\Gammati,A)$ solves the second RT-equation \eqref{RT_2}. By assumption, $(J,B)$ solves the reduced RT-equation \eqref{RT_withB_2}, that is,
\begin{eqnarray} \label{Proof_Thm0_eqn5}
\Delta J   &=& \delta ( J \mm \Gamma ) - B .
\end{eqnarray}
By definition of $A$ in \eqref{A'}, we have 
\beq \label{Proof_Thm0_eqn6}
B = A + \langle dJ ; \Gammati \rangle,
\eeq
so substitution of \eqref{Proof_Thm0_eqn6} into \eqref{Proof_Thm0_eqn5} gives
\beq \nonumber 
\Delta J = \delta ( J \mm \Gamma ) - \langle d J ; \Gammati \rangle - A,
\eeq
which is the sought after RT-equation \eqref{RT_2}. 

We now prove that $(J,\Gammati,A)$ solves the last two RT-equations \eqref{RT_3} - \eqref{RT_4}. By assumption, $(J,B)$ solves \eqref{RT_3} - \eqref{RT_4} for some $w \in W^{m-1,p}(\Omega)$, that is,
\begin{eqnarray}
d \vec{B} &=& \overrightarrow{\text{div}} \big(dJ \wedge \Gamma\big) + \overrightarrow{\text{div}} \big( J\, d\Gamma\big), \label{Proof_Thm0_RT3} \\
\delta \vec{B} &=& w \label{Proof_Thm0_RT4}. 
\end{eqnarray}
Substituting \eqref{Proof_Thm0_eqn6} into \eqref{Proof_Thm0_RT3} and subtracting the resulting equation by $d \overrightarrow{\langle d J ; \Gammati \rangle}$ gives us the equation
\beq \nonumber
d \vec{A} = \overrightarrow{\text{div}} \big(dJ \wedge \Gamma\big) + \overrightarrow{\text{div}} \big( J\, d\Gamma\big) - d \overrightarrow{\langle d J ; \Gammati \rangle}  ,
\eeq
which is the sought after third RT-equation \eqref{RT_3}. Similarly, substituting \eqref{Proof_Thm0_eqn6} into \eqref{Proof_Thm0_RT4} gives 
\beq \nonumber
\delta \vec{A} = w - \delta \langle d J ; \Gammati \rangle , 
\eeq
which is the sought after RT-equation \eqref{RT_4} for $v$ defined by \eqref{v'}, that is, for $v =  w - \delta \langle d J ; \Gammati \rangle$. Taken together, we proved that $(J,\Gammati,A)$ solves the full RT-equations \eqref{RT_1} - \eqref{RT_4} for the gauge freedom in $v$ fixed by the choice \eqref{v'}. This proves (i) of the proposition.

To prove (ii) of Proposition \ref{Thm_gauge-invariance}, we need to show the regularity $\Gammati \in W^{m+1,p}(\Omega)$ together with estimate \eqref{Gammati'_estimate_smooth}. In a first step, we establish the lower regularity $\Gammati \in W^{m,p}(\Omega)$ and $A \in  W^{m,p}(\Omega)$ from their definitions in \eqref{Gammati'} and \eqref{A'}. For this we use that by Morrey's inequality \eqref{Morrey_textbook} the space $W^{m,p}(\Omega)$ is closed under multiplication when $m\geq1$, $p>n$, c.f. \cite{ReintjesTemple_ell1}. Now $\Gammati \in W^{m,p}(\Omega)$, since $\Gamma \in W^{m,p}(\Omega)$, $J^{-1} \in  W^{m+1,p}(\Omega)$ and $ dJ \in W^{m,p}(\Omega)$ so that the closedness of $W^{m,p}(\Omega)$ yields $\Gammati = \Gamma - J^{-1} dJ \in W^{m,p}(\Omega)$ by \eqref{Gammati'}. Moreover, the regularity $A \in  W^{m,p}(\Omega)$ directly follows from \eqref{A'}, since $dJ, \Gammati \in  W^{m,p}(\Omega)$ implies that $A = B - \langle d J ; \Gammati \rangle \; \in   W^{m,p}(\Omega)$. This shows that $\Gammati$ and $A$ are both in $W^{m,p}(\Omega)$. 

We now prove that $\Gammati \in W^{m,p}(\Omega)$ is one derivative more regular, $\Gammati \in W^{m+1,p}(\Omega)$, by establishing estimate \eqref{Gammati'_estimate_smooth}. For this, we use the first RT-equation \eqref{RT_1},
\beq \label{Poisson_Gammati'}
\Delta \Gammati  
=   \delta d \Gamma - \delta \big( dJ^{-1} \wedge dJ \big) + d\big( J^{-1}A\big)  .
\eeq 
But estimate \eqref{Poisson-4} of elliptic regularity theory gives       
\beq \label{Poisson_Gammati'_eqn1}
\| \Gammati \|_{W^{m+1,p}(\Omega')} \leq C_e \big( \| \Delta \Gammati \|_{W^{m-1,p}(\Omega)}  + \| \Gammati \|_{W^{m,p}(\Omega)} \big)
\eeq
for any open set $\Omega'$ compactly contained in $\Omega$, where $C>0$ is some constant depending only on $\Omega', \Omega, p, n, m$. The regularity gain of $\Gammati$ follows once we show $\Delta \Gammati \in W^{m-1,p}(\Omega)$, since $\Gammati  \in W^{m,p}(\Omega)$.  But to derive the sought after estimate \eqref{Gammati'_estimate_smooth} we need a more refined analysis. For this, we begin by substituting for $\Delta \Gammati$ the right hand side of \eqref{Poisson_Gammati'} to obtain
\begin{align} \label{Poisson_Gammati'_eqn2}
&\| \Gammati \|_{W^{m+1,p}(\Omega')}  \cr
&\leq  C_e\Big( \| \delta d \Gamma \|_{W^{m-1,p}(\Omega)} + \|\delta \big( dJ^{-1} \wedge dJ \big) \|_{W^{m-1,p}(\Omega)}  + \| d\big( J^{-1}A \big) \|_{W^{m-1,p}(\Omega)} \Big) \cr
&\leq  C_e \Big( \|d \Gamma \|_{W^{m,p}(\Omega)} + \|\delta \big( dJ^{-1} \wedge dJ \big) \|_{W^{m-1,p}(\Omega)}  + \| J^{-1}A \|_{W^{m,p}(\Omega)} \Big). 
\end{align}
(We subsequently often write $\| \cdot \|_{W^{m,p}}$ instead of $\| \cdot \|_{W^{m,p}(\Omega)}$.) The first term on the right hand side of \eqref{Poisson_Gammati'_eqn2} is bounded by assumption. Using first the product rule and then Morrey's inequality \eqref{Morrey_textbook}, the second term can be bounded by
\begin{align} \nonumber
&\|\delta \big( dJ^{-1} \wedge dJ \big) \|_{W^{m-1,p}(\Omega)}   \cr
&\leq  \big\| |D( dJ^{-1})| \cdot|dJ|  \big\|_{W^{m-1,p}}  +  \big\| |dJ^{-1}| \cdot|D(dJ)| \big\|_{W^{m-1,p}}  \cr
&\leq  \| D( dJ^{-1}) \|_{W^{m-1,p}}  \| dJ  \|_{W^{m-1,\infty}}  +  \| dJ^{-1} \|_{W^{m-1,\infty}} \| D(dJ) \|_{W^{m-1,p}}  \cr
&\leq  \| J^{-1} \|_{W^{m+1,p}}  \| J  \|_{W^{m,\infty}}  +  \| J^{-1} \|_{W^{m,\infty}} \| J \|_{W^{m+1,p}}  \cr
&\overset{\eqref{Morrey_textbook}}{\leq} 2 C_M  \| J^{-1} \|_{W^{m+1,p}(\Omega)}  \| J  \|_{W^{m+1,p}(\Omega)} ,
\end{align}
which is bounded by our incoming assumptions $J^{-1}, J \in W^{m+1,p}(\Omega)$. Using that $dJ=d(J-I)$, the previous estimate gives us
\beq \label{Poisson_Gammati'_eqn3}
\|\delta \big( dJ^{-1} \wedge dJ \big) \|_{W^{m-1,p}(\Omega)}  \leq 2 C_M  \| J^{-1} \|_{W^{m+1,p}(\Omega)}  \| I-J  \|_{W^{m+1,p}(\Omega)}
\eeq  
To estimate the third term on the right hand side of \eqref{Poisson_Gammati'_eqn2}, we use that by definition $\Gammati \equiv \Gammati_J =\Gamma - J^{-1}dJ $ and $A  =  B - \langle d J ; \Gammati \rangle $, to write
$$ 
J^{-1}A = J^{-1} B - J^{-1}\langle d J ; \Gamma\rangle + \langle d J^{-1} ;dJ \rangle  ,
$$ 
where we used the multiplication property \eqref{inner-product_muliplications} to get $J^{-1}\langle dJ ; J^{-1} dJ \rangle = \langle d J^{-1} ;dJ \rangle$. We now estimate the third term on the right hand side of  \eqref{Poisson_Gammati'_eqn2} as 
\begin{align}  \label{Poisson_Gammati'_eqn4}
\| J^{-1}A \|_{W^{m,p}(\Omega)} 
 \leq  \| J^{-1} B \|_{W^{m,p}} + \| J^{-1}\langle d J ; \Gamma\rangle \|_{W^{m,p}}  + \|\langle d J^{-1} ;dJ \rangle \|_{W^{m,p}} .
\end{align}
We now use the closedness of $W^{m,p}(\Omega)$ under multiplication ($m\geq 1$, $n>p$) by Morrey's inequality \eqref{Morrey_textbook}, to estimate the products in \eqref{Poisson_Gammati'_eqn4}, for instance,
\begin{align} \nonumber
\| J^{-1} B \|_{W^{m,p}(\Omega)}& = \| D(J^{-1} B) \|_{W^{m-1,p}} + \| J^{-1} B \|_{W^{m-1,p}} \cr
& \leq \ \| D(J^{-1}) B \|_{W^{m-1,p}} + \| J^{-1} DB \|_{W^{m-1,p}}  + \| J^{-1} B \|_{W^{m-1,p}} \cr
& \leq  \ \| D(J^{-1})\|_{W^{m-1,p}} \| B \|_{W^{m-1,\infty}} + \| J^{-1}\|_{W^{m-1,\infty}} \| DB \|_{W^{m-1,p}} \cr & \ \ \ \ \ \ + \| J^{-1} \|_{W^{m-1,\infty}} \| B \|_{W^{m-1,p}}  \cr
& \leq \ C_M \| J^{-1}\|_{W^{m,p}} \| B \|_{W^{m,p}}.
\end{align}
In this fashion, replacing $dJ$ on the right hand side of \eqref{Poisson_Gammati'_eqn4} by $d(I-J)$, we obtain      
\small
\begin{align} 
\| J^{-1}A \|_{W^{m,p}} 
 \leq  C \| J^{-1} \|_{W^{m+1,p}} \Big( \| B \|_{W^{m,p}} + \| J \|_{W^{m+1,p}} \|\Gamma \|_{W^{m,p}}  +  \|I-J \|_{W^{m+1,p}} \Big) \nonumber \\
 \leq  C \| J^{-1} \|_{W^{m+1,p}} \Big( 1 +  \| J \|_{W^{m+1,p}}\Big)  \Big( \| B \|_{W^{m,p}} + \|\Gamma \|_{W^{m,p}}  +  \|I-J \|_{W^{m+1,p}} \Big),  \label{Poisson_Gammati'_eqn5}
\end{align}
\normalsize
where $C>0$ is some constant depending only on $\Omega, m, n, p$. Substituting \eqref{Poisson_Gammati'_eqn3} and \eqref{Poisson_Gammati'_eqn5} back into the original estimate \eqref{Poisson_Gammati'_eqn2}, we obtain
\small
\beq  \label{Poisson_Gammati'_eqn6}
\| \Gammati \|_{W^{m+1,p}(\Omega')} \leq \mathcal{P}(J) \Big(\|\Gamma \|_{W^{m,p}} + \|d \Gamma \|_{W^{m,p}} + \|I-J \|_{W^{m+1,p}} + \| B \|_{W^{m,p}} \Big) + C\|\Gammati\|_{W^{m,p}},
\eeq
\normalsize
where $\mathcal{P}(J) \equiv C \big( 1+ \| J^{-1} \|_{W^{m+1,p}}\big) \big( 1 +  \| J \|_{W^{m+1,p}}\big)$
for some constant $C>0$ depending only on $\Omega, m, n, p$. By the definition of $\Gammati$ in \eqref{Gammati'}, we bound $\|\Gammati\|_{W^{m,p}(\Omega)}$ using Morrey's inequality as
$$
\|\Gammati\|_{W^{m,p}(\Omega)} \leq C_M \big( \|\Gamma\|_{W^{m,p}(\Omega)} + \|J^{-1}\|_{W^{m+1,p}(\Omega)}  \|J\|_{W^{m+1,p}(\Omega)}   \big),
$$  
from which we obtain the simplified bound
\beq  \label{Poisson_Gammati'_eqn6b}
\| \Gammati \|_{W^{m+1,p}(\Omega')} \leq \mathcal{P}(J) \Big(\|\Gamma \|_{W^{m,p}} + \|d \Gamma \|_{W^{m,p}} + \|I-J \|_{W^{m+1,p}} + \| B \|_{W^{m,p}} \Big),
\eeq
by changing the constant $C>0$ in the definition on $\mathcal{P}$ suitably. Finally, using our incoming assumption \eqref{Gammati_estimate_smooth_assump} to bound the above Sobolev norms on $J, J^{-1}$ and $B$, we obtain the sought after uniform bound \eqref{Gammati'_estimate_smooth}. Clearly, estimate \eqref{Gammati'_estimate_smooth} implies the sought after regularity $\Gammati \in W^{m+1,p}(\Omega)$.\footnote{Note, the gain of one derivative to the required regularity $\Gammati \in W^{m+1,p}(\Omega)$, is entirely based on the cancellation of $\delta \Gamma$-terms in equation \eqref{Proof_Thm_informal_eqn4} of the proof of Lemma \ref{Thm_gauge_informal}.} This completes the proof of Proposition \ref{Thm_gauge-invariance}.
\QED

\subsection{Proof of Theorem \ref{Thm_gauge_existence}} \label{Sec_gauge_lowreg}

We now prove Theorem \ref{Thm_gauge_existence} by adapting the steps in the proof of Proposition \ref{Thm_gauge-invariance} to the weak formulation of the RT-equations to account for the low regularity addressed here. This is significantly more complicated because the substitution of the $J$-equation in the proof of Lemma \ref{Thm_gauge_informal}, (required to get the cancellation of terms involving $\delta\Gamma$, on which our whole theories rests), is not a simple replacement when dealing with the weak form of the equations, due to the problem of multiplying distributions by low regularity functions, c.f. Lemma \ref{Lemma1_Thmgauge} below. Moreover, for the low regularity here, products must be estimated by H\"older inequality instead of Morrey, which we compensate for by putting $J$ in the smaller Sobolev space $W^{1,2p}$ while estimating $\Gammati$ in $W^{1,p}$. Even though we begin here with a given solution of the reduced RT-equations, low regularity products have to be incorporated into the weak formulation of the $\Gammati$ equation, reflecting the non-linear nature of the RT-equations.

So assume $\Gamma  \in L^{2p}(\Omega)$ and $d\Gamma \in L^{p}(\Omega)$ in $x$-coordinates  for $n/2< p < \infty$, and assume that for some constant $M>0$
\beq 
\|(\Gamma,d\Gamma)\|_{L^{2p,p}(\Omega)} \equiv \|\Gamma_x\|_{L^{2p}(\Omega)} + \|d\Gamma_x\|_{L^{p}(\Omega)} \leq M.
\eeq
Assume given some $J \in W^{1,2p}(\Omega)$ invertible and some $B \in L^p(\Omega)$ such that $(J,B)$ solves the reduced RT-equations \eqref{RT_withB_2} - \eqref{RT_withB_4} on $\Omega$, such that
\beq \label{curvature_estimate_soln_again}
\|I-J\|_{W^{1,2p}(\Omega)} + \|I-J^{-1}\|_{W^{1,2p}(\Omega)} + \|B\|_{L^{p}(\Omega)} 
\ \leq \ C_2(M) \|(\Gamma,d\Gamma)\|_{L^{2p,p}(\Omega)}, 
\eeq
where $C_2(M)>0$ is some constant depending only on $\Omega, n, p$ and $M$. Define $\Gammati$ by \eqref{Gammati'} and $A$ by \eqref{A'}, that is,
\beq \label{Gammati&A'_again}
\Gammati = \Gammati_J \equiv \Gamma - J^{-1} dJ \hspace{.5cm} \text{and} \hspace{.5cm} A=A' \equiv B - \langle d J ; \Gammati \rangle .
\eeq   
Again we denote $\Gammati_J$ by $\Gammati$ and $A'$ by $A$ from here on, and we let $\Omega \equiv \Omega_x$ denote the neighborhood $\Omega'_x$ of Theorem \ref{Thm_gauge_existence} in $x$-coordinates, and we denote the compactly contained subset $\Omega''_x$ by $\Omega'$. Theorem \ref{Thm_gauge_existence} $(i)$ states that $(J,\Gammati,A)$ solves the full RT-equations \eqref{RT_1} - \eqref{RT_4} in $\Omega$ for $v=v'$ defined in \eqref{v'}.   Parts $(ii)$ and $(iii)$ of Theorem \ref{Thm_gauge_existence} states that $\Gammati$ is in $W^{1,p}(\Omega')$ and satisfies the uniform bound                
\beq  \label{Gammati'_estimate_again}
\| \Gammati \|_{W^{1,p}(\Omega')} \leq C_3(M) \|(\Gamma,d\Gamma)\|_{L^{2p,p}(\Omega)}, 
\eeq
on any open set $\Omega'$ compactly contained in $\Omega$, and where $C_3(M)>0$ is some constant depending only on $\Omega', \Omega, n, p$ and $M$.\footnote{The version of Theorem \ref{Thm_gauge_existence} applicable to GR shock waves is obtained directly by substituting $\|(\Gamma,d\Gamma)\|_{L^{\infty}}$ for $\|(\Gamma,d\Gamma)\|_{L^{2p,p}}$ everywhere above, since $\|(\Gamma,d\Gamma)\|_{L^{2p,p}(\Omega)} \leq {\rm vol}(\Omega)^\frac{1}{p} \|(\Gamma,d\Gamma)\|_{L^{\infty}(\Omega)}$ for $\Omega$ bounded.}
\vspace{.2cm}

\noindent {\it Proof of Theorem \ref{Thm_gauge_existence}.}  
So assume $J \in W^{1,2p}(\Omega)$ and $B\in L^{p}(\Omega)$ solve the reduced RT-equations \eqref{RT_withB_2} - \eqref{RT_withB_4}. Since only the regularity of the gauge variable $w$ in \eqref{RT_withB_4} is relevant for the proof, we assume here without loss of generality that $w=0$, the case for which we prove existence of solutions in Section \ref{Sec_existence_theory}. So $(J,B)$ is assumed to solve
\begin{eqnarray} 
\Delta J &=& \delta ( J \mm \Gamma ) -  B , \label{reduced_RT_1} \\
d \vec{B} &=& \overrightarrow{\text{div}} \big(dJ \wedge \Gamma\big) + \overrightarrow{\text{div}} \big( J\, d\Gamma\big) ,   \label{reduced_RT_3}\\
\delta \vec{B} &=& 0,  \label{reduced_RT_4}
\end{eqnarray} 
in the weak sense of Definition \ref{Def_weak_reduced_RT}. That is, we assume  
\beq \label{RT_reduced_J_weak_proof}
-\Delta J[\phi] =  \langle  J \mm \Gamma , d \phi \rangle_{L^2} + \langle B , \phi \rangle_{L^2} 
\eeq
for any matrix valued $0$-form $\phi \in W^{1,(2p)^*}_0(\Omega)$, ($\frac{1}{(2p)^*} + \frac{1}{2p} =1$), where $\Delta J[\phi] =- \langle d J, d \phi \rangle_{L^2}$, and we assume
\beq
\begin{cases} \label{RT_reduced_A_weak_proof}
\langle \vec{B}, \delta \psi \rangle_{L^2} = \big\langle ( dJ\wedge \Gamma + J\, d\Gamma), \underleftarrow{\text{div}}(\psi) \big\rangle_{L^2} \cr
\langle \vec{B}, d \varphi \rangle_{L^2}  = 0 ,  
\end{cases}
\eeq
for any vector valued $2$-form $\psi \in W^{1,p^*}_0(\Omega)$ and any vector valued $0$-form $\varphi \in W^{1,p^*}_0(\Omega)$. We here use $C>0$ to denote an universal (``running'') constant depending only on $\Omega$, $n$ and $p$.

The proof of Theorem \ref{Thm_gauge_existence} requires several lemmas. Before we establish these lemmas, we derive the preliminary regularity $\Gammati \in L^{2p}(\Omega)$ and that $A  \in L^{p}(\Omega)$ from the definitions in \eqref{Gammati&A'_again}, regularities we require to bootstrap to the desired regularity $\Gammati \in W^{1,p}(\Omega)$ in the end. For this, recall that $J$ is assumed to be invertible with $J^{-1} \in W^{1,2p}(\Omega)$, so Morrey's inequality \eqref{Morrey_textbook} implies that $J^{-1} dJ \in L^{2p}(\Omega)$, since $p>n/2$. Thus, since $\Gamma \in L^{2p}(\Omega)$ by assumption, $\Gammati \equiv \Gamma - J^{-1} dJ  \in L^{2p}(\Omega)$ follows directly from \eqref{Gammati'}. To show that $A  \in L^{p}(\Omega)$, we first apply H\"older's inequality as in \eqref{Holder_L2p-trick} using that $dJ$ and $\Gammati$ are both in $L^{2p}(\Omega)$ to conclude with $\langle d J ; \Gammati \rangle \in L^{p}(\Omega)$. This implies that $A  \in L^{p}(\Omega)$ by \eqref{Gammati&A'_again} and our incoming assumption $B \in L^{p}(\Omega)$. We have established $\Gammati \in L^{2p}(\Omega)$ and $A \in L^{p}(\Omega)$.  

In Lemma \ref{Lemma3_Thmgauge} below we prove that $\Gammati$ defined in \eqref{Gammati'} solves the first RT-equation \eqref{RT_1} for $A$ defined by \eqref{A'}, in the weak sense \eqref{weak_RT_1}. We then apply elliptic regularity theory to this equation to boost the regularity of $\Gammati$ by essentially one order to $\Gammati \in W^{1,p}(\Omega)$, as asserted by Theorem \ref{Thm_gauge_existence}. The main step in the proof of Lemma \ref{Lemma3_Thmgauge} is accomplished in the following lemma by adapting the computation in the proof of Lemma \ref{Thm_gauge_informal} to the weak formulation required for the low regularity here. 

\begin{Lemma} \label{Lemma1_Thmgauge}
Under the assumption of Theorem \ref{Thm_gauge_existence}, we have
\beq  \label{Lemma1_Thmgauge_eqn}
\delta \Gammati [\phi] = \langle J^{-1}A, \phi \rangle_{L^2}
\eeq
for any matrix valued $0$-form $\phi \in W^{1,p^*}_0(\Omega)$, where $\delta \Gammati [\phi] \equiv - \langle \Gammati, d\phi\rangle_{L^2}$. That is, $\delta \Gammati = J^{-1}A$ in the sense of weak derivatives.\footnote{For the reader familiar with our previous work in \cite{ReintjesTemple_ell1,ReintjesTemple_ell3}, note that establishing \eqref{Lemma1_Thmgauge_eqn} reverses a basic identity in the derivation of the RT-equations from the Riemann-flat condition where we defined $A$ in terms of $J \delta\Gammati$.}
\end{Lemma}

\Proof
Assume $J$ is a solution of the first reduced RT-equation in the weak sense \eqref{RT_reduced_J_weak} for some given $B\in L^{p}(\Omega)$. The proof is based on adapting the computation \eqref{Proof_Thm_informal_eqn3a} - \eqref{Proof_Thm_informal_eqn5} of Lemma \ref{Thm_gauge_informal} to regularities $J\in W^{1,2p}(\Omega)$ and $B \in L^{p}(\Omega)$. This requires care because of the presence of products. Instead of computing  $\delta \Gammati$ directly as in \eqref{Proof_Thm_informal_eqn3a}, which would not yield the weak Laplacian on $J$ (essential for the argument), we begin by taking $\delta$ of $J \Gammati$.\footnote{Note, applying the results of Lemma \ref{Thm_gauge_informal} in this setting at a mollified level, would entail the problem of controlling the zero mollification limit through the second reduced RT-equation \eqref{RT_2}, a system of inhomogeneous elliptic PDE's for which it would be difficult to avoid generalized eigenvalues.} Using that $\Gammati\equiv \Gamma - J^{-1} dJ$ by its definition in \eqref{Gammati'}, taking $\delta$ of $J \Gammati$ gives us 
\begin{eqnarray} \label{proofThmgauge_eqn2}
\delta(J\Gammati)[\phi] 
&=& - \langle J\Gammati , d\phi \rangle_{L^2}   \cr
&=&  - \langle J\Gamma , d\phi \rangle_{L^2} + \langle dJ , d\phi \rangle_{L^2},
\end{eqnarray} 
for any matrix valued $0$-form $\phi \in W^{1,p^*}_0(\Omega)$. Note that the expressions in \eqref{proofThmgauge_eqn2} are finite by H\"older inequality \eqref{Holder}, since $\Gamma \in L^{2p}(\Omega)$ and $J \in W^{1,2p}(\Omega)$, and since $L^{p^*}(\Omega) \subset L^{(2p)^*}(\Omega)$ because $\Omega$ is bounded and $p^* = \frac{p}{p-1}$ is larger than $(2p)^* = \frac{2p}{2p-1}$. For example, by H\"older's inequality \eqref{Holder_matrix-multi},
$$
\big| \langle dJ , d\phi \rangle_{L^2} \big| \leq C\, \| dJ \|_{L^{2p}} \|d\phi\|_{L^{(2p)^*}} \leq C \| dJ \|_{L^{2p}} \, \|d\phi\|_{L^{p^*}}^{\frac{p^*}{(2p)^*}}  < \infty ,
$$  
where $C>0$ is some constant depending only on $\Omega$, $n$ and $p$.  Now, we replace the second term on the right hand side of \eqref{proofThmgauge_eqn2} by the weak reduced RT-equation \eqref{RT_reduced_J_weak_proof}, 
$$
\langle d J, d \phi \rangle_{L^2}=  \langle  J \mm \Gamma , d \phi \rangle_{L^2} + \langle B , \phi \rangle_{L^2}, 
$$ 
which, after the cancellation of the lowest regularity term $\langle  J \mm \Gamma , d \phi \rangle_{L^2}$ crucial to our method, leads to
\beq \label{proofThmgauge_eqn3}
\delta(J\Gammati)[\phi]    =   \langle B , \phi \rangle_{L^2}, 
\hspace{.5cm} \text{that is,}  \hspace{.5cm}
\langle J\Gammati , d\phi \rangle_{L^2}  + \langle B , \phi \rangle_{L^2} =0.
\eeq

Our goal now is to move $d$ to the other side of the first inner product in \eqref{proofThmgauge_eqn3} as $\delta$ on the product $J\Gammati$ and isolate the weak derivative $\delta\Gammati$, from which the sought after equation \eqref{Lemma1_Thmgauge_eqn} then follows. At the start, $J\Gammati$ is not regular enough to apply partial integration nor the Leibniz product rule directly, but the following mollification suffices to complete this last step in the proof.\footnote{The reason why this mollification argument works seems to be that one can anticipate $\Gammati$ to be one order more regular for given solutions of the RT-equations by their consistency.} So consider standard mollifiers $\Gamma_\epsilon \in C^\infty(\Omega)$ and $J_\epsilon \in C^\infty(\Omega)$ of $\Gamma$ and $J$ respectively, so that $\Gamma_\epsilon$ converges to $\Gamma$ in $L^{2p}(\Omega)$ as $\epsilon \rightarrow 0$ and $J_\epsilon$ converges to $J$ in $W^{1,2p}(\Omega)$ as $\epsilon \rightarrow 0$.\footnote{By the compact of the test functions, mollifying does not change the region of integration $\Omega$.}   Define 
$$
\Gammati_\epsilon \equiv \Gamma_\epsilon - J^{-1} dJ_\epsilon,
$$
then $\Gammati_\epsilon \in W^{1,2p}(\Omega)$ holds. To show that $\Gammati_\epsilon \longrightarrow \Gammati$ in $L^{2p}(\Omega)$ as $\epsilon \rightarrow 0$, recall $J^{-1} \in W^{1,2p}(\Omega)$ is bounded in $L^\infty(\Omega)$ by Morrey's inequality, so we have 
$$
\|J^{-1}(dJ_\epsilon - dJ) \|_{L^{2p}} \leq  \|J^{-1}\|_{L^\infty} \|dJ_\epsilon - dJ\|_{L^{2p}} \longrightarrow 0
$$ 
as $\epsilon \rightarrow 0$, and this implies that $\Gammati_\epsilon \longrightarrow \Gammati$ in $L^{2p}(\Omega)$ as $\epsilon \rightarrow 0$. Now, since $\Gammati_\epsilon \longrightarrow \Gammati$ in $L^{2p}(\Omega)$ as $\epsilon \rightarrow 0$, equation \eqref{proofThmgauge_eqn3} implies that
\beq \label{proofThmgauge_eqn4}
\lim_{\epsilon \rightarrow 0} \Big( \langle J\Gammati_\epsilon , d\phi \rangle_{L^2}  + \langle B , \phi \rangle_{L^2} \Big) =0.
\eeq
The regularity $\Gammati_\epsilon \in W^{1,2p}(\Omega)$ allows us now to use partial integration \eqref{partial_integration_matrix}, followed by the Leibniz rule \eqref{ Leibniz-rule-delta}, $\delta(J\Gammati_\epsilon) = J \delta\Gammati_\epsilon +\langle dJ;\Gammati_\epsilon \rangle$, to compute 
\begin{eqnarray}   \label{proofThmgauge_eqn5a}
\langle J\Gammati_\epsilon , d\phi \rangle_{L^2}  + \langle B , \phi \rangle_{L^2} 
&=&  - \langle \delta(J\Gammati_\epsilon) , \phi \rangle_{L^2}  + \langle B , \phi \rangle_{L^2}  \cr
&\overset{\eqref{ Leibniz-rule-delta}}{=}& - \langle J \delta\Gammati_\epsilon , \phi \rangle_{L^2}  + \big\langle (B - \langle dJ;\Gammati_\epsilon \rangle ) , \phi \big\rangle_{L^2}  \cr
&=& - \langle J \delta\Gammati_\epsilon , \phi \rangle_{L^2}+ \langle A_\epsilon,\phi \rangle_{L^2}, 
\end{eqnarray}
where we set $A_\epsilon \equiv B - \langle dJ;\Gammati_\epsilon \rangle$.  Observe that $A_\epsilon \to A$ in $L^p(\Omega)$ as $\epsilon \rightarrow 0$, since $\Gamma_\epsilon \to \Gamma$ and $dJ_\epsilon \to dJ$ in $L^{2p}(\Omega)$ as $\epsilon \rightarrow 0$. We now applying the multiplication property of $\langle\cdot\: ;\cdot\rangle$ in \eqref{inner-product_muliplications} together with cyclic commutativity of matrix multiplication in the trace, to write \eqref{proofThmgauge_eqn5a} as 
\begin{eqnarray}  \label{proofThmgauge_eqn5}
\langle J\Gammati_\epsilon , d\phi \rangle_{L^2}  + \langle B , \phi \rangle_{L^2} 
&=& -\langle\delta\Gammati_\epsilon , J^T \phi  \rangle_{L^2}+ \langle J^{-1} A_\epsilon, J^T\phi \rangle_{L^2} .
\end{eqnarray}
To clarify this step, consider for example  
\begin{align} \nonumber
\langle J \delta\Gammati_\epsilon , \phi \rangle_{L^2} 
&\overset{\eqref{inner-product_L2}}{=} \int_\Omega\tr \langle J \delta\Gammati_\epsilon ; \phi^T \rangle dx 
\overset{\eqref{inner-product_muliplications}}{=} \int_\Omega\tr\big( J \mm \langle \delta\Gammati_\epsilon ; \phi^T \rangle \big) dx \cr
& =   \int_\Omega\tr\big( \langle \delta\Gammati_\epsilon ; \phi^T \rangle \mm J \big) dx   
\overset{\eqref{inner-product_muliplications}}{=}  \int_\Omega\tr \langle \delta\Gammati_\epsilon ; (J^T\phi)^T \rangle 
= \langle \delta\Gammati_\epsilon , J^T\phi  \rangle_{L^2}.
\end{align}
Applying now the integration by parts formula \eqref{partial_integration_matrix} to the first term on the right hand side of \eqref{proofThmgauge_eqn5} and defining $\psi \equiv J^T\phi$, we obtain
\begin{eqnarray} \label{proofThmgauge_eqn6}
\langle J\Gammati_\epsilon , d\phi \rangle_{L^2}  + \langle B , \phi \rangle_{L^2} 
&=&  \langle \Gammati_\epsilon , d\psi \rangle_{L^2} + \langle J^{-1} A_\epsilon,\psi \rangle_{L^2} .
\end{eqnarray}
Since $J \in W^{1,2p}(\Omega)$, (for $p >n/2$, $p>2$, so $1 <p^*<2$), it follows that $\psi \equiv J^T\phi \in W^{1,p^*}_0(\Omega)$ is indeed a test function, (as proven in Lemma \ref{Lemma2_Thmgauge} below).  Thus, since  $\Gammati_\epsilon$ converges to $\Gammati$ in $L^{2p}(\Omega)$  and since $A_\epsilon \rightarrow A$ in $L^p(\Omega)$ as $\epsilon \rightarrow 0$, we conclude that the right hand side in \eqref{proofThmgauge_eqn6} converges as $\epsilon \rightarrow 0$. Moreover, by \eqref{proofThmgauge_eqn4} the left hand side in \eqref{proofThmgauge_eqn6} converges as well and the limit of \eqref{proofThmgauge_eqn6} as $\epsilon \rightarrow 0$ vanishes, which gives
\beq \label{proofThmgauge_eqn7}
- \langle \Gammati , d\psi \rangle_{L^2} = \langle J^{-1} A,\psi \rangle_{L^2},
\eeq
for $\psi =J^T \phi$. Since any $\psi \in W^{1,p^*}_0(\Omega)$ there exists some some $\phi \in W^{1,p^*}_0(\Omega)$ such that $\psi = J^T \phi$, as proven in Lemma \ref{Lemma2_Thmgauge} below, we conclude that the sought after equation \eqref{proofThmgauge_eqn7} holds. This proves Lemma \ref{Lemma1_Thmgauge}.
\QED

We now apply Lemma \ref{Lemma1_Thmgauge} to establish that $\Gammati$ solves the first RT-equation for \eqref{RT_1} for $A=A'$ defined in \eqref{A'}, at the correct level of regularity. This lemma establishes the first statement in part $(i)$ of Theorem \ref{Thm_gauge_existence} together with part $(ii)$.

\begin{Lemma} \label{Lemma3_Thmgauge}
Under the assumption of Theorem \ref{Thm_gauge_existence}, $\Gammati \in L^{2p}(\Omega)$ solves the first RT-equation \eqref{RT_1} with $A$ given by \eqref{A'}, and thus $\Gammati$ has the regularity $\Gammati \in W^{1,p}(\Omega)$.
\end{Lemma}

\Proof
To begin recall that by \eqref{Lemma1_Thmgauge_eqn} of Lemma \ref{Lemma1_Thmgauge}, we have 
\beq \nonumber
\delta \Gammati [\phi] = \langle J^{-1}A, \phi \rangle_{L^2}
\eeq
for any matrix valued $0$-form $\phi \in W^{1,p^*}_0(\Omega)$, where $\delta \Gammati [\phi] \equiv - \langle \Gammati, d\phi\rangle_{L^2}$. Since $A \in L^p(\Omega)$ and $J^{-1} \in W^{1,2p}(\Omega)$, this directly implies that the weak co-derivative $\delta$ of  $\Gammati$ is an $L^p$ function, $\delta \Gammati \in L^p(\Omega)$, and it follows that 
\beq  \label{proofThmgauge_eqn_13}
\delta \Gammati =  J^{-1}A   \ \in L^p(\Omega)
\eeq
holds in the sense of $L^p$ functions. We can now take the exterior derivative $d$ of $\delta \Gammati$ in a weak sense, which gives in light of \eqref{proofThmgauge_eqn_13} that
\beq  \label{proofThmgauge_eqn_14}
\langle \delta \Gammati, \delta \Phi \rangle_{L^2} =  \langle J^{-1}A, \delta \Phi \rangle_{L^2}
\eeq
for any matrix valued $1$-form $\Phi \in W^{1,p^*}_0(\Omega)$. This determines the second term of the weak Laplacian of $\Gammati$ in \eqref{weak_Delta_Gammati}.

To determine the first term of the weak Laplacian $\Delta \Gammati[\Phi]$ in \eqref{weak_Delta_Gammati}, we take the exterior derivative of $\Gammati_\epsilon \equiv \Gamma_\epsilon - J^{-1} dJ_\epsilon$, where $\Gamma_\epsilon$ and $J_\epsilon$ are the mollifications of $\Gamma$ and $J$ introduced in the proof of Lemma \ref{Lemma1_Thmgauge}. This gives us
\begin{eqnarray} \label{proofThmgauge_eqn_14b}
d \Gammati_\epsilon  &=& d \Gamma_\epsilon  - d (J^{-1} dJ_\epsilon)  \cr
&=& d \Gamma_\epsilon  - dJ^{-1} \wedge dJ_\epsilon,
\end{eqnarray}
where we applied the Leibniz rule \eqref{ Leibniz-rule-d} for the last equality.  We now show that the right hand side of \eqref{proofThmgauge_eqn_14b} converges in $L^p(\Omega)$ as $\epsilon \rightarrow 0$. For this recall first that $d\Gamma \in L^{p}(\Omega)$, which implies that $d(\Gamma_\epsilon) = (d\Gamma)_\epsilon \longrightarrow d\Gamma$ in $L^{p}(\Omega)$ as $\epsilon \rightarrow 0$. Moreover, by using H\"older's inequality as in \eqref{Holder_wedge}, we find that       
\begin{eqnarray} \label{proofThmgauge_eqn_15}
\big\| dJ^{-1} \wedge \big( dJ_\epsilon -  dJ \big) \big\|_{L^p} 
\leq  \big\| dJ^{-1} \big\|_{L^{2p}} \big\| dJ_\epsilon - dJ \big\|_{L^{2p}}   .
\end{eqnarray}                
By $L^{2p}$ convergence of $dJ_\epsilon$ to $dJ$, \eqref{proofThmgauge_eqn_15} implies that $dJ^{-1} \wedge dJ_\epsilon$ converges to $dJ^{-1} \wedge dJ$ in $L^p(\Omega)$ as $\epsilon \rightarrow 0$. We conclude that the right hand side of \eqref{proofThmgauge_eqn_14b} converges in $L^p(\Omega)$ as $\epsilon \rightarrow 0$, and this yields that
\begin{eqnarray} \label{proofThmgauge_eqn_16}
d \Gammati  = d \Gamma - dJ^{-1} \wedge dJ \ \in L^p(\Omega). 
\end{eqnarray}
Taking now the co-derivative $\delta$ of \eqref{proofThmgauge_eqn_16} gives
\beq \nonumber
\delta d \Gammati = \delta d \Gamma - \delta \big(d(J^{-1}) \wedge dJ \big) \in W^{-1,p}(\Omega),
\eeq
but of course in the weak sense
\beq \label{proofThmgauge_eqn_17}
\langle d \Gammati, d\Phi \rangle_{L^2} = \langle \big( d \Gamma -  dJ^{-1} \wedge dJ \big), d\Phi \rangle_{L^2} ,
\eeq
for any matrix valued $1$-form $\Phi \in W^{1,p^*}_0(\Omega)$. Combing  now \eqref{proofThmgauge_eqn_14} and \eqref{proofThmgauge_eqn_17}, and using that by our definition in \eqref{weak_Delta_Gammati}
$$
\Delta \Gammati [\Phi]   
=  - \langle \delta \Gammati, \delta \Phi \rangle_{L^2} - \langle d \Gammati, d\Phi \rangle_{L^2},
$$
we finally obtain that $\Gammati$ solves the sought after  first RT-equation,
\beq \label{proofThmgauge_eqn_18}
\Delta \Gammati [\Phi]  = - \langle \big( d \Gamma -  dJ^{-1} \wedge dJ \big), d\Phi \rangle_{L^2} - \langle J^{-1}A, \delta \Phi \rangle_{L^2} ,
\eeq
in the weak form \eqref{weak_RT_1}. 

To complete the proof of Lemma \ref{Lemma3_Thmgauge}, it remains to show $\Gammati$ gains one derivative to $\Gammati \in W^{1,p}(\Omega)$. For this, note that we have already established that $\delta\Gammati$ and $d\Gammati$ are in $L^p(\Omega)$, c.f. \eqref{proofThmgauge_eqn_13} and \eqref{proofThmgauge_eqn_16}. Thus, since the right hand side of \eqref{proofThmgauge_eqn_18} results from taking $d$ of $\delta\Gammati$ and $\delta$ of $d\Gammati$, we conclude that the right hand side of \eqref{proofThmgauge_eqn_18} lies in $W^{-1,p}(\Omega)$, that is,      
\beq \label{proofThmgauge_eqn_18b}
\Delta \Gammati \in W^{-1,p}(\Omega).
\eeq
Applying now Lemma \ref{Lemma_weak_Poisson_equivalence} for the case of regularity $\Gammati \in L^p(\Omega)$ with $d\Gammati, \delta \Gammati \in L^p(\Omega)$, we find that $\Delta\Gammati[\Phi] = - \langle \nabla\Gammati,\nabla\Phi \rangle_{L^2}$. Thus the standard weak Laplacian $- \langle \nabla\Gammati,\nabla\Phi \rangle_{L^2}$ lies in $ W^{-1,p}(\Omega)$ by \eqref{proofThmgauge_eqn_18b}. Applying now basic elliptic regularity theory, c.f. Theorem \ref{Thm_Poisson_interior} in the appendix, we conclude with the sought after regularity $\Gammati\in W^{1,p}(\Omega')$ for any open set $\Omega' \subset \Omega$ compactly contained in $\Omega$. This completes the proof of Lemma \ref{Lemma3_Thmgauge}. 
\QED

In the next Lemma we prove the basic elliptic estimate \eqref{Gammati'_estimate} from which we later derive the curvature bound \eqref{curvature}, using H\"older and Morrey inequalities in combination with estimate \eqref{curvature_estimate_soln} assumed on $J,J^{-1}$ and $B$. Lemma \ref{Lemma_Gammati'_estimate} proves part $(iii)$ of Theorem \ref{Thm_gauge_existence}. 

\begin{Lemma} \label{Lemma_Gammati'_estimate}
Under the assumption of Theorem \ref{Thm_gauge_existence}, in particular assuming the bound \eqref{curvature_estimate_soln_again} on $(J,B)$ and $\|(\Gamma,d\Gamma) \|_{L^{2p,p}}\leq M$, the weak solution $\Gammati \equiv \Gammati_J$ of the first RT-equation \eqref{RT_1}, defined by \eqref{Gammati'}, satisfies
\beq  \label{Gammati'_estimate_proof}
\| \Gammati \|_{W^{1,p}(\Omega')} \leq C(M)  \|(\Gamma,d\Gamma)\|_{L^{2p,p}(\Omega)} ,
\eeq
for any open set $\Omega'$ compactly contained in $\Omega$ and some constant $C(M)>0$ depending only on $\Omega, \Omega', p, n$ and $M$.
\end{Lemma}

\Proof
By Lemma \ref{Lemma3_Thmgauge}, $\Gammati \equiv \Gammati_J$ solves  the weak first RT-equation \eqref{proofThmgauge_eqn_18}, that is,
\beq \label{Gammati'_techeqn1}
\Delta \Gammati [\Phi]  = - \big\langle \big( d \Gamma -  dJ^{-1} \wedge dJ \big), d\Phi \big\rangle_{L^2} - \langle J^{-1}A, \delta \Phi \rangle_{L^2} 
\eeq
holds for any matrix valued $1$-form $\Phi \in W^{1,p^*}_0(\Omega)$. Applying the basic elliptic estimates \eqref{Poisson-4} to \eqref{Gammati'_techeqn1}, we obtain     
\beq \label{Gammati'_techeqn2}
\| \Gammati \|_{W^{1,p}(\Omega')}  \leq C \big( \| \mathcal{F}\|_{W^{-1,p}(\Omega)}  + \| \Gammati \|_{L^p(\Omega)} \big)
\eeq
for any open $\Omega'$ compactly contained in $\Omega$, and where we define the functional
\beq \label{Gammati'_techeqn3}
 \mathcal{F}(\Phi) \equiv - \big\langle \big( d \Gamma -  dJ^{-1} \wedge dJ \big), d\Phi \big\rangle_{L^2} - \langle J^{-1}A, \delta \Phi \rangle_{L^2} 
\eeq
on matrix valued $1$-forms $\Phi \in W^{1,p^*}_0(\Omega)$, and the operator norm $\|\cdot\|_{W^{-1,p}}$ is defined by
\beq \label{Gammati'_techeqn4}
\| \mathcal{F}\|_{W^{-1,p}(\Omega)} \equiv \sup \big\{ |\mathcal{F}(\Phi)| \, \big| \, \Phi \in W^{1,p^*}_0(\Omega), \ \|\Phi\|_{W^{1,p^*}(\Omega)} =1  \big\}.
\eeq
From \eqref{Gammati'_techeqn3} we find
\beq \nonumber 
 \big|\mathcal{F}(\Phi)\big|  \leq \big| \big\langle \big( d \Gamma -  dJ^{-1} \wedge dJ \big), d\Phi \big\rangle_{L^2} \big| + \big| \big\langle J^{-1}A, \delta \Phi \big\rangle_{L^2} \big|,
\eeq
which we further estimate using H\"older's inequality \eqref{Holder} as
\begin{eqnarray} \label{Gammati'_techeqn5}
 \big|\mathcal{F}(\Phi)\big|     
 &\leq & C\, \Big( \big\| \big( d \Gamma -  dJ^{-1} \wedge dJ \big) \big\|_{L^p}  \| d\Phi \|_{L^{p^*}} + \big\| J^{-1}A \big\|_{L^p}  \|\delta \Phi \|_{L^{p^*}} \Big) \cr
& \leq & C\, \Big( \| d \Gamma \|_{L^p(\Omega)}  + \big\| dJ^{-1} \wedge dJ \big\|_{L^p(\Omega)}  + \big\| J^{-1}A \big\|_{L^p(\Omega)} \Big) ,
\end{eqnarray}
where we estimated $\| d\Phi \|_{L^{p^*}} \leq \| \Phi \|_{W^{1,p^*}} = 1$ and $\|\delta \Phi \|_{L^{p^*}} \leq 1$ for the last inequality. Throughout this proof $C>0$ denotes a universal constant depending only on $\Omega, n, p$ We now use H\"older's inequality as in \eqref{Holder_wedge} to estimate the second term in \eqref{Gammati'_techeqn5} as 
\begin{eqnarray} \label{Gammati'_techeqn6a}
\big\| dJ^{-1} \wedge dJ \big\|^p_{L^p(\Omega)} 
&\leq & C\, \big\langle | dJ^{-1} |^p, | dJ|^p  \big\rangle_{L^2} \cr
&\leq & C\, \big\| | dJ^{-1}|^p \big\|_{L^{2}(\Omega)} \big\| | dJ |^p \big\|_{L^{2}(\Omega)}  \cr
&= & C\, \big\| dJ^{-1} \big\|^p_{L^{2p}(\Omega)} \big\| dJ \big\|^p_{L^{2p}(\Omega)} ,
\end{eqnarray}
where we lost a little regularity from $L^{2p}$ to $L^p$, as anticipated in our theory by starting with $J \in W^{1,2p}(\Omega)$. Now taking the $p$-th root of \eqref{Gammati'_techeqn6a} and using that $dJ = d(J-I)$, we obtain
\beq \label{Gammati'_techeqn6}
\big\| dJ^{-1} \wedge dJ \big\|_{L^p(\Omega)} 
\leq \big\| J^{-1} \big\|_{W^{1,2p}(\Omega)}  \big\| I-J \big\|_{W^{1,2p}(\Omega)}.
\eeq
To estimate the third term in \eqref{Gammati'_techeqn5}, we substitute \eqref{Gammati'} and \eqref{A'}, that is, we substitute $\Gammati =\Gamma - J^{-1}dJ $ into $A  =  B - \langle d J ; \Gammati \rangle$. This leads to the identity
$$ 
J^{-1}A = J^{-1} B - J^{-1}\langle d J ; \Gamma\rangle + \langle d J^{-1} ;dJ \rangle ,
$$ 
where we used the multiplication property \eqref{inner-product_muliplications} to write 
$$
J^{-1}\langle dJ ; J^{-1} dJ \rangle = \langle d J^{-1} ;dJ \rangle.
$$ 
We now obtain the bound
\begin{align} \label{Gammati'_techeqn7}
\big\| &J^{-1}A \big\|_{L^p}   \
\leq \  \| J^{-1} B \|_{L^p} + \| J^{-1}\langle d J ; \Gamma\rangle \|_{L^p}  + \|\langle d J^{-1} ;dJ \rangle \|_{L^p} \cr
& \leq \ \| J^{-1} B \|_{L^p} + C_M \| J^{-1}\|_{W^{1,p}}\, \|\langle d J ; \Gamma\rangle \|_{L^p}  + \|\langle d J^{-1} ;dJ \rangle \|_{L^p},
\end{align}
where we applied Morrey's inequality \eqref{Morrey_textbook} in the last step to bound the $L^\infty$-norm of $J^{-1}$. We now estimate the remaining product terms in \eqref{Gammati'_techeqn7} employing H\"older's inequality as in \eqref{Gammati'_techeqn6a}, to obtain
\beq \label{Gammati'_techeqn8a}
\big\| J^{-1}A \big\|_{L^p}   
\leq  C \| J^{-1}\|_{W^{1,p}} \big( \| B\|_{L^{p}} + \| d J \|_{L^{2p}}  \| \Gamma\|_{L^{2p}}  + \| d(I-J) \|_{L^{2p}} \big),
\eeq
where $C>0$ is some constant depending only on $\Omega, n, p$. Applying $dJ = d(J-I)$ we write bound \eqref{Gammati'_techeqn8a} further as 
\beq \label{Gammati'_techeqn8}
\big\| J^{-1}A \big\|_{L^p}   
\leq  C \| J^{-1}\|_{W^{1,p}} \big( 1 + \| J\|_{W^{1,p}} \big) \big( \| B\|_{L^{p}} + \| \Gamma\|_{L^{2p}}  + \| I-J \|_{W^{1,2p}} \big).
\eeq
Combining \eqref{Gammati'_techeqn6} and \eqref{Gammati'_techeqn8} to bound the right hand side of \eqref{Gammati'_techeqn5}, we obtain
\beq \label{Gammati'_techeqn9}
 \big|\mathcal{F}(\Phi)\big|  \leq \mathcal{P}_0 \big( \| d\Gamma\|_{L^{p}} + \| \Gamma\|_{L^{2p}} + \| B\|_{L^{p}}  + \| I-J \|_{W^{1,2p}} \big), 
\eeq 
where $\mathcal{P}_0 = C \big( 1 +\| J^{-1}\|_{W^{1,2p}}\big) \big( 1 + \| J\|_{W^{1,2p}} \big)$. Substituting the bound \eqref{Gammati'_techeqn9} for the right hand side of \eqref{Gammati'_techeqn2}, we find that
\beq \label{Gammati'_techeqn9b}
\| \Gammati \|_{W^{1,p}(\Omega')}  \leq \mathcal{P}_0 \big( \| d\Gamma\|_{L^{2p}} + \| \Gamma\|_{L^{2p}}  + \| B\|_{L^{p}} + \| I-J \|_{W^{1,2p}} \big) +C \| \Gammati \|_{L^p(\Omega)}  .
\eeq
From $\Gammati = \Gamma - J^{-1} dJ$ and H\"older inequality, we bound $\| \Gammati \|_{L^p(\Omega)}$ by    
$$
\| \Gammati \|_{L^p(\Omega)} \leq {\rm vol}(\Omega)^{\frac{1}{2p}} \|\Gamma \|_{L^{2p}(\Omega)} + C\, \| J^{-1}\|_{L^{2p}(\Omega)} \|dJ\|_{L^{2p}(\Omega)},
$$
(where we estimated $\|\Gamma \|_{L^{p}} \leq  {\rm vol}(\Omega)^{\frac{1}{2p}} \|\Gamma \|_{L^{2p}(\Omega)}$ by applying H\"older's inequality to $\Gamma = \Gamma \cdot 1$). Substituting this bound on $\| \Gammati \|_{L^p(\Omega)}$ into \eqref{Gammati'_techeqn9b}, we obtain
\beq \label{Gammati'_techeqn_10}
\| \Gammati \|_{W^{1,p}(\Omega')}  \leq \mathcal{P}_0 \big( \| d\Gamma\|_{L^{2p}} + \| \Gamma\|_{L^{2p}}  + \| B\|_{L^{p}} + \| I-J \|_{W^{1,2p}} \big) ,
\eeq
by modifying the constant $C>0$ in the definition of $\mathcal{P}_0$ suitably.

To derive the sought after bound \eqref{Gammati'_estimate_proof} we now use that $(J,B)$ are assumed to meet the bound \eqref{curvature_estimate_soln_again}, that is, 
\beq \label{curvaturebound_techeqn1}
\|I-J\|_{W^{1,2p}(\Omega)} + \|I-J^{-1}\|_{W^{1,2p}(\Omega)} + \|B\|_{L^{p}(\Omega)} 
\ \leq \ C_2(M)  \|(\Gamma,d\Gamma)\|_{L^{2p,p}(\Omega)}  .
\eeq
So using \eqref{curvaturebound_techeqn1} together with the definition $\|(\Gamma,d\Gamma)\|_{L^{2p,p}(\Omega)}  = \| \Gamma\|_{L^{2p}}  + \| d\Gamma\|_{L^{p}}$ to bound the right hand side of \eqref{Gammati'_techeqn_10}, we conclude that there exists some constant $C_3(M)>0$ depending only on $\Omega, n, p$ and $M$, such that
\beq  \nonumber
\| \Gammati \|_{W^{1,p}(\Omega')} \leq C_3(M) \|(\Gamma,d\Gamma)\|_{L^{2p,p}(\Omega)} 
\eeq
which is the sought after bound \eqref{Gammati'_estimate_again}. This completes the proof of Lemma \ref{Lemma_Gammati'_estimate}. 
\QED

To establish part $(i)$ of Theorem \ref{Thm_gauge_existence} it only remains to verify that $(J,\Gammati,A)$ solves the weak RT-equations \eqref{weak_RT_2} - \eqref{weak_RT_A}. 

\begin{Lemma} \label{Lemma4_Thmgauge}
Under the assumption of Theorem \ref{Thm_gauge_existence}, $(J,\Gammati,A)$ solves the second, third and fourth weak RT-equations \eqref{weak_RT_2} - \eqref{weak_RT_A}  for $v$ defined in \eqref{v'} with $w=0$.
\end{Lemma}

\Proof
By our assumption that $J \in W^{1,2p}(\Omega)$ and $B\in L^{p}(\omega)$ solve the first weak reduced RT-equation \eqref{RT_reduced_J_weak_proof}, that is,
\beq \label{proofThmgauge_eqn_19}
-\Delta J[\phi] =  \langle  J \mm \Gamma , d \phi \rangle_{L^2} + \langle B , \phi \rangle_{L^2} 
\eeq
holds for any matrix valued $0$-form $\phi \in W^{1,(2p)^*}_0(\Omega)$, where $\Delta J[\phi] =- \langle d J, d \phi \rangle_{L^2}$, and where $\langle B , \phi \rangle_{L^2}$ is finite by Lemma \ref{Lemma_Sobolev_embedding}. Substituting $B  =  \langle d J ; \Gammati \rangle + A$, (which follows from the definition of $A$ in \eqref{A'}), into \eqref{proofThmgauge_eqn_19}, we directly obtain
\beq \nonumber
- \Delta J [\phi] 
= \big\langle J \mm \Gamma , \: d\phi \big\rangle_{L^2} + \big\langle \big(\langle d J ; \Gammati\rangle + A\big)  ,\: \phi \big\rangle_{L^2},
\eeq
which is the sought after weak RT-equation \eqref{weak_RT_2}. 

We now show that the third and fourth weak RT-equations \eqref{weak_RT_A} hold. By assumption $J \in W^{1,2p}(\Omega)$ and $B\in L^{p}(\Omega)$ solve the weak reduced RT-equations \eqref{RT_reduced_A_weak_proof}, that is,
\beq \label{proofThmgauge_eqn_20}
\begin{cases} \nonumber 
\langle \vec{B}, \delta \psi \rangle_{L^2} = \big\langle ( dJ\wedge \Gamma + J\, d\Gamma), \underleftarrow{\text{div}}(\psi) \big\rangle_{L^2} \cr
\langle \vec{B}, d \varphi \rangle_{L^2}  = 0 ,  
\end{cases}
\eeq
holds for any vector valued $2$-form $\psi \in W^{1,p^*}_0(\Omega)$ and any vector valued $0$-form $\varphi \in W^{1,p^*}_0(\Omega)$. From the definition of $A$ in \eqref{A'}, we find that $\vec{B} = \vec{A} + \overrightarrow{\langle d J ; \Gammati \rangle}$ and substituting this for $\vec{A}$ in \eqref{proofThmgauge_eqn_20} gives us
\beq
\begin{cases} \nonumber
\langle \vec{A}, \delta \psi \rangle_{L^2} 
= \big\langle ( dJ\wedge \Gamma + J\, d\Gamma), \underleftarrow{\text{div}}(\psi) \big\rangle_{L^2} - \langle \overrightarrow{\langle d J ; \Gammati\rangle} ,\delta \psi \rangle_{L^2} \cr
\langle \vec{A}, d \varphi \rangle_{L^2}  = -\langle v,  \varphi \rangle_{L^2} ,  
\end{cases}
\eeq
for $v = \delta \langle d J ; \Gammati \rangle$. This proves that $(J,\Gammati,A)$ solves the weak RT-equations \eqref{weak_RT_2} - \eqref{weak_RT_A}, and completes the proof of Lemma \ref{Lemma4_Thmgauge}.
\QED

For completeness we prove the following technical lemma which was used in the proof of Lemma \ref{Lemma1_Thmgauge} above. The point was that $\psi \equiv J^T \phi$ can be taken as test function in the same space as the test function $\phi$ used in the argument. It suffices to prove the lemma for products $\phi J$ instead of $J^T \phi$, since only regularity of components is at issue here. It is only in this lemma where we require the assumption $p>2$.

\begin{Lemma} \label{Lemma2_Thmgauge}
Let $J \in W^{1,2p}(\Omega)$ for $p>\max \{n/2,2\}$, $n \geq 2$. Assume $J$ is invertible with inverse $J^{-1} \in W^{1,2p}(\Omega)$. Then $\phi J \in W^{1,p^*}_0(\Omega)$ for any $\phi \in W^{1,p^*}_0(\Omega)$, and for every $\psi \in W^{1,p^*}_0(\Omega)$ there exists some $\phi \in W^{1,p^*}_0(\Omega)$ such that $\psi = \phi J$. That is, $ J \mm W^{1,p^*}_0(\Omega) = W^{1,p^*}_0(\Omega)$.
\end{Lemma}

\Proof
So let $J \in W^{1,2p}(\Omega)$ for $p>\max \{n/2,2\}$. We first show that $\phi J \in W^{1,p^*}_0(\Omega)$ for any $\phi \in W^{1,p^*}_0(\Omega)$. To begin, observe that by the Leibniz rule, we obtain
\begin{eqnarray} \label{proofThmgauge_eqn8}
\|\phi J \|_{W^{1,p^*}} 
&\equiv & \|\phi J \|_{L^{p^*}} + \|d(\phi J)\|_{L^{p^*}}   \cr
&\leq & \|\phi J \|_{L^{p^*}} + \|d\phi\mm J\|_{L^{p^*}}  + \|\phi \, dJ\|_{L^{p^*}}.
\end{eqnarray}
By Morrey's inequality the first term in \eqref{proofThmgauge_eqn8} can be bounded, namely  
\beq \nonumber
\|\phi J \|_{L^{p^*}} \leq C \|\phi\|_{L^{p^*}} \| J \|_{L^{\infty}}  \leq C  \|\phi\|_{L^{p^*}} \| J \|_{W^{1,2p}} 
\eeq 
and similarly
\beq \nonumber
\|d\phi\mm J \|_{L^{p^*}} \leq C \|d\phi\|_{L^{p^*}} \| J \|_{W^{1,2p}} .
\eeq 
So, to prove $\phi J \in W^{1,p^*}_0(\Omega)$, it remains to show the third term in \eqref{proofThmgauge_eqn8} is bounded. 
For this, by \eqref{def_norms}, we first write
\beq \nonumber
\|\phi\, dJ\|_{L^{p^*}}
\overset{\eqref{def_norms}}{=} \sum_{\mu,\nu,\sigma,j} \|\phi^\mu_\sigma\, \partial_j J^\sigma_\nu\|_{L^{p^*}} 
= \sum_{\mu,\nu,\sigma,j} \Big( \big\| |\phi^\mu_\sigma|^{p^*}\, |\partial_j J^\sigma_\nu|^{p^*}\big\|_{L^1}\Big)^{\frac{1}{p^*}}.
\eeq
Applying now H\"older's inequality \eqref{Holder} to each term separately gives
\begin{eqnarray} \nonumber
\|\phi\, dJ\|_{L^{p^*}}
&\overset{\eqref{Holder}}{\leq} & \sum_{\mu,\nu,\sigma,j} \Big(\big\| |\phi^\mu_\sigma|^{p^*}\big\|_{L^q} \, \big\| |\partial_j J^\sigma_\nu|^{p^*}\big\|_{L^{\frac{2p}{p^*}}}  \Big)^{\frac{1}{p^*}} \cr
& = & \sum_{\mu,\nu,\sigma,j} \big\| \phi^\mu_\sigma\big\|_{L^{qp^*}} \, \big\| \partial_j J^\sigma_\nu \big\|_{L^{2p}} \cr
& \leq & \Big(  \sum_{\mu,\nu} \big\| \phi^\mu_\nu\big\|_{L^{qp^*}} \Big) \Big( \sum_{\mu,\nu,j}\big\| \partial_j J^\mu_\nu \big\|_{L^{2p}}\Big), 
\end{eqnarray}
where $q$ is the conjugate exponent of $\frac{2p}{p^*}$,  (i.e., $\frac{1}{q} + \frac{p^*}{2p} =1$). By our definitions of norms in \eqref{def_norms}, the sought after estimate now follows directly,
\beq \label{proofThmgauge_eqn_10}
\|\phi\, dJ\|_{L^{p^*}}  \leq  \| \phi \|_{L^{qp^*}} \, \| dJ \|_{L^{2p}}.
\eeq     
We conclude that to prove $\phi J \in W^{1,p^*}_0(\Omega)$, it suffices to show that $\phi \in W^{1,p^*}_0(\Omega)$ implies $\| \phi \|_{L^{qp^*}}$ to be finite. For this, we apply the Sobolev embedding Theorem for bounded domains \cite[Thm 2, Ch. 5.6]{Evans}, stating that $W^{1,p^*}(\Omega) \subset L^{\frac{np^*}{n-p^*}}(\Omega)$ for $1\leq p^* <n$, (note that $p>2$ implies $p^* \in (1,2)$, so $p^* <n$ holds).  Thus the boundedness of $\| \phi \|_{L^{qp^*}}$ follows from the Sobolev embedding Theorem, as long as 
\beq \label{proofThmgauge_eqn_11}
qp^* \leq  \frac{np^*}{n-p*}. 
\eeq
To verify \eqref{proofThmgauge_eqn_11}, we first compute that 
\beq \label{proofThmgauge_eqn_12}
q = \frac{2}{3-p^*},
\eeq
as follows; inserting into the defining identity $\frac{1}{q} + \frac{p^*}{2p} =1$ that $p^* = \frac{p}{p-1}$ and solving for $q$, we find that $q= \frac{2(p-1)}{2(p-1)-1} = \frac{2}{2-\frac{1}{p-1}} = \frac{2}{3-p^*}$, where the last equality follows from the identity $p^* = \frac{p}{p-1} = 1 + \frac{1}{p-1}$. To continue, we substitute \eqref{proofThmgauge_eqn_12} into the left hand side of \eqref{proofThmgauge_eqn_11}, and show that \eqref{proofThmgauge_eqn_11} holds if and only if $p > n/2$. For this, recall that $p^* \in (1,2)$ and $n\geq 2$, so $n-p^* > 0$ and $3-p^* > 0$. This now allows us to write \eqref{proofThmgauge_eqn_11} equivalently as 
\beq \label{proofThmgauge_eqn_12b}
\frac{p^*}{p^*-1} \geq n/2.
\eeq 
Since $p = \frac{p^*}{p^*-1}$ by $\frac1{p} + \frac1{p^*}=1$, \eqref{proofThmgauge_eqn_12b} is equivalent to $p \geq n/2$ which holds by assumption. This shows that \eqref{proofThmgauge_eqn_11} holds and thereby proves that $\phi J \in W^{1,p^*}_0(\Omega)$ for any $\phi \in W^{1,p^*}_0(\Omega)$, which is the forward implication of Lemma \ref{Lemma2_Thmgauge}.

To prove the backward implication, that for every $\psi \in W^{1,p^*}_0(\Omega)$ there exists a $\phi \in W^{1,p^*}_0(\Omega)$ such that $\psi = \phi J$, we make the ansatz $\phi \equiv \psi J^{-1}$. Since $J^{-1} \in W^{1,2p}(\Omega)$ by assumption, we use the forward implication of Lemma \ref{Lemma2_Thmgauge} to conclude that  $\phi  = \psi J^{-1} \in W^{1,p^*}_0(\Omega)$, while $\phi J = \psi$ holds trivially. This proves the backward implication and completes the proof of Lemma \ref{Lemma2_Thmgauge}.
\QED

\noindent Taken together, Lemmas \ref{Lemma1_Thmgauge} - \ref{Lemma4_Thmgauge} complete the proof of Theorem \ref{Thm_gauge_existence}. To complete the proofs of Theorems \ref{Thm_Smoothing} and \ref{Thm_compactness} it remains to prove Theorem \ref{Thm_Existence_J}, asserting existence of the solutions of the reduced RT-equations precisely as assumed in this section. This is accomplished in Sections \ref{Sec_existence_theory} - \ref{Sec_Lp-extension}.

\section{Existence theory for the reduced RT-equations - Proof of Theorem \ref{Thm_Existence_J}}  \label{Sec_existence_theory}

In this section we prove Theorem \ref{Thm_Existence_J}, regarding existence of solutions to the reduced RT-equations \eqref{RT_withB_2} - \eqref{RT_withB_4} which meet the assumptions of Theorem \ref{Thm_gauge_existence}. This is the final step remaining to complete the proof of Theorems \ref{Thm_Smoothing} and \ref{Thm_compactness}. The proof of Theorem \ref{Thm_Existence_J} is based on an iteration scheme which reduces the problem to known estimates in elliptic PDE theory, recorded in Appendix \ref{Sec_Prelimiaries-elliptic}. To handle the first order system of equations for $B$ \eqref{RT_withB_3} - \eqref{RT_withB_4}, we extend the existence theory for Cauchy-Riemann type equations in \cite{Dac} to the low regularity required here. This extension is presented in Appendix \ref{Sec_A}. The proof of Theorem \ref{Thm_Existence_J} is given in terms of several technical lemmas whose proofs are postponed to Section \ref{Sec_Proofs}. We give the proof of Theorem \ref{Thm_Existence_J} first under the stronger assumption $\Gamma, d\Gamma \in L^{\infty}(\Omega)$, for which our scaling argument below is cleaner, and extend the proof to connection regularity $\Gamma \in L^{2p}, d\Gamma \in L^p$, $p>n/2$, in Section \ref{Sec_Lp-extension}. For simplicity we establish here solutions $B$ in $L^{2p}$ and explain in Section \ref{Sec_Lp-extension} the modification to $B\in L^p$.

So assume $\Gamma, d\Gamma \in L^{\infty}(\Omega)$ in $x$-coordinates, let $M>0$ be a constant such that 
$$
\|(\Gamma,d\Gamma)\|_{L^\infty(\Omega)} \equiv \|\Gamma \|_{L^\infty(\Omega)} + \|d\Gamma \|_{L^\infty(\Omega)} \; \leq\;  M.
$$
Again, we work in fixed $x$-coordinates and we write $\Omega$ for $\Omega_x$ and $\Gamma$ for $\Gamma_x$ throughout the remainder of this paper. Let  $q\in \Omega$ and let $n<p<\infty$. Then, to prove Theorem \ref{Thm_Existence_J}, it suffices to prove that there exists a neighborhood $\Omega' \subset \Omega$ of $x(q)$, depending only on $\Omega, n, p, M$, and there exists $J \in W^{1,p}(\Omega')$ and $B \in L^{p}(\Omega')$ such that $(J,B)$ solves the reduced RT-equations \eqref{RT_withB_2} - \eqref{RT_withB_4} in $\Omega'$ in the weak sense of Definition \ref{Def_weak_reduced_RT}, such that $(J,B)$ satisfies the uniform bound \eqref{curvature_estimate_soln}, $J$ is invertible with $J^{-1} \in W^{1,p}(\Omega')$ and 
\beq \label{integrability_condition_reducedRT}
d\vec{J} \equiv {\rm Curl}(J) \; = \; 0
\eeq 
in $\Omega'$, (implying that $J$ is integrable to coordinates). For ease of notation we show $J \in W^{1,p}$ and $B\in L^{p}$ for $p>n$, instead of $J \in W^{1,2p}$ and $B\in L^{2p}$ for $p>n/2$. For simplicity we assume without loss of generality that $w=0$ in \eqref{RT_withB_4}, that is, we prove existence of a solution $(J,B)$ of 
\begin{eqnarray} 
\Delta J &=& \delta ( J \mm \Gamma ) -  B , \label{RT_reduced_J} \\
d \vec{B} &=& \overrightarrow{\text{div}} \big(dJ \wedge \Gamma\big) + \overrightarrow{\text{div}} \big( J\, d\Gamma\big) ,   \label{RT_reduced_A1}\\
\delta \vec{B} &=& 0,  \label{RT_reduced_A2}
\end{eqnarray}
in the weak sense specified in Definition \ref{Def_weak_reduced_RT}. Without loss of generality we assume that $\Omega$ is the unit ball  in $\R^n$ centered at $x(q)=0$, $\Omega = B_1(0)$. We show below that it suffices to take $\Omega'=B_\epsilon(0)$, the ball of radius $\epsilon >0$ centered at $x=0$, where $\epsilon >0$ is taken sufficiently small for the iteration scheme to converge. We begin the proof of Theorem \ref{Thm_Existence_J} by giving a formal introduction to the iteration scheme on which our existence proof is based.

\subsection{The iteration scheme} \label{Sec_iteration_intro}

Start with $J_0=\1$. For induction, we show that $(B_{k+1},J_{k+1})$ can be constructed from $J_{k}$ for each $k\geq0$. So assume $J_k$ is given for some $k\geq 0$.  Define $B_{k+1}$ as a weak solution of 
\beq \label{iteration_eqnA}
\begin{cases}
d\vec{B}_{k+1} = \overrightarrow{\text{div}} \big(dJ_k \wedge \Gamma\big) + \overrightarrow{\text{div}} \big( J_k\, d\Gamma\big), \cr
\delta \vec{B}_{k+1}=0,
\end{cases}
\eeq    
such that $B_{k+1} \in L^p(\Omega)$ satisfies a uniform bound in the $L^p$-norm. The regularity $L^p$ is too low to impose boundary data in \ref{iteration_eqnA}, and our theory does not require $B_{k+1}$ to meet any boundary conditions, $B_{k+1}$ only needs to satisfy a uniform $L^p$ bound. This is achieved by choosing $B_{k+1}$ to be the zero mollification limit of a solution of the corresponding mollified equation with zero Dirichlet boundary data, c.f. Sections \ref{Sec_iteration_scheme} and \ref{Sec_A}. 

Likewise, the regularity $J\in W^{1,p}(\Omega)$ is too low to impose the boundary condition \eqref{RT_data}, $d\vec{J}=0$ on $\partial \Omega$, a problem we circumvent by imposing $\vec{J} = dy$, for $y$ solving an auxiliary elliptic equation. For this, define auxiliary variables $\Psi_{k+1}$ and $y_{k+1}$ in terms of $J_k$ and $B_{k+1}$, but independent of the previous iterates $\Psi_{k}$ and $y_{k}$. That is, we define the vector valued $0$-form $\Psi_{k+1} \in L^p(\Omega)$ as a weak solution of  
\beq \label{iteration_eqnPsi}
d\Psi_{k+1}  = \overrightarrow{\delta ( J_k \mm \Gamma )} - \overrightarrow{B_{k+1}} , 
\eeq
with uniform $L^p$ bounds, obtained again by mollification in the same manner as in the case for $B_{k+1}$. We then define the vector valued $0$-form $y_{k+1} \in W^{2,p}(\Omega)$ as the solution of  
\beq \label{iteration_eqn_y}
\Delta y_{k+1}=\Psi_{k+1},
\eeq
for zero Dirichlet data. Given $B_{k+1},\Psi_{k+1}$ and $y_{k+1}$, now define $J_{k+1} \in W^{1,p}(\Omega)$ as the weak solution of the following Dirichlet boundary value problem:   
\begin{eqnarray} \label{iteration_eqn_J}
\Delta J_{k+1} &=& \delta ( J_k \mm \Gamma ) -  B_{k+1}  ,\\
\overrightarrow{J_{k+1}} &=& dy_{k+1} \ \ {\rm on}\ \ \partial\Omega. \label{iteration_eqn_y_bdd}
\end{eqnarray}      
Equations \eqref{iteration_eqnA} - \eqref{iteration_eqn_J} define our iteration scheme in a formal way. To prove convergence we need a small parameter $\epsilon >0$. We incorporate $\epsilon$ into the iteration scheme in Sections \ref{Sec_small_parameter} - \ref{Sec_iteration_scheme}, and prove convergence for $\epsilon>0$ sufficiently small in Section \ref{Sec_convegrence}.

Two clarifying remarks are in order. First note that \eqref{iteration_eqnPsi} requires a solvability condition, namely that $d$ of its right hand side must be zero. This condition is meet, because by \eqref{regularity-miracle} we have
\beq \nonumber
d\Big(\overrightarrow{\delta ( J_k \mm \Gamma )}\Big) = \overrightarrow{\text{div}} \big(dJ_k \wedge \Gamma\big) + \overrightarrow{\text{div}} \big( J_k\mm  d\Gamma\big),
\eeq
which implies in light of equation \eqref{iteration_eqnA} for $B_{k+1}$ that
$$
d\Big(\overrightarrow{\delta ( J_k \mm \Gamma )} - \overrightarrow{B_{k+1}} \Big) =0,
$$
so the right hand side of \eqref{iteration_eqnA} has a vanishing exterior derivative. That this consistency condition is necessary and sufficient for the low regularity here is shown in Appendix \ref{Sec_A}.

Secondly, we remark on the role of auxiliary equations \eqref{iteration_eqnPsi} - \eqref{iteration_eqn_y}. The reason for introducing $\Psi_{k}$ and $y_{k+1}$ is that the $W^{1,p}$-regularity of $J_{k+1}$ is too low to impose the boundary data $d\vec{J}_{k+1}=0$ which was required in \cite{ReintjesTemple_ell1} to arrange for the integrability condition of $J$ \eqref{integrability_condition_reducedRT}. Now, augmenting the reduced RT-equations by equations \eqref{iteration_eqnPsi} and \eqref{iteration_eqn_y}, allows us to impose Dirichlet data for $J_{k+1}$ which again gives rise to integrability of $J_{k+1}$ to coordinates, as we show in the following lemma for smooth solutions. In Lemma \ref{Lemma_curl} below we extend this result to the low regularities required by the above iteration scheme. 

\begin{Lemma} \label{Lemma_integrability_J}
Assume $\Gamma$ is smooth and that $B_{k+1}, \Psi_{k+1}, y_{k+1}, J_{k+1}$ are defined by the iteration scheme \eqref{iteration_eqnA} - \eqref{iteration_eqn_y_bdd} and are smooth. Then $dy_{k+1} =\vec{J}_{k+1}$, and hence $J_{k+1}$ is integrable to coordinates $y_{k+1}$ and $d\vec{J}_{k+1}=0$.
\end{Lemma}

\Proof
By \eqref{iteration_eqnA} - \eqref{iteration_eqn_y_bdd} it follows that
\begin{eqnarray}\nonumber
\Delta dy_{k+1}=d\Delta y_{k+1}=d\Psi_{k+1}=\overrightarrow{\delta ( J_{k} \mm \Gamma )} - \overrightarrow{B_{k+1}} = \Delta\vec{J}_{k+1},
\end{eqnarray} 
where the last equality holds, since the operation $vec$ commutes with the Laplacian $\Delta$  (which acts component wise).  Thus,
\begin{eqnarray} \nonumber
&\Delta(\vec{J}_{k+1}-dy_{k+1})=0\ \ {\rm in}\ \ \Omega,&\\
&\vec{J}-dy=0\ \  {\rm on}\ \ \partial\Omega,&  
\end{eqnarray} 
which implies by uniqueness of solutions of the Laplace equation that $\vec{J}_{k+1}=dy_{k+1}$ in $\Omega$. Since second derivatives of $y_{k+1}$ commute, we conclude that 
\beq  \nonumber
d\vec{J}_{k+1}=Curl(\vec{J}_{k+1})=0 \ \ \ \ \text{in} \ \ \Omega.
\eeq 
Moreover, $J_{k+1}$ is the Jacobian of the coordinate system $y_{k+1}$.
\QED

Lemma \ref{Lemma_curl} below generalizes the above result to the low regularity required in this paper. To prove convergence of the iteration scheme, we introduce a small parameter $\epsilon >0$ by restricting to $\Omega'=B_\epsilon(0)$, and prove convergence for $\epsilon>0$ sufficiently small.

\subsection{The $\epsilon$-rescaled reduced RT-equations} \label{Sec_small_parameter} 

We first incorporate the small parameter $\epsilon>0$ into the theory by deriving an $\epsilon$-rescaled version of the reduced RT-equations, required to prove convergence of our iteration scheme. For this we use the fact that regularity is a local problem, so that we can suitably restrict and rescale $\Gamma$ to isolate  the small parameter $\epsilon$, while maintaining the uniform bound \eqref{bound_incoming_ass} assumed in Theorems \ref{Thm_Smoothing} and \ref{Thm_compactness}. This is accomplished in the following lemma. 

\begin{Lemma} \label{rescaling_Gamma}
Assume $\Omega=B_1(0)$ and introduce the coordinate transformation $x \to \tilde{x}(x) = \frac{x}{\epsilon}$. Define $\Gamma^*$ as the restriction of the components of $\Gamma_x$ to $B_\epsilon(0)$, transformed to $\tilde{x}$-coordinates as scalars, $\Gamma^*(\tilde{x}) \equiv \Gamma_x(x(\tilde{x})).$ 
Then, $\Gamma_{\tilde{x}}$ satisfies in $\tilde{x}$-coordinates 
\beq \label{rescaling_Gamma*}
\Gamma_{\tilde{x}}(\tilde{x})=\epsilon\; \Gamma^*(\tilde{x}),
\eeq 
together with the bound
\beq \label{rescaling_bound_Gamma*}
\|(\Gamma^*,d\Gamma^*)\|_{L^\infty(\Omega'_{\tilde{x}})} = \|\Gamma_x\|_{L^\infty(\Omega'_x)} + \epsilon \|d\Gamma_x\|_{L^\infty(\Omega'_x)},
\eeq
where $\Omega'_{\tilde{x}} = B_1(0)$ and $\Omega'_x = B_\epsilon(0)$.
\end{Lemma}

\Proof
By the connection transformation law we have
\beq \label{rescaling_connection_transfo}
(\Gamma_{\tilde{x}})^\sigma_{\mu\nu} 
\ =\ \frac{\partial \tilde{x}^\sigma}{\partial x^k} \Big( \frac{\partial x^i}{\partial \tilde{x}^\mu} \frac{\partial x^j}{\partial \tilde{x}^\nu} (\Gamma_{x})^k_{ij} +  \frac{\partial^2 x^k}{\partial \tilde{x}^\mu \partial \tilde{x}^\nu}   \Big)
\ =\ \epsilon\: (\Gamma_{x})^\sigma_{\mu\nu} ,
\eeq 
since $\frac{\partial x^i}{\partial \tilde{x}^j} = \epsilon\, \delta^i_{\; j}$ under the transformation $\tilde{x}(x) = \frac{x}{\epsilon}$. It follows that for $\tilde{x} \in B_1(0)$, we have component wise
$$
\Gamma_{\tilde{x}}(\tilde{x})= \epsilon \;\Gamma_x(x(\tilde{x})) \equiv \epsilon\; \Gamma^*(\tilde{x}).
$$ 
To prove \eqref{rescaling_bound_Gamma*}, observe that by construction of $\Gamma^*$, as the scalar transformed components of the restriction of $\Gamma_x$ to the ball of radius $\epsilon$, we have $\| \Gamma^* \|_{L^\infty(\Omega'_{\tilde{x}})} = \|\Gamma_x\|_{L^\infty(\Omega_x')}$, and by the chain rule we have $\|d\Gamma^*\|_{L^\infty(\Omega_{\tilde{x}}')} = \epsilon \| d\Gamma_x\|_{L^\infty(\Omega'_x)}$ since $\frac{\partial x}{\partial \tilde{x}} = \epsilon I$. In combination, this gives \eqref{rescaling_bound_Gamma*} and completes the proof.
\QED

By \eqref{rescaling_bound_Gamma*}, and assuming without loss of generality that $\epsilon \leq 1$, we find that $\Gamma^*$ satisfies the original uniform bound \eqref{bound_incoming_ass}, 
$$
\|(\Gamma^*,d\Gamma^*)\|_{L^\infty(\Omega_{\tilde{x}}')} \leq  M .
$$ 
We can thus construct solutions to the reduced RT-equations and apply Theorem \ref{Thm_gauge_existence} in $\tilde{x}$-coordinates, and obtain the uniform curvature bound \eqref{curvature} in $\tilde{x}$-coordinates, without ever scaling back to $x$-coordinates.\footnote{A note might be in order here regarding our proof of Uhlenbeck compactness.   In the proof we implicitly assumed that there is a uniform $\epsilon>0$ which applies uniformly to each connection in the sequence.  In the Proposition\ref{Prop3} below we show that $\epsilon$ can be taken to be on the order of $\frac{1}{M}$, and therefore independent of connections in the sequence.}  

So now we can take the $\tilde{x}$-coordinates to be the original $x$-coordinates, and assume without loss of generality that the connection in $x$-coordinates has the form 
\begin{eqnarray}
\Gamma_x=\epsilon\; \Gamma^* ,  \label{small_Gamma}
\end{eqnarray} 
for some $\Gamma^*$ satisfying
\begin{eqnarray} \label{Gamma-bound}
\|(\Gamma^*,d\Gamma^*)\|_{L^\infty(\Omega_x)}   < M,
\end{eqnarray}   
and we assume without loss of generality that $\Omega_x=B_1(0)$. In light of \eqref{small_Gamma}, we introduce the scaling ansatz 
\beq \label{ansatz_scaling}
J=I+\epsilon \, u , \hspace{1cm} 
B = \epsilon\, a.
\eeq 
Since we only need to prove {\it existence} of a solution to establish optimal connection regularity via the RT-equations, assumption \ref{ansatz_scaling} is made without loss of generality. Note, the variables $\Psi$ and $y$ ``inherit'' their $\epsilon$-scaling from $B$ and $J$, c.f. Section \ref{Sec_iteration_scheme} below. To derive the reduced RT-equations expressed in terms of the rescaled variables, we now substitute \eqref{small_Gamma} and \eqref{ansatz_scaling} into \eqref{RT_reduced_J} - \eqref{RT_reduced_A2} and divide by $\epsilon$. This yields the following equivalent set of equations:

\begin{Lemma}   \label{Lemma_rescaled_RT-eqn}
The reduced RT-equations \eqref{RT_reduced_J_weak} - \eqref{RT_reduced_A_weak} written in terms of the rescaled connection \eqref{small_Gamma} and rescaled variables \eqref{ansatz_scaling} are equivalent to
\begin{eqnarray}
-\Delta u[\phi] = F_u(u,a)[\phi], \label{pde_u}  \\
\begin{cases} \label{pde_a}
\langle \vec{a}, \delta \psi \rangle_{L^2} =  F_a(u)[\psi]  \\
\langle \vec{a}, d \varphi \rangle_{L^2} = 0 ,
\end{cases}
\end{eqnarray}
for any matrix valued $0$-form $\phi \in W^{1,p^*}_0(\Omega)$, any vector valued $2$-form $\psi \in W^{1,p^*}_0(\Omega)$ and any vector valued function $\varphi \in W^{1,p^*}_0(\Omega)$, where we define the linear functionals
\begin{align}  
& F_u(u,a)[\phi] \equiv  \big\langle \Gamma^* , d\phi \big\rangle_{L^2}  + \epsilon\: \big\langle u \mm \Gamma^* , d \phi \big\rangle_{L^2} + \big\langle a, \phi \big\rangle_{L^2} , \label{Def_Fu}  \\  
& F_a(u)[\psi]  \equiv  \big\langle d\Gamma^*, \underleftarrow{\text{div}}(\psi) \big\rangle_{L^2} + \epsilon\: \big\langle ( u \mm d\Gamma^* + du \wedge \Gamma^*), \underleftarrow{\text{div}}(\psi) \big\rangle_{L^2} . \label{Def_Fa}
\end{align}
\end{Lemma}                 

\Proof
Substituting $J= I + \epsilon u$ and $B= \epsilon a$ into \eqref{RT_reduced_J_weak}  and dividing by $\epsilon$, we obtain
\begin{eqnarray}  \nonumber
-\Delta u[\phi] &\overset{\eqref{Laplacian_weak}}{=}& \langle d u, d \phi \rangle_{L^2}  \cr
&=&  \langle  (I + \epsilon u) \mm \Gamma^* , d \phi \rangle_{L^2} + \langle a , \phi \rangle_{L^2}       \cr
 &=&  \langle \Gamma^* , d \phi \rangle_{L^2} + \epsilon \langle  u \mm \Gamma^* , d \phi \rangle_{L^2}  + \langle a , \phi \rangle_{L^2}    \cr 
&=&  F_u(u,a)[\phi]
\end{eqnarray}
which proves the equivalence between \eqref{RT_reduced_J} and \eqref{pde_u}. Similarly, substituting our scaling ansatz $J= I + \epsilon u$ and $B= \epsilon a$ into \eqref{RT_reduced_A_weak}, a division by $\epsilon$ gives
\begin{eqnarray} \nonumber
\langle \vec{a}, \delta \psi \rangle_{L^2}  &=&  \big\langle ( \epsilon du\wedge \Gamma^* + (I+\epsilon u)\mm d\Gamma^*), \underleftarrow{\text{div}}(\psi) \big\rangle_{L^2}           \cr
&=&  \big\langle  d\Gamma^*, \underleftarrow{\text{div}}(\psi) \big\rangle_{L^2} + \epsilon \big\langle ( du\wedge \Gamma^* + u\mm d\Gamma^*), \underleftarrow{\text{div}}(\psi) \big\rangle_{L^2}           \cr
&=&  F_a(u)[\psi]   
\end{eqnarray}
as well as $\langle \vec{a}, d \varphi \rangle_{L^2} = 0$, which proves the equivalence between \eqref{RT_reduced_A_weak} and \eqref{pde_a}. 
\QED

The existence result of Theorem \ref{Thm_Existence_J} is a corollary of the following proposition, the proof of which is topic of Sections \ref{Sec_iteration_scheme} - \ref{Sec_Proofs}.    

\begin{Prop}  \label{Prop2}
Let $\Gamma^*, d\Gamma^* \in L^\infty(\Omega)$ satisfy the bound \eqref{Gamma-bound} and let $n<p<\infty$. Then, for every $\epsilon>0$ sufficiently small, there exists $u \in W^{1,p}(\Omega)$ and $a \in L^p(\Omega)$ which solve the $\epsilon$-rescaled reduced RT-equation \eqref{pde_u} - \eqref{pde_a}.
\end{Prop}

The proof of Proposition \ref{Prop2} is based on the iteration scheme introduced in Section \ref{Sec_iteration_intro}, but adjusted to incorporate the small parameter $\epsilon$. The resulting iteration scheme for the $\epsilon$-rescaled reduced RT-equations is introduced in the next Section \ref{Sec_iteration_scheme}. The proof of our main existence result, Theorem \ref{Thm_Existence_J}, is completed in Section \ref{Sec_Proof_existence_Thm} by applying Proposition \ref{Prop2} together with additional arguments to establish the integrability and invertability of the Jacobian $J=I+\epsilon \, u$, claimed in the theorem, as well as the uniform bound \eqref{curvature_estimate_soln}.

\subsection{The iteration scheme in the $\epsilon$-rescaled variables}   \label{Sec_iteration_scheme}

In this section we define the iterates $(u_{k},a_{k})$, $k\geq0$, for approximating solutions of (\ref{pde_u})-(\ref{pde_a}), and set up the framework for proving convergence of the scheme in the appropriate Sobolev spaces for $\epsilon$ sufficiently small. The iteration scheme we introduce here differs from the iteration scheme in Section \ref{Sec_iteration_intro} in that it is adapted to the rescaled equations (\ref{pde_u})-(\ref{pde_a}). Existence at each stage will be established in Lemma \ref{Lemma_existence_iterates} below. From here on we often omit dependence of norms on $\Omega$, e.g., writing $\|\cdot\|_{L^p}$ for $\|\cdot\|_{L^p(\Omega)}$. We define now the matrix valued $0$-forms $u_{k+1}$ and $a_{k+1}$ by induction as follows.

To start the iteration, set 
$$u_0=a_0=0.$$   
Given $u_k \in W^{1,p}(\Omega)$ and $a_k \in L^p(\Omega)$ for $k\geq0$, we then construct a particular matrix valued $0$-form $a_{k+1} \in L^p(\Omega)$ which solves                    
\beq \label{iterate_a}
\begin{cases} 
\langle \overrightarrow{a_{k+1}}, \delta \psi \rangle_{L^2} = F_a(u_k)[\psi], \cr 
\langle \overrightarrow{a_{k+1}}, d\varphi \rangle_{L^2}=0,
\end{cases}
\eeq
and satisfies the estimate 
\beq \label{iterate_a_Lp-bound}  
\| a_{k+1} \|_{L^p} \leq C \|F_a(u_k)\|_{W^{-1,p}} ,
\eeq 
for some constant $C>0$ independent of $k$, where \eqref{iterate_a} are taken in the weak sense specified in Lemma \ref{Lemma_rescaled_RT-eqn}, i.e., \eqref{iterate_a} shall hold for any vector valued $2$-form $\psi \in W^{1,p^*}_0(\Omega)$ and any vector valued function $\varphi \in W^{1,p^*}_0(\Omega)$. Existence is established Lemma \ref{Lemma_existence_iterates} and relies on the algorithm developed in Appendix \ref{Sec_A} for constructing particular solutions when regularity is too low to impose Dirichlet data in a classical sense. 

We next introduce the vector valued $0$-form $\Psi_{k+1} \in L^p(\Omega)$ as a weak solution of
\beq \label{iterate_psi_strong-form}                
d \Psi_{k+1} = \overrightarrow{\delta ( J_k \mm \Gamma^* )}  - \overrightarrow{a_{k+1}} , 
\eeq
where $J_k \equiv I+\epsilon u_k$, such that $\Psi_{k+1}$ meets the bound 
\beq \label{iterate_psi_bound} 
\| \Psi_{k+1} \|_{L^p} \leq C \|F_a(u_k)\|_{W^{-1,p}} ,
\eeq 
for some constant $C>0$ independent of $k$. That is, $\Psi_{k+1} \in L^p(\Omega)$ meets the bound \eqref{iterate_psi_bound} and satisfies
\beq \label{iterate_psi} 
\langle \Psi_{k+1}, \delta \vec{\phi} \rangle_{L^2} =  F_{\Psi}(u_k,a_{k+1})[\phi],
\eeq
for any matrix-valued $0$-form $\phi \in W^{1,p^*}_0(\Omega)$, where we set       
\beq \label{iterate_psi_DefF} 
F_{\Psi}(u_k,a_{k+1})[\phi]   \equiv  \langle J_k \Gamma^* , d\phi \rangle_{L^2}  + \langle \overrightarrow{a_{k+1}}, \vec{\phi} \rangle_{L^2} .
\eeq 
The definition in \eqref{iterate_psi_DefF} is based on the product rule $\langle \overrightarrow{\delta w} , \vec{\phi} \rangle_{L^2} = - \langle w , d\phi \rangle_{L^2} $ for matrix valued $1$-forms $w\in W^{1,p}(\Omega)$, established in the proof of Lemma \ref{Lemma2_weak_psi_eqn} below. Because of this product rule it is convenient to interpret the test forms in \eqref{iterate_psi} and \eqref{iterate_psi_DefF} as matrix valued $0$-forms instead of vector valued $1$-forms. In Lemma \ref{Lemma2_weak_psi_eqn} below we show that the weak formulation  of \eqref{iterate_psi} is equivalent to the strong formulation \eqref{iterate_psi_strong-form} in the case of smooth solutions. Existence of $\Psi_{k+1}$ also follows from Lemma \ref{Lemma_existence_iterates} by use of the algorithm in Appendix \ref{Sec_A}.

We next define the vector valued $0$-form $y_{k+1} \in W^{2,p}(\Omega)$ as the solution of
\beq \label{iterate_y}
\begin{cases} 
\Delta y_{k+1} = \Psi_{k+1}, \cr 
y_{k+1}\big|_{\partial\Omega} = 0.
\end{cases}
\eeq
Similar to Lemma \ref{Lemma_integrability_J}, equations \eqref{iterate_a}, \eqref{iterate_psi} and \eqref{iterate_y} again arrange for the integrability of $J_{k+1}=I+\epsilon u_{k+1}$. 

Finally, we define $u_{k+1} \in W^{1,p}(\Omega)$ as the unique weak solution satisfying  
\beq \label{iterate_u}
- \Delta u_{k+1}[\phi]=F_u(u_k,a_{k+1})[\phi],
\eeq
for every matrix valued $0$-forms $\phi \in W^{1,p^*}_0(\Omega)$, with Dirichlet boundary data   
\begin{eqnarray} 
u_{k+1}|_{\partial\Omega} &=& dy_{k+1}|_{\partial\Omega}.  \label{bdd_J}
\end{eqnarray}
Equations \eqref{iterate_a} - \eqref{bdd_J} define our iteration scheme. For completeness we show in the next lemma that smooth solutions $\Psi_{k+1}$ of the weak equation \eqref{iterate_psi} are indeed strong solutions of \eqref{iterate_psi_strong-form}. 

\begin{Lemma} \label{Lemma2_weak_psi_eqn}
Let $\Psi_{k+1} \in W^{1,p}(\Omega)$ be a vector valued $0$-form, then $\Psi_{k+1}$ solves \eqref{iterate_psi} if and only if $\Psi_{k+1}$ solves \eqref{iterate_psi_strong-form}.
\end{Lemma}  

\Proof
We first prove the following statement: Let $w\in W^{1,p}(\Omega)$ be a matrix valued $1$-form and $\phi \in W^{1,p^*}_0(\Omega)$ a matrix valued $0$-form, then 
\beq \label{Leipnitz_for_psi-eqn}
\langle \overrightarrow{\delta w} , \vec{\phi} \rangle_{L^2} = - \langle w , d\phi \rangle_{L^2} .
\eeq
To prove \eqref{Leipnitz_for_psi-eqn}, let $w= w^\mu_{\nu i} dx^i$ and $\phi = \phi^\mu_\nu$, then 
\beq \nonumber
\delta w = \sum_{i=1,..,n} \partial_i w^\mu_{\nu i}
\hspace{1cm} \text{and} \hspace{1cm}  
\overrightarrow{\delta w}^\mu = (\delta w)^\mu_\nu dx^\nu =  \sum_{i=1}^{n} \partial_i w^\mu_{\nu i} dx^\nu.
\eeq
Using partial integration component wise, we compute
\begin{eqnarray} \nonumber
\langle \overrightarrow{\delta w} , \vec{\phi} \rangle_{L^2} 
&=& \sum_{\mu,\nu} \int_\Omega {\delta w}^\mu_\nu  \phi^\mu_\nu\: dx  
= \sum_{\mu,\nu,i} \int_\Omega \partial_i w^\mu_{\nu i}\,  \phi^\mu_\nu dx \cr
&=& - \sum_{\mu,\nu,i} \int_\Omega w^\mu_{\nu i}\, \partial_i \phi^\mu_\nu dx 
=  - \langle w , d\phi \rangle_{L^2},
\end{eqnarray}
c.f. the definition of inner products on matrix and vector valued differential forms \eqref{inner-product_L2} and \eqref{inner-product_L2_vec}. This proves the sought after equation \eqref{Leipnitz_for_psi-eqn}.

We now apply \eqref{Leipnitz_for_psi-eqn} to prove Lemma \ref{Lemma2_weak_psi_eqn}. So assume $\Psi_{k+1} \in W^{1,p}(\Omega)$ solves  \eqref{iterate_psi}, that is, 
\beq \label{techeqn1_Lemma2_weak_psi}
\langle \Psi_{k+1}, \delta \vec{\phi} \rangle_{L^2} 
= \langle J_k \Gamma^* , d\phi \rangle_{L^2}  + \langle \overrightarrow{a_{k+1}}, \vec{\phi} \rangle_{L^2} ,
\eeq
for any matrix-valued $0$-form $\phi \in W^{1,p^*}_0(\Omega)$. From the partial integration formula \eqref{partial_integration_matrix} for vector valued forms, we find that
\beq \label{techeqn2_Lemma2_weak_psi}
\langle \Psi_{k+1}, \delta \vec{\phi} \rangle_{L^2} = - \langle d\Psi_{k+1}, \vec{\phi} \rangle_{L^2} ,
\eeq
and by \eqref{Leipnitz_for_psi-eqn}, we have
\beq \label{techeqn3_Lemma2_weak_psi}
\langle J_k \Gamma^* , d\phi \rangle_{L^2}  = - \langle \overrightarrow{\delta (J_k \Gamma^*)} , \vec{\phi} \rangle_{L^2}.
\eeq
Combining \eqref{techeqn2_Lemma2_weak_psi} and \eqref{techeqn3_Lemma2_weak_psi}, we write  \eqref{techeqn1_Lemma2_weak_psi} as
\beq \nonumber
\langle  d\Psi_{k+1} , \vec{\phi} \rangle_{L^2} = \Big\langle \Big(\overrightarrow{\delta (J_k \Gamma^*)} - \overrightarrow{a_{k+1}}\Big) , \vec{\phi} \Big\rangle_{L^2},
\eeq
and since this equation holds for any matrix valued $0$-form $\phi \in W^{1,p^*}_0(\Omega)$, we conclude by Riesz representation that the strong form \eqref{iterate_psi_strong-form} holds. The opposite implication is straightforward. This completes the proof of Lemma \ref{Lemma2_weak_psi_eqn}.
\QED

The iteration scheme on which the existence theory for the reduced RT-equations stated in Proposition \ref{Prop2} is based, is defined in \eqref{iterate_a} - \eqref{bdd_J}. Our strategy for completing the proof of Proposition \ref{Prop2} is to first state the main technical lemmas  regarding the iteration scheme being well-defined and convergent, including elliptic estimate for differences of iterates to establish convergence in suitable Sobolev spaces. The statement of these lemmas is the topic of the next section.

\subsection{Well-posedness and convergence of the iteration scheme}   \label{Sec_convegrence}

In this section we state the main lemmas required for the proof of Proposition \ref{Prop2}, and assuming these, we give the proof of Proposition \ref{Prop2}. Proofs of the supporting lemmas are postponed until Section \ref{Sec_Proofs} below. The first lemma provides an apriori estimate for the source terms.

\begin{Lemma}\label{Lemma_source_estimate_Fa_Fu}
Let $\Gamma^*,d\Gamma^* \in L^\infty(\Omega)$ and assume $u \in W^{1,p}(\Omega)$ and $a\in L^p(\Omega)$, for $n<p<\infty$, then 
\begin{align}                      
\|F_u(u,a)\|_{W^{-1,p}} &\leq    \|a\|_{L^p} \, + \,   \big( |\Omega|^\frac{1}{p} +   \epsilon\: \|u\|_{L^p} \big) \|(\Gamma^*,d\Gamma^*)\|_{L^\infty}, \label{bound_Fu}\\
\|F_a(u)\|_{W^{-1,p}} &\leq  \big(|\Omega|^\frac{1}{p} + \epsilon\: \|u\|_{W^{1,p}} \big)  \|(\Gamma^*,d\Gamma^*)\|_{L^\infty}   .   \label{bound_Fa}
\end{align}
\end{Lemma} 

\noindent Lemma \ref{Lemma_source_estimate_Fa_Fu} is proven in Section \ref{Sec_source_estimates} below.  Our second lemma gives the elliptic estimates required to establish that the iteration scheme is well-defined.

\begin{Lemma} \label{Lemma_existence_iterates}
Assume $u_k\in W^{1,p}(\Omega)$ is given, $n < p < \infty$. Then there exists $a_{k+1}\in L^p(\Omega)$ which solves \eqref{iterate_a}, there exists the auxiliary iterates $\Psi_{k+1} \in L^p(\Omega)$ and $y_{k+1} \in W^{2,p}(\Omega)$ which solve \eqref{iterate_psi} - \eqref{iterate_y}, and there exists $u_{k+1}\in W^{1,p}(\Omega)$ which solves \eqref{iterate_u} with boundary data \eqref{bdd_J}. In addition,  the iterates satisfy the following elliptic estimates:  
\begin{eqnarray}
\|a_{k+1}\|_{L^p(\Omega)}  &\leq & C_e \; \|F_a(u_k)\|_{W^{-1,p}(\Omega)}, \label{existence_est1} \\
\|\Psi_{k+1}\|_{L^p(\Omega)}  &\leq & C_e \; \|F_u(u_k,a_{k+1})\|_{W^{-1,p}(\Omega)}, \label{existence_est2} \\
\|y_{k+1}\|_{W^{2,p}(\Omega)} &\leq & C_e \; \|F_u(u_k,a_{k+1})\|_{W^{-1,p}(\Omega)}, \label{existence_est3}\\
\|u_{k+1}\|_{W^{1,p}(\Omega)} &\leq & C_e \; \|F_u(u_k,a_{k+1})\|_{W^{-1,p}(\Omega)}, \label{existence_est4} 
\end{eqnarray}
for some constant $C_e >0$ depending only on $n, p$ and $\Omega$. 
\end{Lemma}

\noindent The proof of Lemma \ref{Lemma_existence_iterates}, given in Section \ref{Sec_existence}, is based on the $L^p$ elliptic estimate \eqref{Poisson-3} of Theorem \ref{Thm_Poisson} and Gaffney's inequality \eqref{Gaffney} of Theorem \ref{Thm_CauchyRiemann}.\footnote{Note that the boundary data \eqref{bdd_J} for $J$, i.e. $J=dy$ on $\partial\Omega$, does not enter estimate \eqref{existence_est4}, since it can be bounded by $\|F_u(u_k,a_{k+1})\|_{W^{-1,p}(\Omega)}$ using \eqref{existence_est3}.} Lemma \ref{Lemma_existence_iterates} directly implies the following corollary.

\begin{Corollary}
The iteration scheme is well-defined.
\end{Corollary}

Before we establish convergence of the iteration scheme, we show that each Jacobian $J_k=I + \epsilon\: u_k$ is integrable to coordinates for each $k \in \mathbb{N}$. This is the subject of the next lemma, proven in Section \ref{Sec_curl} below, which extends Lemma \ref{Lemma_integrability_J} to the low regularities here.

\begin{Lemma} \label{Lemma_curl}
Let $u_{k+1} \in W^{1,p}(\Omega)$ be a solution of \eqref{iterate_u} with boundary data \eqref{bdd_J}, and let $y_{k+1} \in W^{2,p}(\Omega)$ be a solution of \eqref{iterate_y}. Then 
\beq \label{curl_J_iterates}
d\overrightarrow{u_{k+1}} =0
\eeq
in $\Omega$ and $J_{k+1} \equiv I + \epsilon\, u_{k+1}$ is the Jacobian of the coordinate transformation $x \to x +\epsilon\, y_{k+1}(x)$.
\end{Lemma}

We now discuss convergence of the iteration scheme. Lemma \ref{Lemma_existence_iterates} yields a sequence of iterates $(u_k,a_k)_{k\in \mathbb{N}}$. To establish convergence of this sequence in $W^{1,p}\times L^p$, we require estimates on the differences 
\beq
\begin{aligned} \label{diff1}
\overline{a_{k}} &\equiv  a_k- a_{k-1}   
  \ \ \ \ \ \ \text{and} \ \ \ \ \ \
\overline{u_{k}} &\equiv  u_{k}-u_{k-1} 
\end{aligned}
\eeq 
in terms of the corresponding previous difference of iterates, $\overline{a_{k-1}}$ and $\overline{u_{k-1}}$. This is accomplished in the following lemma, proven in Section \ref{Sec_decay}. The proof of the lemma combines the elliptic estimates \eqref{existence_est1} - \eqref{existence_est2} with suitable bounds on differences of source terms by previous differences of iterates in the fashion of the estimates of Lemma \ref{Lemma_source_estimate_Fa_Fu}. 

\begin{Lemma} \label{Lemma_decay}                         
Assume $\Gamma^*, d\Gamma^* \in L^\infty(\Omega)$, then
\begin{eqnarray}
\|\overline{a_{k+1}}\|_{L^p} 
&\leq & \epsilon\: C_d\;\|(\Gamma^*, d\Gamma^*)\|_{L^\infty} \:\|\overline{u_{k}}\|_{W^{1,p}} , \label{decay_a} \\
\|\overline{u_{k+1}}\|_{W^{1,p}}  
&\leq & \epsilon\: C_d \; \|(\Gamma^*, d\Gamma^*)\|_{L^\infty} \:\|\overline{u_{k}}\|_{W^{1,p}}, \label{decay_u} 
\end{eqnarray} 
where $C_d \equiv  C_e (1+C_e) >0$ depends only on $n$, $p$, $\Omega$, through the constant  $C_e>0$  of Lemma \ref{Lemma_existence_iterates}.
\end{Lemma}

\noindent Convergence of the iteration scheme will follow from Lemma \ref{Lemma_decay}, because
$$
\|(\Gamma^*, d\Gamma^*)\|_{L^\infty(\Omega)} \leq M,
$$ 
by \eqref{Gamma-bound}. This is proven in the following proposition, which completes the proof of Proposition \ref{Prop2}, assuming Lemmas \ref{Lemma_source_estimate_Fa_Fu}, \ref{Lemma_existence_iterates} and \ref{Lemma_decay} hold. 

\begin{Prop}   \label{Prop3}
Assume Lemmas \ref{Lemma_source_estimate_Fa_Fu}, \ref{Lemma_existence_iterates} and \ref{Lemma_decay} hold. Let $\Gamma^*, d\Gamma^* \in L^\infty(\Omega)$ satisfy the initial bound \eqref{Gamma-bound}, $\|(\Gamma^*,d\Gamma^*)\|_{L^\infty} < M$ for some constant $M>0$, and assume 
\beq \label{epsilon_bound_2}
0\ < \ \epsilon \ < \ \frac{1}{C_d M } ,
\eeq 
where $C_d>0$ is the constant from Lemma \ref{Lemma_decay}. Then the sequence of iterates $(u_k,a_k)_{k\in \mathbb{N}}$ defined by \eqref{iterate_a} - \eqref{bdd_J} converges in $W^{1,p}(\Omega) \times L^p(\Omega)$, and the corresponding limits 
\begin{eqnarray} \nonumber
& u \equiv \lim\limits_{k\rightarrow\infty} u_k \ \in W^{1,p}(\Omega), & \cr
&a \equiv \lim\limits_{k\rightarrow\infty} a_k \ \in L^p(\Omega), & 
\end{eqnarray}
solve the reduced RT-equations \eqref{pde_u} - \eqref{pde_a} and satisfy the bound
\beq \label{curvature_bound_(u,a)}
\|u\|_{W^{1,p}(\Omega)} + \|a\|_{L^p(\Omega)} \leq C_2(M) \|(\Gamma^*,d\Gamma^*)\|_{L^\infty(\Omega)} , 
\eeq 
for some constant $C_2(M)>0$ depending only on $\Omega, n, p$ and $M$.
\end{Prop}

\Proof
We prove Proposition \ref{Prop3} under the assumption that Lemmas \ref{Lemma_source_estimate_Fa_Fu}, \ref{Lemma_existence_iterates} and \ref{Lemma_decay} are valid, and postpone their proofs to Section \ref{Sec_Proofs}. So by Lemma \ref{Lemma_existence_iterates} there exist a sequence of iterates $(u_k)$. Given two such iterates $u_k,u_l \in W^{1,p}(\Omega)$, ($k\geq l$), estimate \eqref{decay_u} of Lemma \ref{Lemma_decay} in combination with our incoming bound $\|(\Gamma^*, d\Gamma^*)\|_{L^\infty} \leq M$, implies 
\begin{eqnarray} \nonumber
\|u_{k}-u_{l}\|_{W^{1,p}} 
& \leq & \sum_{j=l+1}^{k}\|\overline{u_j}\|_{W^{1,p}} \ \leq \ \ \|\overline{u_{l+1}}\|_{W^{1,p}} \sum_{j=l+1}^{k}(\epsilon\, C_d\, M)^j .
\end{eqnarray} 
By the bound \eqref{epsilon_bound_2} on $\epsilon$, the above geometric series converges as $k \rightarrow \infty$. This implies that $(u_k)_{k\in \mathbb{N}}$ is a Cauchy sequence in the Banach space $W^{1,p}(\Omega)$. Therefore, $(u_k)_{k\in \mathbb{N}}$ converges to some $u$ in $W^{1,p}(\Omega)$. Similarly, \eqref{decay_a} together with the bound \eqref{Gamma-bound} implies
\begin{eqnarray} \nonumber
\|a_{k}-a_{l}\|_{L^p} 
& \leq & \sum_{j=l+1}^{k}\|\overline{a_j}\|_{L^p} \ \leq \ \ \|\overline{u_{l+1}}\|_{L^p} \sum_{j=l+1}^{k}(\epsilon\, C_d\, M)^j ,
\end{eqnarray} 
which in light of \eqref{epsilon_bound_2} is a convergent geometric series, and we conclude with convergence of $(a_k)_{k \in \mathbb{N}}$ to some $a$ in the Banach space $L^p(\Omega)$. 

The limit $(u,a)$ solves \eqref{pde_u} and \eqref{pde_a} because each term in the equations \eqref{iterate_a} and \eqref{iterate_u} converge to the corresponding terms in \eqref{pde_u} and \eqref{pde_a} with respect to the $L^p$-norm on $\Omega$. For example, using H\"older inequality we find from \eqref{iterate_u} that
\begin{eqnarray} \nonumber
\Delta u[\phi] + F_u(u,a)[\Phi]  \ = \ \lim\limits_{k\to \infty} \Big(\Delta u_{k+1}[\phi] + F_u(u_k,a_{k+1})[\Phi] \Big) = 0,
\end{eqnarray}
which shows that $u = \lim\limits_{k\rightarrow\infty} u_k$ is indeed a solution of \eqref{pde_u}.

To derive estimate \eqref{curvature_bound_(u,a)}, we use the bounds on source terms of Lemma \ref{Lemma_source_estimate_Fa_Fu} in combination with the above convergence $(u_k,a_k) \to (u,a)$. That is, using that we initiated the iteration with $a_0=0$ and $u_0=0$, we find  
\begin{eqnarray} \nonumber
\|a\|_{L^p} =  \|a - a_0\|_{L^p} \leq \sum_{k=1}^\infty \|a_{k+1}-a_k\|_{L^p}  + \|a_1\|_{L^p}.
\end{eqnarray}
Using first \eqref{decay_a} and then \eqref{decay_u} successively, we estimate the above sum as
\begin{eqnarray} \label{proof_Prop3_eqn1}
\|a\|_{L^p} \leq \sum_{k=1}^\infty \big(\epsilon\: C_d\,\|(\Gamma^*, d\Gamma^*)\|_{L^\infty}\big)^k  + \|a_1\|_{L^p}.
\end{eqnarray}
We now use the elliptic estimate \eqref{existence_est1} in combination with the bound \eqref{bound_Fa} on $F_a(u_0)$  and $u_0=0$ to obtain 
\begin{eqnarray} \label{proof_Prop3_eqn2}
\|a_1\|_{L^p} \leq C_e\, {\rm vol}(\Omega) \, \|(\Gamma^*,d\Gamma^*)\|_{L^\infty}
\end{eqnarray}
 Substituting this back into \eqref{proof_Prop3_eqn1} and using \eqref{Gamma-bound} to estimate $\|(\Gamma^*,d\Gamma^*)\|_{L^\infty}$ by $M>0$, we obtain
\begin{eqnarray} \nonumber
\|a\|_{L^p} \leq \left( C_e {\rm vol}(\Omega) + \epsilon\: C_d \sum_{k=1}^\infty \big(   \epsilon\: C_d\,M \big)^{k-1}  \right) \|(\Gamma^*, d\Gamma^*)\|_{L^\infty}, 
\end{eqnarray}
and our $\epsilon$-bound \eqref{epsilon_bound_2} implies the above infinite sum converges, so we conclude that
\beq \label{proof_Prop3_eqn3}
\|a\|_{L^p} \leq C_2(M) \|(\Gamma^*, d\Gamma^*)\|_{L^\infty}
\eeq
for some constant $C_2(M)>0$ depending only on $\Omega, n, p$ and $M$.

We now derive an estimate on $u$ in a similar way. Using $u_0=0$, we begin by writing 
\begin{eqnarray} \nonumber
\|u\|_{W^{1,p}} =  \|u - u_0\|_{W^{1,p}} \leq \sum_{k=1}^\infty \|u_{k+1}-u_k\|_{W^{1,p}}  + \|u_1\|_{W^{1,p}},
\end{eqnarray}
and applying \eqref{decay_u} together with our initial bound \eqref{Gamma-bound} yields
\begin{eqnarray} \label{proof_Prop3_eqn4}
\|u\|_{W^{1,p}} \leq \epsilon\, C_d \sum_{k=1}^\infty \big( \epsilon\, C_d  \, M \big)^{k-1} \|(\Gamma^*, d\Gamma^*)\|_{L^\infty}  + \|u_1\|_{W^{1,p}},
\end{eqnarray}
where the sum is finite by our $\epsilon$-bound \eqref{epsilon_bound_2}. Using the elliptic estimate \eqref{existence_est4} in combination with the bound \eqref{bound_Fu} on $F_u(u_0,a_1)$, we obtain 
\begin{eqnarray} \nonumber
\|u_1\|_{W^{1,p}} 
&\leq & C_e  \big( \|a_1\|_{L^p} \, + \,   {\rm vol}(\Omega) \big) \|(\Gamma^*,d\Gamma^*)\|_{L^\infty}     \cr
&\overset{\eqref{proof_Prop3_eqn2}}{\leq} &  C_e ( C_e +1 ) \, {\rm vol}(\Omega) \|(\Gamma^*,d\Gamma^*)\|_{L^\infty}.
\end{eqnarray}
Substituting this estimate into \eqref{proof_Prop3_eqn4}, we obtain the estimate
\begin{eqnarray} \label{proof_Prop3_eqn5}
\|u\|_{W^{1,p}} \leq C_2(M) \, \|(\Gamma^*, d\Gamma^*)\|_{L^\infty}  ,
\end{eqnarray}
for some constant $C_2(M)>0$ depending only on $\Omega, n, p$ and $M$. Adding \eqref{proof_Prop3_eqn3} and \eqref{proof_Prop3_eqn5} yields the sought after estimate \eqref{curvature_bound_(u,a)}. 
\QED

Theorem \ref{Prop3} is a refined restatement of Proposition \ref{Prop2}, and thereby completes the proof of Proposition \ref{Prop2}, once we give the proofs of Lemmas \ref{Lemma_source_estimate_Fa_Fu}, \ref{Lemma_existence_iterates} and \ref{Lemma_decay}. This is accomplished in Section \ref{Sec_Proofs} below.

\subsection{Proof of Theorem \ref{Thm_Existence_J}}   \label{Sec_Proof_existence_Thm}

We now give the proof of our main existence result, Theorem \ref{Thm_Existence_J}, for connection $\Gamma \in L^\infty(\Omega)$ with $d\Gamma \in L^\infty(\Omega)$, under the assumption that Proposition \ref{Prop3} and Lemma \ref{Lemma_curl} hold. So given a weak solution $u \in W^{1,p}(\Omega)$ and $a \in L^p(\Omega)$ of the rescaled reduced RT-equations \eqref{pde_u} - \eqref{pde_a} constructed in Proposition \ref{Prop3}, we obtain  a solution $J \in W^{1,p}(\Omega)$ and $B  \in L^p(\Omega)$ of the reduced RT-equations \eqref{RT_reduced_J} - \eqref{RT_reduced_A2} by setting
\beq \label{ansatz_scaling_proof}
J= I + \epsilon\, u   \hspace{1cm} \text{and}  \hspace{1cm}    B= \epsilon\, a,
\eeq
as can be verified by inspection, c.f. our scaling ansatz \eqref{ansatz_scaling} and Lemma \ref{Lemma_rescaled_RT-eqn}. It remains to prove that $(J,B)$ satisfy estimate \ref{curvature_estimate_soln}, and that $J$ is integrable to coordinates as well as invertible for any $\epsilon> 0$ subject to some upper bound depending only on $\Omega, n, p$ and $M$.

We first prove that $J$ is invertible assuming $\epsilon >0$ meets \eqref{epsilon_bound_2} together with the upper bound  
\beq \label{epsilon_uniform_bound}
\epsilon < \frac{1}{2 C_M C_2(M) M},
\eeq
where $C_M>0$ is the constant from Morrey's inequality \eqref{Morrey_textbook}, $C_2(M)$ is the constant from estimate \eqref{curvature_bound_(u,a)} of Proposition \ref{Prop3}, and $M>0$ is our incoming bound on $\|(\Gamma,d\Gamma)\|_{L^\infty}$ in \eqref{curvature_estimate_soln}. For this we use the following lemma, which was proven in \cite[Lemma 6.1]{ReintjesTemple_ell2}.

\begin{Lemma} \label{Lemma_J_inverse}  \label{Lemma2_J_inverse} 
Let $J= I + \epsilon u$ for some matrix valued $0$-form $u\in W^{1,p}(\Omega)$, $p>n$, and assume 
\beq \label{epsilon_for_J-inv}
0 < \epsilon <  \frac{1}{2 C_M \|u\|_{W^{1,p}}} ,
\eeq                                 
where $C_M>0$ is the constant from Morrey's inequality \eqref{Morrey_textbook}. Then $J$ is invertible and there exists a matrix valued $0$-form $u^{-}\in W^{1,p}(\Omega)$ such that 
\beq \label{J_inverse_eqn}
J^{-1}= I + \epsilon \, u^{-}
\eeq
and there exists a constant $C_->0$  depending only on $\Omega, n, p$ such that
\beq \label{J_inverse_bound}
\| u^{-}\|_{W^{1,p}(\Omega)} \leq \: C_- \: \|u\|_{W^{1,p}(\Omega)}.
\eeq
\end{Lemma} 

To apply Lemma \ref{Lemma_J_inverse} to the matrix valued $0$-form $u \in W^{1,p}(\Omega)$ constructed in Proposition \ref{Prop3}, it suffices to show that the $\epsilon$-bound \eqref{epsilon_uniform_bound} implies the $\epsilon$-bound \eqref{epsilon_for_J-inv} of Lemma \ref{Lemma_J_inverse}. By estimate \eqref{curvature_bound_(u,a)} of Proposition \ref{Prop3} and our initial bound \eqref{ansatz_scaling} on $\Gamma^*$, we find
$$
\|u\|_{W^{1,p}(\Omega)} \ \leq \ C_2(M) \|(\Gamma^*,d\Gamma^*)\|_{L^\infty(\Omega)} \ \leq \ C_2(M)\, M ,
$$
which implies
$$
\frac{1}{2 C_M C_2(M) M} \leq  \frac{1}{2 C_M \|u\|_{W^{1,p}}},
$$
and this shows that our $\epsilon$-bound \eqref{epsilon_uniform_bound} implies \eqref{epsilon_for_J-inv}. We conclude that the Jacobian $J= I + \epsilon\, u$ is invertible with $J^{-1} \in W^{1,p}(\Omega)$, c.f. \eqref{J_inverse_eqn}. Moreover, by \eqref{J_inverse_eqn} and \eqref{J_inverse_bound} it follows that 
\beq \label{bound_J-inv}
\|I-J^{-1}\|_{W^{1,p}(\Omega)} \leq \epsilon \, C_- \|u\|_{W^{1,p}(\Omega)},
\eeq
where $C_->0$ is the constant from Lemma \ref{Lemma_J_inverse}, which depends only on $\Omega, n, p$.

We now prove estimate \eqref{curvature_estimate_soln}. For this, observe first that by \eqref{ansatz_scaling_proof}, we have $\|I-J\|_{W^{1,p}} = \epsilon \, \|u\|_{W^{1,p}}$ and $\|B\|_{L^p} = \epsilon \, \|a\|_{L^p}$.  Applying now estimate \eqref{curvature_bound_(u,a)} of Proposition \ref{Prop3}, we obtain
\begin{align} \nonumber
\|I-J^{-1}\|_{W^{1,p}} + \|I-J\|_{W^{1,p}} + \|B\|_{W^{1,p}}  
\leq & \ \epsilon\, (1+C_-)  \big( \|u\|_{W^{1,p}} + \|a\|_{L^p} \big) \cr
\leq & \ C_2(M) \epsilon\, \|(\Gamma^*,d\Gamma^*)\|_{L^\infty(\Omega)},
\end{align}
absorbing $(1+C_-)>0$ into the constant $C_2(M)>0$. Now our scaling assumption $\Gamma_x= \epsilon \, \Gamma^*$ in \eqref{small_Gamma} directly gives
\beq \nonumber 
\|I-J^{-1}\|_{W^{1,p}(\Omega)} + \|I-J\|_{W^{1,p}(\Omega)} + \|B\|_{W^{1,p}(\Omega)}   \leq C(M) \|(\Gamma_x,d\Gamma_x)\|_{L^\infty(\Omega)} , 
\eeq 
which is the sought after estimate \ref{curvature_estimate_soln} of Theorem \ref{Thm_Existence_J}.

We finally show that $J \equiv I + \epsilon\: u$ is indeed a Jacobian which is integrable to coordinates. For this recall that by Lemma \ref{Lemma_curl}, for each $k \geq 1$, the Jacobian $J_{k} \equiv I + \epsilon\, u_{k}$ is integrable to coordinates, that is,
\beq \label{curl_J_iterates_proof}
d\overrightarrow{J_{k}} =0
\eeq
holds, c.f. \eqref{curl_J_iterates}, where $u_{k+1} \in W^{1,p}(\Omega)$ is defined by \eqref{iterate_u} - \eqref{bdd_J} of the iteration scheme. By the convergence $u_k \to u$ in $W^{1,p}$ as $k \to \infty$, it follows that  $J_k$ converges to $J$ in $W^{1,p}$ as $k \to \infty$ as well. Thus 
$d\overrightarrow{J_{k}}$ converges to $d\overrightarrow{J}$ in $L^p,$ and this implies $d\overrightarrow{J}=0$ by \eqref{curl_J_iterates_proof}. That is, $J$ is integrable to some coordinate system $y$. It then follows directly that $y_k$ defined by our iteration scheme converges to some $y$ in $W^{2,p}(\Omega)$, and that $J$ is the Jacobian of the coordinate transformation $x \to x+ \epsilon\, y(x)$. This completes the proof of Theorem \ref{Thm_Existence_J}.    \hfill $\Box$  
\vspace{.2cm}

It remains only to prove Lemmas \ref{Lemma_source_estimate_Fa_Fu}, \ref{Lemma_existence_iterates} and \ref{Lemma_decay}, used to prove Proposition \ref{Prop3}, and to prove Lemma \ref{Lemma_curl}, which together with Proposition \ref{Prop3} was used to prove Theorem \ref{Thm_Existence_J}.

\section{Proof of Lemmas \ref{Lemma_source_estimate_Fa_Fu}, \ref{Lemma_existence_iterates}, \ref{Lemma_curl} and \ref{Lemma_decay}} \label{Sec_Proofs}

The proof of Theorem \ref{Thm_Existence_J} in Section \ref{Sec_Proof_existence_Thm} above followed from  Lemma \ref{Lemma_curl} and Proposition \ref{Prop3}, which assumed Lemmas \ref{Lemma_source_estimate_Fa_Fu}, \ref{Lemma_existence_iterates} and \ref{Lemma_decay} to be valid. In this section we prove these lemmas and thereby complete the proof of Theorem \ref{Thm_Existence_J}.

\subsection{Proof of Lemma \ref{Lemma_source_estimate_Fa_Fu}  (Estimates of the source terms)} \label{Sec_source_estimates}

Lemma \ref{Lemma_source_estimate_Fa_Fu} provides the basic estimates for the terms on the right hand side of equations \eqref{pde_u} - \eqref{pde_a}, and is required for the proofs of Lemmas \ref{Lemma_existence_iterates} and \ref{Lemma_decay}.  Let $\Gamma^*,d\Gamma^* \in L^\infty(\Omega)$, and assume $u \in W^{1,p}(\Omega)$ and $a\in L^p(\Omega)$, for $n < p < \infty$. Then Lemma \ref{Lemma_source_estimate_Fa_Fu} states that \eqref{bound_Fu} and \eqref{bound_Fa} hold, namely
\begin{align}   \nonumber
\|F_u(u,a)\|_{W^{-1,p}} \leq &   \|a\|_{L^p} \, + \, \big( |\Omega|^\frac{1}{p} +   \epsilon\: \|u\|_{L^p} \big) \|(\Gamma^*,d\Gamma^*)\|_{L^\infty(\Omega)} \cr
\|F_a(u)\|_{W^{-1,p}} \leq &  \big( |\Omega|^\frac{1}{p} + \epsilon\: \|u\|_{W^{1,p}} \big)  \|(\Gamma^*,d\Gamma^*)\|_{L^\infty(\Omega)}   ,  
\end{align}
where $F_u(u,a)$ and $F_a(u)$ are defined in \eqref{Def_Fu} and \eqref{Def_Fa}.
 
\Proof  
Recall that the operator norm on a linear functional $F \in W^{-1,p}(\Omega)$, $F : \; W^{1,p^*}_0(\Omega) \longrightarrow \R$, is defined as
\beq \label{techeqn0_estimates_Fu}
\| F\|_{W^{-1,p}} \equiv     \sup_{\phi \in \mathcal{T}}     \big|  F[\phi] \big| ,
\eeq
where
$$
\mathcal{T} \equiv \big\{\phi \in W^{1,p^*}_0(\Omega) \big| \; \|\phi\|_{W^{1,p^*}} =1 \big\},
$$
and $p^*$ is the conjugate of $p$, $\frac1{p} + \frac1{p^*}=1$. We first derive \eqref{bound_Fu}. For this, recall the definition of $F_u(u,a)$ in \eqref{Def_Fu},
\beq \nonumber
F_u(u,a)[\phi] 
\equiv  \big\langle \Gamma^* , d\phi \big\rangle_{L^2}  + \epsilon\: \big\langle u \mm \Gamma^* , d \phi \big\rangle_{L^2} + \big\langle a, \phi \big\rangle_{L^2} , 
\eeq
for any matrix valued $0$-form $\phi \in W^{1,p^*}_0(\Omega)$. From this together with \eqref{techeqn0_estimates_Fu}, we directly obtain that 
\begin{align}  \label{techeqn1_estimates_Fu}
\| F_u(u,a) \|_{W^{-1,p}} 
&\leq  \sup_{\phi \in \mathcal{T}} \Big( \big|\big\langle \Gamma^* , d\phi \big\rangle_{L^2}\big|  + \epsilon\:\big| \big\langle u \mm \Gamma^* , d \phi \big\rangle_{L^2}\big| + \big|\big\langle a, \phi \big\rangle_{L^2}\big| \Big).
\end{align}
We now estimate the right hand side of \eqref{techeqn1_estimates_Fu} term by term. For the first term we apply H\"older's inequality \eqref{Holder} component-wise to obtain 
\begin{eqnarray} \label{techeqn2a_estimates_Fu}
\big|\big\langle \Gamma^* , d\phi \big\rangle_{L^2}\big|  
& \leq &  \| \Gamma^* \|_{L^p} \|d\phi\|_{L^{p^*}} \cr
& \leq &  |\Omega|^\frac{1}{p} \, \| \Gamma^* \|_{L^\infty} \|\phi\|_{W^{1,p^*}}  \cr
&\leq &  |\Omega|^\frac{1}{p} \, \|(\Gamma^*,d\Gamma^*)\|_{L^\infty} ,
\end{eqnarray}
for $|\Omega| \equiv {\rm vol}(\Omega)$, and where the last estimate follows from $\|\phi\|_{W^{1,p^*}}=1$ for $\phi \in \mathcal{T}$. Likewise, using H\"older's inequality, we estimate the second term in \eqref{techeqn1_estimates_Fu} as
\begin{eqnarray} \label{techeqn2b_estimates_Fu}
\big| \big\langle u \mm \Gamma^* , d \phi \big\rangle_{L^2}\big|
&\leq &  \| u\mm \Gamma^* \|_{L^p} \|d\phi\|_{L^{p^*}} \cr
& \leq &  \| \Gamma^* \|_{L^\infty}  \| u \|_{L^p}  \|\phi\|_{W^{1,p^*}}  \cr
&\leq & \| u \|_{L^p} \, \|(\Gamma^*,d\Gamma^*)\|_{L^\infty} 
\end{eqnarray}
and the third term as
\beq \label{techeqn2c_estimates_Fu}
\big|\big\langle a, \phi \big\rangle_{L^2}\big| 
\ \leq \  \| a \|_{L^p}    \|\phi\|_{L^{p^*}} 
\ \leq \  \| a \|_{L^p} .
\eeq
Substituting \eqref{techeqn2a_estimates_Fu} - \eqref{techeqn2c_estimates_Fu} into \eqref{techeqn1_estimates_Fu}, we obtain
\beq \nonumber
\| F_u(u,a) \|_{W^{-1,p}} 
\leq   \big( |\Omega|^\frac{1}{p} + \epsilon\:  \| u \|_{L^p} \big) \|(\Gamma^*,d\Gamma^*)\|_{L^\infty} \, +\, \| a \|_{L^p} 
\eeq
which implies the sought after estimate \eqref{bound_Fu}.

We next prove \eqref{bound_Fa}. The functional $F_a$ is defined in \eqref{Def_Fa} as
\beq  \nonumber 
F_a(u)[\psi]  \equiv  \big\langle d\Gamma^*, \underleftarrow{\text{div}}(\psi) \big\rangle_{L^2} + \epsilon\: \big\langle ( u \mm d\Gamma^* + du \wedge \Gamma^*), \underleftarrow{\text{div}}(\psi) \big\rangle_{L^2}  ,
\eeq
for any vector valued $2$-form $\psi \in W^{1,p^*}_0(\Omega)$, where 
$$
\big(\underleftarrow{\text{div}}(\psi)\big)^\mu_{\nu}  = \partial_\nu \psi^\mu_{ij} \; dx^i dx^j
$$ 
by \eqref{vec-div_adjoint}. Thus we have
\beq   \label{techeqn1_estimates_Fa}
\|F_a(u)\|_{W^{-1,p}}  \leq  
\sup_{\psi \in \mathcal{T}} \Big( \big| \big\langle d\Gamma^*, \underleftarrow{\text{div}}(\psi) \big\rangle_{L^2} \big| + \epsilon\: \big|\big\langle ( u \mm d\Gamma^* + du \wedge \Gamma^*), \underleftarrow{\text{div}}(\psi) \big\rangle_{L^2} \big| \Big),
\eeq
where $\mathcal{T}$ is now taken as the space of vector valued $2$-forms in $W^{1,p^*}_0(\Omega)$ having unit length with respect to the $W^{1,p^*}$-norm. Applying again H\"older's inequality, we estimate the first term in \eqref{techeqn1_estimates_Fa} by
\begin{eqnarray}   \label{techeqn2a_estimates_Fa}
\big| \big\langle d\Gamma^*, \underleftarrow{\text{div}}(\psi) \big\rangle_{L^2} \big| 
& \leq &  \| d\Gamma^*\|_{L^p} \, \|\underleftarrow{\text{div}}(\psi)\|_{L^{p^*}} \cr
& \leq &  |\Omega|^\frac{1}{p} \, \| d\Gamma^*\|_{L^\infty} \, \|\psi\|_{W^{1,p^*}} \cr
& \leq &  |\Omega|^\frac{1}{p}\, \|(\Gamma^*,d\Gamma^*)\|_{L^\infty} ,
\end{eqnarray}
since $\|\psi\|_{W^{1,p^*}}=1$ for all $\psi \in \mathcal{T}$. Likewise, we estimate the second term in \eqref{techeqn1_estimates_Fa} using H\"older's inequality by
\begin{align}   \label{techeqn2b_estimates_Fa}
 \big|\big\langle ( u \mm d\Gamma^* + du \wedge \Gamma^*),& \underleftarrow{\text{div}}(\psi) \big\rangle_{L^2} \big| 
 \leq   \big\| u \mm d\Gamma^* + du \wedge \Gamma^* \big\|_{L^p} \|\underleftarrow{\text{div}}(\psi)\|_{L^{p^*}} \cr
\leq &  \big( \| u \|_{L^p} \| d\Gamma^* \|_{L^\infty} + \| du\|_{L^p} \| \Gamma^* \|_{L^\infty} \big) \|\psi\|_{W^{1,p^*}} \cr
\leq &    \| u\|_{W^{1,p}} \, \|(\Gamma^*,d\Gamma^*)\|_{L^\infty}  ,
\end{align}
again using $\|\psi\|_{W^{1,p^*}} =1$.  Substituting \eqref{techeqn2a_estimates_Fa} and \eqref{techeqn2b_estimates_Fa} into \eqref{techeqn1_estimates_Fa}, we finally obtain
\beq   \label{techeqn3_estimates_Fa}
\|F_a(u)\|_{W^{-1,p}}  
\leq  
 \big(|\Omega|^\frac{1}{p} + \epsilon\:  \| u\|_{W^{1,p}} \big) \|(\Gamma^*,d\Gamma^*)\|_{L^\infty}   ,
\eeq
which is the sought after estimate \eqref{bound_Fa}. This completes the proof.
\QED

\subsection{Proof of Lemma \ref{Lemma_existence_iterates} (Well-posedness of the iteration scheme)} \label{Sec_existence}

We now prove Lemma \ref{Lemma_existence_iterates} regarding well-posedness of the iteration scheme. For this, assume $u_k\in W^{1,p}(\Omega)$ is given and let $n < p < \infty$, $n\geq 2$.  Lemma \ref{Lemma_existence_iterates} then states that there exists $a_{k+1}\in L^p(\Omega)$ which solves \eqref{iterate_a}, there exists $\Psi_{k+1} \in L^p(\Omega)$ and $y_{k+1} \in W^{2,p}(\Omega)$ which solve \eqref{iterate_psi} - \eqref{iterate_y}, and there exists $u_{k+1}\in W^{1,p}(\Omega)$ which solves \eqref{iterate_u} with boundary data \eqref{bdd_J}, and these solutions satisfy the elliptic estimates \eqref{existence_est1} - \eqref{existence_est4},
\begin{eqnarray}
\|a_{k+1}\|_{L^p(\Omega)}  &\leq & C_e \; \|F_a(u_k)\|_{W^{-1,p}(\Omega)}, \label{ell_est_a_proof} \\
\|\Psi_{k+1}\|_{L^p(\Omega)}  &\leq & C_e \; \|F_u(u_k,a_{k+1})\|_{W^{-1,p}(\Omega)}, \label{ell_est_psi_proof} \\
\|y_{k+1}\|_{W^{2,p}(\Omega)} &\leq & C_e \; \|F_u(u_k,a_{k+1})\|_{W^{-1,p}(\Omega)}, \label{ell_est_y_proof}\\
\|u_{k+1}\|_{W^{1,p}(\Omega)} &\leq & C_e \; \|F_u(u_k,a_{k+1})\|_{W^{-1,p}(\Omega)}, \label{ell_est_u_proof} 
\end{eqnarray}
for some constant $C_e >0$ depending only on $n, p$ and $\Omega$.  \\

\noindent {\it Proof.}  
We begin by proving existence of a weak solution $a_{k+1}$ to the first order system \eqref{iterate_a}, namely
\beq \nonumber 
\begin{cases} 
\langle \overrightarrow{a_{k+1}}, \delta \psi \rangle_{L^2} = F_a(u_k)[\psi], \cr 
\langle \overrightarrow{a_{k+1}}, d\varphi \rangle_{L^2}=0,
\end{cases}
\eeq
subject to the bound \eqref{existence_est1}, by applying Proposition \ref{Lemma1_appendix} of Appendix \ref{Sec_A}. Proposition \ref{Lemma1_appendix} gives the existence of solutions to Cauchy Riemann type systems in a scalar variable at low level of regularity $a_{k+1} \in L^p(\Omega)$. We obtain such solutions by solving mollified equations with classical Dirichlet data, and then taking the zero mollification limit to obtain solutions $a_{k+1} \in L^p(\Omega)$.   (Note that $a_{k+1}\in L^p$ is too weak to impose Dirichlet data directly.) To start, note that the incoming assumption $u_k\in W^{1,p}(\Omega)$ together with the source estimates of Lemma \ref{Lemma_source_estimate_Fa_Fu} show that $F_a(u_k) \in W^{-1,p}(\Omega)$, which is the regularity assumed in Proposition \ref{Lemma1_appendix}. We now show that each vector component of \eqref{iterate_a} is a Cauchy Riemann type system in scalar variables, each component satisfying the assumptions of Proposition \ref{Lemma1_appendix}. The right hand side of the second equation in \eqref{iterate_a} is zero, and hence of the form assumed in Proposition \ref{Lemma1_appendix}. To apply Proposition \ref{Lemma1_appendix} to the first equation in \eqref{iterate_a}, it suffices to show that there exists a vector valued $1$-form $w \in W^{-1,p}(\Omega)$ such that $F_a(u)=dw$ in a weak sense, which then also implies the standard consistency conditions $dF_a(u)=0$ a weak sense. This is accomplished in the next lemma.

\begin{Lemma} \label{Lemma_dF=0_weakly}
Assume $u \in W^{1,p}(\Omega)$ is given, then there exists a vector valued $1$-form $w \in W^{-1,p}(\Omega)$ such that $F_a(u)=dw$ in the weak sense $F_a(u)[\varphi]= - w[\delta\varphi]$ for any vector valued $2$-form $\varphi \in W^{2,p^*}_0(\Omega)$. Moreover,  $dF_a(u)=0$ holds in the weak sense that $F_a(u)[\delta\varphi]=0$ for any vector valued $3$-form $\varphi \in W^{2,p^*}_0(\Omega)$. 
\end{Lemma}

\Proof
By definition \eqref{Def_Fa}, we have
\begin{align}  \nonumber
F_a(u)[\psi] & \equiv  \big\langle d\Gamma^*, \underleftarrow{\text{div}}(\psi) \big\rangle_{L^2} + \epsilon\: \big\langle ( u \mm d\Gamma^* + du \wedge \Gamma^*), \underleftarrow{\text{div}}(\psi) \big\rangle_{L^2} \cr
& =  \Big\langle \big((I+\epsilon u) \mm d\Gamma^* + d(I+\epsilon u) \wedge \Gamma^*\big), \underleftarrow{\text{div}}(\psi) \Big\rangle_{L^2} 
\end{align} 
for any vector valued $2$-form $\psi \in W^{1,p^*}_0(\Omega)$. Let $\Gamma^*_\rho$ denote a standard mollifier of $\Gamma^* \in L^\infty(\Omega)$ and $u_\rho$ a mollifier of $u \in W^{1,p}(\Omega)$, then $\Gamma^*_\rho \rightarrow \Gamma^*$ in $L^\infty$ and $d\Gamma^*_\rho \rightarrow d\Gamma^*$ in $L^\infty$ as $\rho \rightarrow 0$, while $u_\rho \rightarrow u$ in $W^{1,p}(\Omega)$ as $\rho \rightarrow 0$. As a consequence, setting 
\beq \nonumber
\mathcal{F}_\rho[\psi] \equiv \lim_{\rho \rightarrow 0} \Big\langle \big((I+\epsilon u_\rho) \mm d\Gamma^*_\rho + d(I+\epsilon u_\rho) \wedge \Gamma^*_\rho\big), \underleftarrow{\text{div}}(\psi) \Big\rangle_{L^2} ,
\eeq
H\"older inequality \eqref{Holder} implies convergence     
\begin{eqnarray} \label{existence_iterates_techeqn0}
F_a(u)[\psi]  &=&  \lim_{\rho \rightarrow 0}  \mathcal{F}_\rho[\psi] .
\end{eqnarray}
We now show that $\mathcal{F}_\rho[\delta\varphi]=0$. For this we begin by using the  Leibniz rule for differential forms \eqref{ Leibniz-rule-d} to compute 
\begin{eqnarray} \nonumber
\mathcal{F}_\rho[\psi]  
&=& \Big\langle d\big((I+\epsilon u_\rho) \mm \Gamma^*_\rho\big), \underleftarrow{\text{div}}(\psi) \Big\rangle_{L^2} ,
\end{eqnarray} 
Application of the adjoint property \eqref{vec-div_adjoint_eqn1} gives   
\begin{eqnarray} \nonumber
\mathcal{F}_\rho[\psi]  =   \Big\langle \overrightarrow{\text{div}}\big(d\big((I+\epsilon u_\rho) \mm \Gamma^*_\rho\big)\big), \psi \Big\rangle_{L^2}  .
\end{eqnarray}
We now apply \eqref{regularity-miracle} to commute $d$ and $\overrightarrow{\text{div}}$, from which we obtain
\begin{eqnarray} \nonumber
\mathcal{F}_\rho[\psi]  &=&   \Big\langle d\; \overrightarrow{\delta \big( (I+\epsilon u_\rho) \mm \Gamma^*_\rho\big)}, \psi \Big\rangle_{L^2}    \\
&=&  \Big\langle \overrightarrow{\delta \big( (I+\epsilon u_\rho) \mm \Gamma^*_\rho\big)}, \delta\psi \Big\rangle_{L^2},   \label{existence_iterates_techeqn1}
\end{eqnarray}
where the last equality follows from partial integration for differential forms \eqref{partial_integration_matrix}.  Now, since $\Gamma^*_\rho \rightarrow \Gamma^*$ in $L^\infty$ and $u_\rho \rightarrow u$ in $W^{1,p}(\Omega)$ as $\rho \rightarrow 0$, it follows that the expression on the right hand side converges in $W^{-1,p}(\Omega)$ and defines the vector valued $1$-form $w \in W^{-1,p}(\Omega)$ as
\beq \nonumber
w[\psi] \equiv \lim_{\rho \to 0}\Big\langle \overrightarrow{\delta \big( (I+\epsilon u_\rho) \mm \Gamma^*_\rho\big)}, \delta\psi \Big\rangle_{L^2}.
\eeq
Combining \eqref{existence_iterates_techeqn0} with \eqref{existence_iterates_techeqn1} imply that $ F_a(u)[\psi] = w[\delta\psi]$ for any vector valued $2$-form $\varphi \in W^{2,p^*}_0(\Omega)$, which is the sought after equation.

To prove the supplement, substitute $\psi = \delta \varphi$ into \eqref{existence_iterates_techeqn1}, the identity $\delta^2=0$ then gives us
\beq \nonumber
\mathcal{F}_\rho[\delta \varphi] = \Big\langle \overrightarrow{\delta \big( (I+\epsilon u_\rho) \mm \Gamma^*_\rho\big)}, \delta\delta\varphi \Big\rangle_{L^2} =0,
\eeq
which implies by \eqref{existence_iterates_techeqn0} that 
$$
F_a(u)[\delta\varphi]= \lim_{\rho \rightarrow 0}  \mathcal{F}_\rho[\delta\varphi] = 0
$$ 
for any vector valued $3$-form $\varphi \in W^{2,p^*}_0(\Omega)$. This proves Lemma \ref{Lemma_dF=0_weakly}.
\QED

By Lemma \ref{Lemma_dF=0_weakly}, the desired condition $dF_a(u)=0$ holds for each vector component in the weak sense, since there exists of a vector valued $1$-form $w \in W^{-1,p}(\Omega)$ such that $F_a(u)=dw$. We conclude that Proposition \ref{Lemma1_appendix} applies component wise and yields the existence of a solution $a_{k+1} \in L^p(\Omega)$ to \eqref{iterate_a}. Moreover, the solution constructed in Proposition \ref{Lemma1_appendix} meets the $L^p$-bound \eqref{appendix_eqn2}, which by application to each vector component directly implies the sought after $L^p$-bound \eqref{existence_est1} on $a_{k+1}$. 

Next, we prove existence of a weak solution $\Psi_{k+1} \in L^p(\Omega)$ of \eqref{iterate_psi}, namely of
\beq \nonumber 
\langle \Psi_{k+1}, \delta \vec{\phi} \rangle_{L^2} =  F_{\Psi}(u_k,a_{k+1})[\phi]
\eeq 
for any matrix-valued $0$-form $\phi \in W^{1,p^*}_0(\Omega)$, subject to the $L^p$ bound \eqref{iterate_psi_bound} by applying Proposition \ref{Lemma2_appendix}, which is a version of Proposition \ref{Lemma1_appendix} applying to the simpler case of $0$-forms. For this, we need to verify that each vector component of \eqref{iterate_psi} meets the consistency condition $df=0$, in the weak sense $f(\delta \psi)=0$, of Proposition \ref{Lemma2_appendix}, which is achieved in the next Lemma. 

\begin{Lemma} \label{Lemma_dF_Psi=0_weakly}
Assume $a_{k+1} \in L^p(\Omega)$ solves \eqref{iterate_a} for some $u_k \in W^{1,p}(\Omega)$, then $
F_\Psi$, defined in \eqref{iterate_psi}, satisfies the weak consistency condition 
\beq \label{dF_Psi=0_weakly}
F_\Psi(u_k,a_{k+1})[\delta\psi] =0
\eeq
for any vector valued $2$-form $\psi \in W^{2,p^*}_0(\Omega)$, (so $\psi|_{\partial\Omega} =0$ and $\delta\psi|_{\partial\Omega}=0$).\footnote{Note, $\delta \psi$ is a vector valued $1$-form, and any such form can always be interpreted as a matrix valued $0$-form. So $\delta\psi$ is an admissible argument for $F_\Psi$.}
\end{Lemma}

\Proof            
By \eqref{iterate_psi_DefF}, $F_\Psi$ is defined as
\beq \nonumber
F_\Psi(u_k,a_{k+1})[\phi]  = \langle J_k \Gamma^* , d\phi \rangle_{L^2}  + \langle \overrightarrow{a_{k+1}}, \overrightarrow{\phi} \rangle_{L^2},
\eeq
for matrix valued $0$-forms $\phi \in W^{1,p^*}_0(\Omega)$, where $J_k \equiv I+ \epsilon u_k$. In order for $F_\Psi$ to act on  the vector valued $1$-form $\delta\big(\psi^\mu_{ij}dx^i\wedge dx^j\big) = (\delta\psi)^\mu_\nu dx^\nu$, we express $\delta\psi$ as the associated matrix valued $0$-form $(\delta \psi)^\mu_\nu$, then
\begin{eqnarray}   \label{techeqn1_dF=0_psi_weak}
F_\Psi(u_k,a_{k+1})[\delta\psi]  
= \langle J_k \Gamma^* , d(\delta\psi) \rangle_{L^2}  + \langle \overrightarrow{a_{k+1}}, \overrightarrow{\delta\psi} \rangle_{L^2}  ,
\end{eqnarray}
where $d(\delta\psi)= \partial_i (\delta\psi)^\mu_\nu dx^i$ is the exterior derivative of the matrix valued $0$-form $\delta\psi$, and where $\overrightarrow{\delta\psi}$ denotes the original vector valued $1$-form $\delta\psi^\mu = (\delta\psi)^\mu_\nu dx^\nu$. The main technical step of this proof is to show by a mollification argument that the first term on the right hand side of \eqref{techeqn1_dF=0_psi_weak} equals
\begin{eqnarray} \label{techeqn2_dF=0_psi_weak}
\langle J_k \Gamma^* , d(\delta\psi) \rangle_{L^2}   
= - \big\langle \big(J_k \mm d\Gamma^* + dJ_k \wedge \Gamma^*\big) , \underleftarrow{\text{div}}(\psi) \big\rangle_{L^2}  .
\end{eqnarray}
Assuming for the moment \eqref{techeqn2_dF=0_psi_weak} holds, we substitute $J_k = I + \epsilon \: u_k$  to write \eqref{techeqn2_dF=0_psi_weak} as 
\begin{eqnarray}  \label{techeqn3_dF=0_psi_weak}
\langle J_k \Gamma^* , d(\delta\psi) \rangle_{L^2}   
&=& - \big\langle d\Gamma^*, \underleftarrow{\text{div}}(\psi) \big\rangle_{L^2} - \epsilon\: \big\langle ( u_k \mm d\Gamma^* + du_k \wedge \Gamma^*), \underleftarrow{\text{div}}(\psi) \big\rangle_{L^2} \cr
&=& - \langle \overrightarrow{a_{k+1}} , \overrightarrow{\delta\psi} \rangle_{L^2} ,
\end{eqnarray}
where the last equality follows from \eqref{iterate_a}, the equation for $a_{k+1}$. Substituting \eqref{techeqn3_dF=0_psi_weak} into \eqref{techeqn1_dF=0_psi_weak} gives the sought after consistency condition, $F_\Psi(u_k,a_{k+1})[\delta\psi] =0$ for any vector valued $2$-form $\psi \in W^{1,p^*}_0(\Omega)$, which completes the proof of Lemma \ref{Lemma_dF_Psi=0_weakly} once we prove equation \eqref{techeqn2_dF=0_psi_weak} holds.

To verify \eqref{techeqn2_dF=0_psi_weak}, we consider a standard mollifier $\Gamma^*_\rho$ of $\Gamma^*$ together with a mollifier $(u_{k})_\rho$ of $u_{k}$, as in the proof of Lemma \ref{Lemma_dF=0_weakly}. For ease of notation we omit writing out the mollifier $(u_k)_\rho$ in the subsequent argument, that is, whenever $\Gamma^*_\rho$ appears we assume $J_k$ denotes the mollification $(J_k)_\rho= I +\epsilon (u_k)_\rho$. Now, since $\Gamma^*_\rho \to \Gamma^*$ in $L^\infty(\Omega)$ and $(u_k)_\rho \to u_k$ in $W^{1,p}(\Omega)$, it follows that
\begin{eqnarray} \nonumber
\langle J_k \Gamma^* , d(\delta\psi) \rangle_{L^2}   
= \lim_{\rho \rightarrow 0} \langle J_k \Gamma^*_\rho , d(\delta\psi) \rangle_{L^2} .
\end{eqnarray}
Using the partial integration formula \eqref{partial_integration_matrix}, we obtain
\begin{eqnarray} \nonumber
\langle J_k \Gamma^*_\rho , d(\delta\psi) \rangle_{L^2}  
&=&  -  \langle \delta(J_k \Gamma^*_\rho) , (\delta\psi) \rangle_{L^2}   \cr
&=& -  \langle \overrightarrow{\delta(J_k \Gamma^*_\rho)} , \overrightarrow{\delta\psi} \rangle_{L^2}  ,
\end{eqnarray}
where  for the last equality we used the inner product identity \eqref{tracestuff} for matrix and valued forms, using again the notation $\overrightarrow{\delta\psi}=(\delta\psi)^\mu_\nu dx^\nu$. Applying now partial integration \eqref{partial_integration_matrix} for vector valued $1$-forms, we get
\begin{eqnarray}  \nonumber
\langle J_k \Gamma^*_\rho , d(\delta\psi) \rangle_{L^2}   
= \langle d\overrightarrow{\delta(J_k \Gamma^*_\rho)} , \psi \rangle_{L^2}  ,
\end{eqnarray}
and using \eqref{regularity-miracle} to commute $d$ and $\overrightarrow{\text{div}}$ as
$$
d\overrightarrow{\delta(J_k \Gamma^*_\rho)} = \overrightarrow{\text{div}}\big(d(J_k \Gamma^*_\rho)\big),
$$
we find that
\begin{eqnarray} \label{techeqn4_dF=0_psi_weak}
\langle J_k \Gamma^*_\rho , d(\delta\psi) \rangle_{L^2}
&=&  \langle \overrightarrow{\text{div}}\big(d(J_k \Gamma^*_\rho)\big) , \psi \rangle_{L^2}  \cr
&\overset{\eqref{vec-div_adjoint_eqn1}}{=}& - \langle d(J_k \Gamma^*_\rho) , \underleftarrow{\text{div}}(\psi) \rangle_{L^2} ,
\end{eqnarray}
using the adjoint property \eqref{vec-div_adjoint_eqn1} for $\overrightarrow{\text{div}}$ in the last step. Now, by the  Leibniz rule, the $L^\infty$ convergence of $\Gamma^*_\rho$ and $d\Gamma^*_\rho$ and the $W^{1,p}$ convergence of $(J_k)_\rho$, it follows that
$$
d\big((J_k)_\rho \Gamma^*_\rho\big) 
= d(J_k)_\rho \wedge \Gamma^*_\rho + (J_k)_\rho \mm d\Gamma^*_\rho 
\ \overset{\rho \to 0}{\longrightarrow} \ dJ_k \wedge \Gamma^* + J_k \mm d\Gamma^*
$$
converges in $L^p(\Omega)$. Thus the left and right hand sides in \eqref{techeqn4_dF=0_psi_weak} both converge (as can be shown using H\"older inequality) and yield
\beq \nonumber
\langle J_k \Gamma^* , d(\delta\psi) \rangle_{L^2}   
= - \big\langle \big(dJ_k \wedge \Gamma^* + J_k \mm d\Gamma^*\big) , \underleftarrow{\text{div}}(\psi) \big\rangle_{L^2},
\eeq
which is the sought after identity \eqref{techeqn2_dF=0_psi_weak}. This completes the proof of Lemma \ref{Lemma_dF_Psi=0_weakly}.
\QED

Lemma \ref{Lemma_dF_Psi=0_weakly} establishes the consistency condition required by Proposition \ref{Lemma2_appendix} for existence of a solution to the first order Cauchy Riemann type system. To apply Proposition \ref{Lemma2_appendix} and conclude with the sought after existence of a vector valued $0$-form $\Psi_{k+1} \in L^p(\Omega)$ which solves \eqref{iterate_psi}, it remains only to show that $F_\Psi(u_k,a_{k+1}) \in W^{-1,p}(\Omega)$. For this, recall that by \eqref{iterate_psi_DefF}, $F_\Psi$ is defined as
\beq \nonumber
F_\Psi(u_k,a_{k+1})[\phi]  = \langle J_k \Gamma^* , d\phi \rangle_{L^2}  + \langle \overrightarrow{a_{k+1}}, \overrightarrow{\phi} \rangle_{L^2},
\eeq
for any matrix valued $0$-form $\phi$, where $J_k \equiv I+ \epsilon u_k$. Comparing this $F_u$ in \eqref{Def_Fu},  
\beq \nonumber
F_u(u,a)[\phi] \equiv    \big\langle (I+\epsilon\:u) \mm \Gamma^* , d \phi \big\rangle_{L^2} + \big\langle a, \phi \big\rangle_{L^2} ,
\eeq
where $\phi \in W^{1,p}(\Omega)$ can be any matrix valued $0$-form, we conclude that
\beq \nonumber
\|F_\Psi(u_k,a_{k+1})\|_{W^{-1,p}(\Omega)}  = \|F_u(u_k,a_{k+1})\|_{W^{-1,p}(\Omega)}.
\eeq
which is finite by the source estimate \eqref{bound_Fu} of Lemma \ref{Lemma_source_estimate_Fa_Fu}. We can now apply Proposition \eqref{Lemma2_appendix} and conclude with existence of a vector valued $0$-form $\Psi_{k+1} \in L^p(\Omega)$ which solves \eqref{iterate_psi} and satisfies the sought after estimate \eqref{ell_est_psi_proof}.

We now prove the existence of a solution $y_{k+1} \in W^{2,p}(\Omega)$ to the Dirichlet problem \eqref{iterate_y},
\beq \nonumber
\begin{cases} 
\Delta y_{k+1} = \Psi_{k+1}, \cr 
y_{k+1}\big|_{\partial\Omega} = 0,
\end{cases}
\eeq
together with the elliptic estimate \eqref{ell_est_y_proof}. By Lemma \ref{Lemma_source_estimate_Fa_Fu}, $F_u(u_k,a_{k+1})$ is in $W^{-1,p}(\Omega)$ and we can apply the basic existence result for the Poisson equation with $L^p$ sources, Theorem \ref{Thm_Poisson}. This yields the existence of a solution $y_{k+1} \in W^{2,p}(\Omega)$. To prove estimate \eqref{existence_est4}, we now apply the elliptic estimate \eqref{Poisson-3_Lp} of Theorem \ref{Thm_Poisson} component wise to \eqref{iterate_y}, which gives us
\begin{eqnarray} \nonumber
\|y_{k+1}\|_{W^{2,p}(\Omega)} 
&\leq &   C \:  \| \Psi_{k+1} \|_{L^p(\Omega)}.
\end{eqnarray}
Using now estimate \eqref{existence_est3} on $\| \Psi_{k+1} \|_{L^p(\Omega)}$, we obtain
\begin{eqnarray} \nonumber
\|y_{k+1}\|_{W^{2,p}(\Omega)} 
& \leq &   C \: \|F_u(u_k,a_{k+1})\|_{W^{-1,p}(\Omega)} ,\label{existence_iterates_techeqn3}
\end{eqnarray}
where we have absorbed the constant from the estimate on $\|\Psi_{k+1}\|_{L^p}$ into the universal constant $C>0$. This is the sought after estimate \eqref{ell_est_y_proof}.

We now prove existence of a solution $u_{k+1} \in W^{1,p}(\Omega)$ of \eqref{iterate_u} with boundary data \eqref{bdd_J}, that is, 
\beq \label{iterate_u_proof}
- \Delta u_{k+1}[\phi]=F_u(u_k,a_{k+1})[\phi],
\eeq
for any matrix valued $0$-form $\phi \in W^{1,p^*}_0(\Omega)$, and with Dirichlet boundary data $u_{k+1}|_{\partial\Omega} = dy_{k+1}|_{\partial\Omega}$. By Lemma \ref{Lemma_source_estimate_Fa_Fu} we have $F_u(u_k,a_{k+1}) \in W^{-1,p}(\Omega)$, so existence of a solution $u_{k+1} \in W^{1,p}(\Omega)$ of \eqref{iterate_u_proof} follows directly from Theorem \ref{Thm_Poisson}. To prove estimate \eqref{existence_est2}, we apply estimate \eqref{Poisson-3} of Theorem \ref{Thm_Poisson}  component wise to equation \eqref{iterate_u_proof} and obtain
\beq \label{elliptic_estimate_techeqn1}
\|u_{k+1}\|_{W^{1,p}(\Omega)} 
\leq  C \Big( \|F_u(u_k,a_{k+1})\|_{W^{-1,p}(\Omega)} + \|dy_{k+1}\|_{W^{1,p}(\Omega)}  \Big),
\eeq
where the second terms on the right hand side results from the boundary data, $u_{k+1} = dy_{k+1}$ on $\partial\Omega$. 
Applying now estimate \eqref{ell_est_y_proof} to bound the boundary term by 
$$
\|dy_{k+1}\|_{W^{1,p}(\Omega)}  \leq \|y_{k+1}\|_{W^{2,p}(\Omega)} \leq  C\, \|F_u(u_k,a_{k+1})\|_{W^{-1,p}(\Omega)}     ,
$$ 
we obtain  
\beq \nonumber
\|u_{k+1}\|_{W^{1,p}(\Omega)} \leq  C \; \|F_u(u_k,a_{k+1})\|_{W^{-1,p}(\Omega)},
\eeq
which is the sought after estimate \eqref{existence_est2}. We now choose the maximum over all constants in the above estimates as the constant $C_e>0$ stated in Lemma \ref{Lemma_existence_iterates}. This completes the proof of Lemma \ref{Lemma_existence_iterates}.
\hfill $\Box$

\subsection{Proof of Lemma \ref{Lemma_decay} (Bounds on differences of iterates)} \label{Sec_decay}
             
We prove the closeness of subsequent iterates required to conclude with convergence of the iteration scheme in the proof of Proposition \ref{Prop3}. So assume $\Gamma^*, d\Gamma^* \in L^\infty(\Omega)$, and let $C_e>0$ denote the constant from the elliptic estimates of Lemma \ref{Lemma_existence_iterates}, which depends only on $n$, $p$, $\Omega$. Then, to prove Lemma \ref{Lemma_decay}, it suffices to show that differences of iterates satisfy
\begin{eqnarray}
\|\overline{a_{k+1}}\|_{L^p} 
&\leq & \epsilon\: C_e\, \|(\Gamma^*, d\Gamma^*)\|_{L^\infty} \:\|\overline{u_{k}}\|_{W^{1,p}} , \label{decay_a_proof} \\
\|\overline{u_{k+1}}\|_{W^{1,p}}  
&\leq & \epsilon\: C_e (1+C_e)\, \|(\Gamma^*, d\Gamma^*)\|_{L^\infty} \:\|\overline{u_{k}}\|_{W^{1,p}}, \label{decay_u_proof} 
\end{eqnarray} 
for any $k \in \mathbb{N}$. To prove Lemma \ref{Lemma_decay}, we require the following lemma which gives bounds on differences of source terms, 
\beq
\begin{aligned}
\overline{F_a(u_{k})} &\equiv  F_a(u_{k}) - F_a(u_{k-1}), \cr 
\overline{F_{u}(u_{k},a_{k+1})} &\equiv  F_{u}(u_{k},a_{k+1}) - F_{u}(u_{k-1},a_{k}),\label{diff2}
\end{aligned}
\eeq 
which by linearity of $F_a$ and $F_u$ is a straightforward modification of the proof of Lemma \ref{Lemma_source_estimate_Fa_Fu}.

\begin{Lemma} \label{Lemma_sources_difference}
Assume $(u_k,a_k)$ are defined by the iteration scheme \eqref{iterate_a} - \eqref{iterate_u}, then the differences of source terms defined in \eqref{diff2} satisfy
\begin{align} 
\big\|\overline{F_u(u_k,a_{k+1})}\big\|_{W^{-1,p}} 
 \leq &    \|\overline{a_{k+1}}\|_{L^p}   \, +\,      \epsilon\: \|\overline{u_{k}}\|_{L^p} \, \|(\Gamma^*,d\Gamma^*)\|_{L^\infty},  \label{bound_diff_Fu}  \\
\big\|\overline{F_a(u_{k})}\big\|_{W^{-1,p}} 
 \leq & \epsilon \,  \|\overline{u_{k}}\|_{W^{1,p}} \, \|(\Gamma^*,d\Gamma^*)\|_{L^\infty} . \label{bound_diff_Fa}
\end{align} 
\end{Lemma}

\Proof
We prove the lemma by using linearity of the source terms $F_u$ and $F_a$ and following the steps in the proof of Lemma \ref{Lemma_source_estimate_Fa_Fu}. In more detail, we find from the definition of $F_u(u,a)$ in \eqref{Def_Fu}, that
\beq \nonumber
\overline{F_u(u_k,a_{k+1})[\phi]} 
=   \epsilon\: \big\langle \overline{u_{k}} \mm \Gamma^* , d \phi \big\rangle_{L^2} + \big\langle \overline{a_{k+1}}, \phi \big\rangle_{L^2} , 
\eeq
for any matrix valued $0$-form $\phi \in W^{1,p^*}_0(\Omega)$. Following the steps in the proof of Lemma \ref{Lemma_source_estimate_Fa_Fu}, then yields \eqref{bound_diff_Fu}. Similarly, from the definition of $F_a$ in \eqref{Def_Fa} that
\beq  \nonumber
\overline{F_a(u_k)[\psi]}  \equiv  \epsilon\: \big\langle ( \overline{u_k} \mm d\Gamma^* + d\overline{u_k} \wedge \Gamma^*), \underleftarrow{\text{div}}(\psi) \big\rangle_{L^2}  ,
\eeq
for any vector valued $2$-form $\psi \in W^{1,p^*}_0(\Omega)$, and following the steps in the proof of Lemma \ref{Lemma_source_estimate_Fa_Fu} gives us \eqref{bound_diff_Fa}. This completes the proof of Lemma \ref{Lemma_sources_difference}.
\QED

Lemma \ref{Lemma_decay} now follows from the elliptic estimates \eqref{existence_est1} and \eqref{existence_est4} together with the bounds on differences of sources in Lemma \ref{Lemma_sources_difference}. That is, by linearity of \eqref{existence_est1}, we have
\begin{eqnarray}  \label{techeqn1_decay_Lemma}
\|\overline{a_{k+1}}\|_{L^p(\Omega)}  
&\leq & C_e \; \|\overline{F_a(u_k)}\|_{W^{-1,p}(\Omega)}  \cr
&\overset{\eqref{bound_diff_Fa}}{\leq} & \epsilon \, C_e \, \|\overline{u_{k}}\|_{W^{1,p}} \, \|(\Gamma^*,d\Gamma^*)\|_{L^\infty} .
\end{eqnarray}
Likewise, by linearity of \eqref{existence_est4}, we find
\begin{eqnarray} \label{techeqn2_decay_Lemma}
\|\overline{u_{k+1}}\|_{W^{1,p}(\Omega)} 
&\leq & C_e \; \|\overline{F_u(u_k,a_{k+1})}\|_{W^{-1,p}(\Omega)}   \cr
&\overset{\eqref{bound_diff_Fu}}{\leq} & C_e \; \big(    \|\overline{a_{k+1}}\|_{L^p}   \, +\,      \epsilon\: \|\overline{u_{k}}\|_{L^p} \, \|(\Gamma^*,d\Gamma^*)\|_{L^\infty}  \big)  \cr
&\overset{\eqref{techeqn1_decay_Lemma}}{\leq} & \epsilon \, C_e (1+C_e) \,  \|\overline{u_{k}}\|_{W^{1,p}} \, \|(\Gamma^*,d\Gamma^*)\|_{L^\infty} .
\end{eqnarray}
This completes the proof of Lemma \ref{Lemma_decay}. \hfill $\Box$

\subsection{Proof of Lemma \ref{Lemma_curl} (Integrability of $J$)}   \label{Sec_curl}

On smooth $k$-forms the Laplacian acts component wise, (i.e., on components of matrix-, vector- and differential forms separately), and the relation between vector and matrix valued solutions of the Poisson equations in a classical sense is straightforward. That is, we have $\overrightarrow{\Delta u} = \Delta \vec{u}$ in a classical sense. This is used in Lemma \ref{Lemma_integrability_J} to prove that the Jacobian $J$ produced by the iteration scheme is integrable to coordinates.  The next lemma establishes the relation $\overrightarrow{\Delta u} = \Delta \vec{u}$ for the weak Laplacian.

\begin{Lemma}  \label{Lemma_Laplacian_equiv_matrix_vec}
Let $u \in W^{1,p}(\Omega)$ be a matrix valued $0$-form, then
\beq \label{Laplacian_equiv_vec_matrix}
\Delta(u)\big[\phi\big] = \Delta\vec{u}\big[\vec{\phi}\,\big]
\eeq
for any matrix valued $0$-form $\phi \in W^{1,p^*}_0(\Omega)$.
\end{Lemma}

\Proof
From the weak form of the Laplacian in \eqref{Laplacian_weak}, using that $\delta u =0$ and $du = \nabla u$ for matrix valued $0$-forms, (where again $\nabla u$ denotes the gradient acting on each component of $u$), we find
\begin{eqnarray} \nonumber
-\Delta(u)[\phi] 
&=&  \langle du, d\phi \rangle_{L^2} + \langle \delta u, \delta \phi \rangle_{L^2} 
\ = \ \langle du, d\phi \rangle_{L^2}  \cr
&=&  \langle \nabla u, \nabla\phi \rangle_{L^2} \ 
= \  \sum_j \langle \partial_j u, \partial_j \phi \rangle_{L^2},
\end{eqnarray}
where the last equality follows from the definition in \eqref{inner-product_Du_matrix}, c.f. Lemma \ref{Lemma_weak_Poisson_equivalence}. Using now that for fixed $j$ the inner product is invariant under vectorization for matrix valued $0$-forms, c.f. \eqref{tracestuff}, we obtain
\begin{eqnarray} \nonumber
-\Delta(u)[\phi] 
&=&  \langle \nabla\vec{u}, \nabla\vec{\phi} \rangle_{L^2} .
\end{eqnarray}
Now, let $u^\epsilon$ be a standard mollification of $u$. Then $u^\epsilon \rightarrow u$ in $W^{1,p}(\Omega)$ as $\epsilon \rightarrow 0$, which allows us to compute
\begin{eqnarray} \nonumber
-\Delta(u)[\phi] 
&=&  \lim_{\epsilon \rightarrow 0}  \langle \nabla\vec{u}^\epsilon, \nabla\vec{\phi} \rangle_{L^2} \cr
&=& - \lim_{\epsilon \rightarrow 0}  \langle \Delta \vec{u}^\epsilon, \vec{\phi} \rangle_{L^2} \cr
&=& \lim_{\epsilon \rightarrow 0} \big( \langle d\vec{u}^\epsilon, d\vec{\phi} \rangle_{L^2} + \langle \delta \vec{u}^\epsilon, \delta \vec{\phi} \rangle_{L^2} \big) \cr
&=& \langle d\vec{u}, d\vec{\phi} \rangle_{L^2} + \langle \delta \vec{u}, \delta \vec{\phi} \rangle_{L^2} \cr
&=& - \Delta\vec{u}[\vec{\phi}].
\end{eqnarray}
This completes the proof.
\QED

We now prove Lemma \ref{Lemma_curl}, which states that 
\beq \label{curl_J_iterates_proof_2}
d\overrightarrow{u_{k+1}} =0,
\eeq
where $u_{k+1} \in W^{1,p}(\Omega)$ is a solution of \eqref{iterate_u} with boundary data \eqref{bdd_J}. Equation \eqref{curl_J_iterates_proof_2} implies directly that $J_{k+1} \equiv I + \epsilon\, u_{k+1}$ is integrable to coordinates. Moreover, Lemma \ref{Lemma_curl} states that $J_{k+1}$ is the Jacobian of the coordinate transformation $x \to x +\epsilon\, y_{k+1}(x)$, where $y_{k+1} \in W^{2,p}(\Omega)$ is the solution of \eqref{iterate_y}.\\

\noindent \textit{Proof of Lemma \ref{Lemma_curl}.} 
The idea of proof is similar to that of Lemma \ref{Lemma_integrability_J}, but adapted to the weak formulation of \eqref{iterate_psi_strong-form}, to take into account the regularity of $\Psi_{k+1} \in L^p(\Omega)$ and $u_{k+1} \in L^p(\Omega)$. That is, we need to show that 
\beq \label{techeqn1_Lemma_curl}
\Delta \big(\overrightarrow{u_{k+1}} - dy_{k+1} \big)\big[\vec{\phi}\big] =0,
\eeq
for any matrix valued $0$-form $\phi \in W^{1,p^*}_0(\Omega)$. Assume for the moment \eqref{techeqn1_Lemma_curl} is true. Then, since $\overrightarrow{u_{k+1}} - dy_{k+1}$ vanishes on $\partial\Omega$ by the boundary condition \eqref{bdd_J}, equation \eqref{techeqn1_Lemma_curl} implies that 
\beq \label{techeqn2_Lemma_curl}
\overrightarrow{u_{k+1}} - dy_{k+1} =0
\eeq
in $\Omega$, which is the sought after equation \eqref{curl_J_iterates}. Moreover, \eqref{techeqn2_Lemma_curl} directly implies that 
\beq \nonumber
d(x + \epsilon y_{k+1}) = \overrightarrow{J_{k+1}}, 
\eeq 
where $J_{k+1}\equiv I + \epsilon\, u_{k+1}$. Thus $J_{k+1}$ is in fact the Jacobian of the coordinate transformation $x \to x +\epsilon\, y_{k+1}(x)$. This proves Lemma \ref{Lemma_curl} once we establish \eqref{techeqn1_Lemma_curl}.

To prove \eqref{techeqn1_Lemma_curl}, recall that $\Psi_{k+1}$ satisfies \eqref{iterate_psi},
\beq \label{iterate_psi_Lemma_curl} 
\langle \Psi_{k+1}, \delta \vec{\phi} \rangle_{L^2} = \langle J_k \Gamma^* , d\phi \rangle_{L^2}  + \langle \overrightarrow{a_{k+1}}, \vec{\phi} \rangle_{L^2} ,
\eeq
for any matrix-valued $0$-form $\phi \in W^{1,p^*}_0(\Omega)$, where $J_k \equiv I+\epsilon u_k$. Moreover, $y_{k+1}  \in W^{2,p}(\Omega)$ solves 
\beq \label{iterate_y_Lemma_curl}
\Delta y_{k+1} = \Psi_{k+1},
\eeq
with boundary data $y_{k+1}\big|_{\partial\Omega} = 0$, c.f. \eqref{iterate_y}. 

Combining \eqref{iterate_psi_Lemma_curl} and \eqref{iterate_y_Lemma_curl}, and using $d^2y_{k+1} =0$, we obtain from the definition of the weak Laplacian in \eqref{Laplacian_weak} that
\begin{eqnarray}  \label{techeqn3_Lemma_curl}
-\Delta(dy_{k+1})[\vec{\phi}] 
&=&  \big\langle \delta d y_{k+1}, \delta \vec{\phi} \big\rangle_{L^2}  +  \big\langle  d^2 y_{k+1}, d \vec{\phi} \big\rangle_{L^2}  \cr
&=&  \big\langle \delta d y_{k+1}, \delta \vec{\phi} \big\rangle_{L^2}     \cr
&=&   \big\langle \Delta y_{k+1}, \delta \vec{\phi} \big\rangle_{L^2} ,
\end{eqnarray}
since $y_{k+1}$ is a vector valued $0$-form, so that $\delta y_{k+1}=0$ and $\Delta y_{k+1} = (\delta d + d \delta) y_{k+1} = \delta d y_{k+1}$.  Substituting now \eqref{iterate_y_Lemma_curl} for $\Delta y_{k+1}$, we write \eqref{techeqn3_Lemma_curl} as
\begin{eqnarray}  \label{techeqn4_Lemma_curl}
-\Delta(dy_{k+1})[\vec{\phi}] 
&\overset{\eqref{iterate_y_Lemma_curl}}{=}& \big\langle \Psi_{k+1}, \delta \vec{\phi} \big\rangle_{L^2}  \cr
&\overset{\eqref{iterate_psi_Lemma_curl}}{=}&  \langle J_k\, \Gamma^* , d\phi \rangle_{L^2}  + \langle \overrightarrow{a_{k+1}}, \vec{\phi} \rangle_{L^2}.
\end{eqnarray}
Now, recall that $u_{k+1}$ solves \eqref{iterate_u}, that is,
\beq \label{techeqn5_Lemma_curl}
- \Delta u_{k+1}[\phi]=  \big\langle J_k \, \Gamma^* , d \phi \big\rangle_{L^2} + \big\langle a_{k+1} , \phi \big\rangle_{L^2}.
\eeq
By definition of the inner products we have $\big\langle a_{k+1} , \phi \big\rangle_{L^2} = \big\langle \overrightarrow{a_{k+1}} , \overrightarrow{\phi} \big\rangle_{L^2}$, c.f. \eqref{tracestuff}. Thus, 
\eqref{techeqn4_Lemma_curl} in combination with \eqref{techeqn5_Lemma_curl} gives us
\beq \label{techeqn7_Lemma_curl}
\Delta(dy_{k+1})\big[\vec{\phi}\big] = \Delta u_{k+1}\big[\phi\big]
\eeq
for any matrix valued $0$-form $\phi \in W^{1,p^*}_0(\Omega)$. Finally, applying Lemma \ref{Lemma_Laplacian_equiv_matrix_vec} to the right hand side of \eqref{techeqn7_Lemma_curl}, we obtain 
\beq \nonumber
\Delta(dy_{k+1})\big[\vec{\phi}\big] = \Delta \overrightarrow{u_{k+1}}\big[\vec{\phi}\big],
\eeq
which directly gives the sought after equation \eqref{techeqn1_Lemma_curl}. This completes the proof of Lemma \ref{Lemma_curl}. \hfill $\Box$  \vspace{.2cm}

This finishes the proof of Theorem \ref{Thm_Smoothing} and \ref{Thm_compactness}, thereby establishing optimal regularity and Uhlenbeck compactness for $L^\infty$ connections.

\section{Extension of the existence theory to $L^p$ connections}     \label{Sec_Lp-extension}

We now extend the existence theory developed in Sections \ref{Sec_existence_theory} - \ref{Sec_Proofs} for $L^\infty$ connections with $L^\infty$ curvature to the setting of connections $\Gamma \in L^{2p}(\Omega)$ with $d\Gamma \in L^p(\Omega)$, $n/2 <p < \infty$, addressed in Theorem \ref{Thm_Existence_J}, for which we now seek solutions $J \in W^{1,2p}(\Omega)$ and $B \in L^{p}(\Omega)$ of the reduced RT-equations \eqref{reduced_RT_1} - \eqref{reduced_RT_4}, 
\begin{eqnarray} 
\Delta J &=& \delta ( J \mm \Gamma ) - B , \label{Lp-ext_red_RT_2} \\
d \vec{B} &=& \overrightarrow{\text{div}} \big(dJ \wedge \Gamma\big) + \overrightarrow{\text{div}} \big( J\, d\Gamma\big) ,   \label{Lp-ext_red_RT_3} \\
\delta \vec{B} &=& 0 .  \label{Lp-ext_red_RT_4}
\end{eqnarray}
Consistency of the reduced RT-equations at the level of regularity here is established in Section \ref{Sec_consistency} by application of the H\"older and Morrey inequalities as well as Sobolev embedding in the form of Lemma \ref{Lemma_Sobolev_embedding}, which states that $L^{p}(\Omega) \subset W^{-1,2p}(\Omega)$ for $p > n/2$ and that
\beq \label{Lp-ext_Sobolev_embedding_estimate}
\| B \|_{W^{-1,2p}(\Omega)} \leq C \| B \|_{L^p(\Omega)} ,
\eeq
where $C>0$ is some constant depending only on $\Omega, p, n$. The consistency of the auxilliary system of equations
\begin{eqnarray}
d\Psi & = & \overrightarrow{\delta(J \Gamma)} - B, \label{Lp-ext_aux1}  \\
\Delta y &=& \Psi, \label{Lp-ext_aux2}
\end{eqnarray}
follows analogously, since $B \in W^{-1,2p}(\Omega)$ together with $\delta(J \Gamma) \in W^{-1,2p}(\Omega)$ place $\Psi$ in the desired space $L^{2p}(\Omega)$, which in turn places $y$ in the sought after space $W^{2,2p}(\Omega)$.
The argument for obtaining consistency of the augmented reduced RT-equations, \eqref{Lp-ext_red_RT_2} - \eqref{Lp-ext_red_RT_4} and \eqref{Lp-ext_aux1} - \eqref{Lp-ext_aux2}, extends rather directly to the well-posedness and convergence proof of our iteration scheme in Section \ref{Sec_existence_theory}, by using estimate \eqref{Lp-ext_Sobolev_embedding_estimate}. This, as we show below, yields the existence of solutions in the sought after spaces $B \in L^p(\Omega)$ and $J \in W^{1,2p}(\Omega)$, invertible and integrable to coordinates.  

The main novel step is to adapt the $\epsilon$-scaling \eqref{small_Gamma} of the connection to the regularities here, and prove that our incoming $L^p$-bound \eqref{bound_incoming_ass}, i.e.,
\beq \label{bound_incoming_SecLp}
 \|(\Gamma,d\Gamma)\|_{L^{2p,p}(\Omega)}  \equiv \|\Gamma\|_{L^{2p}(\Omega)} + \|d\Gamma \|_{L^p(\Omega)} \; \leq \; M,
\eeq    
is maintained when $p<\infty$, analogous to Lemma \ref{rescaling_Gamma}. For this, we no longer introduce $\Gamma^*$ as the restriction of the components of $\Gamma_x$ to $B_\epsilon(0)$, transformed to $\tilde{x}$-coordinates as scalars, $\tilde{x}(x) \equiv \frac{x}{\epsilon}$. The reason is that transformation of the components of $\Gamma^*_{\tilde{x}}$ to $\Gamma_x$ as scalars (not as connections) is problematic, since  
\beq \label{Lp_scaling_bound_eqn1}
\|\Gamma^*_{\tilde{x}}\|_{L^{2p}(B_1(0))} = \epsilon^{-\frac{n}{2p}} \|\Gamma_{x}\|_{L^{2p}(B_\epsilon(0))},
\eeq
which blows up as $\epsilon >0$ approaches zero. On the other hand, the differentiated connection components (transformed as scalars) give
\beq \label{Lp_scaling_bound_eqn2}
\|d_{\tilde{x}}\Gamma^*_{\tilde{x}}\|_{L^p(B_1(0))} = \epsilon^{1-\frac{n}{p}} \|d_x \Gamma_{x}\|_{L^p(B_\epsilon(0))},
\eeq
and hence decrease with $\epsilon >0$ whenever $p\geq n$. An easy way to circumvent the problematic blow-up in \eqref{Lp_scaling_bound_eqn1} is to measure the undifferentiated connection in the $L^\infty$-norm, (i.e., taking $q=\infty$), and measuring $d\Gamma$ in the $L^p$-norm for $p\geq n$. This is exactly what we assume for our Uhlenbeck compactness result, Theorem \ref{Thm_compactness}, precisely to avoid the aforementioned problem. 

Now, to adapt the $\epsilon$-scaling \eqref{small_Gamma} to connections $\Gamma \in L^{2p}(\Omega)$ with $d\Gamma \in L^p(\Omega)$ for $p>n/2$, $p<\infty$, as in Theorem \ref{Thm_Smoothing}, we do the following: By locality of the problem of optimal regularity, it suffices to restrict to arbitrarily small neighborhoods $B_{\delta}(0)$, so we can exploit that
\beq \label{delta_limit}
\|(\Gamma_x,d_x\Gamma)\|_{L^{2p,p}(B_{\delta}(0))} \longrightarrow 0
\eeq
as $\delta\to0$. To implement this smallness into the reduced RT-equations, denote with $\cutoff$ the characteristic function which vanishes outside $B_\delta(0)$ and is identical to $1$ on $B_\delta(0)$. We then introduce $\Gamma^*$ and $d\Gamma^*$ by 
\beq \label{Gamma-star_Lp}
\epsilon\; \Gamma^* \equiv \cutoff \Gamma 
\hspace{1cm} \text{and} \hspace{1cm}
\epsilon\; d\Gamma^* \equiv \cutoff d\Gamma ,
\eeq
and address in our existence theory the RT-equations with $\cutoff\Gamma$ and $\cutoff d\Gamma$ in place of $\Gamma$ and $d\Gamma$. The true geometric solution of the reduced RT-equations is then recovered by restriction to $B_\delta(0)$. Note that $d\Gamma^*$ is the exterior derivative of $\Gamma^*$ only on the interior of $B_\delta(0)$ but not at the boundary. This does not affect our existence theory, because $\Gamma$ and $d\Gamma$ only enter our arguments through their $L^p$ bounds, for which one may consider $\Gamma$ and $d\Gamma$ as functions independent of each other. Now, substituting \eqref{Gamma-star_Lp} for $(\cutoff\Gamma, \cutoff d\Gamma)$ in the reduced RT-equations yields again the sought-after $\epsilon$-rescaled RT-equations \eqref{pde_u} - \eqref{pde_a} in terms of $\Gamma^*$ and $d\Gamma^*$. Now our adapted existence theory (summarized below) applies to these $\epsilon$-rescaled RT-equations \eqref{pde_u} - \eqref{pde_a}, because $\Gamma^*$ and $d\Gamma^*$ satisfy our incoming $L^p$ bound for $\delta>0$ sufficiently small. That is, given some $\epsilon>0$ at the start, we choose $\delta>0$ small enough such that
\beq 
\|(\cutoff \Gamma,\cutoff d\Gamma)\|_{L^{2p,p}(\Omega)} = \|(\Gamma,d\Gamma)\|_{L^{2p,p}(B_{\delta}(0))}\leq  \epsilon M ,  
\eeq
which by \eqref{Gamma-star_Lp} implies the sought-after bound
\beq \label{Lp-ext_initial_bound}
\| (\Gamma^*, d\Gamma^*) \|_{L^{2p,p}(\Omega)} \leq M.
\eeq
Note, since our existence theory establishes an upper bound for $\epsilon>0$ in terms of $M$ alone (c.f., \eqref{epsilon_uniform_bound} in Section \ref{Sec_Proof_existence_Thm}), it follows that $M$ determines $\epsilon$ and thereby $\delta$ at the start.  However, this choice of $\delta>0$ depends also on the shape of $\Gamma$, and thereby might fail to provide a uniform $\delta$ for sequences of connections, by which $B_\delta(0)$ might collapse to a single point. Because of this, our version of Uhlenbeck compactness in Theorem \ref{Thm_compactness} requires currently the stronger bound $\|(\Gamma_x,d\Gamma_x)\|_{L^{\infty,p}(\Omega)} < M$ for $p>n$. 

We now explain how to prove existence of solutions $(J,B)$ of the reduced RT-equations when $\Gamma \in L^{2p}(\Omega)$, $d\Gamma \in L^p(\Omega)$, $p>n/2$, subject to bound \eqref{bound_incoming_SecLp}. The above scaling argument again yields the $\epsilon$-rescaled reduced RT-equations \eqref{pde_u} - \eqref{pde_a}, and one can again introduce the iteration scheme as in Section \ref{Sec_iteration_scheme}, but adjusted to the sought after solution spaces $J \in W^{1,2p}(\Omega)$ and $B \in L^p(\Omega)$, by using precisely the regularities for testing prescribed in Definition \ref{Def_weak_reduced_RT} of the weak formulation. That is, set $u_0=a_0=0$ to start the iteration, and assume $u_k \in W^{1,2p}(\Omega)$ and $a_k \in L^p(\Omega)$ given for some $k\geq0$. (Recall that the Jacobian associated to $u_k$ is $J_k \equiv I+\epsilon u_k$.) We again introduce the $0$-form $a_{k+1} \in L^p(\Omega)$ as the weak solution of
\beq \label{Lp-ext_iterate_a}
\begin{cases} 
\langle \overrightarrow{a_{k+1}}, \delta \psi \rangle_{L^2} = F_a(u_k)[\psi], \cr 
\langle \overrightarrow{a_{k+1}}, d\varphi \rangle_{L^2}=0,
\end{cases}
\eeq
(for vector valued $2$-forms $\psi \in W^{1,p^*}_0(\Omega)$ and vector valued $0$-forms $\varphi \in W^{1,p^*}_0(\Omega)$), subject to the bound 
\beq \label{Lp-ext_iterate_a_Lp-bound}  
\| a_{k+1} \|_{L^p} \leq C \|F_a(u_k)\|_{W^{-1,p}} ,
\eeq 
where $C>0$ denotes again a universal constant independent of $k$. We then introduce the vector valued $0$-form $\Psi_{k+1}$, now in $L^{2p}(\Omega)$, as the weak solution of
\begin{eqnarray} 
\langle \Psi_{k+1}, \delta \vec{\phi} \rangle_{L^2} &=&  F_{\Psi}(u_k,a_{k+1})[\phi] \label{Lp-ext_iterate_psi}  
\end{eqnarray}
for any matrix-valued $0$-form $\phi \in W^{1,(2p)^*}_0(\Omega)$, again subject to the bound
\beq \label{Lp-ext_iterate_psi_bound} 
\| \Psi_{k+1} \|_{L^p} \leq C \|F_a(u_k)\|_{W^{-1,p}} ,
\eeq 
and the vector valued $0$-form $y_{k+1} \in W^{2,2p}(\Omega)$ as the solution of
\beq \label{Lp-ext_iterate_y}
\begin{cases} 
\Delta y_{k+1} = \Psi_{k+1}, \cr 
y_{k+1}\big|_{\partial\Omega} = 0.
\end{cases}
\eeq
In the last step, we introduce $u_{k+1} \in W^{1,2p}(\Omega)$ as the weak solution of  
\beq \label{Lp-ext_iterate_u}
\begin{cases} 
- \Delta u_{k+1}[\phi] = F_u(u_k,a_{k+1})[\phi], \cr
\ \ \ u_{k+1}|_{\partial\Omega} = dy_{k+1}|_{\partial\Omega}
\end{cases}
\eeq
for matrix valued $0$-forms $\phi \in W^{1,(2p)^*}_0(\Omega)$. To prove well-posedness and convergence of the iteration scheme \eqref{Lp-ext_iterate_a} - \eqref{Lp-ext_iterate_u} we proceed as in Section \ref{Sec_convegrence}, by adapting Lemmas \ref{Lemma_source_estimate_Fa_Fu} - \ref{Lemma_decay} to the regularities here. The first lemma gives the adapted source estimates of Lemma \ref{Lemma_source_estimate_Fa_Fu}.

\begin{Lemma}\label{Lp-ext_Lemma_source_estimate_Fa_Fu}
Let $\Gamma^* \in L^{2p}(\Omega)$, $d\Gamma^* \in L^{p}(\Omega)$, for $n/2<p <\infty$, and assume $u \in W^{1,2p}(\Omega)$ and $a\in L^p(\Omega)$. Then there exists a constant $C_s >0$ depending only on $\Omega, n, p$, such that
\begin{align}      \nonumber                 
\|F_u(u,a)\|_{W^{-1,2p}} &\leq    \|a\|_{L^p} \, + \,  C_s \, \big( {\rm vol}(\Omega) +   \epsilon\: \|u\|_{L^{2p}} \big) \|(\Gamma^*,d\Gamma^*)\|_{L^{2p,p}} \cr
\|F_a(u)\|_{W^{-1,p}} &\leq   C_s \,\big({\rm vol}(\Omega) + \epsilon\: \|u\|_{W^{1,2p}} \big)  \|(\Gamma^*,d\Gamma^*)\|_{L^{2p,p}}   .   
\end{align}
\end{Lemma} 

\Proof
The lemma follows by the proof of Lemma \ref{Lemma_source_estimate_Fa_Fu} in Section \ref{Sec_source_estimates}, using estimate \eqref{Lp-ext_Sobolev_embedding_estimate} in suitable places and H\"older's inequality to bound products involving $\Gamma$ (similarly to the consistency proof in Section \eqref{Sec_consistency}). For example, the bound $\big|\big\langle a, \phi \big\rangle_{L^2}\big| \leq C \|a\|_{L^p} \|\phi\|_{W^{1,(2p)^*}}$ follows from \eqref{Sobolev_embedding_estimate} of Lemma \ref{Lemma_Sobolev_embedding}, while 
$$
\big| \big\langle u \Gamma^* , d \phi \big\rangle_{L^2}\big|  
\leq  C \| u \Gamma^* \|_{L^{2p}} \|d\phi\|_{L^{(2p)^*}} 
 \leq  C\,C_M  \| u\|_{W^{1,2p}} \|\Gamma^* \|_{L^{2p}}  \| u \|_{L^p}  \|\phi\|_{W^{1,p^*}} 
$$
follows by  first using H\"older's and then Morrey's inequality.
\QED

\noindent The next lemma gives the well-posedness of the adapted iteration scheme.

\begin{Lemma} \label{Lp-ext_Lemma_existence_iterates}
Assume $u_k\in W^{1,2p}(\Omega)$ is given, $n/2 < p < \infty$. Then there exists $a_{k+1}\in L^p(\Omega)$ which solves \eqref{iterate_a}, there exists the auxiliary iterates $\Psi_{k+1} \in L^{2p}(\Omega)$ and $y_{k+1} \in W^{2,2p}(\Omega)$ which solve \eqref{iterate_psi} - \eqref{iterate_y}, and there exists $u_{k+1}\in W^{1,2p}(\Omega)$ which solves \eqref{iterate_u} with boundary data \eqref{bdd_J}. In addition, the iterates satisfy the following elliptic estimates,  
\begin{eqnarray}
\|a_{k+1}\|_{L^p(\Omega)}  &\leq & C_e \; \|F_a(u_k)\|_{W^{-1,p}(\Omega)}, \label{Lp-ext_existence_est1} \\
\|\Psi_{k+1}\|_{L^{2p}(\Omega)}  &\leq & C_e \; \|F_u(u_k,a_{k+1})\|_{W^{-1,2p}(\Omega)}, \label{Lp-ext_existence_est2} \\
\|y_{k+1}\|_{W^{2,2p}(\Omega)} &\leq & C_e \; \|F_u(u_k,a_{k+1})\|_{W^{-1,2p}(\Omega)}, \label{Lp-ext_existence_est3}\\
\|u_{k+1}\|_{W^{1,2p}(\Omega)} &\leq & C_e \; \|F_u(u_k,a_{k+1})\|_{W^{-1,2p}(\Omega)}, \label{Lp-ext_existence_est4} 
\end{eqnarray}
for some constant $C_e >0$ depending only on $n, p$ and $\Omega$. 
\end{Lemma}

\Proof
The elliptic estimates and resulting well-posedness of the iteration scheme follows exactly as in the proof of Lemma \ref{Lemma_existence_iterates} by simply adapting the regularity of the source function to those in Lemma \ref{Lp-ext_Lemma_source_estimate_Fa_Fu}, $F_u (u_k,a_{k+1}) \in W^{-1,2p}(\Omega)$ and $F_a(u_k) \in W^{-1,p}(\Omega)$.
\QED

\noindent Lemma \ref{Lemma_curl}, asserting the integrability of each Jacobian $J_k=I + \epsilon\: u_k$ to coordinates, applies again at the level of regularity here and the proof carries over directly. The convergence of the iteration scheme is based again on the estimate on difference in the following lemma.

\begin{Lemma} \label{Lp-ext_Lemma_decay}                         
Assume $\Gamma^* \in L^{2p}(\Omega), d\Gamma^* \in L^p(\Omega)$, $n/2 < p <\infty$, then
\begin{eqnarray}
\|\overline{a_{k+1}}\|_{L^p(\Omega)} 
&\leq & \epsilon\: C_d\;\|(\Gamma^*, d\Gamma^*)\|_{L^{2p,p}(\Omega)} \:\|\overline{u_{k}}\|_{W^{1,2p}(\Omega)} , \label{Lp-ext_decay_a} \\
\|\overline{u_{k+1}}\|_{W^{1,2p}(\Omega)}  
&\leq & \epsilon\: C_d \; \|(\Gamma^*, d\Gamma^*)\|_{L^{2p,p}(\Omega)} \:\|\overline{u_{k}}\|_{W^{1,2p}(\Omega)}, \label{Lp-ext_decay_u} 
\end{eqnarray} 
where $C_d \equiv C_s  C_e (1+C_e) >0$ depends only on $n$, $p$, $\Omega$, where $C_s>0$ and $C_e>0$ are the constants of Lemmas \ref{Lp-ext_Lemma_source_estimate_Fa_Fu} and \ref{Lp-ext_Lemma_existence_iterates} respectively.
\end{Lemma}

\Proof
This follows as in the proof of Lemma \ref{Lemma_decay} in Section \ref{Sec_decay}, by combining the elliptic estimates \eqref{Lp-ext_existence_est1} - \eqref{Lp-ext_existence_est2} with bounds on differences of source terms by previous differences of iterates.  As for Lemma \ref{Lp-ext_Lemma_source_estimate_Fa_Fu}, the proof of Lemma \ref{Lemma_decay} can be adapted to the regularities here by suitable use of estimate \eqref{Lp-ext_Sobolev_embedding_estimate} to bound $B$ in $W^{-1,2p}$ and by using H\"older's inequality to bound products involving $\Gamma$.  
\QED

\noindent Analogous to Section \ref{Sec_convegrence}, Lemma \ref{Lp-ext_Lemma_decay} implies convergence of the iteration scheme, as recorded in the next proposition.

\begin{Prop}   \label{Lp-ext_Prop3}
Assume $\Gamma^*  \in L^{2p}(\Omega), d\Gamma^* \in L^p(\Omega)$, $n/2 < p < \infty$, satisfy the initial bound \eqref{Lp-ext_initial_bound}, and assume $0\ < \ \epsilon \ < \ \frac{1}{C_d M }$, where $C_d>0$ is the constant from Lemma \ref{Lemma_decay}. Then the sequence of iterates $(u_k,a_k)_{k\in \mathbb{N}}$ defined by \eqref{iterate_a} - \eqref{bdd_J} converges in $W^{1,2p}(\Omega) \times L^p(\Omega)$, and the corresponding limits 
\begin{eqnarray} \nonumber
& u \equiv \lim\limits_{k\rightarrow\infty} u_k \ \in W^{1,2p}(\Omega), & \cr
&a \equiv \lim\limits_{k\rightarrow\infty} a_k \ \in L^p(\Omega), & 
\end{eqnarray}
solve the reduced RT-equations \eqref{pde_u} - \eqref{pde_a} and satisfy the bound
\beq \label{Lp-ext_curvature_bound_(u,a)}
\|u\|_{W^{1,2p}(\Omega)} + \|a\|_{L^p(\Omega)} \leq C_2(M) \|(\Gamma^*,d\Gamma^*)\|_{L^{2p,p}(\Omega)} , 
\eeq 
for some constant $C_2(M)>0$ depending only on $\Omega, n, p$ and $M$.
\end{Prop}

\Proof
This is a straightforward extension of the proof of Proposition \ref{Prop3} in Section \ref{Sec_convegrence}.
\QED

Theorem \ref{Thm_Existence_J} follows now from Proposition \ref{Lp-ext_Prop3} by the argument given in Section \ref{Sec_Proof_existence_Thm}. This completes the existence theory for the reduced RT-equations at the sought after level of regularity $\Gamma  \in L^{2p}(\Omega), d\Gamma \in L^p(\Omega)$, $n/2 < p < \infty$, and completes the proof of Theorem \ref{Thm_Existence_J}.

\section{Proof of Theorems \ref{Thm_Smoothing} and \ref{Thm_compactness}} \label{Sec_Proof_optimal_regularity}  

In this section we complete the proofs of our main results stated in Section \ref{Sec_MainResults}, Theorem \ref{Thm_Smoothing} on optimal regularity and Theorem \ref{Thm_compactness} on Uhlenbeck compactness, by applying Theorems \ref{Thm_gauge_existence} and \eqref{Thm_Existence_J}. 

\subsection{Proof of Theorem \ref{Thm_Smoothing} - Optimal Regularity}   

Let $q$ be some point in $\Omega \subset \R^n$ and let $p>\max\{ n/2, 2\}$, $n\geq 2$. Assume $\Gamma \in L^{2p}(\Omega)$ and $d\Gamma \in L^{p}(\Omega)$ in $x$-coordinates, and assume the initial bound \eqref{bound_incoming_ass} holds, i.e. $ \|(\Gamma,d\Gamma)\|_{L^{2p,p}(\Omega)} \leq M$ for some constant  $M>0$. {\it Theorem \ref{Thm_Smoothing} now asserts that there exists a neighborhood $\Omega' \subset \Omega$ of $q$ and a coordinate transformation $x \to y$ with Jacobian $J=\frac{\partial y}{\partial x}\, \in W^{1,2p}(\Omega'_x),$ such that the connection components $\Gamma_y$ in $y$-coordinates exhibit optimal regularity $\Gamma_y \in W^{1,p}(\Omega''_y)$, (precisely one derivative above its curvature), on every open set $\Omega''$ compactly contained in $\Omega'$, where $\Omega''_y \equiv y(\Omega'')$.  }

We prove this assertion by combining Theorem \ref{Thm_Existence_J} and Theorem \ref{Thm_gauge_existence}. The existence of the neighborhood $\Omega'$ together with the coordinate transformation with Jacobian $J\in W^{1,2p}(\Omega'_x)$ follows by Theorem \ref{Thm_Existence_J}, which asserts that there exists a solution $(J,B)$ of the reduced RT-equations \eqref{RT_withB_2} - \eqref{RT_withB_4} defined in $\Omega'_x$ containing $q$, such that $J \in W^{1,2p}(\Omega'_x)$, $J^{-1} \in W^{1,2p}(\Omega'_x)$, $B \in L^{2p}(\Omega'_x)$, $d\vec{J}=0$ in $\Omega'_x$. This $J$ is indeed integrable to coordinates, $J=\frac{\partial y}{\partial x}$, since $d\vec{J}=0$, (c.f. Theorem \ref{Thm_CauchyRiemann}). Moreover, since this Jacobian $J$ solves the reduced RT-equations and meets the properties assumed in Theorem \ref{Thm_gauge_existence}, Corollary \ref{Thm_equivalence} of  Theorem \ref{Thm_gauge_existence} implies that $J$ is indeed the Jacobian to coordinates $y$ such that $\Gamma_y$ has optimal regularity, $\Gamma_y \in W^{1,p}(\Omega''_y)$. Namely,  by part (iii) of Theorem \ref{Thm_gauge_existence}, we have that $\Gammati_J \equiv \Gamma - J^{-1} dJ \in W^{1,p}(\Omega''_x)$ on every open set $\Omega''$ compactly contained in $\Omega'$, using then that the connection of optimal regularity is given by 
\beq \label{Gamma_y_formula}
(\Gamma_y)^\gamma_{\alpha\beta} = J_k^\gamma (J^{-1})^i_\alpha  (J^{-1})^j_\beta   (\Gammati_J)^k_{ij},
\eeq
it follows that $\Gamma_y \in W^{1,p}(\Omega''_y)$, (c.f. the proof of Corollary \ref{Thm_equivalence}); below we show in detail how to control the regularity of the products in \eqref{Gamma_y_formula}. This proves the first assertion of Theorem \ref{Thm_Smoothing}.

{\it Theorem \ref{Thm_Smoothing} asserts further that the connection $\Gamma_y$ satisfies the uniform bound   
\beq \label{curvature_again} 
\|\Gamma_y \|_{W^{1,p}(\Omega''_y)}  \leq C_1(M)\; \|(\Gamma,d\Gamma)\|_{L^{2p,p}(\Omega')} , 
\eeq
and the Jacobian $J$ satisfies 
\beq \label{curvature_on_J_again}
\|J\|_{W^{1,2p}(\Omega''_x)}  + \|J^{-1}\|_{W^{1,2p}(\Omega''_x)} \leq C_1(M)\; \|(\Gamma,d\Gamma)\|_{L^{2p,p}(\Omega')} ,  
\eeq    
for some constant $C_1(M) > 0$ depending only on $\Omega''_x, \Omega'_x, p, n, q$ and $M$. Moreover, Theorem \ref{Thm_Smoothing} states that the neighborhood $\Omega'$ can be taken as $\Omega'_x = \Omega_x \cap B_r(q)$, for $B_r(q)$ the Euclidean ball of radius $r$ in $x$-coordinates, where $r$ depends only on $\Omega_x, p, n$ and $\Gamma$ near $q$; if $\|(\Gamma,d\Gamma)\|_{L^{\infty,2p}(\Omega)}~\leq~M$, then $r$ depends only on $\Omega_x, p, n$ and $M$.}

The uniform bound \eqref{curvature_on_J_again} on $J$ directly follows from the elliptic estimate \eqref{curvature_estimate_soln} of Theorem \ref{Thm_Existence_J} on the solution $(J,B)$ of the reduced RT-equations. Moreover, the dependence of $\Omega'$ also follows from Theorem \ref{Thm_Existence_J}. Note that $\Omega'$ arises from our scaling argument in Section \ref{Sec_existence_theory}, and this scaling only depends on $M>0$, (instead of $\Gamma$ near $q$), in case that the stronger initial bound $\|(\Gamma,d\Gamma)\|_{L^{\infty,2p}(\Omega)}~\leq~M$ holds, (as assumed for Uhlenbeck compactness). So it only remains to prove estimate \eqref{curvature_again} on the connection of optimal regularity. For this, we used that part $(iii)$ of Theorem \ref{Thm_gauge_existence} implies the uniform bound \eqref{Gammati'_estimate} on  $\tilde{\Gamma}_J$, and this gives the bound (\ref{curvature_again}) on $\Gamma_y$ by the relation between $\Gammati_J$ and $\Gamma_y$ given in \eqref{Gamma_y_formula}. That is, \eqref{Gamma_y_formula} implies                       
\begin{eqnarray} \label{Gamma_y_estimate_proof}
\|\Gamma_y \|_{W^{1,p}(\Omega''_y)}  
&\leq &  C \: \|J\|_{W^{1,p}(\Omega''_x)} \|J^{-1}\|_{W^{1,p}(\Omega''_x)}^2 \| \Gammati_J\|_{W^{1,p}(\Omega''_x)},
\end{eqnarray}
which can be shown by applying Morrey's inequality to each product in the case that $p>n$.     However, in the case that $n/2 < p \leq n$ a little more care has to be taken, since Morrey's inequality cannot be used to bound the $L^\infty$ norm of $\Gammati_J \in W^{1,p}$. To address the case $n/2 < p \leq n$, note first that Morrey's inequality still applies to $J,J^{-1} \in W^{1,2p}(\Omega'_x)$, and it follows that $f\equiv  J_k^\gamma (J^{-1})^i_\alpha  (J^{-1})^j_\beta$ is again in $W^{1,2p}(\Omega'_x)$, (we use $f$ here formally to represent any of the components, since only the regularity is of relevance at this stage). Writing now \eqref{Gamma_y_formula} symbolically as $\Gamma_y = f \cdot \Gammati_J$, we have 
\beq \label{Gamma_y_estimate_techeqn1}
\|\Gamma_y \|_{W^{1,p}} \leq  \|f \cdot \Gammati_J \|_{L^p} +  \|f \cdot D\Gammati_J \|_{L^p} +  \|Df \cdot \Gammati_J \|_{L^p}.
\eeq
The first two terms on the right hand side of \eqref{Gamma_y_estimate_techeqn1} can be bounded by applying Morrey's inequality to bound $\|f\|_{L^\infty}$, 
\begin{eqnarray} \label{Gamma_y_estimate_techeqn2}
\|f \cdot \Gammati_J \|_{L^p} +  \|f \cdot D\Gammati_J \|_{L^p}  
&\leq &  \|f\|_{L^\infty} \Big( \|\Gammati_J \|_{L^p} +  \|D\Gammati_J \|_{L^p} \Big)  \cr 
&\leq & C_M \|f\|_{W^{1,2p}} \; \|\Gammati_J \|_{W^{1,p}}.
\end{eqnarray}
On the other hand, the third term can be bounded by first applying H\"older's inequality as in \eqref{Holder_L2p-trick},
\beq \label{Gamma_y_estimate_techeqn2b}
\|Df \cdot \Gammati_J \|_{L^p} \leq \|Df\|_{L^{2p}} \|\Gammati_J\|_{L^{2p}},
\eeq
and then by applying Sobolev embedding \cite[Thm 2, Ch. 5.6]{Evans}, in $(*)$, to bound
\beq \label{Gamma_y_estimate_techeqn3}
\|\Gammati_J\|_{L^{2p}} \leq {\rm vol}(\Omega)^\frac{p'}{2p} \|\Gammati_J\|_{L^{p'}} 
\overset{(*)}{\leq} C_S \|\Gammati_J\|_{W^{1,p}},
\eeq
where $C_S>0$ is the constant of Sobolev embedding and $p'$ is the Sobolev conjugate, $p'=\frac{np}{n-p}$. Note that the first inequality in \eqref{Gamma_y_estimate_techeqn3} holds since $2p \leq p'$ for $n/2\leq p<n$ (with equality at $p=n/2$). Note further that the Sobolev embedding in \eqref{Gamma_y_estimate_techeqn3} holds for any $p$ with $n/2 < p <n$. The analogous bound to \eqref{Gamma_y_estimate_techeqn3} in the special case $p=n$ is obtained by applying Sobolev embedding to $W^{1,\frac{2}{3}n}$, that is,
\beq \label{Gamma_y_estimate_techeqn4}
\|\Gammati_J\|_{L^{2n}}    \leq C_S \|\Gammati_J\|_{W^{1,\frac{2}{3}n}} 
\leq  {\rm vol}(\Omega)^\frac{1}{3} \|\Gammati_J\|_{W^{1,p}},
\eeq
where we used that $\Gammati_J \in W^{1,p}(\Omega''_x)$ lies also in $W^{1,\frac{2}{3}n}(\Omega''_x)$ when $p=n$, and that $2n$ is the Sobolev conjugate of $\frac{2}{3}n$. Estimates \eqref{Gamma_y_estimate_techeqn1} - \eqref{Gamma_y_estimate_techeqn4} taken together now establishes \eqref{Gamma_y_estimate_proof} in the remaining case $n/2 < p \leq n$. The sought after estimate (\ref{curvature_again}) now follows by estimating $\Gammati_J$ in \eqref{Gamma_y_estimate_proof} by \eqref{Gammati'_estimate} and $J, J^{-1}$ by (\ref{curvature_estimate_soln}).  This completes the proof of Theorem \ref{Thm_Smoothing}.

\subsection{Proof of Theorem \ref{Thm_compactness} - Uhlenbeck Compactness}

For the proof we employ Morrey's inequality, the Banach-Alaoglu Theorem, Sobolev compactness and Lemma  \ref{technical} below, (a technical lemma which states that products of strongly and weakly converging functions converge weakly).  Recall that Morrey's inequality tells us that, when $p>n$ (our assumption here),  functions uniformly bounded in $W^{1,p}$ are also uniformly bounded in $L^\infty$, so $W^{1,p}$ is closed under products, and uniform bounds in $W^{1,p}$ norms extend to uniform bounds on products.   Banach-Alaoglu tells us that the closed unit ball in $L^p$  is weakly compact \cite{Rudin_FunctionalAnalysis}.  These together with Sobolev compactness and the boundedness of $\Omega$ tell us that sequences of functions uniformly bounded in $W^{1,p}$ and $L^\infty$ admit subsequences which converge weakly in $W^{1,p}$, strongly in $L^p$, with uniform bounds given by the original uniform $W^{1,p}$ and $L^{\infty}$ bounds.  
 
We now give the proof of Theorems \ref{Thm_compactness}. So assume  $\{(\Gamma_i)_x\}_{i\in \mathbb{N}}$ are the $x$-components of a sequence of connections $\Gamma_i$ defined on the tangent bundle $T\mathcal{M}$ of an $n$-dimensional manifold $\mathcal{M}$ in a fixed coordinate system $x$, let $n < p <\infty$, and assume $(\Gamma_i)_x \in L^{\infty}(\Omega_x)$, $d\big((\Gamma_i)_x\big) \in L^{p}(\Omega_x)$ such that 
\beq \label{uniform_bound_again}
\|(\Gamma_i,d\Gamma_i)\|_{L^{\infty,p}(\Omega_x)}   \leq M,
\eeq
for some constant $M>0$ independent of $i \in \mathbb{N}$. We need to prove that for each $q \in \Omega$ there exists a fixed neighborhood $\Omega' \subset \Omega$ of $q$, and for each $(\Gamma_i)_x$ there exists a coordinate transformation $x \to y_i(x)$ taking $\Omega'_x$ to $\Omega'_{y_i}$, such that the components $\Gamma_{y_i} \equiv (\Gamma_i)_{y_i}$ of $\Gamma_i$ in $y_i$-coordinates exhibit optimal regularity $\Gamma_{y_i} \in W^{1,p}(\Omega'_{y_i})$, both in $y_i$-coordinates, and when expressed in $x$-coordinates $\Gamma_{y_i}(x) \equiv \Gamma_{y_i}(y_i(x)) \in W^{1,p}(\Omega'_x)$. We need to prove further that a subsequence of $y_i(x)$ converges to some $y(x)$ weakly in $W^{2,2p}(\Omega'_x)$, strongly in $W^{1,2p}(\Omega'_x)$, and that a further subsequence $\Gamma_{y_i}(x)$ converges to $\Gamma_y(x)$ weakly in $W^{1,p}(\Omega_x')$, strongly in $L^p(\Omega_x')$, and that $\Gamma_y$ is the connection $\Gamma_x$ in $y$-coordinates, where $\Gamma_x$ is the weak $L^p$-limit of $(\Gamma_i)_x$. 

By Theorem \ref{Thm_Smoothing}, there exists a single neighborhood $\Omega'$ depending only on $M$, (assuming $n,p,\Omega$ fixed), on which a coordinate transformation  $x\to y_i(x)$ exists for each $i \in \mathbb{N}$, taking $\Omega'_x$ to $\Omega'_{y_i},$ which maps $(\Gamma_i)_x$ to optimal regularity on $\Omega'$, so $\Gamma_{y_i} \in W^{1,p}(\Omega'_{y_i})$. Moreover, by estimate \eqref{curvature} of Theorem  \ref{Thm_Smoothing}, $\|\Gamma_{y_i}\|_{W^{1,p}(\Omega'_{y_i})}$ are uniformly bounded.  This proves $(i)$ of Theorem \ref{Thm_compactness}.    

For each $i \in \mathbb{N}$, the coordinate transformations $y_i(x)$ are obtained from the Jacobians constructed in Theorem \ref{Thm_Existence_J}. To obtain a uniform $L^p$-bound on $y_i(x)$, we choose $y_i(x(q))=0$ for each $i\in\mathbb{N}$. It follows from estimate \eqref{curvature_on_J} of Theorem \ref{Thm_Smoothing}, (or alternatively \eqref{curvature_estimate_soln} of Theorem \ref{Thm_Existence_J}), that the Jacobians $J_i$ of $x \to y_i(x)$ satisfy the uniform bound $\|J_i\|_{W^{1,2p}(\Omega'_x)}\leq C_2(M)$ and $\|J^{-1}_i\|_{W^{1,2p}(\Omega'_x)}\leq C_2(M)$.  Since the Jacobians bound the derivatives of the coordinate maps $x\to y_i(x)$, and $y_i(x(q))=0$ bounds the $L^p$-norm (and supnorm), it follows that each map $y_i(x)$ as a function of $x$ is uniformly bounded in $W^{2,2p}(\Omega'_x)$ by some constant $C_4(M)$, again depending only on $M$.    It now follows from the basic compactness theorem for Sobolev spaces (Banach Alaoglu) that there exists a subsequence, also denoted $y_i(x)$, on which $y_i(x)$ converges to $y(x)$  weakly in $W^{2,2p}(\Omega'_x)$, strongly in $W^{1,2p}(\Omega'_x)$, such that $\|y\|_{W^{2,2p}(\Omega'_x)}<C_4(M)$.    In particular,  $J_i$ converges to $J$ weakly in $W^{1,2p}(\Omega'_x)$, strongly in $L^{2p}(\Omega'_x)$, and the uniform bound on $J^{-1}_i$ implies invertibility of $J$. This proves $(iii)$ of Theorem \ref{Thm_compactness}.  

Now since $J_i\in W^{1,2p}(\Omega'_x)$, $\Gamma_{y_i}(x)$ are uniformly bounded in $W^{1,p}(\Omega'_x)$ by the chain rule. That is, by Morrey's inequality one can estimate products $\Gamma_{y_i}(x)$ times $J_i$ to lie in $W^{1,p}(\Omega'_x)$, with norm bounded by some $C_5(M)$ depending only on $M$. This proves $(ii)$ of Theorem \ref{Thm_compactness}. 

By the uniform $W^{1,p}$-bound on $\Gamma_{y_i}(x)$, it follows that a further subsequence of $\Gamma_{y_i}(x)$ converges weakly in $W^{1,p}(\Omega'_x)$ to a connection $\Gamma_{y}(x)$ which satisfies the same bound $C_5(M)$ in $W^{1,p}(\Omega'_x)$.   Thus the coordinate map $x\to y$ is in $W^{2,2p}(\Omega'_x)$, and so $\Gamma_y$ exhibits optimal regularity in $y$-coordinates.   

Finally, by taking a further subsequence, $(\Gamma_i)_x$ converges to some $\Gamma_x$ weakly in $L^p(\Omega_x)$ by the Banach Alaoglu Theorem, (i.e., the uniform $L^\infty$-bound \eqref{uniform_bound_again} directly implies a uniform $L^p$-bound because $\Omega$ is bounded). To show that $\Gamma_y$ is indeed the connection $\Gamma_x$ in $y$-coordinates, we use that for each $i$, $\Gamma_{y_i}$ is the connection $(\Gamma_i)_x$ in $y_i$-coordinates, so by the transformation law for connections (written in shorthand, suppressing indices) we have
\beq \label{compactness_techeqn1}
J_i^{-1}\mm J_i \mm J_i \cdot \Gamma_{y_i} =  (\Gamma_i)_x - J_i^{-1}\, dJ_i .
\eeq
Since $J_i$ converges to $J$ weakly in $W^{1,2p}(\Omega'_x)$, $J_i$ converges to $J$ strongly in $L^{2p}$ and $dJ_i$ converges to $dJ$ weakly in $L^{2p}$.  Similarly, $J^{-1}_i$ converges to $J^{-1}$ strongly in $L^{2p}$ and weakly in $W^{1,2p}(\Omega'_x)$, and $\Gamma_{y_i}(x)$ converges to $\Gamma_y(x)$  strongly in $L^p$ and weakly in $W^{1,p}(\Omega'_x)$. Thus by Lemma \ref{technical} below, (taking $J_i^{-1}\mm J_i \mm J_i$ for $f_i$ and $\Gamma_{y_i}$ for $g_i$), the left hand side of \eqref{compactness_techeqn1} converges to $J^{-1}\mm J \mm J \mm \Gamma_{y}$ weakly in $L^p$, and the right hand side of \eqref{compactness_techeqn1} converges to $\Gamma_x - J\, dJ$ weakly in $L^p$.   Taken on whole,  (\ref{compactness_techeqn1}) and Lemma \ref{technical} imply that the connection $\Gamma_y$ is the connection $\Gamma_x$ transformed to $y$-coordinates as $L^p$ functions, which proves $(iv)$ and completes the proof of Theorem \ref{Thm_compactness} once we prove the following lemma.
 
\begin{Lemma}\label{technical}
Let $f_i,g_i$ be sequences of functions on a bounded set $\Omega$ such that $f_i\to f$ in $L^p(\Omega)$ with $\|f_i\|_{L^\infty},\|f\|_{L^\infty}\leq M$,  and such that $g_i\to g$ weakly in $L^p$ with $\|g_i\|_{L^p},\|g\|_{L^p}\leq \tilde{M},$ $p>n.$   Then $f_ig_i\to fg$ weakly in $L^p(\Omega).$
\end{Lemma}

\Proof  
Since $\|f_i\|_{\infty},\|f\|_{\infty}\leq M$,  and $f_i\to f$ in $L^p$ on a bounded set, it follows that $f_i\to f$ in every $L^p$, $p\geq1$.   This follows by measure theory because the measure of the set on which $|f_i-f|>\epsilon$ tends to zero as $i\to\infty$ for every $\epsilon>0$.  Recall now that the dual space of $L^p$ is $L^{p^*}$ with $1/p+1/{p^*}=1$.   Thus to prove that  $f_ig_i-fg\to0$ weakly in $L^p,$ we must show that
\begin{eqnarray}\label{start0}
\langle f_ig_i-fg,\phi \rangle_{L^2}\equiv\int_{\Omega} (f_ig_i-fg)\phi \to 0
\end{eqnarray}
for every $\phi\in L^{p^*}$, where $\langle \cdot , \cdot \rangle_{L^2}$ is the $L^2$ inner product.   But
\begin{eqnarray}\label{start1}
\langle f_ig_i-fg,\phi\rangle_{L^2}=\langle f(g_i-g),\phi\rangle_{L^2}+\langle(f_i-f)g_i,\phi\rangle_{L^2}.
\end{eqnarray}  
But $f\in L^\infty$ implies $f\phi\in L^{p^*}$, so the first term in (\ref{start1}) satisfies
 \begin{eqnarray} \nonumber
\langle f(g-g_i),\phi\rangle_{L^2}=\langle(g-g_i),f\phi\rangle_{L^2}\to0
\end{eqnarray}  
because $g_i-g$ tends to zero weakly in $L^p$.   

Consider now the second term in (\ref{start1}).   Since $f_i-f\in L^\infty$, we have $(f_i-f)\phi\in L^{p^*}$, so we can apply Holder's inequality twice to obtain the estimate
 \begin{eqnarray}\label{start2}
|\langle(f_i-f)g_i,\phi\rangle_{L^2}|&=&|\langle g_i,(f_i-f)\phi\rangle_{L^2}|\\\nonumber
&\leq&\|g_i\|_{L^p}\|(f-f_i)\phi\|_{L^{p^*}}\\\nonumber
&\leq& \tilde{M}\|(f-f_i)\phi\|_{L^{p^*}}.
\end{eqnarray} 
Now let $E_N=\left\{x\in\Omega: |\phi|^{p^*}\geq N\right\}$.   Then since $\phi^{p^*}\in L^1$, it follows that $\int_{E_N}|\phi|^{p^*}\to 0$ as $N\to\infty$.   Thus
\begin{eqnarray}\label{start3}
 \|(f-f_i)\phi\|^{p^*}_{L^{p^*}}&=&\int_{\Omega}|f-f_i|^{p^*}|\phi|^{p^*}\,d\mu\\\nonumber
 &=&\int_{E_N}|f-f_i|^{p^*}|\phi|^{p^*}\,d\mu+\int_{E^c_N}|f-f_i|^{p^*}|\phi|^{p^*}\,d\mu\\\nonumber
 &\leq&\|(f-f_i)^{p*}\|_{L^\infty}\int_{E_N}|\phi|^{p^*}\,d\mu+N\int_{E^c_N}|f-f_i|^{p^*}\,d\mu\\\nonumber
 &\leq&(2M)^{p^*}\int_{E_N}|\phi|^{p^*}\,d\mu+N\int_{\Omega}|f-f_i|^{p^*}\,d\mu.
\end{eqnarray} 
 Now we can make the first term arbitrarily small by choosing $N$ sufficiently large, and the second term tends to zero with $i$ because $f_i\to f$ in $L^{p^*}(\Omega)$.  It follows that $\|(f-f_i)\phi\|_{L^{p^*}}\to0$, and by this we conclude from (\ref{start2}) that the second term in (\ref{start1}) tends to zero as well.   Thus $f_ig_i\to fg$ weakly in $L^p$ as claimed.  
\QED

\appendix

\section{Sobolev norms and inequalities} \label{Sec_appendix_norms}

We first give an overview of the norms used in this paper. These norms are coordinate dependent, so we assume at the start a given coordinate system $x$ defined on an open set $\Omega \subset \mathcal{M}$ such that $\Omega_x \equiv x(\Omega) \subset \R^n$ is bounded. In this section $\Omega$ always refers to $\Omega_x$. We denote by $\|\cdot \|_{W^{m,p}(\Omega)}$ the standard $W^{m,p}$-norm, defined as the sum of the $L^p$-norms of derivatives from order zero up to $m$ \cite{Evans}. When applied to matrix or vector valued differential forms $\omega$,  $\|\omega\|_{W^{m,p}(\Omega)}$ denotes the sum of the $W^{m,p}$-norms of all components, i.e., summation over all matrix (or vector) and differential form components.  That is, for matrix valued $k$-forms $\omega= \omega^\mu_{\nu\, i_1 ... i_k} dx^{i_1} \wedge ... \wedge dx^{i_k}$ we define 
\begin{eqnarray} \label{def_norms}
\| \omega \|_{L^p(\Omega)} &\equiv &  \sum_{\mu,\nu,i_1,...,i_k} \big\|\omega^\mu_{\nu\, i_1 ... i_k}\|_{L^p(\Omega)}   \\
\| \omega \|_{W^{m,p}(\Omega)} &\equiv & \sum_{\mu,\nu,i_1,...,i_k} \big\|\omega^\mu_{\nu\, i_1 ... i_k}\|_{W^{m,p}(\Omega)}
\ = \  \sum_{|l|\leq m} \| \partial^l \omega \|_{L^p(\Omega)} ,
\end{eqnarray}
where $1\leq p \leq \infty$, $l$ is a multi-index, (i.e., $l=(l_1,...,l_n)$, $|l|=l_1+... +l_n$ and  $\partial^l \omega \equiv (\partial_{l_1}\omega,...\partial_{l_n}\omega)$ taken component wise).    We further define the $L^2$-inner product on matrix valued forms $\omega$ and $u$,
\beq \label{Def_L2-inner-product}
\langle \omega , u  \rangle_{L^2} 
\equiv \int_\Omega \tr \big(\langle \omega\; ; u^T \rangle \big) dx
= \sum_{\nu, \sigma=1}^n  \sum_{i_1<...<i_k} \int_\Omega  \omega^\nu_{\sigma\: i_1...i_k} u^\nu_{\sigma\: i_1...i_k}  dx ,
\eeq
c.f. \eqref{inner-product_L2}, where $\tr(\cdot)$ is the matrix trace  and $\langle\cdot \; ;\: \cdot \rangle$ the matrix valued inner product \eqref{def_inner-product}, and $dx$ is the Lebesgue measure in a fixed coordinate system $x$. By this we introduce the Hilbert-Schmidt inner product on the matrix components of matrix valued differential forms. Note that, (by Young's inequality), the $L^2$ norm associated to \eqref{Def_L2-inner-product} is equivalent to \eqref{def_norms} when $p=2$.  When convenient we drop the dependence of norms on $\Omega$, writing $\| \cdot \|_{W^{m,p}}$ instead of $\| \cdot \|_{W^{m,p}(\Omega)}$.

We now summarize the basic integral inequalities we apply in this paper, see \cite{Evans} for details. The space $W^{1,p}$ for $p>n$, is embedded in the space of H\"older continuous functions $C^{0,\alpha}(\overline{\Omega})$. Namely, for $p>n$ Morrey's inequality gives
\beq \label{Morrey_textbook}
\| f\|_{C^{0,\alpha}(\overline{\Omega})}  \leq C_M \|f\|_{W^{1,p}(\Omega)},
\eeq
where $\alpha \equiv 1 - \frac{n}{p}$ and $C_M>0$ is a constant depending  only on $n$, $p$ and $\Omega$ \cite{Evans}.\footnote{In Section \ref{Sec_gauge_smooth}, we absorb combinatorial factors in $C_M$ when applying \eqref{Morrey_textbook} to higher derivatives.}  Morrey's inequality (\ref{Morrey_textbook}) extends unchanged to components of matrix valued differential forms and hence to the norms in \eqref{def_norms}. By Morrey's inequality we can estimate products of $W^{1,p}$ functions $f$ and $g$ on bounded domains $\Omega$ as        
\beq \label{Morrey_algebra_prop}
\| fg \|_{W^{1,p}(\Omega)}  \leq C_M \|f\|_{W^{1,p}(\Omega)} \|g\|_{W^{1,p}(\Omega)},
\eeq 
by pulling $L^\infty$ norms of undifferentiated functions out of $L^p$ norms and applying \eqref{Morrey_textbook} to bound the resulting $L^\infty$ norms. This shows that $W^{1,p}(\Omega)$ is closed under multiplication on bounded domains.

To handle products in the RT-equations at the lowest order of regularity we employ H\"older's inequality, which states
\beq \label{Holder}
\| fg\|_{L^1(\Omega)} \leq  \| f\|_{L^p(\Omega)} \| g\|_{L^{p^*}(\Omega)},
\eeq
where $p$ and $p^*$ are conjugate exponents, i.e., $\frac{1}{p} + \frac{1}{p^*} =1$. Now, assuming $f,g \in L^{2p}(\Omega)$, \eqref{Holder} implies the estimate
\beq \label{Holder_L2p-trick}
\| fg\|_{L^p(\Omega)} =  \big\| |fg|^p \big\|^{\frac1{p}}_{L^1(\Omega)}  \overset{\eqref{Holder}}{\leq}  \big\| |f|^p \big\|^{\frac1{p}}_{L^{2}(\Omega)} \, \big\| |g|^p \big\|^{\frac1{p}}_{L^{2}(\Omega)} 
= \| f \|_{L^{2p}(\Omega)} \ \| g \|_{L^{2p}(\Omega)},
\eeq
which shows in particular that $fg \in L^p(\Omega)$. Estimate \eqref{Holder_L2p-trick} allows us to control the gradient product $dJ^{-1} \wedge dJ$ in \eqref{RT_1} in the proof of Theorem \ref{Thm_gauge_existence}, which is a key step in our analysis. H\"older's inequality \eqref{Holder} and estimate \eqref{Holder_L2p-trick} extend to matrix valued differential forms.  For example, for matrix valued $0$-forms $A$ and matrix valued $k$-forms $B$, we have  
\begin{eqnarray}  \label{Holder_matrix-multi}
\| A \cdot B\|_{L^1(\Omega)}   
&\equiv & \sum_{\mu,\nu} \| (A \cdot B)^\mu_\nu\|_{L^1(\Omega)}   
\ \leq \  \sum_{\mu,\nu,\sigma} \| A^\mu_\sigma  B^\sigma_\nu\|_{L^1(\Omega)}     \cr
&\overset{\eqref{Holder}}{\leq} & \sum_{\mu,\nu,\sigma} \| A^\mu_\sigma\|_{L^p(\Omega)}  \| B^\sigma_\nu\|_{L^{p^*}(\Omega)}   \cr
&\leq &  \Big( \sum_{\mu,\nu} \| A^\mu_\nu\|_{L^p(\Omega)} \Big) \Big( \sum_{\mu,\nu} \| B^\mu_\nu\|_{L^{p^*}(\Omega)} \Big)\cr
&= & \| A\|_{L^p(\Omega)} \| B\|_{L^{p^*}(\Omega)},   
\end{eqnarray}              
and by applying \eqref{Holder_L2p-trick} component-wise, we obtain in a similar fashion
\beq
\| A \cdot B\|_{L^p(\Omega)}  \leq \| A \|_{L^{2p}(\Omega)} \ \| B \|_{L^{2p}(\Omega)}.
\eeq
This extends to general matrix valued differential forms $A$ and $B$ as
\begin{eqnarray}  \label{Holder_wedge}
\| A \wedge B\|_{L^1(\Omega)}  &\leq & \| A\|_{L^p(\Omega)} \| B\|_{L^{p^*}(\Omega)}, \cr
\| A \wedge B\|_{L^p(\Omega)}  &\leq &  \| A\|_{L^{2p}(\Omega)} \| B\|_{L^{2p}(\Omega)}.
\end{eqnarray}

\section{Elliptic PDE theory} \label{Sec_Prelimiaries-elliptic}

We now summarize the estimates we use from elliptic PDE theory.  It suffices to assume here that $1< p<\infty$, $n\geq 2$. We assume throughout that $\Omega \subset \mathbb{R}^n$ is a bounded open domain, simply connected and with smooth boundary. Our estimates are based on the following two theorems, which directly extend to matrix valued and vector valued differential forms since the Laplacian acts component wise, c.f. Lemma \ref{Lemma_weak_Poisson_equivalence}. That is, we take the weak Laplacian here as $\Delta u[\phi] = -\langle \nabla u, \nabla\phi\rangle_{L^2}$ for scalar functions $u \in W^{1,p}(\Omega)$ and for test functions $\phi \in W_0^{1,p^*}(\Omega)$, where $W^{1,p^*}_0(\Omega)$ is the closure of $C^\infty_0(\Omega)$ with respect to the $W^{1,p^*}$-norm (so $\phi|_{\partial\Omega}=0$).  Our first theorem is based on Theorem 7.2 in \cite{Simader}, but adapted to the case of solutions to the Poisson equation with non-zero Dirichlet data.    

\begin{Thm}\label{Thm_Poisson}
Let $\Omega\subset R^n$ be a bounded open set with smooth boundary $\partial\Omega$, assume $f\in W^{-1,p}(\Omega)$ and $u_0\in W^{1,p}(\Omega)\cap C^0(\overline{\Omega})$ for $1< p<\infty$, $n\geq 2$.   Then the Dirichlet boundary value problem
\begin{eqnarray}
 \Delta u[\phi]&=&f[\phi],\ \ \text{in}\ \Omega  \label{Poisson-1}\\
u&=&u_0\ \  \text{on}\ \partial\Omega, \label{Poisson-2}
\end{eqnarray}
for any $\phi \in W_0^{1,p^*}(\Omega)$, has a unique weak solution $u \in W^{1,p}(\Omega)$ with boundary data $u-u_0 \in W_0^{1,p}(\Omega)$. Moreover, any weak solution\footnote{It suffices to assume that $u$ is regular enough to make sense of the weak formulation of the Laplacian, for example, $du, \delta u \in L^p(\Omega)$ for a differential form $u$, as in Section \ref{Sec_gauge_lowreg}.} $u$ of \eqref{Poisson-1} - \eqref{Poisson-2} satisfies
\begin{eqnarray}\label{Poisson-3}
\| u\|_{W^{1,p}(\Omega)}\leq C\left( \|f\|_{W^{-1,p}(\Omega)} + \|u_0\|_{W^{1,p}(\Omega)} \right),
\end{eqnarray}
for some constant $C$ depending only on $\Omega, n, p$, and if $f\in L^p(\Omega)$ and  $u_0\in W^{2,p}(\Omega)$, then the solution $u$ satisfies
\begin{eqnarray}\label{Poisson-3_Lp}
\| u\|_{W^{2,p}(\Omega)}\leq C\left( \|f\|_{L^p(\Omega)} + \|u_0\|_{W^{2,p}(\Omega)} \right).
\end{eqnarray}
\end{Thm}

\Proof
Theorem 7.2 in \cite{Simader} yields existence of a unique solution $u \in W^{1,p}(\Omega)$ to \eqref{Poisson-1} - \eqref{Poisson-2} satisfying estimate \eqref{Poisson-3} in the case of zero Dirichlet data, i.e. when $u_0=0$ in $\Omega$. Note, Theorem 7.2 in \cite{Simader} applies since the weak Laplacian is a strongly uniformly elliptic operator in the sense of equation (1.8) of \cite[Def 1.3]{Simader}.\footnote{That in fact any such solution satisfies estimate \eqref{Poisson-3} follows from Theorem 6.1 in \cite{Simader}, equation (6.2), where we can take $C_2=0$ since $\Delta$ is strongly uniformly elliptic.}  To extend this result to non-zero Dirichlet data, let $\tilde{u}\in W^{1,p}(\Omega)$ be the solution of the Laplace equation $\Delta \tilde{u}=0$ with boundary data $\tilde{u}=u_0$ on $\partial\Omega$ in the sense that $u_0-\tilde{u} \in W^{1,p}_0(\Omega)$; note that $u$ can be constructed via Green's representation formula \cite[Eqn. (2.21)]{GilbargTrudinger} for $W^{1,p}$-data. Assume now $w \in W^{1,p}$ is the solution of \eqref{Poisson-1} with zero Dirichlet data satisfying \eqref{Poisson-3}, $\Delta w=f$ in a weak sense and $w \in W^{1,p}_0(\Omega)$, which exists by Theorem 7.2 in \cite{Simader}. Then $u \equiv w +\tilde{u}$ solves  \eqref{Poisson-1} - \eqref{Poisson-2}, since $\Delta u = \Delta w=f$ (in a weak sense) and $u-u_0 \in W^{1,p}_0(\Omega)$. To show that estimate \eqref{Poisson-3} holds, we begin by using the triangle inequality twice to get
\begin{eqnarray} \label{GilTru_3}
\|u\|_{W^{1,p}}  
& \leq &  \|w\|_{W^{1,p}}  + \|\tilde{u}-u_0\|_{W^{1,p}}  + \|u_0\|_{W^{1,p}}. 
\end{eqnarray}
We can now apply the established case of estimate \eqref{Poisson-3}, for the case of zero Dirichlet data, to the first two terms, since $w,\: \tilde{u}-u_0 \in W^{1,p}(\Omega)$. This yields
\begin{eqnarray} \label{GilTru_4}
\|w\|_{W^{1,p}}    & \leq &  C \|f\|_{W^{-1,p}(\Omega)},  
\end{eqnarray}
while the second term is bounded by
\begin{eqnarray} \label{GilTru_5}
\|\tilde{u}-u_0\|_{W^{1,p}(\Omega)} & \leq & C \|\Delta(\tilde{u}-u_0)\|_{W^{-1,p}(\Omega)} \cr
& \leq & C \|u_0\|_{W^{1,p}(\Omega)},
\end{eqnarray}
where we used that $\Delta(\tilde{u}-u_0) = \Delta u_0$ and $\|\Delta u_0\|_{W^{m-1,p}(\Omega)} \leq \|u_0\|_{W^{m+1,p}(\Omega)}$. Substitution of estimates \eqref{GilTru_4} and \eqref{GilTru_5} into \eqref{GilTru_3} yields the sought after estimate \eqref{Poisson-3} in the general case of non-zero Dirichlet data. We proved that there exists a solution to \eqref{Poisson-1} - \eqref{Poisson-2} which satisfies estimate \eqref{Poisson-3}.  

To complete the proof, note that estimate \eqref{Poisson-3_Lp} in the case of zero Dirichlet data ($u_0=0$) is already proven in \cite[Thm 7.2]{Simader}, (c.f. Lemma 9.17 in \cite{GilbargTrudinger}). The case of estimate \eqref{Poisson-3_Lp} for non-zero Dirichlet data follows by an argument analogous to \eqref{GilTru_3} and \eqref{GilTru_5}. Namely, let $\tilde{u}$ be the solution of $\Delta \tilde{u}=0$ with boundary data $u_0-\tilde{u} \in W^{1,p}_0(\Omega)$, and let $w \in W^{2,p}(\Omega)$ be the solution of $\Delta w=f$ with $w \in W^{1,p}_0(\Omega)$ established in \cite[Thm 7.2]{Simader}. Then setting again $u \equiv w + \tilde{u}$ and applying estimate \eqref{Poisson-3_Lp} in the case of vanishing Dirichlet data ($y=0$) to $w$ and $\tilde{u}-u_0$ yields the sought after estimate \eqref{Poisson-3_Lp}:
\begin{eqnarray} \nonumber
\|u\|_{W^{2,p}(\Omega)} 
& \leq &  \|w\|_{W^{2,p}}  + \|\tilde{u}-u_0\|_{W^{2,p}}  + \|u_0\|_{W^{2,p}}   \cr
& \leq & C \big( \|f\|_{L^p(\Omega)} + \|\Delta(\tilde{u}-u_0)\|_{L^p(\Omega)}  \big)+ \|u_0\|_{W^{2,p}(\Omega)} \cr
& \leq & C \big( \|f\|_{L^p(\Omega)} + \|u_0\|_{W^{2,p}(\Omega)}  \big),
\end{eqnarray}
where $C>0$ was again used as a running constant. This completes the proof of Theorem \ref{Thm_Poisson}.   
\QED

We require the following interior elliptic estimates in the proof of Theorem \ref{Thm_gauge_existence} in Section \ref{Sec_gauge} in the case $m=0$, and for higher regularities $m\geq 1$ to prove Proposition \ref{Thm_gauge-invariance}. Note that interior elliptic estimates usually are established earlier in the development of elliptic PDE theory, but for completeness we derive the interior estimate from \eqref{Poisson-3}.

\begin{Thm}\label{Thm_Poisson_interior}
Let $f\in W^{m-1,p}(\Omega)$, for $m\geq 0$ and $1< p<\infty$, $n\geq 2$. Assume $u$ is a weak solution of \eqref{Poisson-1}. Then $u \in W^{m+1,p}(\Omega')$ for any open set $\Omega'$ compactly contained in $\Omega$ and there exists a constant $C$ depending only on $\Omega, \Omega', m, n, p$ such that
\begin{eqnarray}\label{Poisson-4}
\| u \|_{W^{m+1,p}(\Omega')} \leq C \big( \| f \|_{W^{m-1,p}(\Omega)}  + \|u\|_{W^{m,p}(\Omega)} \big).
\end{eqnarray}
\end{Thm}

\Proof
We only need to prove the case $m=0$. The case for $m\geq 1$ can easily be obtained by differentiating and applying the estimate for the case $m=0$; (c.f. Appendix A in \cite{ReintjesTemple_ell2}.)

We apply estimate \eqref{Poisson-3} to $\phi u$ where $\phi$ is a standard smooth cutoff function, $\phi=1$ in $\Omega'$, $\phi=0$ on $\partial\Omega$.  Then    
 \begin{eqnarray}\label{Poisson-5}
\Delta(\phi u)=\phi\Delta u+2\nabla{\phi}\cdot \nabla{u}+ u\Delta\phi \equiv \hat{f}.
\end{eqnarray}
Then applying \eqref{Poisson-3} together with the assumption that we have a solution of the Poisson equation \eqref{Poisson-1}, we have
 \begin{eqnarray}\label{Poisson-6}
\|u\|_{W^{1,p}(\Omega')}&=&\|\phi u\|_{W^{1,p}(\Omega')}\leq \|\phi u\|_{W^{1,p}(\Omega)} \leq C\|\hat{f}\|_{W^{1,p}(\Omega)}\\ \nonumber
&=&C\|\phi\|_{C^2}\left( \|f\|_{W^{-1,p}(\Omega)}+\|\nabla u\|_{W^{-1,p}(\Omega)}+\|u\|_{W^{-1,p}(\Omega)}\right),
\end{eqnarray}
from which (\ref{Poisson-4}) follows, since by definition of the operator norm we have the bounds $\|\nabla u\|_{W^{-1,p}(\Omega)}\leq \|u\|_{L^p}$ and $\| u\|_{W^{-1,p}(\Omega)}\leq \|u\|_{L^p}$. This completes the proof.
\QED

\section{Cauchy Riemann type equations at low regularities} \label{Sec_A}

In this appendix we prove Propositions \ref{Lemma1_appendix} and \ref{Lemma2_appendix} which give existence of weak solutions to Cauchy Riemann type equation for scalar valued differential forms, required in the proof of Lemma \ref{Lemma_existence_iterates} for well-posedness of the iteration scheme. For this, in Theorems \ref{Thm_Gaffney}, \ref{Thm_CauchyRiemann} and \ref{Thm_Hogde_Morrey} below, we first collect the theorems from \cite{Dac} regarding classical $W^{1,p}$ solutions of first order Cauchy-Riemann type equations
\beq \label{Cauchy_Riemann_prelim}
du=f  \hspace{.5cm}   \text{and} \hspace{.5cm}  \delta u =0, \hspace{.5cm} \text{in} \ \ \ \Omega ,
\eeq 
where the Cartan algebra of differential forms is determined by the Euclidean metric in $\mathbb{R}^n$. We extend these theorems in Proposition \ref{Lemma1_appendix} and \ref{Lemma2_appendix} below to prove existence of weak $L^p$ solutions, the lower regularity required for well-posedness of our iteration scheme, in the special case when boundary data is free to assign. 

To begin, we state the following partial integration formula for non-zero boundary data,  
\beq \label{partial_int_forms}
\int_{\Omega} \langle d u, w \rangle dx + \int_{\Omega} \langle  u, \delta w \rangle dx 
= \int_{\partial\Omega}  \langle N \wedge u, w \rangle
= \int_{\partial\Omega}  \langle u, N\cdot w \rangle,
\eeq
where $u$ is a $k$-form and $w$ a $(k+1)$-form, $N$ denotes the outward-pointing unit normal of $\partial \Omega$ and $N\cdot w$ denotes the contraction of $N$ and $w$, c.f. Theorem 3.28 in \cite{Dac}, (and \eqref{partial_integration_no_bdd} for the case of vanishing boundary data).

We now state the basic elliptic estimate for \eqref{Cauchy_Riemann_prelim}, which mirrors estimate \eqref{Poisson-3} for the Poisson equation, the so-called Gaffney inequality, (c.f. Theorem 5.21 in \cite{Dac}).  The Gaffney inequality shows that $d$ and $\delta$ control all derivatives of $u$.

\begin{Thm}  \label{Thm_Gaffney}
{\bf (Gaffney Inequality):}  Let $u \in W^{m+1,p}(\Omega)$ be a $k$-form, where $m\geq 0$, $1\leq k\leq n-1$, $1<p<\infty$ and $n\geq2$. Then there exists a constant $C>0$ depending only on $\Omega$, $m,n,p$, such that
\beq \label{Gaffney}
\|u\|_{W^{m+1,p}(\Omega)} \leq C \Big( \|du\|_{W^{m,p}(\Omega)} + \|\delta u\|_{W^{m,p}(\Omega)}+\| u\|_{W^{m+\frac{p-1}{p},p}(\partial\Omega)}\Big). 
\eeq 
\end{Thm} 

The following special case of Theorem 7.4 in \cite{Dac},  provides the existence theorem sufficient for our purposes, and contains a refinement of Gaffney's inequality (\ref{Gaffney}) for $1$-forms and $0$-forms.  \\

\begin{Thm} \label{Thm_CauchyRiemann}
$(i)$ Let $f\in W^{m,p}(\Omega)$ be a $2$-form with $df=0$, where $m\geq 0$, $n\geq2$, $1<p< \infty$. Assume further that $f=d v$ for some $1$-form $v \in W^{m,p}(\Omega)$.\footnote{Since $d^2=0$, the assumption $f=d v$ implies $df=0$, and is a slightly stronger assumption than $df=0$, convenient for our purposes.}  Then there exists a $1$-form $u=u_i\, dx^i \in W^{m+1,p}(\Omega)$ which solves 
\beq \label{CauchyRiemann_eqn_Thm}
du=f  \hspace{.5cm}   \text{and} \hspace{.5cm}  \delta u =0 \hspace{1cm} \text{in} \ \ \ \Omega,
\eeq 
together with the boundary condition 
\beq \label{CauchyRiemann_bdd_Thm}
u \cdot N=0 \hspace{1cm} \text{on} \ \ \  \partial\Omega,
\eeq
where $N$ is the unit normal on $\partial\Omega$ and $u \cdot N \equiv u_i N^i$.  Moreover, there exists a constant $C>0$ depending only on $\Omega$, $m,n,p$, such that
\beq \label{Gaffney_2}
\| u \|_{W^{m+1,p}(\Omega)}  \leq C \| f \|_{W^{m,p}(\Omega)} .
\eeq
$(ii)$ Let $f\in W^{m,p}(\Omega)$ be a $1$-form with $df=0$. Then there exists a $0$-form $u \in W^{m+1,p}(\Omega)$ such that $u$ solves $du=f$, has zero average $\int_\Omega u dx=0$ and satisfies estimate \eqref{Gaffney_2}.
\end{Thm}

\Proof
Theorem \ref{Thm_CauchyRiemann} and its proof are taken from \cite{ReintjesTemple_ell2}, and the proof is included for completeness, c.f. Theorem 2.4 in \cite{ReintjesTemple_ell2}.  Part (i) is a special case of Theorem 7.4 in \cite{Dac} for $1$-forms with zero boundary conditions. Namely, our assumption $df=0$ together with zero boundary data, ($\omega_0 =0$, following notation in \cite{Dac}), directly gives condition (C1) of \cite[Thm 7.4]{Dac}. The first equation of condition (C2) of \cite[Thm 7.4]{Dac} follows trivially from our assumptions; $g=0$ and $\omega_0=0$ in the notation of \cite{Dac}. The second equation in (C2), that $\int_\Omega \langle f; \Psi\rangle =0$ for any harmonic form $\Psi$ (i.e. $\delta \Psi=0$) with vanishing normal components (i.e. $N\cdot \Psi=0$) on the boundary ($\Psi \in \mathcal{H}_N$ in the notation of \cite{Dac}), follows by application of the integration by parts formula \eqref{partial_int_forms} for differential forms to $f=d v$,
$$
\langle f,\Psi \rangle_{L^2} = - \langle v,\delta \Psi \rangle_{L^2} + \langle v,N\cdot \Psi \rangle_{L^2} =0.
$$ 
Theorem 7.4 in \cite{Dac} now yields the existence of a solution $u\in W^{m+1,p}(\Omega)$ to \eqref{CauchyRiemann_eqn_Thm} - \eqref{CauchyRiemann_bdd_Thm} satisfying estimate \eqref{Gaffney_2}.

Part (ii) of Theorem \ref{Thm_CauchyRiemann}, can be thought of as a version of Theorem \cite[Thm 7.4]{Dac}, in the special case of $0$-forms, which does not require condition (C2) by abandoning boundary data. That is, we seek a $0$-form $u$ solving the gradient equation $du=f$ such that estimate \eqref{Gaffney_2} holds. (No boundary data is required for our purposes). To begin the proof, observe that a solution $u \in W^{m+1,p}(\Omega)$ of $du=f$, in the case $m\geq 1$, is given by the path integral 
\beq \label{CauchyRiemann_int-formula}
u(x) = \int_{x_0}^x f \cdot d\vec{r} \; +\; u_0
\eeq
along any differentiable curve connecting $x_0$ and $x$, where $x_0 \in \Omega$ is some point we fix, and the constant $u_0$ is the value of $u$ at $x_0$, which is free to be chosen. Note, since $df=0$, the integral \eqref{CauchyRiemann_int-formula} is path independent, as can be shown by applying Stokes Theorem to integration of $df$ over the region enclosed by two curves connecting $x_0$ and $x$. We now choose $u_0$ such that the average of $u$ is zero, $\int_\Omega u dx=0$.  Then Poincar{\'e}'s inequality \cite[Eqn. (7.45)]{GilbargTrudinger} implies that $\| u\|_{L^p(\Omega)} \leq C\| f \|_{L^p(\Omega)}$ for a suitable constant $C>0$. Thus, since $\| du \|_{L^p(\Omega)} = \| f \|_{L^p(\Omega)}$ follows directly from $du=f$, we have 
\beq\label{CauchyRiemann_0estimate} 
\| u \|_{W^{1,p}(\Omega)}  \leq C \| f \|_{L^p(\Omega)}.
\eeq 
Estimate \eqref{Gaffney_2} follows by suitable differentiation of $du=f$ and application of estimate  \eqref{CauchyRiemann_0estimate}. Existence of a solution $u$ to $du=f$ in the case $m=0$ follows again from \eqref{CauchyRiemann_int-formula} by mollifying $f$, and using that this mollification is controlled by estimate \eqref{CauchyRiemann_0estimate}. This completes the proof of Theorem \ref{Thm_CauchyRiemann}. 
\QED

We finally require the so-called Hodge-Morrey decomposition, taken from Theorem 6.12 in \cite{Dac}:

\begin{Thm} \label{Thm_Hogde_Morrey}
{\bf (Hodge-Morrey decomposition):} (i) Let  $\Phi \in L^p(\Omega)$ be a $1$-form for $1< p < \infty$. Then there exists $1$-forms $w_1,w_2 \in W^{2,p}(\Omega)$ such that 
\beq \label{Hodge-Morrey-decomp}
\Phi = d\alpha + \delta \beta + h,
\eeq
where $\alpha = \delta w_1$ and $\beta = d w_2$ such that $N \wedge \alpha\big|_{\partial\Omega} =0$ and $N \cdot \beta\big|_{\partial\Omega}=0$, where $N$ is interpreted as either a $1$-form or a vector normal to $\partial\Omega$, and where $h$ is a harmonic $1$-form in the sense that $dh=0=\delta h$. Moreover, there exists a constant $C>0$ depending only on $\Omega,n,p$ such that
\beq \label{Hodge-Morrey_est}
\| w_1\|_{W^{2,p}(\Omega)} + \| w_2\|_{W^{2,p}(\Omega)} + \|h\|_{L^p(\Omega)} \leq C \| \Phi\|_{L^p(\Omega)}.
\eeq
(ii) Let $\Phi \in L^p(\Omega)$ be a $0$-form, $1<p<\infty$, then there exist $0$-forms $w \in W^{2,p}(\Omega)$ and a constant $h_0$ such that 
\beq \label{Hodge-Morrey-decomp_0}
\Phi = \delta \beta + h_0,
\eeq 
where $\beta = dw$ and $N \cdot \beta\big|_{\partial\Omega}=0$, and exists a constant $C>0$ depending only on $\Omega,n,p$ such that
\beq \label{Hodge-Morrey_est_0}
\| w\|_{W^{2,p}} \leq C \| \Phi\|_{L^p}.
\eeq
\end{Thm}

\Proof
Part (i) of Theorem \ref{Thm_Hogde_Morrey} is the case of Theorem 6.12 (iii) in \cite{Dac} for $1$-forms $\Phi$. Part (ii) follows from (iii) of \cite[Thm 6.12]{Dac} for $0$-forms $\Phi$, by observing that any harmonic $0$-form $h$ is constant, (since $dh=0$ is the vanishing gradient condition for $h$), so $h=h_0$.\footnote{One can understand Theorem \ref{Thm_Hogde_Morrey} (ii) quite easily from the point of view of the Poisson equation. Namely the sought after function $w$ is the solution to the Poisson equation $\Delta w = \Phi - h_0$ with Neumann data $N\cdot dw =0$ on $\partial\Omega$, where $h_0$ is a constant chosen such that $\Phi - h_0$ satisfies the consistency condition $\int_\Omega(\Phi-h_0) dx =0$  existence of $w$, (required by the divergence theorem applies to the equation). The solution $w$ is unique up to addition by a constant, and we choose this constant for $w$ to have zero average $\int_\Omega w dx=0$. Now, the Poincar\'e inequality implies the $L^p$-norm of $w$ to be bounded by the $L^p$ norm of $dw$, and from this estimate \eqref{Hodge-Morrey_est_0} follows from standard elliptic estimates. (Compare also with Theorem 9.2 in \cite{Dac} and its proof.)}
\QED

We are now prepared to establish the existence theorems for $1$-forms and $0$-forms required in our iteration scheme in Section \ref{Sec_existence}, Proposition \ref{Lemma1_appendix} and \ref{Lemma2_appendix} below. We begin with the case of $1$-forms. That is, given $f \in W^{-1,p}(\Omega)$ for $1<p<\infty$, we prove existence of a $1$-form $a \in L^p(\Omega)$ which is a weak solution               
\beq \label{appendix_eqn1}
\begin{cases}  
da = f \cr
\delta a=0,  
\end{cases}
\eeq
such that 
\beq \label{appendix_eqn2}
\|a\|_{L^p} \leq C  \|f\|_{W^{-1,p}}
\eeq
for some constant $C>0$ depending only on $n,p,\Omega$. No boundary data is imposed. Here $a$ is a scalar valued $1$-form and $f$ a linear functional over the space of $2$-forms with components in $W^{1,p^*}_0(\Omega)$, $f: W^{1,p^*}_0(\Omega) \longrightarrow \mathbb{R}$, where $W^{1,p^*}_0(\Omega)$ is the closure of $C^\infty_0(\Omega)$ with respect to the $W^{1,p^*}$-norm, and where $\frac{1}{p}+\frac{1}{p^*}=1$. We refer to such a linear functional again as a $2$-form in $W^{-1,p}(\Omega)$. Equations \eqref{appendix_eqn1} are interpreted in the following weak sense,
\beq \label{appendix_eqn1'}
\begin{cases}  
\langle a, \delta \phi \rangle_{L^2} =- f(\phi) \cr
\langle a, d \psi \rangle_{L^2}=0,  
\end{cases}
\eeq
for all $2$-forms $\phi$ with components in $W^{1,p^*}_0(\Omega)$ and all $0$-forms $\psi \in W^{1,p^*}_0(\Omega)$, (so $\phi|_{\partial\Omega} = 0$ and $\psi|_{\partial\Omega} = 0$), where $\langle\cdot,\cdot\rangle_{L^2}$ denotes the standard $L^2$-inner product on differential forms.            

\begin{Prop} \label{Lemma1_appendix}
Let $f \in W^{-1,p}(\Omega)$ be a $2$-form satisfying $df=0$ in the weak sense that $f(\delta \psi)=0$ for all $3$-forms $\psi$ with components in $W^{2,p^*}_0(\Omega)$; assume further that $f=d v$ for some $1$-form $v \in W^{-1,p}(\Omega)$ in the sense that $f[\phi] = -v(\delta \phi)$ for any $2$-forms $\phi \in W^{1,p^*}_0(\Omega)$.\footnote{As in Theorem \ref{Cauchy_Riemann}, assuming $f=d v$ is a slightly stronger assumption than $df=0$, convenient in our proof of well-posedness of the iteration scheme, c.f. Lemma \ref{Lemma_existence_iterates}.}
Then there exists a solution $a \in L^p(\Omega)$ of  \eqref{appendix_eqn1'} satisfying \eqref{appendix_eqn2}.
\end{Prop}

\Proof
The proof consists of the following three steps: (1) Construct approximate solutions $a^\epsilon$. (2) Derive an $\epsilon$-independent bound on the approximate solutions which implies existence of a convergent subsequence. (3) Prove that the limit of this convergent subsequence is a solution of \eqref{appendix_eqn2} which satisfies estimate \eqref{appendix_eqn1'}. 

To implement step (1), we mollify the functional $f$, that is, we introduce $f^\epsilon(x) \equiv f(\varphi^\epsilon(\cdot -x))$, where $\varphi^\epsilon \equiv \varphi^\epsilon_{ij}dx^i\wedge dx^j$ is a $2$-form with components $\varphi^\epsilon_{ij} \in C^\infty_0(\Omega)$ that are a standard mollifier function. So $f^\epsilon \in C^\infty(\Omega)$, and $f^\epsilon$ converges to $f$ in $W^{-1,p}$ component-wise. For each $\epsilon >0$, we now introduce $a^\epsilon$ as the solution of              
\beq \label{appendix_eqn3}
\begin{cases}
da^\epsilon = f^\epsilon \cr
\delta a^\epsilon=0,  
\end{cases}
\eeq
with boundary data $N^j a^\epsilon_j =0$ on $\partial \Omega$, where $N$ is the outward pointing unit normal of $\partial\Omega$. 
The solution $a^\epsilon$ does indeed exist by Theorem \ref{Thm_CauchyRiemann} $(i)$, since $f^\epsilon=dv^\epsilon$ for the $1$-form $v^\epsilon \equiv v[\varphi^\epsilon] \in C^\infty(\Omega)$. Namely, our assumption $f[\phi] = -v(\delta \phi)$ for any $2$-forms $\phi \in W^{1,p^*}_0(\Omega)$ implies that
$$
f^\epsilon= f(\varphi^\epsilon) = -v[\delta \varphi^\epsilon] = d v[\varphi^\epsilon] = dv^\epsilon,
$$ 
by definition of the distributional derivative $dv$. Clearly, $df^\epsilon = d^2 v^\epsilon =0$. Thus Theorem \ref{Thm_CauchyRiemann} $(i)$ applies and yields a solution $a^\epsilon  \in W^{1,p}(\Omega)$ for each $\epsilon >0$, establishing step (1).

To establish step (2), we now derive a uniform bound on $\| a^\epsilon \|_{L^p}$ in order to conclude convergence of a subsequence to the sought after solution $a$. The uniform bound we derive can be thought of as a version of Gaffney's inequality at the lower level of $L^p$ regularity, when boundary data cannot be imposed strongly. To begin, since the operator norm is equivalent to the $L^p$ norm, we find that
\beq \label{appendix_eqn4}
\| a^\epsilon \|_{L^p} = \sup_{\Phi \in \mathcal{F}} \big| \langle a^\epsilon , \Phi \rangle_{L^2} \big|,
\eeq
where 
$$
\mathcal{F} \equiv \Big\{ \Phi \in L^{p^*}(\Omega) \; \text{a $1$-form with}  \;  \|\Phi\|_{L^{p^*}} =1 \Big\}
$$
is the space of test functions. Now, fix $\Phi \in \mathcal{F}$ and apply the Hodge-Morrey decomposition of Theorem \ref{Thm_Hogde_Morrey} to write
\beq \label{appendix_eqn4'}
\Phi = d\alpha + \delta \beta + h,
\eeq
where $\alpha = \delta w_1$ and $\beta = d w_2$ for $1$-forms $w_1,w_2 \in W^{2,p^*}(\Omega)$, such that $N \wedge \alpha\big|_{\partial\Omega} =0$ and $N \cdot \beta\big|_{\partial\Omega}=0$, and where $h$ is a harmonic $1$-form. Next, applying the existence theory of Theorem \ref{Thm_CauchyRiemann} $(ii)$, we define the $0$-form $\Psi \in W^{1,p^*}_0$ as a solution of                                
\beq  \label{appendix_eqn5}
d\Psi = h
\eeq
which exists, since  $d h=0$ for $h$ harmonic; no boundary data imposed. We now substitute the decomposition \eqref{appendix_eqn4'} for $\Phi$ to write $\langle a^\epsilon , \Phi \rangle_{L^2}$ in \eqref{appendix_eqn4} equivalently as 
\begin{eqnarray} \label{appendix_eqn6}
\langle a^\epsilon , \Phi \rangle_{L^2} 
&=& \langle a^\epsilon , (d\alpha +\delta\beta +h)\rangle_{L^2}   \cr
&=& \langle a^\epsilon , d\alpha \rangle_{L^2} + \langle a^\epsilon , \delta \beta \rangle_{L^2} + \langle a^\epsilon , h \rangle_{L^2}  .
\end{eqnarray}
Applying now the partial integration formula \eqref{partial_int_forms} to each term, we obtain
\begin{eqnarray} \label{appendix_eqn6a}
\langle a^\epsilon , d\alpha \rangle_{L^2} = - \langle \delta a^\epsilon , \alpha \rangle_{L^2} + \langle a^\epsilon , N\wedge \alpha \rangle_{L^2(\partial\Omega)} = - \langle \delta a^\epsilon , \alpha \rangle_{L^2},
\end{eqnarray}
where the last equality follows from $N \wedge \alpha\big|_{\partial\Omega} =0$, c.f. Theorem \ref{Thm_Hogde_Morrey}. Similarly, partial integration together with $N \cdot \beta\big|_{\partial\Omega}=0$ gives
\begin{eqnarray}\label{appendix_eqn6b}
\langle a^\epsilon , \delta \beta \rangle_{L^2} 
= - \langle da^\epsilon , \beta \rangle_{L^2} + \langle a^\epsilon , N\cdot \beta \rangle_{L^2(\partial\Omega)} 
= - \langle da^\epsilon , \beta \rangle_{L^2},
\end{eqnarray}
and by \eqref{appendix_eqn5},
\begin{eqnarray} \label{appendix_eqn6c}
\langle a^\epsilon , h \rangle_{L^2}   
&=& \langle a^\epsilon , d\Psi \rangle_{L^2} \cr
&=& - \langle \delta a^\epsilon , \Psi \rangle_{L^2} + \langle N\cdot a^\epsilon , \Psi \rangle_{L^2(\partial\Omega)} \cr
&=& - \langle \delta a^\epsilon , \Psi \rangle_{L^2} ,
\end{eqnarray}
since $N\cdot a^\epsilon=0$ on $\partial\Omega$ by assumption. Now, substituting \eqref{appendix_eqn6a} - \eqref{appendix_eqn6c} into \eqref{appendix_eqn6}, and using that $a^\epsilon$ solves \eqref{appendix_eqn3}, we obtain 
\begin{eqnarray}  \label{appendix_eqn7a}
\langle a^\epsilon , \Phi \rangle_{L^2} 
&=& - \langle \delta a^\epsilon , \alpha \rangle_{L^2}    - \langle d a^\epsilon , \beta \rangle_{L^2} -\langle \delta a^\epsilon , \Psi \rangle_{L^2} \cr
&=& - \langle f^\epsilon , \beta \rangle_{L^2}.
\end{eqnarray}                         
Now \eqref{appendix_eqn7a} and the definition of the operator norm $\| \cdot \|_{W^{-1,p}}$ imply
\begin{eqnarray} \label{appendix_eqn7}
|\langle a^\epsilon , \Phi \rangle_{L^2} |
&\leq &  \| f^\epsilon\|_{W^{-1,p}} \: \| \beta \|_{W^{1,p^*}} .
\end{eqnarray}
By Theorem \ref{Thm_Hogde_Morrey}, we have $\beta = dw_2$ so estimate \eqref{Hodge-Morrey_est} gives us
\beq  \label{appendix_eqn8}
\| \beta \|_{W^{1,p^*}} =  \| dw_2 \|_{W^{1,p^*}} \leq \|w_2\|_{W^{2,p^*}}  \overset{\eqref{Hodge-Morrey_est}}{\leq} C \|\Phi\|_{L^{p^*}} ,
\eeq
where $C>0$ is some constant only depending on $p,n,\Omega$. Substituting \eqref{appendix_eqn8} into \eqref{appendix_eqn7} and using that $\| \Phi \|_{L^{p^*}}=1$ for any $\Phi \in \mathcal{F}$, we obtain the estimate
\begin{eqnarray} \label{appendix_eqn9}
|\langle a^\epsilon , \Phi \rangle_{L^2} | 
\ \leq \ C \| f^\epsilon\|_{W^{-1,p}} 
\ \leq \ 2 C \| f\|_{W^{-1,p}}  ,
\end{eqnarray}
for all $\epsilon>0$ sufficiently small, since $f^\epsilon$ converges to $f$ in $W^{-1,p}$ by standard mollification. Finally, substituting \eqref{appendix_eqn9} into \eqref{appendix_eqn4}, we obtain the sought after uniform bound
\beq \label{appendix_eqn10}
\| a^\epsilon \|_{L^p} \leq C \| f\|_{W^{-1,p}},
\eeq
where $C>0$ is some constant only depending on $p,n,\Omega$.

We now complete step (3). By \eqref{appendix_eqn10}, $\|a^\epsilon\|_{L^p}$ is bounded independent of $\epsilon$, so the Banach Alaoglu Theorem implies convergence of a subsequence to some differential form $a\in L^p$  weakly in $L^p$. We now show that this limit $a$ solves \eqref{appendix_eqn1}. For this, let $\epsilon_k >0$ such that $\epsilon_k \rightarrow 0$ as $k\rightarrow \infty$  and assume $a_k=a^{\epsilon_k}$ is the convergent subsequence, so $a_k \rightarrow a$  weakly in $L^p$ as $k\rightarrow\infty$. By \eqref{partial_int_forms}, we have for any $\phi$ and $\psi \in W^{1,p^*}_0(\Omega)$ that
\begin{eqnarray} \label{appendix_eqn11}
\langle d a^\epsilon , \phi \rangle_{L^2} &=& - \langle a^\epsilon , \delta\phi \rangle_{L^2}, \cr
\langle \delta a^\epsilon , \psi \rangle_{L^2} &=& - \langle a^\epsilon , d\psi \rangle_{L^2} .
\end{eqnarray}
So using that $a_k$ solves \eqref{appendix_eqn3}, we write \eqref{appendix_eqn11} as 
\beq \label{appendix_eqn12}
\begin{cases}  
\langle a_k, \delta \phi \rangle_{L^2} =- \langle f^{\epsilon_k},\phi\rangle_{L^2} \cr
\langle a_k, d \psi \rangle_{L^2}=0,  
\end{cases}
\eeq
which converges to the sought after equation \eqref{appendix_eqn1'}. We conclude that $a$ is the sought after weak solution of \eqref{appendix_eqn1'}. Moreover, the sought after estimate 
\eqref{appendix_eqn2} follows from the uniform bound \eqref{appendix_eqn10}, since
\begin{eqnarray} \label{appendix_eqn14}
\|a\|_{L^p} &=& \sup_{\psi\in L^{p^*}} \big|\langle a,\psi\rangle\big| 
= \lim\limits_{k\rightarrow\infty} \sup_{\psi\in L^{p^*}} \big|\langle a_k ,\psi\rangle\big| \cr
&=& \lim\limits_{k\rightarrow\infty} \|a_k\|_{L^p} 
 \overset{\eqref{appendix_eqn10}}{\leq} \,  C \| f\|_{W^{-1,p}} .
\end{eqnarray}
This completes the proof of Proposition \ref{Lemma1_appendix}.
\QED

Our final existence result for $0$-forms $u$ is required to extend Theorem \ref{Thm_CauchyRiemann} $(ii)$ to solutions in the space $L^p(\Omega)$. It is an extension of Poincar\'e's Lemma to linear functionals. We seek weak solutions $u \in L^p(\Omega)$ of the first order equation
\beq \label{Appendix_eqn1}
du=f,
\eeq
satisfying
\beq \label{Appendix_eqn2}
\|u\|_{L^p} \leq C \|f\|_{W^{-1,p}},
\eeq
for any $1$-form $f$ with components in $W^{-1,p}(\Omega)$, where $C>0$ is a constant depending only on $n,p,\Omega$. That is, we prove existence of weak solutions $u \in L^p(\Omega)$ of \eqref{Appendix_eqn1} in the sense that 
\beq \label{Appendix_eqn1weak}
\langle u,\delta \phi \rangle = - f(\phi),
\eeq
for any $1$-form $\phi$ with components in $W^{1,p^*}_0(\Omega)$ subject to estimate \eqref{Appendix_eqn2}. 

\begin{Prop} \label{Lemma2_appendix}
Assume the $1$-form $f \in W^{-1,p}(\Omega)$ satisfies $df=0$ in the sense that $f(\delta \phi)=0$ for any $2$-form $\phi$ with components in $W^{2,p^*}_0(\Omega)$. Then there exists a solution $u \in L^p(\Omega)$ of \eqref{Appendix_eqn1weak} satisfying \eqref{Appendix_eqn2}.     
\end{Prop}

\Proof
The proof is similar to that of Proposition \ref{Lemma1_appendix}, consisting of the same three steps. To begin with the first step, we mollify the functional $f$, setting again $f^\epsilon(x) \equiv f(\varphi^\epsilon(\cdot -x))$, where $\varphi^\epsilon \in C^\infty_0(\Omega)$ is a $1$-form whose components are standard mollifier functions. So $f^\epsilon \in C^\infty(\Omega)$, and $f^\epsilon$ converges to $f$ in $W^{-1,p}$ component wise. For each $\epsilon >0$, we have $df^\epsilon = -f(\delta \varphi^\epsilon) =0$ by assumption. Thus Theorem \ref{Thm_CauchyRiemann} $(ii)$ applies, and yields the existence of a $0$-form $u^\epsilon  \in W^{1,p}(\Omega)$  solving              
\beq \label{Appendix_eqn3}
du^\epsilon = f^\epsilon 
\eeq
such that $u^\epsilon$ has zero average, $\int_\Omega u^\epsilon dx =0$.
 
In the next step we derive a uniform bound on $\| u^\epsilon \|_{L^p}$. That is, we express the $L^p$-norm in terms of the operator norm, 
\beq \label{Appendix_eqn4}
\| u^\epsilon \|_{L^p} = \sup_{\Phi \in \mathcal{F}} \big| \langle u^\epsilon , \Phi \rangle_{L^2} \big|,
\eeq
where 
$$
\mathcal{F} \equiv \Big\{ \Phi \in L^{p^*}(\Omega) \; \text{a function with}  \;  \|\Phi\|_{L^{p^*}} =1 \Big\}
$$
is the space of test functions and $\langle\cdot,\cdot \rangle_{L^2}$ denotes the standard $L^2$ inner product. We fix some $\Phi \in \mathcal{F}$ and apply the Hodge-Morrey decomposition in Theorem \ref{Thm_Hogde_Morrey} (ii) to write   
\beq \label{Appendix_eqn6}
\Phi = \delta \beta +h_0,
\eeq
where $\beta = dw$ for some $0$-form $w \in W^{2,p^*}(\Omega)$ and $N \cdot \beta\big|_{\partial\Omega}=0$. From \eqref{Appendix_eqn6} we find that
$$
\langle u^\epsilon , \Phi \rangle_{L^2} 
= \langle u^\epsilon , \delta\beta \rangle_{L^2} +  \langle u^\epsilon , h_0 \rangle_{L^2}  
= \langle u^\epsilon , \delta\beta \rangle_{L^2} ,
$$
since  $\langle u^\epsilon , h_0 \rangle_{L^2} = h_0 \int_\omega u^\epsilon dx =0$ by our zero average assumption on $u^\epsilon$. Integration by parts \eqref{partial_int_forms} gives further  
\begin{eqnarray} \label{Appendix_eqn7}
\langle u^\epsilon , \Phi \rangle_{L^2} 
&=& -\langle du^\epsilon , \beta \rangle_{L^2}   +  \langle u^\epsilon , N\cdot  \beta \rangle_{L^2(\partial\Omega)} \ = \  \langle f^\epsilon , \beta \rangle_{L^2},
\end{eqnarray}
where the last equality follows by substituting \eqref{Appendix_eqn1} for $u^\epsilon$ and $N\cdot \beta|_{\partial\Omega} =0$.  From \eqref{Appendix_eqn7} and the definition of the operator norm $\| \cdot \|_{W^{-1,p}}$, we obtain
\begin{eqnarray} \label{Appendix_eqn8}
|\langle u^\epsilon , \Phi \rangle_{L^2} |
&\leq &  \| f^\epsilon\|_{W^{-1,p}} \: \| \beta \|_{W^{1,p^*}} .
\end{eqnarray}
Using now that $\beta = dw$ in combination with estimate \eqref{Hodge-Morrey_est_0} of the Hodge-Morrey decomposition, we obtain
\beq  \label{Appendix_eqn9}
\| \beta \|_{W^{1,p^*}}  \leq \|w\|_{W^{2,p^*}}  \leq C \|\Phi\|_{L^{p^*}} ,
\eeq
for some constant $C>0$ only depending on $p,n,\Omega$. Substituting now \eqref{Appendix_eqn9} into \eqref{Appendix_eqn8} and using that $\| \Phi \|_{L^{p^*}}=1$ for any $\Phi \in \mathcal{F}$, we obtain the uniform bound
\begin{eqnarray} \label{Appendix_eqn10}
|\langle u^\epsilon , \Phi \rangle_{L^2} | 
\leq  C \| f^\epsilon\|_{W^{-1,p}} 
\leq  2 C \| f\|_{W^{-1,p}}  ,
\end{eqnarray}
for all $\epsilon>0$ sufficiently small, because $f^\epsilon$ converges to $f$ in $W^{-1,p}$ by standard mollification. Finally, substituting \eqref{Appendix_eqn10} into \eqref{Appendix_eqn4}, we obtain the sought after uniform bound
\beq \label{Appendix_eqn11}
\| u^\epsilon \|_{L^p} \leq C \| f\|_{W^{-1,p}},
\eeq
where $C>0$ is some constant only depending on $p,n,\Omega$. 

The uniform bound \eqref{Appendix_eqn11} implies the existence of a subsequence which converges to some function $u \in L^p(\Omega)$ subject to the $L^p$-bound \eqref{Appendix_eqn2}, and an argument similar to that of step (3) in the proof of Proposition \ref{Lemma1_appendix} shows that $u$ solves \eqref{Appendix_eqn3}. This completes the proof of Proposition \ref{Lemma2_appendix}. 
\QED

\section*{Declarations and Statements}

Authors declare that there is no conflict of interest and that there are no competing interests.

\section*{Funding}

M. Reintjes is currently supported by CityU Start-up Grant for New Faculty (7200748) and by CityU Strategic Research Grant (7005839); and was supported by the German Research Foundation, DFG grant FR822/10-1, from June 2019 until July 2021; and by FCT/Portugal, (GPSEinstein) PTDC/MAT-ANA/1275/2014 and UID/MAT/04459/2013, from January 2017 until December 2018.

\section*{Acknowledgements}
We thank Craig Evans for directing us to references \cite{Dac} and \cite{Simader} which contain the basic elliptic estimates which are the starting point for our analysis of convergence of the iteration scheme here.

\providecommand{\bysame}{\leavevmode\hbox to3em{\hrulefill}\thinspace}
\providecommand{\MR}{\relax\ifhmode\unskip\space\fi MR }
\providecommand{\MRhref}[2]{  \href{http://www.ams.org/mathscinet-getitem?mr=#1}{#2}}
\providecommand{\href}[2]{#2}

\end{document}